 \documentclass[final,3p,times,sort&compress]{elsarticle}

\usepackage{amssymb,bm}
\usepackage{xcolor}
\usepackage{multirow}

\journal{Physics Reports}

\usepackage{hyperref}
\usepackage{amsmath}
\usepackage{soul} % for strike out used in macro cut see below
\usepackage[utf8]{inputenc}  %needed for some of the bibtex entries

\newcommand{\be}{\begin{eqnarray}}
\newcommand{\ee}{\end{eqnarray}}
\newcommand{\non}{\nonumber\\}

\newcommand{\ave}[1]{\left\langle #1 \right\rangle}

\newcommand{\mev}{{\rm \, MeV}}
\newcommand{\gev}{{\rm \, GeV}}

\newcommand{\cum}[1]{\kappa_{#1}}

 \newcommand\op{{\sigma}}

 \newcommand\p{{\bm{p}}}
 \newcommand\la{\langle}
 \newcommand\ra{\rangle}
 \newcommand\xis{\xi}

\graphicspath{{figures/}}   % better than figpath

\begin{document}

\begin{frontmatter}



\title{Mapping the Phases of Quantum Chromodynamics with Beam Energy Scan}


\author[add1]{Adam Bzdak}
\author[add2]{ShinIchi Esumi}
\author[add3]{Volker Koch}
\author[add4,add5,add6]{Jinfeng Liao}
\author[add7]{Mikhail Stephanov}
\author[add3,add5]{Nu Xu}

\address[add1]{AGH University of Science and Technology,
Faculty of Physics and Applied Computer Science, 30-059 Krakow, Poland}
\address[add2]{Tomonaga Center for the History of the Universe, University of Tsukuba, Tsukuba, Ibaraki 305, Japan}
\address[add3]{Nuclear Science Division, Lawrence Berkeley National Laboratory, Berkeley, CA 94720, USA}
\address[add4]{Physics Department and CEEM,
Indiana University, 2401 N Milo B. Sampson Lane, Bloomington, IN 47408, USA}
\address[add5]{Institute of Particle Physics and Key Laboratory of Quark \& Lepton Physics (MOE), Central China Normal University, Wuhan 430079, China}
\address[add6]{Guangdong Provincial Key Laboratory of Nuclear Science, Institute of Quantum Matter, South China Normal University, Guangzhou 510006, China.}
\address[add7]{Physics Department, University of Illinois at Chicago, 845 W. Taylor St., Chicago IL 60607-7059, USA}

\begin{abstract}
We review the present status of the search for a  phase transition and
critical point as well as anomalous transport
phenomena in Quantum Chromodynamics (QCD), with an emphasis on the Beam Energy Scan program at the Relativistic Heavy Ion Collider
 at Brookhaven National Laboratory. We present the conceptual framework and
discuss the observables  deemed most sensitive to a phase transition, QCD critical point, and
 anomalous transport, focusing  on fluctuation and correlation measurements. 
 Selected experimental results for these observables together with those characterizing the global properties
of the systems created in heavy ion collisions are presented. We then discuss what can be
already learned from the currently available data about the QCD critical point and anomalous
transport as well as what additional measurements and theoretical developments are needed in order to
discover these phenomena.

%%% Local Variables:
%%% mode: latex
%%% TeX-master: "BES_Main_current"
%%% End:

\end{abstract}


\end{frontmatter}



\tableofcontents

\newpage

%-------------Section 1-----------------------------------
%
%=====================================================================================
%=====================================================================================
\section{Introduction}
\label{sec1}

The understanding of matter and its phases has always been of fundamental interest in
science. Matter in our physical world is governed by a hierarchical microscopic structure,
reflecting various length scales: ``normal'' matter such as water or crystals is made out of atoms
and molecules which by way of their interaction arrange themselves in various ways, such as liquid
or gas or crystalline structures. At a shorter length scale, molecules consist of atoms and again
the interactions among them result in a variety of configurations ranging from diatomic molecules to
complex polymers. The atoms themselves are made of a positively charged nuclei surrounded by
negatively charged electrons. The nucleus itself consists of protons and neutrons, which themselves
are made of even more fundamental particles, quarks and gluons which are fundamental degrees of freedom of the
Standard Model. In this review we are concerned with matter which is governed by the strong
interactions. It therefore involves nucleons or more generally hadrons as well as quarks and gluons.
The properties of strongly interacting matter is not only of fundamental interest but it also important for our understanding of the Universe~\cite{BraunMunzinger:2008tz,Braun-Munzinger:2015hba}. Indeed a few milliseconds after the ``Big Bang'' the entire Universe was occupied by strongly interacting matter of quarks and gluons. Also, highly compressed strongly interacting matter at densities of several times nuclear saturation density ($\rho_0\sim 0.16/{\rm fm}^3$) is expected to exist inside the compact stars and thus affects their properties~\cite{Lattimer:2015nhk,Page:2006ud}. 

Matter can occur in various states, or phases. Most well known are probably the liquid, gas
and solid phases of water. So the obvious question arises, if strongly interacting matter can also
exhibit different 
phases. And indeed, soon after the discovery of Quantum Chromodynamics (QCD) \cite{Fritzsch:1973pi}, the theory of the strong interactions, and following the realization that QCD exhibits asymptotic freedom
\cite{Gross:1973id,Politzer:1973fx}, it was recognized that QCD likely predicts a transition from a
hadronic phase to one dominated by quarks and gluons, the so called quark-gluon plasma (QGP)
\cite{Collins:1974ky,Cabibbo:1975ig,Shuryak:1977ut,Shuryak:1978ij,Freedman:1976ub,Kapusta:1979fh}.  However, already before the advent of QCD the
possibility of a limiting temperature of the hadron gas was discussed and a quantitative prediction of $T_{limit}\simeq 170 \mev$ was obtained in the statistical bootstrap model of Hagedorn 
\cite{Hagedorn:1965st}. The existence of a new phase was confirmed 
in the first calculations using the lattice formulation of QCD, initially 
for pure $SU(2)$ gauge theory \cite{Creutz:1980zw,Creutz:1979dw,McLerran:1980pk}. 

These results soon sparked ideas to create and study the quark-gluon plasma in high energy heavy ion
collisions (see e.g. \cite{Shuryak:1978ij,Lee:1974kn,Chin:1978gj} ) extending early considerations of creating
thermodynamically equilibrated matter in high energy hadronic collisions by Fermi
\cite{Fermi:1950jd}, Landau \cite{Landau:1953gs}, and Hagedorn \cite{Hagedorn:1965st}. These 
early initiatives have meanwhile led to very active and large research programs at the  
Bevatron of LBL, the Alternating Gradient Synchrotron (AGS) and the Relativistic Heavy Ion Collider (RHIC) of Brookhaven as well as the Super Proton Synchrotron (SPS) and the Large Hadron Collider (LHC) of CERN. The central discoveries at RHIC,
and later confirmed and refined at the LHC, include  a large azimuthal asymmetry, known as elliptic flow $v_2$, in the particle yields
\cite{Ackermann:2000tr}, as well as a strong suppression of high energy jets and heavy quarks
\cite{Adcox:2001jp}.  The observed elliptic flow, which is consistent with predictions from
nearly ideal hydrodynamics, together with the strong suppression of jets suggested that   the system
created in these collisions is strongly  {coupled}. These findings required a paradigm shift, away
from the expectations, based on asymptotic freedom, of a weakly coupled quark-gluon plasma
\cite{Back:2004je,Arsene:2004fa,Adcox:2004mh,Adams:2005dq} (for a recent review see, e.g., \cite{Braun-Munzinger:2015hba}). 
 {Of course at very high temperatures and/or baryon density QCD will eventually become weakly coupled due to asymptotic freedom. To which extent these conditions can be reached in experiments at even higher energies or in the interior of neutron stars is at present not known.}    

Together with the advances in experiment, important developments occurred on the theoretical side. In
particular, lattice QCD methods became sufficiently refined and powerful to perform continuum
extrapolated calculations in full QCD with physical quark masses. These simulations for example
showed that the transition at vanishing baryon density is a crossover \cite{Aoki:2006we} with a
pseudo-critical temperature of $T\simeq 150 \mev$
\cite{Aoki:2006br,Aoki:2009sc,Bazavov:2011nk,Bazavov:2014pvz}. They also showed that the
(approximate) chiral symmetry of QCD, which is spontaneously broken in the vacuum, is restored at
high temperatures. 

\begin{figure}[hbt] 
\begin{center}
\includegraphics[height=20em]{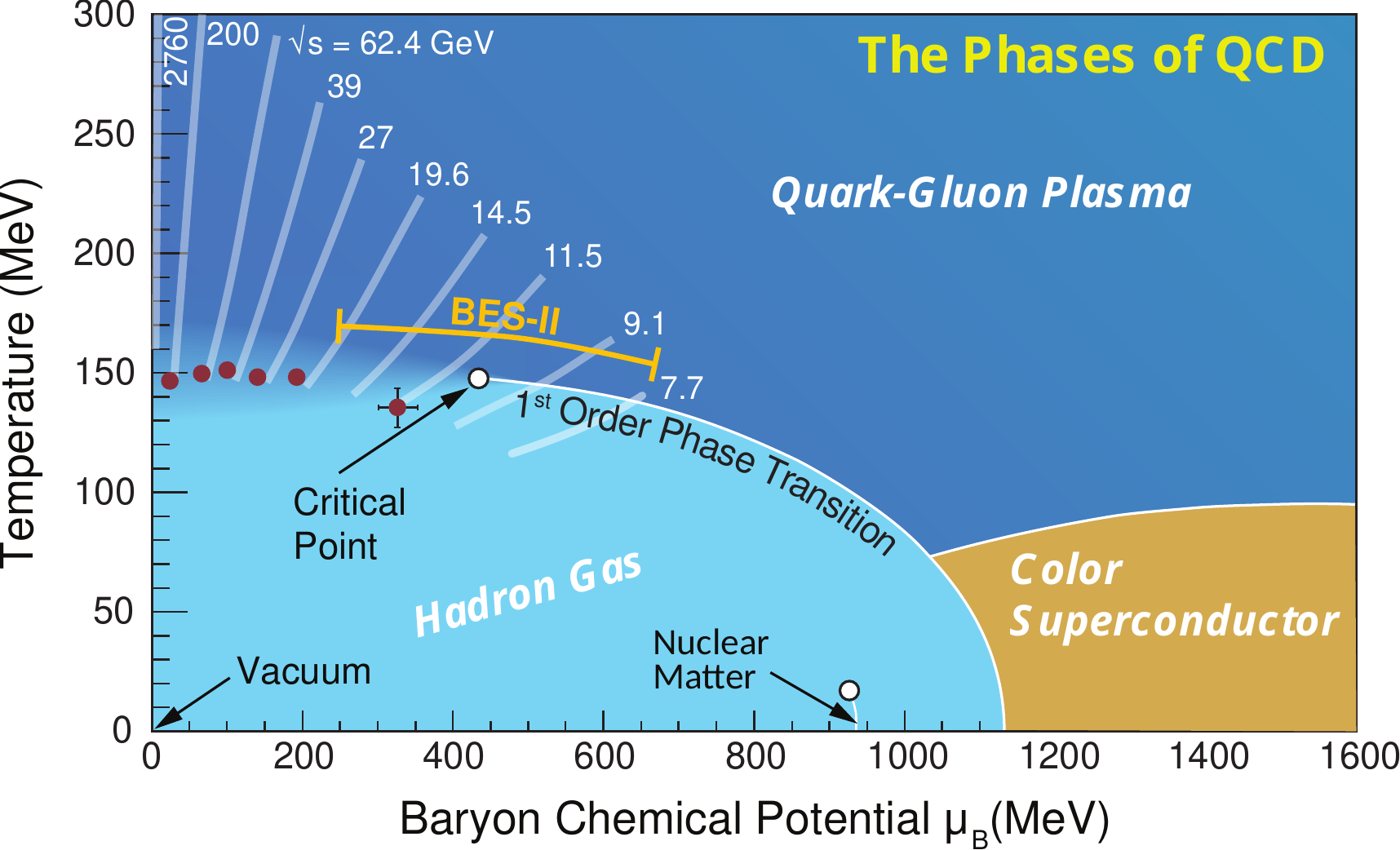}
%\vspace{-0.5in}
\caption{
A schematic QCD phase diagram in the  thermodynamic parameter space spanned by the temperature $T$ and  baryonic chemical potential $\mu_B$.  
The corresponding  (center-of-mass) collision energy ranges for different accelerator facilities, especially the RHIC beam energy scan program, are indicated in the figure.  {Figure adapted from \cite{Geesaman:2015fha}. }
}
 \label{fig:qcdpd}
%\vspace{-0.15in}
\end{center}
\end{figure}

Due to the fermion sign problem\footnote{In the  presence of a baryon-number chemical potential, the
  fermion determinant in the path integral of the partition function becomes complex, making
  importance sampling difficult if not impossible.},
lattice QCD methods are restricted to the region of vanishing or small baryon density. Also most
experiments both at the highest energies at RHIC as well as at the LHC explore this region of nearly
vanishing baryon density.  Therefore, very little is known about the properties of strongly
interacting matter at large baryon density. Many model calculations, see, e.g. Refs. \cite{Braun-Munzinger:2015hba,Stephanov:2004wx,Stephanov:2007fk,BraunMunzinger:2008tz,Fukushima:2013rx,Baym:2017whm}
for recent reviews, predict a first order phase transition at large baryon
density or equivalently baryon number chemical potential. If true, this phase transition will end at
a critical point since we know the transition to be a crossover at small baryon chemical
potential. These model ideas are typically summarized in a (conjectured) phase diagram of QCD
matter shown in Fig.~\ref{fig:qcdpd}. Clearly such a phase diagram has to be rather schematic.  The
only regions where we have firm knowledge are: { (a) at low temperature and baryon chemical potential, where we have a dilute gas of hadrons, which are predominantly pions, where interactions are small corrections which can be systematically described in chiral perturbation theory and the experimental knowledge of hadronic interactions. (b) At small values of the baryon number chemical
potential and finite temperatures ($T \gtrsim 130 \mev$)   from lattice QCD, and, (c), for small temperatures close to the nuclear matter saturation
density from the extrapolation of well tested nuclear forces and experiments of nuclear
fragmentation \cite{Chomaz:2003dz,Das:2004az,Moretto_2011} .}

In order to experimentally explore the QCD phase diagram at finite net-baryon density, one needs to
create systems with finite net-baryon density in heavy ion collisions. Since baryon number is conserved, the only way to
increase the net-baryon density is to ensure that some of the baryons from the colliding nuclei are transported to the
mid-rapidity region. This can be achieved by lowering the beam energy, and available particle
production systematics \cite{Andronic:2014zha} confirm that this strategy indeed works. Therefore,
a systematic scan of heavy ion measurements over a range of beam energies enables the exploration of
the high baryon density region of the QCD phase diagram and the search for the existence of a
first order phase transition and its associated critical point. Such a beam energy scan (BES) program has been started at
RHIC in 2010 (see e.g. \cite{Odyniec:2013aaa}) and its next phase with improved beam quality,  {such as increased luminosity and smaller beam packages,} and
detector capability has just started in 2019. Also, several experiments at other facilities
such as NA61/SHINE and HADES are able to measure some of the same observables at energies even lower
than those achievable at RHIC. In addition, new experiments extending the reach of the RHIC beam
energies towards lower energies are planned at the FAIR facility in Darmstadt (CBM), at NICA in
Dubna (MPD), as well as at the CSR in Lanzhou (CEE). 
 {Finally, a fixed-target experimental program enabling much lower energy collisions at RHIC is underway, which allows measurements at center of mass energies down to $\sqrt{s_{\rm NN}} = 3 \gev$, although with somewhat limited acceptance.}  These current and future programs provide  unique opportunities for exploring 
and mapping the phases of QCD across a wide range of conditions in the laboratory.
 
The first set of measurements resulting from the RHIC beam energy scan made a number of intriguing
observations, such as a non-monotonic dependence on the beam energy of some of the key
observables and the disappearance at low energy of certain key signals observed at high energy. 
These observations underline the discovery potential for locating the Critical Point (CP) as a landmark of QCD phase diagram, as well as for unambiguously observing the anomalous chiral transport effects thus experimentally verifying chiral symmetry restoration. 
% Notably there emerge opportunities of potential discovery for possibly locating the Critical End Point (CEP) as a landmark of QCD phase diagram, as well as for unambiguously observing the anomalous chiral transport effects thus experimentally verifying chiral symmetry restoration. 
Therefore, and in view of the second phase of the beam energy scan, it appears to be
appropriate to provide a review of the key physics motivations together with the experimental
results of the RHIC beam energy scan so far.  Special emphasis in this review will be given to observables 
which are sensitive to the QCD critical point, namely fluctuations of conserved charges, and to
chiral restoration, namely various correlation functions which measure the induced currents from the
chiral anomaly.

% We note in passing   that  the laboratory study of QCD phase diagram fosters fruitful connections between the physics of the smallest scales with that of the largest scales. Indeed the quark-gluon plasma,  occurring at a temperature as hot as a trillion degrees or more,  was the cosmic primordial matter occupying the whole Universe at about a few microseconds after the ``Big Bang''. Highly compressed baryonic matter at a densities of several times  nuclear saturation
% density ($\rho_0\sim 0.16/{\rm fm}^3$), on the other hand, is expected to exist inside the compact
% stars.  Its equation of state manifests itself in  the properties of such astrophysical objects that
% are accessible through a wide range  of electromagnetic observations as well as through to the detection of neutrinos and gravitational waves.  The investigation  of hot and dense phases of QCD matter would therefore not only help unravel the mysteries of how QCD operates in nature but also bear far-reaching implications for cosmic and astrophysical matter. 

The rest of this review is organized as follows. 
In the next section we provide an overview of those aspects of
QCD pertinent to the topics discussed in this review. Next we provide a more detailed discussion
of the QCD phase diagram followed by a section on the physics of the critical point. We then turn to
the dynamics induced by the chiral anomaly and how it can be used to extract information about
chiral symmetry restoration. After that we present an overview of the various measurements, the implications of  which will then be discussed  in the context of the QCD phase diagram and chiral restoration. Finally we will
summarize the current status and provide an outlook towards future experiments and developments.

%=====================================================================================
%=====================================================================================
%

%%% Local Variables:
%%% mode: latex
%%% TeX-master: "BES_Main"
%%% End:

%-------------Section 2-----------------------------------
\section{QCD and Chiral Symmetry}
\label{sec:vk:qcd_overview}

In this section we will discuss the properties of QCD, the theory of the strong interaction,
which are relevant to this review. The Lagrangian of QCD is given by 
\begin{align}
{\cal L}_{QCD} = \bar{\psi}_{f} \left( i \, \gamma^{\mu} D_{\mu}
  \, - m   \right) \psi_{f} -\frac{1}{4}G_{\mu\nu}^{a}G^{\mu\nu}_{a} 
  \label{eq:vk:QCD_lagrange}
\end{align}
Here $D_{\mu} = \partial_{\mu}- i g A_{\mu}^{a} \lambda_{a}$ is the
co-variant derivative, and $A_\mu^a$ is the gluon field of $SU(3)$ color symmetry with $a$ the color index, $\lambda_{a}$ the Gell-Mann matrices, and $G_{\mu\nu}^{a}$ the  corresponding non-Abelian gauge field strength tensor~\cite{Peskin:1995ev}. In the following we will consider only the light (up and down) quarks and assume their mass, $m$, to be the same. The quark fields $\psi_{f}$ are then
two-component spinors in flavor space, denoted by the index $f$, and we have suppressed the 
additional color indices.
One of the most widely known properties of QCD
is asymptotic freedom, i.e., the fact that the strength of the strong coupling constant,
$\alpha_{s}=\frac{g^2}{4\pi}$, decreases at short distances or, equivalently, at large momentum scales. This property,
first pointed out by Gross, Wilczek, and Politzer \cite{Gross:1973id,Politzer:1973fx} 
has been widely tested and confirmed in experiment and has been awarded the Nobel prize in 2004. Asymptotic
freedom allows for a perturbative treatment of processes at high energies, such as deep inelastic
scattering etc.   { Asymptotic freedom also implies an emergent energy scale $\Lambda_{QCD}\simeq 200\ \rm MeV$ at which the QCD running coupling becomes large and the theory gets strongly coupled. Equivalently this implies an emergent distance scale $1/\Lambda_{QCD} \simeq 1\  \rm fm$ that is empirically found to set the size of all hadrons containing light flavor valence quarks.} At such low energy scales or long distances, QCD is highly non-perturbative, and its
treatment requires either numerical methods such as lattice QCD or effective field theories which
respect the symmetries of QCD. It is the low energy regime of QCD which is most relevant for the
proprieties of QCD probed in heavy ion collisions, such as the phase structure, its collective
behavior etc. Especially relevant for the topics discussed in this review are chiral symmetry, the
axial anomaly as well as the topological structure of the QCD vacuum, which we will briefly discuss 
in the following.~\footnote{We note that there are other highly important non-perturbative features of QCD such as confinement, which are not much discussed in the present review. We refer interested readers to  e.g.~\cite{Greensite:2011zz}.}  Given the quark fields one can define the (iso-vector) vector and axial vector currents, ${\mathcal V}_{\mu}^{i}$ and ${\mathcal A}_{\mu}^{i}$, given by: 
\begin{align}
{\mathcal  V}_\mu^i  &= \bar{\psi} \, \gamma_\mu \tau^i \,  \psi
  \label{eq:vk:vector_current}
\\
{\mathcal  A}_{\mu }^i &= \bar{\psi} \, \gamma_\mu  \gamma_{5} \tau^{i} \psi
  \label{eq:vk:axial_current}
\end{align}
which play an important role in the context of chiral symmetry which we will discuss below.

\subsection{Chiral symmetry }
\label{sec:vk:chiral_symmetry}

Chiral symmetry is a symmetry of QCD in the limit of vanishing (light)
quark masses. In reality, however, the current masses of the up and
down quarks are finite, $m_{u,d}\simeq 10 \mev$, but very small compared to the 
scale of QCD $\Lambda_{QCD}\simeq 200 \mev$. Therefore, chiral
symmetry may be considered an approximate symmetry of the strong
interactions. Before we provide a proper definition of chiral symmetry in QCD
let us briefly review its history.

Long before QCD was known to be the theory of the strong interactions,
phenomenological indications for the existence of chiral symmetry came
from the study of the nuclear beta decay. There one finds that the
weak coupling constants for the vector and axial-vector
hadronic-currents, $C_V$ and $C_A$, did not (in case of $C_V$) or only
slightly ($25 \%$ in case of $C_A$) differ from those for the leptonic
counterparts.  Consequently strong interaction `radiative' corrections
to the weak vector and axial vector `charge' are essentially
absent. This absence of radiative corrections indicates the presence of 
conserved currents and thus symmetries, similar to the familiar case
of the electric charge, which is protected  from radiative corrections
as consequence of electric charge conservation.  In case of the vector
current, the underlying symmetry is the well known isospin symmetry of
the strong interactions and the hadronic vector current is
identified with the isospin current. The identification of the axial
current, on the other hand is not so straightforward. Contrary to the
vector or iso-spin symmetry, which is clearly visible in the hadronic
mass spectrum -- degeneracy of proton and neutron or the three pion
states, for example -- no such symmetry exists for states related by 
axial symmetry. For instance, the difference between the masses of the $\rho$ and
$a_{1}$ mesons, $m_{a_{1}}-m_{\rho}\simeq 500\mev $, is of the order
of QCD scale, $\Lambda_{QCD}$, rather than the electromagnetic or weak
scale.

However there is additional phenomenological evidence that the
axial current is (almost) conserved. Since the pion is a pseudo-scalar, the weak decay of the
pion with four-momentum $q$ involves
the axial current $\mathcal{A}_{\mu}^{j}$ in Eq.~\eqref{eq:vk:axial_current}. Since  $\mathcal{A}_{\mu}^{j}$ is a
Lorentz vector and since we are considering a pion with given four-momentum,  the matrix element has to be proportional to 
$q_{\mu} e^{-iq \cdot x}$. Thus the matrix element can be written as
\begin{align}
\langle 0 |\, {\mathcal A}_\mu^j(x) \,| \pi^k(q) \rangle = i f_\pi q_\mu \delta^{jk} e^{-iq \cdot x}
\label{eq:vk:pcac1}
\end{align} 
where $f_{\pi} = 93\mev $ is the pion decay constant and $j,k$ are
isospin indices. From its divergence 
\begin{align}
\langle 0 |\, \partial^\mu \mathcal{A}_\mu^j(x) \,| \pi^k(q) \rangle = 
- f_\pi q^2 \delta^{jk} e^{-iq \cdot x} = - f_\pi m_\pi^2 \delta^{jk} 
e^{-iq \cdot x}
\label{eq:vk:pcac2}
\end{align}
we find that $ \partial^\mu \mathcal{A}_\mu^j(x) \sim m_{\pi}^{2}$. Given that
the mass of the pion is very small, the divergence of the axial
current is also small, and thus the axial current is ``almost''
conserved.  This observation led to the so called partially conserved axial current
(PCAC) hypothesis, which typically refers to the relation
\eqref{eq:vk:pcac2}. The above relations, Eqs.\eqref{eq:vk:pcac1} and 
\eqref{eq:vk:pcac2}, suggest that the pion contributes to the axial
current and that the divergence of the axial current may  serve as an
interpolating pion field. 
%{\bf (JL: not clear about this phrase ``interpolating'' pion field?)}
 This observation led to further
developments, such as the Goldberger-Treiman relation
\cite{Goldberger:1958zz}, which relates the {\em strong} pion nucleon
coupling constant $g_{\pi}$ with the {\em weak} pion decay constant,
$f_{\pi}$, a nice example of the power of symmetries. Also, the
so-called Weinberg sum rules \cite{Weinberg:1967kj} which are
concerned with moments of the spectral functions of the vector and axial currents,
predict the mass difference of the $\rho$ and $a_{1}$ mesons to
reasonable accuracy. The mass splitting is  simply a result of the
contribution  of the pion to the axial channel, which has no
equivalent in the vector channel.

\begin{figure}[ht]
  \centering
  \includegraphics[width=0.9 \textwidth]{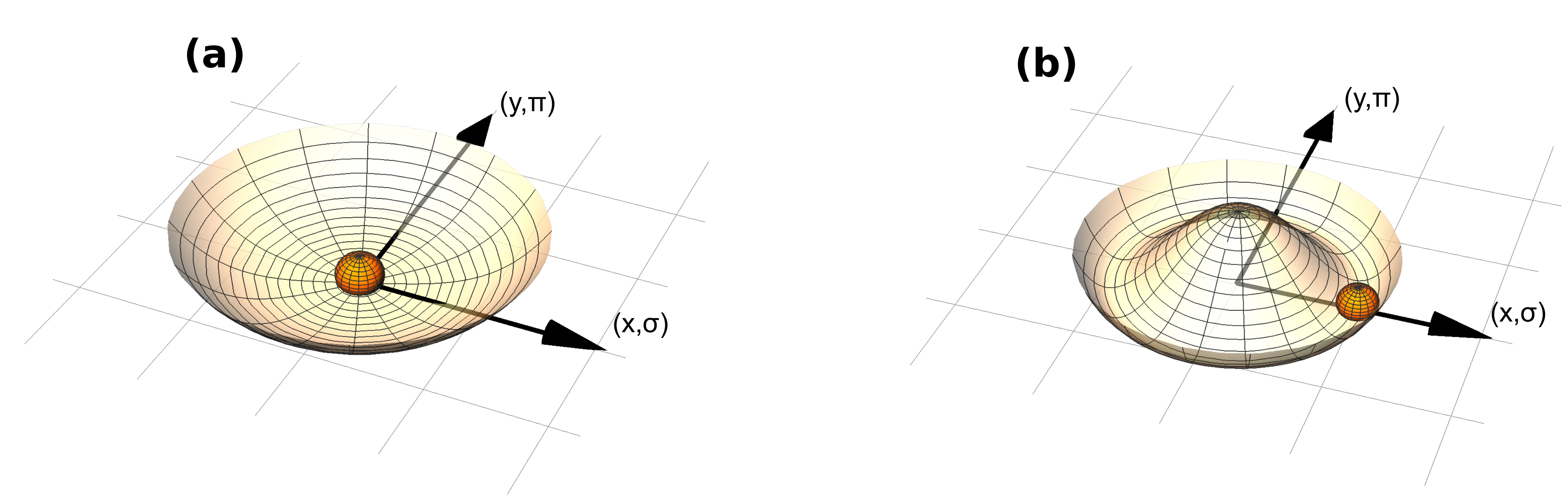}
  \caption{\label{fig:vk:mexican_hat} (a) Potential and ground state
    indicated by small sphere are rotationally symmetric. (b)
    Potential is rotationally symmetric but the ground state breaks the symmetry
    spontaneously.}
\end{figure}

The above examples, which were obtained prior to the advent of QCD,
show that there is indeed a (partial) axial symmetry in the strong
interaction. But why is this symmetry not visible in the  mass
spectrum? The answer to this is the mechamism of  ``spontaneous 
broken symmetry'' first discovered by Y. Nambu
\cite{Nambu:1960xd,Nambu:1961tp} in the context of the
conserved axial current of the strong interaction, and he was awarded the 2008 Nobel prize for this work. 
The essential
concept of spontaneous symmetry breaking is the following:  while the
Hamiltonian possesses the symmetry, its ground state does not. A
familiar example is a ferromagnet, where the interaction between the
spins is invariant under rotation, but the ground state, which has a
magnetic moment pointing in a certain direction, obviously breaks
rotation symmetry.  An important consequence of the spontaneous
symmetry breaking, as expressed in the Goldstone
theorem \cite{Goldstone:1961eq}, is the existence of a massless mode,
the  so-called Goldstone mode. While a more formal proof of the 
Goldstone theorem  may be found in textbooks such as
\cite{Peskin:1995ev}, its essence may be understood in the simple
example shown in Fig.~\ref{fig:vk:mexican_hat}. In both (a) and (b) we see a
potential (Hamiltonian) which is symmetric under rotations. However,
in (a) the ground state indicated by the little ball, is in the
middle, so that both the potential and ground state are invariant
under rotations. In (b), on the other hand, the ground state, i.e., the
position of the  little ball obviously breaks the rotational
symmetry. Thus the symmetry is spontaneously broken. However because
of the rotational symmetry of the potential, rotational excitations,
or small excitations in the $(y,\pi)$ direction,
do not cost any energy. Or in other words, we have a ``massless''
mode. Excitations in the radial or $(x,\sigma)$ direction, on the
other hand, cost energy and are thus ``massive''. Compare this with
the situation in (a). There excitations in both the  $(x,\sigma)$ and
$(y,\pi)$ direction are massive.  And, due to the symmetry 
of the potential, they are degenerate. Thus spontaneous breaking leads to a non-degenerate
excitation spectrum and the symmetry reveals itself in the appearance
of a massless Goldstone mode. 
Indeed, the very existence of a ``flat direction'' of the potential, and the resulting Goldstone mode, in the spontaneously broken vacuum is guaranteed by the original symmetry of the Hamiltonian.  
In case of the strong interaction, the Goldstone bosons are the
pions. The massive radial excitation is often referred to as the
$\sigma$ meson, although such a meson has not been identified as a
well defined state in experiment. The reason is that the
$\sigma$, which carries the quantum number of the vacuum, is likely to
mix with many other states, and thus  broadens considerably due to, e.g., decay   {into two or more pions.}
\footnote{{For example in the pion-pion phase shift, a broad feature occurs at an invariant mass of about $500\  \rm MeV$ that could be plausibly due to the $\sigma$ meson.}}   
% If chiral symmetry were a perfect symmetry of QCD, the pion
% should be massless. Since chiral symmetry is only approximate, we
% expect the pion to have a finite but small (compared to all other
% hadrons) mass. This is indeed the case!  
Finally, similar to the
ferromagnet, where the magnetization disappears above the Curie
temperature, one expects the chiral symmetry to be restored at sufficiently high 
temperatures.

After this historical introduction, we next turn to QCD. Chiral symmetry,
is a symmetry of the quark and flavor sector of QCD and thus it is sufficient to consider only the
fermionic part of the QCD Lagrangian \eqref{eq:vk:QCD_lagrange}, 
\begin{align}
\delta {\cal L}_{quarks} = \bar{\psi}_{f} \left[ i \left( \gamma_{\mu} D^{\mu}
  \right) - m   \right] \psi_{f} 
  \label{eq:vk:QCD_quarks}
\end{align}
where we  restrict ourselves again to the light (up and down) flavors. 

Chiral symmetry is related to the following vector and axial-vector transformations of
the quark fields
\begin{align}
\Lambda_V:& \,\, \psi \longrightarrow e^{-i \vec{\tau}\cdot\vec{\Theta } }
\psi 
\non
\Lambda_A: &\,\, \psi \longrightarrow 
e^{-i \gamma_{5} \vec{\tau}\cdot \vec{\Theta } } \psi 
  \label{eq:vk:vector_rotation}
\end{align}
where $\vec{\tau}=(\tau_{1},\tau_{2},\tau_{3})$ represents the isospin
generators, which are Pauli matrices acting upon the flavor indices of
the quark fields, and $\vec{\Theta }=(\theta_1,\theta_2,\theta_3)$ are the corresponding transformation angles. 
The above transformations do not involve the gluon
fields, and  it is easy to see that, in the limit of massless quarks,
$m =0$, often referred to as the {\em chiral limit}, the QCD
Lagrangian is invariant under both $\Lambda_{V}$ and
$\Lambda_{A}$. The conserved Noether currents (see, e.g., Ref. \cite{Peskin:1995ev}) 
associated with these symmetries are the axial and vector currents,
Eqs.~\eqref{eq:vk:vector_current} and \eqref{eq:vk:axial_current}. 

Before we discuss chiral symmetry further let us step back and
examine the special case of massless quarks or, generally, massless fermions a
bit closer. In the absence of any fermion masses, the  Dirac
Hamiltonian reduces to $\hat{H} \to
\gamma^0\vec{\gamma}\cdot \mathbf{p}$ and it commutes with  $\gamma_{5}$.
Therefore, we can characterize the eigenstates of $\hat{H}$ in
terms of eigenstates of $\gamma_{5}$. This is done simply by the
following decomposition of a given spinor into its left-handed and
right-handed component
\begin{eqnarray} \label{eq:jl:PsiRL}
 \psi_R = \frac{1+\gamma_{5}}{2} \psi  \,\,\, , \,\,\,   \psi_L = \frac{1- \gamma_{5}}{2} \psi \,\, .
 \end{eqnarray} 
with $\gamma^5 \Psi_{R/L} = (+/-) \Psi_{R/L}$.  So instead of
classifying the fermions for example by spin-up and spin-down, in case
of massless fermions it is more appropriate to sort them into left- and
right-handed states, as they are eigenstates of both Hamiltonian and $\gamma_5$. For free fermions, where
one frequently works with helicity states, the helicity of a
left/right handed fermion is indeed left/right handed, which explains
the origin of the concept of handedness or chirality in Greek. And as
long as we are dealing with massless fermions the handedness or
chirality is conserved classically. That is,  a right-handed (left-handed) fermion with its spin parallel (antiparallel) to its momentum remains right-handed (left-handed). 

Returning to QCD, in the limit of vanishing quark masses, both vector and
axial vector are conserved and thus we can form left and right handed
combinations which are both conserved 
\begin{eqnarray}
  J^{\mu,i}_{R}= \frac{1}{2} \left( V^{\mu,i}  + A^{\mu,i} \right) 
  = \bar{\psi}_{R} \, \gamma^\mu
  \tau^i \,  \psi_{R}, \,\,\,\, \partial_{\mu}J^{\mu,i}_{R}=0 
\non
  J^{\mu,i}_{L}= \frac{1}{2} \left( V^{\mu,i} - A^{\mu,i} \right) 
  = \bar{\psi}_{L} \, \gamma^\mu
  \tau^i \,  \psi_{L}, \,\,\,\, \partial_{\mu}J^{\mu,i}_{L}=0 
  \label{eq:vk:left_right_current}
\end{eqnarray}
Thus, both left and right handed quarks form individually conserved
(iso-vector) currents. The combined symmetries for  right handed and
left handed quarks is, therefore, referred to as chiral symmetry, which is governed
by the group $SU(2)_{L}\times SU(2)_{R}$.

It is easy to see that the scalar combination of the quark fields,
$\bar{\psi}\psi$, which comes with the mass term in the Lagrangian, is
not invariant under the axial vector transformation. The vector
transformation, $\Lambda_{V}$, on the on the other hand, leaves the
mass term unchanged. Therefore, for real QCD, where even the light
quarks have a finite but small mass, the axial vector symmetry is
(slightly) broken, whereas the vector symmetry remains
intact. However, as already discussed, the dominant effect is the
spontaneous breaking of the chiral symmetry, where the ground state
does not posses the symmetries of the Hamiltonian. In QCD this happens 
due to the dynamics in the gluon fields. In particular the so-called instantons,
which characterize certain topological fluctuation of the gluon field,
are believed to be the main source for the spontaneous chiral symmetry
breaking \cite{Schafer:1996wv}. Therefore, even in case of vanishing
quark masses chiral symmetry would be (spontaneously) broken, and the {\em explicit}
breaking, due to the small current quark masses is only a small
correction. This is evidenced by the fact that in the chiral limit the
pion mass would be zero while in nature $m_{\pi}\simeq 139.6\mev$, which is
still very small compared to that of other hadrons. As a result of the
spontaneous symmetry breaking, the vacuum expectation value of the
scalar bi-linear quark operator, $\ave{\bar{q}q}$ obtains a finite
value. The chiral condensate, $\ave{\bar{q}q}$, is not invariant under the chiral 
transformation, and   may serve as an order parameter for the spontaneous symmetry breaking. 

\begin{figure}[ht]
  \centering
  \includegraphics[width=0.7 \textwidth]{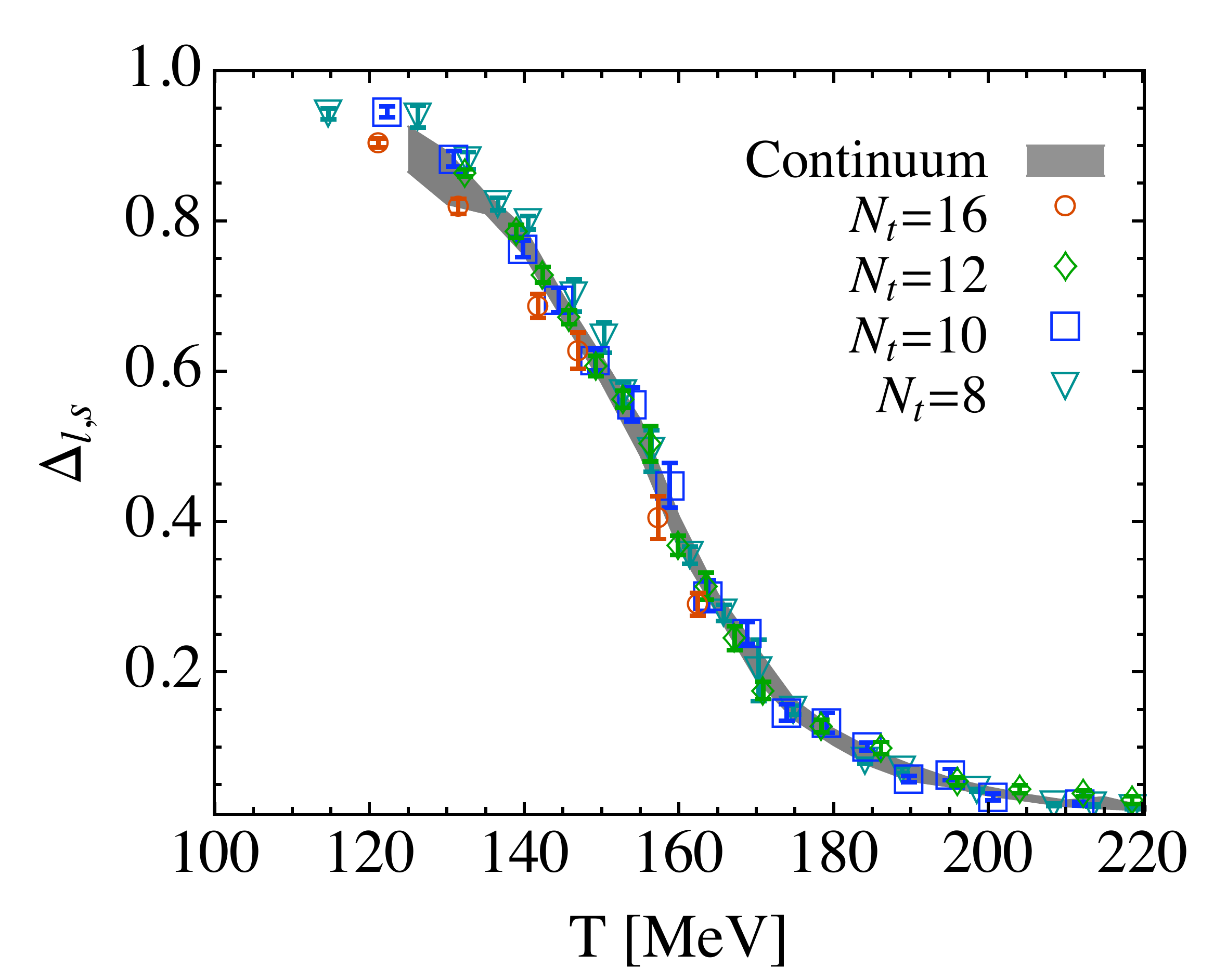}
  \caption{Chiral condensate $\ave{\bar{q}q}$ as function of
    temperature from Lattice QCD (see text and footnote \ref{ftn:1} for details). $N_t$ is the temporal lattice size used in a calculation. Figure adopted from \cite{Borsanyi:2010bp}.  }
\label{fig:vk:lattice_chiral_condesate}
\end{figure}

As one increases the temperature, thermal fluctuations suppress the
instantons and as a result the chiral condensate $\ave{\bar{q}q}$ will
vanish and chiral symmetry is restored, in analogy to the a
ferro-magnet, where above the Curie temperature the (spontaneous)
magnetization vanishes. That this is indeed the case has been
established by lattice QCD calculations
\cite{Bazavov:2011nk,Borsanyi:2010bp} and in
Fig.~\ref{fig:vk:lattice_chiral_condesate} we show the temperature
dependence of the chiral condensate.\footnote{What is plotted is
  actually the ratio of the chiral condensate at finite temperature
  over that at $T=0$. In addition some subtractions had to be made in
  order to properly renormalize this quantity. For details see Ref.
  \cite{Borsanyi:2010bp}.\label{ftn:1}} We see that at a temperature of
$T \gtrsim 180 \mev$ the chiral condensate has essentially vanished
and thus chiral symmetry is restored. As a consequence the chirality
of the light quarks becomes well defined, which is an essential feature for
the possible occurrence of the so-called chiral magnetic effect which
we will discuss in detail in Section \ref{sec5}. 

The system created in heavy ion collisions at RHIC and LHC is expected to 
reach the chiral transition temperature and it would be interesting and
important to devise a measurement which would confirm the restoration
of chiral symmetry. Chiral symmetry predicts that the vector and
axial-vector spectral function should be the same. The vector spectral
function is accessible via dilepton measurements. These have been
carried out over a wide range of collision energies, from
$\sqrt{s_{NN}}\simeq 2.2 \gev$ up to $\sqrt{s_{NN}} = 200 \gev $. The analysis
of these data \cite{Rapp:2009yu,vanHees:2007th} indicate that the
spectral function flattens, i.e. the peak associated with the
$\rho$-meson disappears. Unfortunately there is no easy access to the
axial spectral function and one has to rely on model estimates to see
if chiral symmetry is restored. 
One such attempt 
\cite{Hohler:2013eba}  finds  chiral symmetry
would be restored and that both the vector and axial vector spectral
functions become featureless, similar to those of free quarks.

{Another approach is to look for the so-called Chiral Magnetic Effect (CME), an anomalous chiral transport phenomenon whose presence  
requires that the chirality of the quarks in the hot matter is
well defined. The correlation between the spin and momentum of a chiral fermion, which would be completely spoiled by the vacuum chiral condensate mixing up right-/left-handed sectors, is an essential ingredient of CME. } Thus experimental proof for its existence would immediately imply
that chiral symmetry is restored in these collisions. This idea and
the present status of experimental searches and their interpretation is
discussed in detail in Sections~\ref{sec5} and \ref{sec7c}.

So far we have concentrated on the 
vector and axial vector transformations $\Lambda_{V}$ and
$\Lambda_{A}$ in the flavor $SU(2)$ iso-vector channels, Eq.~\eqref{eq:vk:vector_rotation}, since they
constitute what is commonly referred to as chiral symmetry. However QCD is also
invariant under the analogous $U(1)$ transformations:   

\begin{align}
U(1):& \,\, \psi \longrightarrow e^{-i \Theta } \psi 
\non
U(1)_A: &\,\, \psi \longrightarrow 
e^{-i \gamma_{5} \Theta  } \psi 
  \label{eq:vk:U1_symmetry}
\end{align}
The former, $U(1)$, which holds even in case of finite (and
non-equal) quark-masses, gives rise to the conservation of baryon number (or
equivalently quark number).  The latter, $U(1)_{A}$, which appears to  be a symmetry of QCD in
the chiral limit, is actually broken by quantum effects, the so-called
anomaly which we discuss next.

\subsection{Anomaly}

In the classical limit, the axial $U(1)_{A}$ symmetry would lead to a conserved current (i.e. its corresponding Noether current)  
$J^{\mu}_{5}=\bar{\psi} \gamma^\mu \gamma_{5} \psi$  in the case of massless quarks.  In quantum field theory,
however, the {\em anomaly} gives rise to yet another type of symmetry breaking, in addition to the
explicit or the spontaneous breaking discussed previously.  Even if the quarks are
massless, upon the quantization of the Dirac field when
coupled to any (Abelian or non-Abelian) gauge theory,  the axial symmetry is explicitly broken and the axial
current is thus not conserved. 
This phenomenon, which is different from the aforementioned spontaneous symmetry breaking,  is the well-known {\em chiral anomaly}, often also
referred to as axial anomaly or triangle anomaly. It was first discovered in the 1960's in the
effort to try to understand the neutral pion decay, $\pi^0\to 2\gamma$,  with pioneering works from Adler as well
as from Bell and Jackiw (thus sometimes also called ABJ anomaly) in the QED
context~\cite{Adler:1969gk,Bell:1969ts}. It was shown later in an insightful paper by
Fujikawa in the path integral formalism that the chiral anomaly is an essential and general feature
for the quantum description of chiral fermions with gauge interaction~\cite{Fujikawa:1979ay}. The
chiral anomaly has now become an indispensable and integral ingredient in many basic aspects of quantum field theories and string theories. 

Let us examine the chiral anomaly in a more concrete situation, with a single species of massless Dirac  fermions of electric charge $q e$ coupled to an electromagnetic field $A_\mu$. The anomaly relation reads: 
\begin{eqnarray} \label{eq:jl:CA}
\partial_\mu J^\mu_5 = \left(\frac{1}{2\pi^2} \right) (qe\mathbf{E}) \cdot (qe\mathbf{B}) 
%= \frac{q^2e^2}{4\pi^2}  F^{\mu\nu} \tilde{F}_{\mu\nu} \,\, ,
\end{eqnarray}
where $\mathbf{E}$ and $\mathbf{B}$ are the electric and magnetic fields. The constant  in front
of the right-hand side is the anomaly coefficient $C_A =
\frac{1}{2\pi^2}$.  
%The last equality simply rewrites it in terms of the covariant field strength tensor
%$F^{\mu\nu}=\partial^\mu A^\nu -\partial^\nu A^\mu$ and its dual 
%$\tilde{F}_{\mu\nu}=\frac{1}{2}\epsilon_{\mu\nu\rho\sigma}F^{\rho\sigma}$. 
Clearly this implies that
the axial charge, $N_{5}\equiv \int_{V} d^3 \mathbf{r} \, J_5^{0}$,  
is no longer conserved. This is evident upon integrating the above equation over a certain spatial volume $V$ (with vanishing current on the boundary): 
\footnote{We note that in condensed matter physics a nonzero axial charge is generated in samples of the so-called Dirac or Weyl semimetals~\cite{Li:2014bha} by simply applying parallel electric and magnetic fields.}
\begin{eqnarray} \label{eq:jl:dN5dt}
\frac{d N_5}{dt} = \int_V d^3 \mathbf{r}\, \frac{\partial J_5^{0}}{\partial t} =\left( \frac{1}{2\pi^2} \right) \int_V d^3 \mathbf{r} \,  (qe\mathbf{E}) \cdot (qe\mathbf{B})  = C_A  \int_V d^3 \mathbf{r} \,  (qe\mathbf{E}) \cdot (qe\mathbf{B})  \,\, .
\end{eqnarray} 
It deserves mentioning that the anomaly relation may be viewed as occurring for the left-handed and
right handed quarks separately: 
\begin{eqnarray} \label{eq:jl:ARL}
\partial_\mu J^\mu_R = + \frac{C_A}{2}  (qe\mathbf{E}) \cdot (qe\mathbf{B})    \,\,\, , \,\,\, 
 \partial_\mu J^\mu_L =  - \frac{C_A}{2}   (qe\mathbf{E}) \cdot (qe\mathbf{B})  
\end{eqnarray}
Each sector ``suffers'' half of the axial symmetry violation but they together maintain vector current conservation $\partial_\mu J^\mu=0$ which of course is required by gauge invariance.  

A unique feature of the violation of the axial symmetry via the anomaly, is that the anomaly
coefficient $C_A  $ is {\em universal}, i.e. independent of the coupling strength and  other
dynamical details of the system. 
For example, in the context of perturbative computation of the
anomaly relation, it first arises at one-loop level via the famous triangle diagram where one
immediately obtains the full value of $C_A$. Naively one would expect that there will be
all kinds --- in fact an infinite number --- of higher order diagrams that would contribute and
thus modify the coefficient into a series of expansion ordered by the increasing power of the
coupling constant (as is typically the case for a perturbative calculation). Interestingly, all
the higher order contributions simply add to zero, thus leaving the coefficient $C_A$ intact. That
is, the anomaly coefficient, $C_A$, is dictated by the one-loop contribution. Calculations based on
nonperturbative methods show the same value for $C_A$ as well. As we will see later, such a universal
value of the anomaly coefficient has profound consequences for its macroscopic manifestation. 

Let us now turn to the anomaly in QCD in the chiral limit, $m_{q}\simeq 0$. In this case the
non-Abelian anomaly occurs similarly through the triangle diagram and it is analogous to the Abelian
anomaly Eq.~(\ref{eq:jl:CA}). One simply replaces  the QED coupling with the QCD coupling and
electromagnetic fields with the gluon fields as well as sums over color index. Thus one obtains  the following anomaly relation (for each flavor):  
\begin{eqnarray} \label{eq:jl:dJ5}
\partial_\mu J^\mu_5 = - \frac{g^2 }{16\pi^2} G^{\mu\nu}_a \tilde{G}_{\mu\nu}^a 
= \frac{1 }{4\pi^2} (g \mathbf{E}^a) \cdot (g \mathbf{B}_a) = \frac{C_A}{2} \, (g \mathbf{E}^a) \cdot (g \mathbf{B}_a)   \,\,  
\end{eqnarray}
with $g$ the color charge and $\mathbf{E}^a,  \mathbf{B}_a$ the chromo electric and magnetic fields. 

Clearly this implies that the axial charge (per flavor) in a certain spatial volume could be changed
by gluon field configurations according to
$\frac{dN_5}{dt}  =
\frac{C_A}{2} \, \int_V d^3\mathbf{r} \, (g \mathbf{E}^a) \cdot (g \mathbf{B}_a)  $. That is, chromo
electric and magnetic fields in parallel (or anti-parallel) configurations will induce the
generation of nonzero axial charges. As it turns out, such configurations naturally occur during the
very initial states, often referred as the glasma, in relativistic heavy ion collisions~\cite{Gelis:2012ri,Gelis:2010nm}.  The strong chromo electric and magnetic fields in the glasma
organize themselves into many flux tubes with collinear $\mathbf{E}^a$ and $\mathbf{B}_a$ fields
extending along the beam direction and with a typical transverse size determined by the so-called
saturation scale $Q_{s}^2=(1\sim 2)\gev^2$, which characterizes the gluon wave function of a nucleus~\cite{Gelis:2012ri,Gelis:2010nm}.  
Consequently nonzero axial charge will be generated inside each flux tube with fluctuating signs depending
upon whether the $\mathbf{E}^a, \mathbf{B}_a$ are parallel or anti-parallel. Rough estimates as well
as recent real-time lattice simulations have suggested that sizable axial charge fluctuations can be
generated through the glasma flux tubes in heavy ion
collisions~\cite{Kharzeev:2001ev,Mace:2016svc,Mace:2016shq}.

 A much more nontrivial question is whether axial charge could be generated {\em globally}. To appreciate this issue, one may perform integration of Eq.~(\ref{eq:jl:dJ5})  over the whole spacetime and notice that its right-hand side   can actually be rewritten as a full derivative:  
 \begin{eqnarray} \label{eq:jl:N5Qw}
N_5(t\to +\infty) - N_5(t\to -\infty) = - \frac{g^2 }{16\pi^2}  \int dt d^3\mathbf{r} \,  G^{\mu\nu}_a \tilde{G}_{\mu\nu}^a  
=  C_A   \int d^4x \, \partial^\mu K_\mu  \,\, ,
\end{eqnarray}
 with the topological current $K_\mu\equiv \frac{1}{8} \epsilon_{\mu\nu\rho\sigma}  \left( \tilde{A}_a^\nu \partial^\rho
   \tilde{A}_a^\sigma - \frac{1}{3} f_{abc} \tilde{A}_a^\nu \tilde{A}_b^\rho \tilde{A}_c^\sigma \right)$ for a given gluon
 configuration $\tilde{A}_a^\mu = g A_a^\mu$. This integral would evaluate to zero, unless there is
 a nonzero contribution from the spacetime boundary due to gluonic configurations with highly
 nontrivial boundary topology. These do exist in QCD, known as the instantons and
 sphalerons,\footnote{We note that these configurations are fluctuations of the QCD vacuum such that
 $\ave{N_5}=0$, i.e. the CP symmetry is not violated by QCD.} which
 we shall discuss next.

\subsection{Topology}
Topology is a mathematical concept pertaining to the properties of shapes (in its general yet
abstract sense) that would not change under arbitrary continuous deformations. It has wide
applications in many branches of physics. In the context of field theories, there exist
configurations with nontrivial topological structures in various spacetime dimensions. In
particular, such nontrivial topological configurations are found to play important roles in
non-Abelian gauge theories including QCD.

\begin{figure}[ht]
  \centering
  \includegraphics[width=0.95 \textwidth]{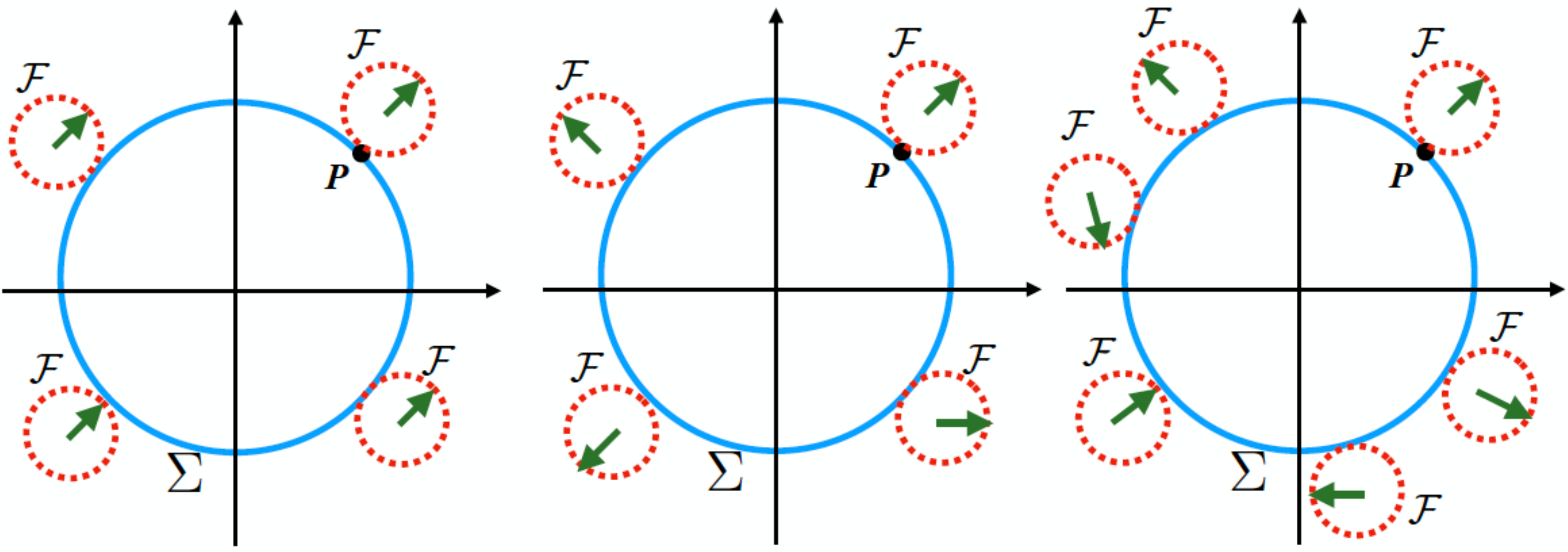}
  \caption{ An illustration of topological configurations based on a vortex in two dimensions. (see text for details). }
\label{fig:jl:vortex}
\end{figure}

Let us first use a simple example, the vortex in two spatial dimension, to illustrate the emergence
of topological configurations. This is shown in Fig.~\ref{fig:jl:vortex}.  Let us consider
possible configurations of a $U(1)$ gauge field $A_\mu$ on the boundary ${\Sigma}$ at spatial
infinity which is a circle or $S(1)$ (represented by the blue solid circle in
Fig.~\ref{fig:jl:vortex}). The possible values of the field $A_\mu$ on ${\Sigma}$ must be pure gauge~\footnote{By ``pure gauge'' one means field configurations $A_\mu$ with zero field strength $F_{\mu\nu}=0$ and thus carrying no energy or action.}
in order to ensure a finite action. In other words, $A_{\mu}$ should be zero up to a gauge
transformation characterized by an internal $U(1)$ phase angle $\theta(x)$ (represented by the green
arrows in Fig.~\ref{fig:jl:vortex}).  Therefore at each point on ${\Sigma}$, the field $A_\mu$ takes
a value from a group of gauge-equivalent $A_\mu$ labeled by the internal angle $\theta$. The space
of gauge-equivalent $A_\mu$ or angles, which is represented by the dashed red circles at the
boundary in Fig.~\ref{fig:jl:vortex} is often referred to as the functional space $\mathcal{F}$. In other words, we have a map from the configuration space, $\Sigma$ to the functional space,
$\mathcal{F}$, which in our simple example are both circles.  Now let start from an arbitrary point
P (with and initial value for the field $\theta_0$) on the boundary ${\Sigma}$ and go around for a full circle
back to the point P while watching the change of the internal phase angle $\theta$ along the
way. Upon returning to P, $\theta$ must return to its initial value $\theta_0$: this however could
be realized in many ways, e.g. allowing $\theta$ to stay constant\footnote{It actually does not have
  to be constant, it may fluctuate around the initial angles while going around the blue circle.}
along $\Sigma$ (left configuration of Fig.~\ref{fig:jl:vortex}), or changing for a full $2\pi$ along
$\Sigma$ (middle configuration of Fig.~\ref{fig:jl:vortex}), or changing for a full $4\pi$ along
$\Sigma$ (right configuration of Fig.~\ref{fig:jl:vortex}), etc.  Clearly, in these distinctive
configurations, the gauge field winds around its internal space by zero, one, or two times while
going around the spatial boundary once. This brings the concept of an integer number $Q_w$ whose
magnitude simply counts the number of times the gauge field winds around. One can further assign a positive or
negative sign to $Q_w$, depending on whether one is winding in a counterclockwise or clockwise direction
(while assuming always going counterclockwise on the spatial boundary $\Sigma$). This intuitive
observation can be made more precise by the following mathematical definition:~\footnote{We note that if one considers this 2D system as actually embedded in a 3D space, then this definition is  counting the quantized magnetic flux through the vortex.}  
\begin{eqnarray}
Q_w = \frac{1}{2\pi} \int_{\Sigma} d{\vec l} \cdot (e \vec{A})
\label{eq:sec2:simple_topology}
\end{eqnarray}
which is a line-integral of the gauge field over the spatial boundary $\Sigma$. 
Let us briefly show that this definition actually works. As already said, on the boundary the gauge field is pure gauge, given by the local gauge phase angle $\theta(x)$ via $\vec{A} = \frac{1}{e} \nabla \theta(x) $. It is most transparent if one performs the above integral with polar coordinates $(r,\phi)$, in which case $d\vec l \to Rd\phi$  (with $R$ the radius of the boundary $\Sigma$) while $(e \vec{A}) \to \frac{1}{R}\partial_\phi (\theta)$. It shall now become obvious that, going around $\Sigma$ from point P back to P, the above integral yields $Q_w=[\theta(P)_{final}-\theta_0]/(2\pi)$ where $\theta(P)_{final}$ must end up as $(\theta_0+2\pi \, n)$ (where $n$ is any integer) for periodicity. Thus the so-defined $Q_w$ indeed counts the number of windings. 

%Going around $\Sigma$ from point P back to P, the integral runs from $\phi=\phi_0$ to 

It is not difficult to recognize that configurations with the same $Q_w$ could be continuously
deformed into each other while configurations with different $Q_w$ could never be. That is precisely
a statement about topology: configurations with the same $Q_w$ are topologically equivalent, while
configurations with different $Q_w$ are topologically distinctive. The number $Q_w$ is the
topological number labeling various topological sectors.  Mathematically speaking, one defines a
mapping from the spacetime manifold $\Sigma$ to the field space manifold $\mathcal{F}$. Such a mapping
could be topologically trivial or nontrivial depending upon the specific $\Sigma$ and
$\mathcal{F}$. There are a number of famous examples for topologically nontrivial configurations in
field theories, including the kink (domain wall) in one spatial dimension, the vortex in two spatial
dimensions discussed above, the monopole in three spatial dimensions, as well as the instantons and
sphalerons in four spacetime dimensions. Interested readers are referred to a number of excellent
and pedagogical review articles on this topic, see
e.g. \cite{Schafer:1996wv,tHooft:1999cgx,Tong:2005un}.

Focusing our discussions on QCD, the relevant topological configurations are the instantons and
sphalerons arising from the gluon field. It is most convenient to discuss them in the four
dimensional Euclidean spacetime. In this case, the boundary is the $O(3)$ hypersphere at infinity in
4D (Euclidean) spacetime. As in our example, any gluon configuration $A_\mu(x)$ with finite action
can only take values on the boundary which are pure gauge. Therefore  the field space $\mathcal{F}$
would be the collection of all pure gauge field configurations connected with each other through
gauge transformations which in the case of QCD would be $\mathrm{SU(3)}$. This mapping may be topologically nontrivial, and gluon field configurations can be labelled by a topological winding number $Q_w$ (in analogy to the winding number $Q_w$ of the 2D vortex example): 
\begin{eqnarray} \label{eq:jl:Qw}
Q_w =   - \frac{1 }{32\pi^2}  \int d^4 x  \,  (gG^{\mu\nu}_a) \cdot (g \tilde{G}_{\mu\nu}^a) 
= \frac{C_A}{2} \int_{boundary} d\Sigma^\mu \cdot K_\mu  \,\, ,
\end{eqnarray} 
where $g$ is QCD gauge coupling, $G$ and $\tilde{G}$ are gluon field strength tensor and dual tensor, and $a=1,2,...8$ is the adjoint color index to be summed. Explicit solutions with nonzero integer $Q_w$, i.e. the instantons, were discovered a while ago~\cite{Belavin:1975fg}.  They were found to play vital roles in understanding many fundamental aspects of QCD from vacuum to finite temperature and density: see reviews in \cite{Schafer:1996wv,Diakonov:2002fq}.

Next, let us highlight a few important features of such topological configurations with nonzero
integer $Q_w$. We first note that, given Eq.~\eqref{eq:jl:Qw}, the gluon field is {\em strong} and
thus non-perturbative, in
particular if the gauge coupling $g$ is weak. Indeed, Eq.~\eqref{eq:jl:Qw} suggests that the gluon
field strength scales as $G\sim \sqrt{Q_w}/g$ and its contribution to the action is
$S\sim G^2 \sim Q_w/g^2$.  A precise calculation reveals the action of such a configuration, which is a solution to the classical equation of motion, is given
by $Q_w (8\pi^2/g^2)$. This has nontrivial implications: as the gauge coupling becomes stronger, the
action ``cost'' of such configurations becomes less. As a result these configurations become more
important at strong coupling or equivalently at the low energy scale in QCD.  Furthermore the exact solution shows that the total action of such
configurations, instead of spreading widely over large spacetime volume, is more concentrated across
a domain of certain size, typically on the order of a fraction of $1\,\rm fm$ in QCD.\footnote{The
  scale is actually set by the running of the coupling constant as the action $S\sim 1/g^{2}$.}
Furthermore, let us emphasize that the above topological winding number of gluon field is
${\hat{\mathcal P}}$-odd and ${\hat{\mathcal C}}{\hat{\mathcal P}}$-odd, which perhaps becomes
apparent upon the rewriting
$G^{\mu\nu}_a \tilde{G}_{\mu\nu}^a \to \mathbf{E}^a \cdot \mathbf{B}_a$.  In short, the gluonic
topological transitions are non-perturbative phenomena naturally occurring in QCD and are actually
important in the regime where the gauge coupling is strong. Accompanying such transitions are the
creation of ${\hat{\mathcal P}}$-odd and ${\hat{\mathcal C}}{\hat{\mathcal P}}$-odd local
domains. Needless to say, an experimental verification of such topological transitions would be of
fundamental interest.

This however is nontrivial as the gluons do not carry the kind of quantum numbers (such as electric
charge or baryon number) that could be readily measured.  The gluonic topological fluctuations could
be better ``seen'' if their unique features would translate into the quark sector. And indeed, such
a ``translation'' is readily provided by the chiral anomaly, as discussed in the previous
section. By comparing Eq.~\eqref{eq:jl:N5Qw} with Eq.~\eqref{eq:jl:Qw} one immediately recognizes a key relation between the change of chirality in the quark sector and the gluonic field topology: 
\begin{eqnarray}
N_R-N_L = N_5 = 2Q_w \ . 
\end{eqnarray} 
  {The above relation bears a deep connection to mathematics, being essentially a special case of the Atiyah-Singer index theorem~\cite{Atiyah:1968mp} for the Dirac operator in the instanton background fields.}  
Therefore,  {\em every gluonic topological transition induces  a corresponding fluctuation in
  the quark chirality imbalance.} Consequently,   topological transitions in the gluon sector
are always accompanied by considerable fluctuations of the net chirality in local domains of matter,
and an experimental measurement of the latter would be a unique and direct means for probing the
former. To which extent this is possible we will discuss in Section~\ref{sec5}.

%-------------Section 3-----------------------------------

\section{Phase Diagram of QCD }
\label{sec3}

% \subsection{Phase diagram of QCD}
% \label{PD}

\subsection{Chiral symmetry restoration and deconfinement}

The phase diagram of QCD is a map of thermodynamic states of QCD in the
plane of temperature $T$ and baryon chemical potential $\mu_B$. It is
shaped by the fundamental properties of QCD: spontaneous chiral
symmetry breaking and confinement. If quark masses were zero, the QCD
Lagrangian would possess exact chiral symmetry, which would be
spontaneously broken in the vacuum state at $T=\mu_B=0$ and restored
at sufficiently high temperatures.  On the other hand, if there were
no light quarks at all, the color charges would be linearly confined
in the vacuum by color flux tubes or QCD strings, while the string
tension would vanish at sufficiently high temperatures leading to deconfinement.

Strictly speaking, in our exquisitely imperfect world, neither of
these two fundamental properties are realized exactly. The chiral
symmetry is, of course, explicitly broken by non-zero quark masses. Meanwhile,
confinement, in the rigorous sense of the infinite energy required to
separate two opposite color sources, does not hold exactly unless
quarks are absent (or infinitely heavy): string breaking via creation
of quark-anti-quark pairs limits the range of the linear growth of the
confining potential. 

   {An alternative characterization of confinement in the vacuum state %(useful even for the case with dynamical quarks) can be 
is based on the fact that the physical spectrum of QCD contains only color-singlet states, i.e. the various hadrons. This definition, however, is not directly useful for studying the phase diagram. Rather than being a consequence of dynamics,
for a non-abelian gauge theory such as QCD, the Hilbert space of physical states must contain only color-singlet states {\em by definition}, as a necessary condition for quantization of the theory.}
~\footnote{   {
For many subtle facets of confinement, especially in a theory like QCD with dynamical quarks, see,  e.g., Ref.~\cite{Greensite:2011zz}.}}
%Of course, we are not defining
 % confinement as the absence of non-singlet %color states in the
 % physical Hilbert space of QCD. We view the absence of color states
 % as part of the {\em definition\/} of QCD as a non-Abelian gauge
  %theory. As such it is not a dynamical property which could
 % distinguish thermodynamic states.

   {Both idealized limits discussed above -- massless quarks and infinitely massive (or absent) quarks have their own order parameters.  In the massless limit, the chiral condensate (as discussed in Sec. 2.1) is a local order parameter identifying different phases at finite temperature and density: being zero in the symmetric phase and being non-zero in the symmetry-breaking phase. Somewhat similarly in the limit without dynamical quarks (i.e. in a  pure non-Abelian gauge theory), one can define an order parameter by considering finite-temperature theory as a 4-dimensional theory in Euclidean space (with compactified imaginary time direction):  Polyakov loop 
$L\equiv \left\langle \frac{1}{N_c} {\rm Tr}\left\{\hat{P} \ \exp\left(i \int_0^{1/T} dx_4 A_4  \right) \right\}\right\rangle \ $ where $N_c=3$ is the number of colors and $\hat{P}$ signifies path-ordering for the integration of gauge field component $A_4$ along temporal direction $x_4$.
As shown in \cite{Polyakov:1978vu} there is a $Z_{N_c}$ symmetry of the theory for which $L$ is an order parameter. This symmetry is present at zero temperature (vacuum) but  is broken at high temperature signifying deconfinement transition in pure gauge theory. 
Thus the change of the Polyakov loop from zero (in vacuum) to nonzero value at finite temperature signifies deconfinement transition in a pure-gauge theory.

The $Z_3$ symmetry is, however, explicitly broken in QCD because of quarks and the Polyakov loop is nonzero both in vacuum and at high-temperatures and thus cannot be used as an order parameter for rigorous definition of deconfinement in QCD. Likewise the chiral symmetry is also explicitly broken by nonzero physical quark masses in QCD so the chiral condensate is not an order parameter in QCD. 
 Nevertheless, examining the behavior of the chiral condensate and the Polyakov loop at finite temperature and density is found to be very useful for studying QCD phase diagram, as we shall discuss further. }

Notwithstanding chiral symmetry's approximate nature, the smallness of
quark masses relative to the relevant QCD scale makes this symmetry a
useful guide to the states of QCD at finite temperature and
density. Lattice calculations at zero baryon density show the rapid
drop of the measure of the spontaneous breaking of the chiral symmetry
-- the chiral condensate -- as a function of temperature (see
Fig. \ref{fig:vk:lattice_chiral_condesate}). This behavior is expected due to the
restoration of the chiral symmetry at high temperature. However, consistent with the
{\em approximate} nature of the symmetry, the chiral condensate does not drop
to zero and no thermodynamic singularity or discontinuity has been
found in lattice calculations (at
$\mu_B=0$)~\cite{Brown:1990ev,Aoki:2006we}.

\begin{figure}
  \centering
  \includegraphics[height=22em]{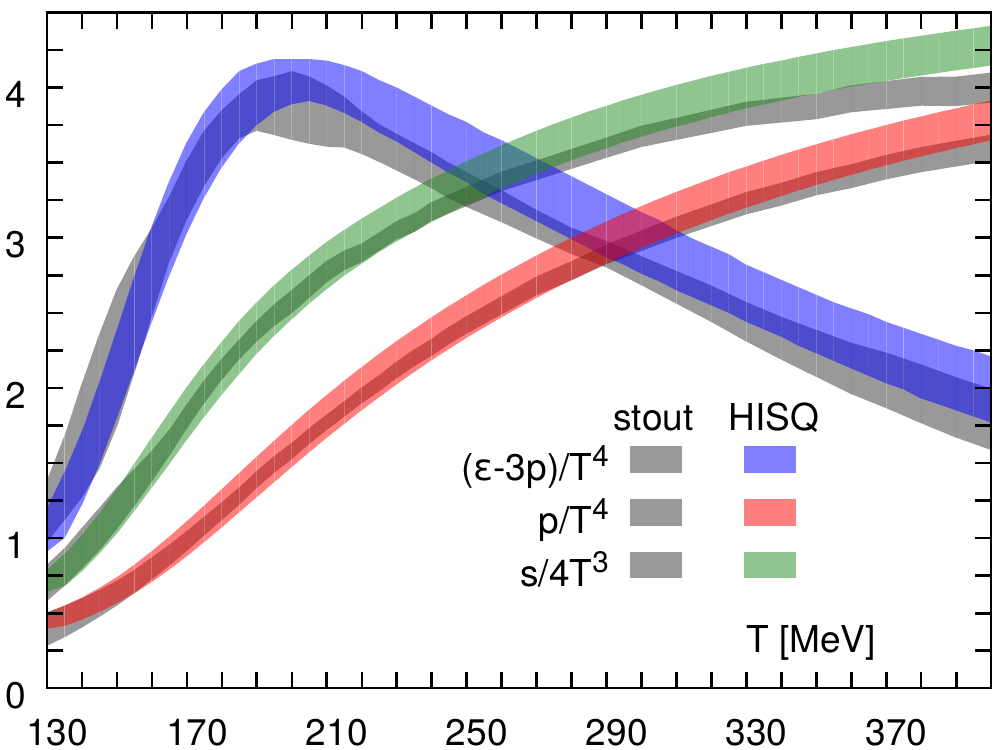}
  \caption{The dependence of the QCD thermodynamic quantities  in
    the transition temperature (crossover) region at zero baryon
    density.  The dramatic rise
    of the effective number of degrees of freedom per unit volume, as
    quantified, for example, by the ratio $s/T^3$, demonstrates color
    deconfinment. The transition is rapid but
    smooth, i.e., no thermodynamic singularity is found. 
    {The labels ``stout''~\cite{Borsanyi:2013bia} and HISQ~\cite{Bazavov:2014pvz} refer to computations  based on two different approaches of lattice discretization that were performed by two different groups.} The figure is taken from Ref.~\cite{Soltz:2015ula}.   
% or \cite{Borsanyi:2010cj}.
  }
  \label{fig:sT}
\end{figure}

As a function of increasing temperature QCD undergoes the transition
from a thermal state of hadron gas, which is a pion-dominated, dilute and nearly ideal gas  for sufficiently low temperatures, to a much
denser state of quark-gluon plasma, which is nevertheless also nearly ideal
at sufficiently high temperatures due to asymptotic freedom. The temperature of the
transition region is determined largely by the
confinement scale $\Lambda_{\rm QCD}\sim 200$ MeV (inverse of the
confinement radius). The fact that this transition is associated with
deconfinement can be seen from the first-principle lattice
calculations of thermodynamic functions. The main feature, largely due to liberation of colored gluons, is a
dramatic rise of the effective number of the degrees of freedom,
measured, e.g., by entropy density in units of temperature, $s/T^3$ as
seen in Fig. \ref{fig:sT}.

Lattice calculations show that this color liberation transition is not
abrupt, and is not accompanied by a thermodynamic
singularity. Rather, it is a smooth
crossover~\cite{Brown:1990ev,Aoki:2006we}. This is in agreement with
the observation that the confinement-deconfinement transition is not
well-defined in the presence of quarks, i.e., there is no order
parameter, such as string tension or the Polyakov loop expectation value, whose non-analytic behavior would
signal a true phase transition \cite{Polyakov:1978vu}. Similarly, because of the explicit breaking
of the chiral symmetry by the quark masses, there is no order
parameter for the chiral symmetry breaking which would have to be strictly zero
when the system is in the state where the symmetry is restored.

In contrast, in the idealized chiral limit $m_q\to0$, the
thermodynamic realization of the chiral symmetry (in this case exact on the level of
the QCD Lagrangian)
qualitatively changes at a certain
transition temperature (from being broken to being restored).  This, by a standard textbook argument
\cite{Landau:1980mil}, has to lead to a true thermodynamic singularity
-- a phase transition. The chiral symmetry plays a central role in
this argument as it enforces the vanishing of the order parameter, the chiral condensate,   in the high-temperature phase. An
analytic function cannot remain constant on any finite interval unless
it is constant everywhere. Therefore, the high-temperature phase must be
separated from the low-temperature phase by a 
singularity.

Using symmetry, universality and renormalization group
considerations, it can be further argued~\cite{Pisarski:1983ms} that
due to the SU(3)$\times$ SU(3) chiral symmetry and the U(1)$_A$ anomaly in
an idealized world with {\em three} lightest quarks ($u$, $d$ and $s$)
being massless, the finite-temperature transition must be of the first
order. By continuity, this must remain true for a range of quark
masses near the point $m_{u}=m_d=m_s=0$. The size and shape of this parameter
region, usually drawn on the $m_s$ vs $m_{u,d}\equiv m_u=m_d$ diagram known as
``Columbia plot''~\cite{Brown:1990ev}, has been and still remains a subject of
lattice QCD studies, with the main problem currently being strong
cutoff dependence of the result~\cite{deForcrand:2017cgb}. All lattice
results, however, are in agreement on the conclusion that the region
of the first-order transition (at $\mu_B=0$) on the Columbia
plot does not extend to the physical point given by the real-world
values of quark masses $m_s$ and $m_{u,d}$. In other words, the finite
temperature transition in QCD with physical quark masses is not a
first-order transition but an analytic crossover.

\subsection{Phase transition at finite baryon density}

\begin{figure}
  \centering
  \includegraphics[height=22em]{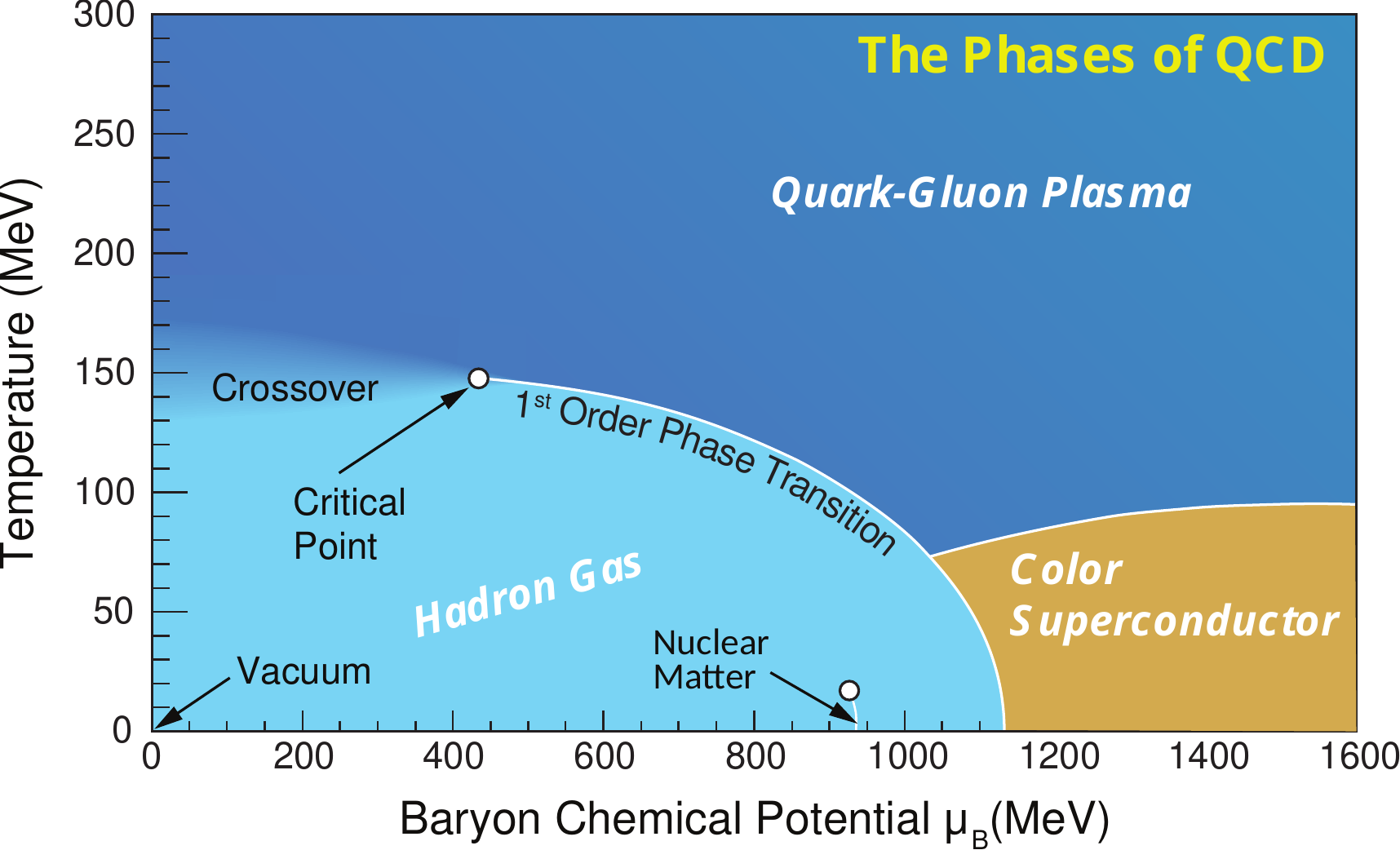}
  \caption{A sketch of the QCD phase diagram in the plane of
    temperature and baryon chemical potential illustrating theoretical
    expectations (as well as empirically known coexistence between
    nuclear matter and vacuum). The finite temperature
    crossover at zero or small $\mu_B$  switches to   the large-$\mu_B$ first-order transition
    at a critical point. Higher density phases, not covered by this
    review, are also sketched.}
  \label{fig:pd}
\end{figure}

As we increase the baryon chemical potential $\mu_B$ the nature of the
finite-temperature transition from hadron gas to quark-gluon plasma
can change from the continuous (second-order transition or crossover)
to discontinuous (first-order) transition. In the idealized world with
{\em massless} $u$ and $d$ quarks and sufficiently heavy $s$ quark, where
the transition is second order at $\mu_B=0$, the change of the order of
the transition at finite $\mu_B$ would correspond to a tricritical
point on the phase diagram found in models of QCD~\cite{Berges:1998rc,Halasz:1998qr,Stephanov:1998dy,Stephanov:2004wx}. In the world where $u$ and $d$ quarks have small masses and the second-order transition at
$\mu_B=0$ is replaced by a crossover, the first-order transition would
have to begin at a critical
point as shown in the sketch
of the QCD phase diagram in Fig.~\ref{fig:pd}.  Many models of QCD show such a phenomenon
(see, e.g., Refs.~\cite{Stephanov:2004wx,Stephanov:2007fk} for a review). The experimental search
for the signatures of the QCD critical point and associated
first-order transition is underway and is a main  subject of this review.

First-principle lattice QCD calculations provide a powerful method for
studying QCD thermodynamics at $\mu_B=0$~\cite{Ding:2015ona}.  However, reliable lattice
calculations at {\em finite} chemical potentials are impeded by the
notorious sign problem (see, e.g., Ref.~\cite{deForcrand:2010ys} for a
review). Estimates based on various techniques
circumventing the sign problem such as reweighting or Taylor expansion,
have yielded results which are consistent with the existence of the
critical point at $\mu_B\gtrsim200-300$ MeV, while the imaginary
$\mu_B$ extrapolation does not show a critical point at least until
$\mu_B\gtrsim 500$ MeV. Taken together at face value, these results
disfavor the existence of the critical point for
$\mu_B\lesssim 200$ MeV. It should be emphasized that some of these results are
obtained with lattice spacing  in the range where the cutoff
dependence is still very strong and makes the extrapolation to continuum
limit uncertain. There has been also much effort and progress
in addressing the sign problem recently which will have to be reviewed
elsewhere. This review will focus on the {\em experimental\/} search for
the QCD critical point and the corresponding first order phase
transition.

\subsection{First-order transition and the critical point}

Many physical properties of the critical point can be better understood 
as a consequence of the fact that this point terminates the
coexistence curve in the $T$ vs $\mu_B$ plane. On this curve the two
coexisting phases are separated by a first-order phase transition and
the densities of conserved quantities, such as energy or baryon
density, as well as other quantities, such as the chiral condensate,
are discontinuous. The discontinuities vanish at the critical point,
where the two coexisting phases become one and the same. The
increasing similarity between the two phases, and the diminishing thermodynamic barrier between them are at the origin of the large
fluctuations at the critical point. These qualitative features are universal,
i.e., independent of the physical nature of the underlying system or substance 
with the critical point.

The discontinuities across the first-order phase transition line
reflect the difference in the physics of the phases. Mechanical,
thermal and chemical equilibrium on the coexistence line require
continuity of pressure $p$, $T$ and $\mu_B$ respectively. However, the entropy
density, $s$, is higher on the QGP side, i.e., $\Delta s>0$, due to the liberation of color
degrees of freedom from confinement. Similarly, the baryon number
density $n_B$ is higher on the QGP side, i.e., $\Delta n_B>0$, because the degrees of freedom
carrying the baryon number -- quarks -- are lighter due to restoration
of the chiral symmetry. 
These discontinuities across the first-order phase transition curve are
related by the Clasius-Clapeyron relationship to the slope
of the curve: $dT/d\mu_B= - \Delta n_B/\Delta s$, which follows from
equilibrium conditions $\Delta p = \Delta T = \Delta\mu_B=0$ and
$dp=sdT+n_Bd\mu_B $. The negative slope $dT/d\mu_B<0$ of the transition line on the phase
diagram is consistent with the physics associated with the
discontinuity: $\Delta s>0$ and $\Delta n_B>0$~\cite{Halasz:1998qr}.

As we shall discuss in more detail in the next section, a critical point
terminating the first-order phase transition is a ubiquitous phenomenon
which occurs in physically very different systems such as ferromagnets
and liquids. Despite completely different microscopic nature, the
collective effects near such critical points are remarkably
universal. After a certain remapping the leading singularity of the
equation of state near all such critical points (within the same universality class) is the same. This is a
consequence of the fact that critical phenomena are caused by the {\em
  long-range\/} correlations which emerge in the thermal fluctuations
of the system. Qualitatively, these fluctuations are related to the
transitions between two coexisting phases of the system which become
increasingly similar, i.e., close to each other as the first-order
transition approaches its terminal (critical) point. In the limit of
infinite correlation length such long-range fluctuations are described
by a conformal (scale-invariant) field theory defined as the infrared
fixed point of a renormalization group evolution (coarse-graining) of
the system. The universality of critical phenomena directly follows
from the independence of the fixed-point theory on the initial point
of the evolution (within a universality class).    {The scaling and universality of the critical phenomena is the key to our ability to make predictions about the QCD critical point and its signatures in heavy-ion collisions despite having only incomplete information about the QCD equation of state. We shall discuss this in more detail in Section~\ref{sec4}.}

% for a typical critical point in fluids there is no
% symmetry between the two phases separated by the coexistence
% curve. This is also true for the QCD critical point.% ~\footnote{
% % In the chiral limit, the $O(4)$ symmetry relates different states
% % below the first-order phase transition curve.}
% The {\em leading}
% singularity of the equation of state at the critical point, however,
% is universal and possesses the $Z_2$ symmetry of the Ising model. The
% universality allows to draw qualitative and semiquantitative
% predictions without precise knowledge of the equation of state.

\subsection{High baryon density regime}

At still larger baryon chemical potentials the QCD phase diagram is believed
to be very rich. Much of the physics at {\em asymptotically large}
$\mu_B$ can be addressed perturbatively, where color-flavor locking
determines the ground state of QCD matter~\cite{Alford:2007xm}. The
most interesting regime of {\em intermediate\/} $\mu_B$ relevant to neutron
stars~\cite{Lattimer:2015nhk,Page:2006ud} and, possibly, low-energy heavy-ion collisions is a subject of
current theoretical research. Many interesting possible phases and 
regimes have been proposed, with predictions often depending on
the interplay between non-perturbative effects and dynamical
assumptions made in calculations. The equation of state of QCD in
this regime is, unfortunately, inaccessible to present lattice
first-principle calculations due to the sign problem. 

It should also be mentioned that gravitational wave astronomy and the recent discovery of
neutron star mergers have opened a new exciting avenue for exploring the
QCD equation of state in this domain. We shall leave these interesting
topics to surveys elsewhere.  The present review will  focus on the investigation of the QCD
phase diagram in the regime readily accessible by heavy-ion collisions.

%=====================================================================================
%=====================================================================================
%

%%% MS: this is useful for emacs:
%%% Local Variables: 
%%% TeX-PDF-mode: t
%%% TeX-master: "BES_Main_current"
%%% End: 

%-------------Section 4-----------------------------------
%\graphicspath{{figures/}}

% %% Macros used in this section
% \newcommand\op{{\sigma}}
% \newcommand\pd{\partial}
% \newcommand\LambdaQCD{\Lambda_{\rm QCD}}
% \newcommand\paper{Letter}
% \newcommand\p{{\bm{p}}}
% \newcommand\la{\langle}
% \newcommand\ra{\rangle}
% \newcommand\xis{\xi}
% \newcommand\cO{{\cal O}}

\section{Theory and Phenomenology of the Critical Point}
\label{sec4}

\subsection{Critical phenomena}

The critical phenomena which we shall focus on occur at an end-point
of a first-order transition in a thermodynamic system. The first-order
transition corresponds to a situation when a thermodynamic system
under given external conditions (such as $T$ and $\mu$, for example)
can be in equilibrium in two distinct thermodynamic states. Such a
two-phase coexistence can occur only for special values of external
parameters, typically, on a manifold of one less dimension than the
space of external parameters. E.g., in the $T-\mu$ plane this manifold
is a first-order transition line. One of the two states is
thermodynamically stable on one side of the first-order phase
transition, and the other -- on the other side. By adjusting
parameters along the phase-coexistence line one could arrive at a
special point where the difference between the two coexisting phases
disappears. This is a critical point, also known as a second-order
phase transition. This point is characterized by critical phenomena which
manifest themselves in singular thermodynamic and hydrodynamic
properties.

The two most common examples of such critical points are the end point
of the liquid-gas coexistence curve and the Curie point in a
(uniaxial) ferromagnet. Although the two systems in which these two
examples occur are different on a fundamental, microscopic level, the
physics near the critical point is remarkably similar on qualitative
as well as quantitative level. This observation is the basis of the
concept of universality of the second-order phase transitions.

\begin{figure}[b]
  \centering
  \includegraphics[height=15em]{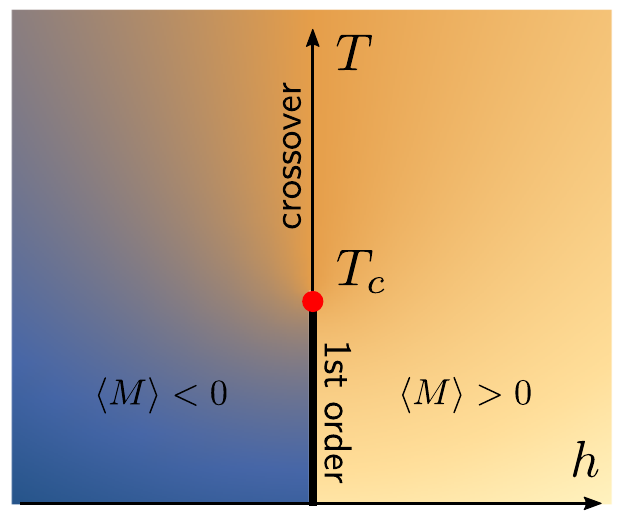}
  \caption{The phase diagram of the Ising ferromagnet. The transition
    between the phases with positive and negative magnetization is a
    first-order transition for $T<T_c$ and a continuous crossover at
    $T>T_c$. The transition changes its character at the critical point.}
  \label{fig:pd-ising}
\end{figure}

The uniaxial, or Ising, ferromagnet is the simplest of such systems. It
can be modeled by a lattice of spins $s_i=\pm1$, or two-state systems,
with local (e.g., nearest neighbor) interaction favoring the alignment
of spins in the same direction. There are two ground states, with all
the spins pointing in one of the two possible directions. The
degeneracy is lifted if one applies a magnetic, or ordering, field
$h$, which changes the energy of the spins by $h \sum_i s_i$. The two
ordered states are distinguished by the value of the magnetization
\begin{equation}\label{eq:M}
M=\frac1N\sum_{i=1}^N s_i
\end{equation}
which equals $+1$ or $-1$ depending on the sign of $h$, or more
precisely, by its thermal average $\langle M\rangle$. At finite, low
enough temperature the ordering persists and $\langle M\rangle$ plays the role of the
order parameter which flips sign at $h=0$. The two ordered phases
coexist on the line $h=0$ in the $T-h$ plane as shown in Fig.~\ref{fig:pd-ising}

The magnetization $M$ along the coexistence line, $h=0$, decreases with
increasing temperature due to thermal fluctuations. At the Curie
temperature, $T_c$, the magnetization completely vanishes and remains
zero for all higher temperatures. The coexistence line (the
first-order phase transition) ends at $T=T_c$ -- the critical
point. There is only one phase at and above the Curie point temperature.

Similarly, liquids (e.g., water) coexists with their vapour at given
pressure $p$ at the boiling temperature $T$, which defines a line in
the $T$ vs $p$ plane.  At any of the coexistence points on this line
the molecules making up the substance can be arranged in two possible
ways, or phases, with the same pressure and temperature. The liquid phase
has higher density, but the kinetic energy of molecules in it is
smaller due to the molecular attraction. The resulting pressure is
the same as that of the gas phase of a lower density.

As the temperature and pressure increase
along the coexistence curve the density difference $\Delta
\rho\equiv\rho_\mathrm{liquid}-\rho_\mathrm{gas}$ decreases and
vanishes at the critical point, where the coexistence line
ends.

In order to enhance the similarity between the two examples of the
phase transitions, it is instructive to think of the line of vanishing
ordering (magnetic) field, $h=0$, between $T=0$ and $T_\mathrm{c}$ on
the $T$ vs $h$ plane of an Ising ferromagnet as a coexistence line
where two possible arrangements of spins with equal magnitude and
opposite sign of magnetization coexist. The discontinuity of the
magnetization $\Delta M = 2M$ is then the Ising ferromagnet analog of
the density discontinuity $\Delta \rho$ in the liquid-gas
transition. Even though $\rho$ is not a measure of ``order'' in the
same sense as $M$ is, the similarity is enough to justify common
terminology ``order parameter'' in reference to density $\rho$.

In the following we will review the most important properties of
criticality pertinent to the experimental search for the QCD critical
point, such as the divergences of the fluctuations and their
relation with the correlation length~$\xi$. For many other aspects
of critical behavior of the phase transitions we refer the
reader to standard textbooks and reviews on the subject (e.g., Refs.~\cite{Amit:1984ms,ZinnJustin:2002ru,Pelissetto:2000ek}).

%Ferromagnet

%Liquid gas

%Critical opalescence...

\subsection{The role of fluctuations}
\label{sec:fluctuations-role}

A useful way to understand the state of a system at a critical point
is to think of it as a point where the difference between two
coexisting states, or phases, has just vanished. 
These states
coexist along the first-order transition which terminates at the critical
point. In the Landau-Ginzburg picture \cite{Landau:1980mil} of the transition each of the
states corresponds to a minimum of the Landau-Ginzburg free energy (as
a function of an order parameter). At any point on the first order transition line the two
alternative (or coexisting) states are equally deep and are separated
by the Landau-Ginzburg free-energy
barrier. The path from one state to the other lies through {\em
  non-equilibrium} states (or more precisely, through inhomogeneous states).  At the critical point this barrier shrinks
and disappears (the lowest free energy cost, which is proportional to interface
tension, vanishes).

The approach to the critical point is characterized by an increase
of fluctuations, which is the manifestation of the ease of changing
the state of the system  from one state to the
other. Away from the critical point, on the first order line, these
fluctuations are suppressed by the barrier of non-equilibrium states,
while on the crossover side there is simply only one state.

\begin{figure}
  \centering
  \includegraphics[height=100pt]{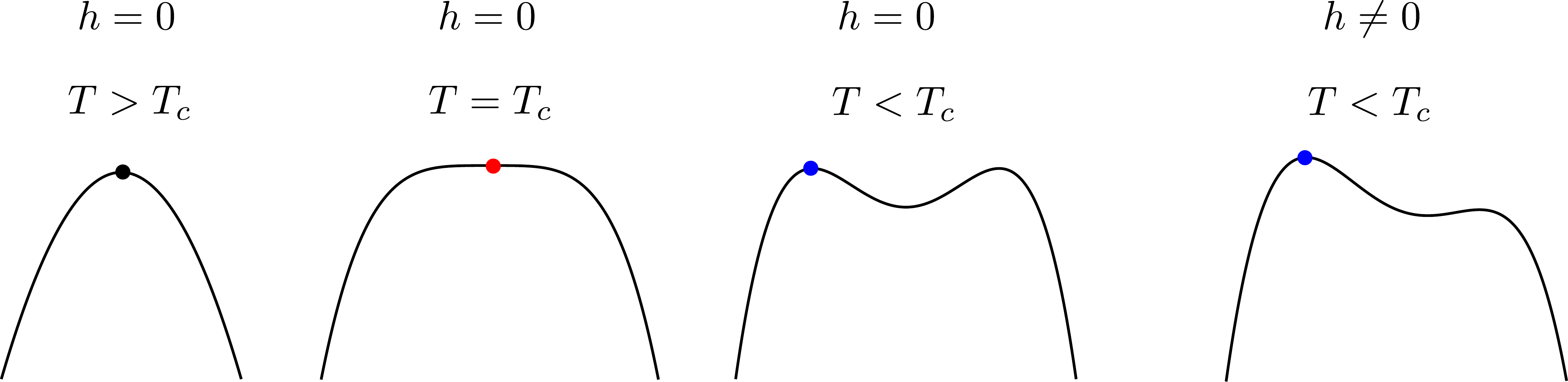}
  \caption{A schematic representation of the probability $P(M)$  distribution
    of the order parameter $M$ of the Ising model, plotted as $\log P(M)/N$, at three
    different temperatures -- above, at, and below critical point --
    at zero ordering field $h=0$, so that $P(M)$ is symmetric
    around $M=0$, as well as at $T<T_c$ at nonzero~$h$. The filled
    circle indicates the maximum of $P(M)$, i.e., equilibrium value of
    $M$. At $h=0$, $T<T_c$ the system has to choose between two equally probable maxima, breaking the symmetry spontaneously. The magnitude of fluctuations around the maximum,
    $\langle(\delta M)^2\rangle^{1/2}$, is of order $1/\sqrt N$
    according to Eq.~(\ref{eq:F''}), except at $T=T_c$ when the
    susceptibility $(\partial M/\partial h)_T$ diverges. For $T<T_c$,
    when the two maxima of $P(M)$ coexist, an arbitrary nonzero ordering
    field $h$ leads to exponential (in $N$) suppression of the height of the
    lower peak by a factor $\exp(-N \langle M\rangle h/T)$, making the
    contribution of the lower peak of $P(M)$ to fluctuations negligible as
    $N\to\infty$. This suppression disappears as $T\to T_c$, since
    $\langle M\rangle\to 0$, which is one of the ways to understand
    the enhancement of fluctuations at the critical point. It can be
    viewed as a result of the large fluctuations between the two
    coexisting maxima. This also helps explain why critical point
    exists only at $h=0$. }
  \label{fig:3P}
\end{figure}

More quantitatively, one can describe the fluctuations by the
probability distribution $P$ of the quantities characterizing a
macroscopic state of the system. For the Ising model -- magnetization
$M$ is a natural choice, since the coexisting states differ by the
value of $M$. For a large, but finite, system, $1\ll N<\infty$, the
probability distribution $P(M)$ at finite temperature $T$ and given
$h$ is a two-peak function, with the two peaks reaching equal height
at the coexistence point $h=0$, as shown schematically in Fig.~\ref{fig:3P}. Above $T_c$ the probability
distribution $P(M)$ has only one peak. The critical point is
characterized by the probability distribution $P$, which has a single
peak, with a very flat (i.e., zero curvature) top -- the two coexisting
states (peaks) have just merged together. 

The probability distribution of $M$ is related to the Helmholtz free
energy $F(M)$:
\begin{equation}\label{eq:PF}
P(M)\sim\exp\left\{-\frac NT \left(\,F(M)-hM\,\right)\right\}\,. 
\end{equation}
Indeed, for large $N$ the average (equilibrium) value of $M$, $\langle
M\rangle$, given
by maximum of $P(M)$, satisfies
\begin{equation}
  \label{eq:<M>}
  \left(\frac{\partial F}{\partial M}\right)_T = h \quad\mbox{at $M=\langle M \rangle$}. 
\end{equation}
A qualitatively useful but, due to fluctuations, quantitatively
incorrect % (except in a space of more than four dimensions)
 expression
for $F(M)$ is given by the mean-field (or Landau-Ginzburg) free energy:
\begin{equation}
  \label{eq:LG}
  F(M) = \frac12 a(T) M^2 + \frac 14 b(T) M^4,
\quad\mbox{(mean-field free energy)}
\end{equation}
where $b(T)>0$ at all temperatures while $a(T)>0$ for $T>T_c$ and
 $a(T)<0$ for $T<T_c$. Thus $F$ has two minima at $T<T_c$, and
correspondingly, $P(M)$ has two peaks (see Fig.~\ref{fig:3P}).

The critical point is characterized by the vanishing of the second
derivative of $F$ at its minimum, which is related to the magnitude of
the fluctuations of $M$ around the equilibrium,
$\delta M \equiv M - \langle M\rangle$, as well as isothermal magnetic
susceptibility $(\partial \langle M\rangle /\partial h)_T$:
\begin{equation}
  \label{eq:F''}
  \langle(\delta M)^2\rangle
= \frac TN\,\left(\frac{\partial^2 F}{\partial M^2}\right)_T^{-1} =
  \frac TN\, \left(\frac{\partial \langle M\rangle}{\partial h}\right)_{T} \,.%\equiv T \chi\,,
\end{equation}
Thus the divergence of the susceptibility, i.e., flatness of $F$ as a
function of $M$ at the critical point, is directly related to the fact
that the fluctuations are large at the critical point (see
Fig.~\ref{fig:3P}).

The magnitude of fluctuations is of order $\delta M\sim1/\sqrt N$. The
subject of thermodynamic fluctuations is usually peripheral in the
study of condensed matter systems because the typical number of
degrees of freedom $N$ in such systems is huge (e.g., $10^{24}$).
There is one notable exception. Near the critical point the
correlation length $\xi$ of fluctuations grows arbitrarily large. The
actual measure of the importance of the fluctuations is the ratio of
the volume of the system $V$ to the {\em typical} correlation volume, i.e., the cube of the correlation length, $\xi^3$. Although it is
still impractical to fine-tune the system to achieve correlation
volumes of {\em macroscopic} sizes, in the phenomenon of critical
opalescence the relevant size (for light scattering) is only the wavelength of visible
light. This length is few orders of magnitude longer than the typical
correlation length, but near the critical point the correlation length
can achieve such large values, resulting in the observable loss of
transparency due to scattering of light on the density fluctuations.

Contrary to condensed matter systems, fluctuations in heavy-ion collision experiments do play a significant role even away
from a critical point. These systems are large enough to be treated
thermodynamically (or hydrodynamically), and yet not too large
($10^{2-4}$ particles) for the fluctuations to become unimportant. As
in condensed matter systems, the fluctuations should be even larger near the
critical point.

Another fruitful example of a system where fluctuations are important
is provided by the statistical lattice models and lattice field
theory simulated on computers, since they require the number of
degrees of freedom to be not too large for practical reasons.

\subsection{Long-range correlations}
\label{sec:long-range-corr}

It is important to keep in mind that the divergence of fluctuations at
the critical point is not a microscopically local phenomenon. Rather,
it is a collective phenomenon -- the result of the correlation between
fluctuations of many degrees of freedom. One can say that it is
related to the divergence of the number of the degrees of freedom in a
correlation volume $\xi^3$, not to the magnitude of the fluctuations
of each individual degree of freedom (as is obvious in the case of the
Ising model, where $s_i=\pm1$).

This can be quantified by considering the correlation function of an
order parameter. To make the discussion unified, we shall call this
order parameter $\op(\bm x)$. For the ferromagnet $\op(\bm x)$ is $M(\bm x)$
-- magnetization of a small patch of the lattice centered around point
$\bm x$, and for the liquid-gas transition it could be chosen as density $\rho(\bm x)$.

To describe spatial correlations we consider the probability distribution
as a functional, ${\cal
  P}[\op(\bm x)]$, for a spatially varying parameter $\op(\bm x)$. To determine the correlation length $\xi$ we
measure the thermal expectation value of $\delta\op(\bm x)\delta\op(\bm y)$,
where $\delta\op=\op-\langle\op\rangle$:
\begin{equation}
  \label{eq:MM-expectation}
  \langle \delta\op(\bm x)\delta\op(\bm y)\rangle 
= \int {\cal D}\op\, {\cal P}[\op]\,\delta\op(\bm x)\delta\op(\bm y)
\sim {\exp(-|\bm x - \bm y|/\xi)},  
\qquad{\rm for}\quad |\bm x - \bm y|\to\infty.
\end{equation}
where the functional integral is taken over all configurations of the
fluctuating field $\sigma$ with weights given by $\mathcal P[\sigma]$.
The logarithm of the probability density ${\cal P}$ defines the corresponding 
{\em Ginzburg-Landau free energy functional},
$\Omega[\op]$:
\begin{equation}
  \label{eq:P-Omega-sigma}
  P[\sigma] \sim \exp\left\{ - \Omega[\sigma]/T\right\},
\end{equation}
 also known as effective
Hamiltonian or effective action, depending on the context,
which we shall expand in the gradients of $\sigma$:
\begin{equation}
  \label{eq:Omega-sigma}
  \Omega[\sigma] 
=\! \int\!d^3\bm x\left[U(\sigma)+
\frac{1}{2}(\bm\nabla\sigma)^2 +
% \frac{m_\sigma^2}2 \sigma^2 
% + \frac{\lambda_3}{3}\sigma^3
% + \frac{\lambda_4}{4}\sigma^4 +
 \ldots
\right]
\,.
\end{equation}
For a given field $\sigma$, the coefficient of the gradient term
$(\bm\nabla\sigma)^2$ will not necessarily have the canonical value
$1/2$. We can, however, always redefine the field $\sigma$,
multiplying it by an appropriate constant, to reduce the gradient term
to the canonical form. 

Note that for a uniform field ($\sigma=M$) $\Omega[\sigma]$ is directly
related to the free energy $F - h M$ according to Eq.~(\ref{eq:PF}).

\subsection{Mean-field critical behavior}

The Gaussian approximation (also known as saddle-point or mean-field
approximation) to the path integral in Eq.~(\ref{eq:MM-expectation})
amounts to expanding $\Omega[\op(\bm x)]$ around its minimum,
$\sigma_0$, or
\begin{equation}
  \label{eq:U0}
  U(\sigma) = U_0 + \frac12m_\sigma^2 (\sigma - \sigma_0)^2 + \ldots
\end{equation}
and dropping
all terms beyond quadratic. The Gaussian path integral is then
easy to take. One finds $\langle\sigma\rangle=\sigma_0$ and, for 
fluctuation
\begin{equation}
    \delta\sigma =\sigma -\langle\sigma\rangle,
\end{equation}
the correlator
\begin{equation}
  \label{eq:MM-mean-field}
  \langle \delta \sigma(\bm x)\delta\sigma(\bm y)\rangle
=  \frac{T }{4\pi |\bm x - \bm
  y|}\exp\left(-\frac{|\bm x - \bm y|}{\xi}\right)\,,
\quad\mbox{(Gaussian approximation)}
\end{equation}
where the correlation length $\xi$ is related to the curvature of the
potential $U(\sigma)$ at its minimum:
\begin{equation}
  \label{eq:corr-length}
  \xi = (U''(\sigma))^{-1/2}\Big|_{\sigma=\sigma_0} = m_\sigma^{-1}.
\end{equation}

At the critical point, two
coexisting minima of $U$ continuously merge into one, 
and the curvature $U''(\sigma_0)$ vanishes. Thus $\xi=(U'')^{-1/2}$ diverges.
% We can redefine (by a constant shift) the order parameter field
% $\sigma$ in such a way that $\sigma_0=0$. We shall do that now to
% simplify the discussion that follows.
The divergence of the correlation length Eq.~(\ref{eq:corr-length}) is the
source of the non-analyticity of all thermodynamic functions at the
critical point. 

Indeed, let us define the volume
integral of the fluctuation of the order parameter field
\begin{equation}
  \label{eq:}
  \sigma_V \equiv \int d^3\bm x\, \delta\sigma(\bm x)\,.
\end{equation}
This extensive quantity fluctuates around its expectation value 
$\langle\sigma_V\rangle=0$. The (square) magnitude of these fluctuations is
measured by the second moment $\langle\sigma_V^2\rangle$ which is
related to the integral of the correlator of $\sigma(\bm x)$
\begin{equation}
  \label{eq:kappa2-sV-ss}
\kappa_2[\sigma_V]= \langle\sigma_V^2\rangle 
= V \int d^3{\bm x}\, \langle \delta\op(\bm x)\delta\op(\bm 0)\rangle 
\end{equation}
This quantity is also extensive, i.e., $\kappa_2\sim V$, provided the
correlator $\la\sigma\sigma\ra$ is local, i.e., $\xi$ is finite, or
more precisely, $V\gg\xi^3$. In the
mean-field approximation
\begin{equation}
  \label{eq:kappa2-xi}
  \kappa_2[\sigma_V]= VT\,\xi^2\,.
\end{equation}
This result is easy to understand as the zero-momentum value of the
propagator of the field $\sigma$, $(q^2+m_\sigma^2)^{-1}$. It
is also useful to view the divergence $\xi^2$ as a result of
the divergence of the integral of $1/|\bm x|$ being cut off at $|\bm x|\sim\xi$ according to
Eq.~(\ref{eq:MM-mean-field}). I.e., the divergence of the magnitude of
the fluctuations is due to the increase of the {\em range} of the
correlations, not their local magnitude.

Going beyond the mean-field approximation is a non-trivial task
requiring the application of renormalization group     {(see, e.g., \cite{ZinnJustin:2002ru})}, outside of the scope
of this review. This typically
results in small corrections to the power laws in
Eqs. (\ref{eq:MM-mean-field}) and~(\ref{eq:kappa2-xi}):
$\langle\op\op\rangle\sim 1/|\bm x - \bm y|^{1+\eta}$ for $|\bm x - \bm y|\ll\xi$ and
correspondingly $\chi\sim \xi^{2-\eta}$, with very small $\eta\approx
0.04$.    {\footnote{    {
The smallness can be understood as a consequence of the fact the corrections to Gaussian approximation for $\eta$ appear only at two-loop order. In terms of $\epsilon=4-d$ expansion: $\eta=\mathcal O(\epsilon^2)$. There are other power laws, and corresponding exponents, which differ from the mean-field ($\epsilon=0$) values more significantly, i.e., at one-loop order. Most notably, the correlation length diverges as $\xi\sim |T-T_c|^{-\nu}$, where $\nu=1/2$ in mean-field approximation, but, in fact, is closer to $2/3$ in $d=3$ dimensions. {This means that the specific heat exponent $\alpha$ in $C_V\sim|T-T_c|^{-\alpha}$ is close to zero  $\alpha=2-d\nu\approx0.11$ in $d=3$ (note that it is exactly zero in $d=2$ and $d=4$),  i.e., the specific heat is only weakly divergent.}
}}}
These corrections become visible only for very large
correlation lengths, and for finite non-static systems such as
heavy-ion collisions are negligible compared to other more important effects.

\subsection{Higher-order moments and cumulants of the fluctuations}
\label{sec:high-moments-cumul}

\begin{figure}
  \centering
  \includegraphics[height=6em]{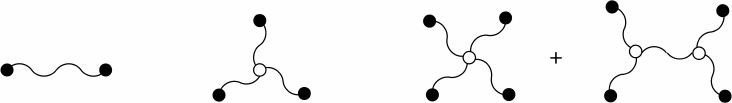}
  \caption{Tree diagrams corresponding to Eq. \eqref{eq:sigma-moments}.}
  \label{fig:tree-graphs}
\end{figure}

 The divergence of the correlation length at the critical point has
even stronger effect on the measures of the ``shape'' of the probability distribution
$\mathcal P[\sigma]$, as pointed out in Ref.~\cite{Stephanov:2008qz}. As the critical point is approached the
probability distribution becomes less Gaussian, i.e., flatter at the
top if the symmetry is maintained as in the case $h=0$ (see
Fig.~\ref{fig:3P}) and, in general, less symmetric around a maximum
for a given $h\neq 0$. These non-Gaussian properties can be quantified
using higher-order cumulants of fluctuations. Extending the expansion
of the potential $U(\sigma)$ to quartic order in $\sigma$
\begin{equation}
  \label{eq:U34}
U(\sigma_0+\delta\sigma) = U(\sigma_0) +  \frac{m_\sigma^2}2 \delta\sigma^2 
+ \frac{\lambda_3}{3}\delta\sigma^3
+ \frac{\lambda_4}{4}\delta\sigma^4 + \ldots
\,,
\end{equation}
and performing the path integrals in the saddle point approximation one finds
\begin{equation}
  \label{eq:sigma-moments}
  \begin{split}
  &  \kappa_2=\la \sigma_V^2 \ra = V T\,\xi^2\,;
\\
  &
%\qquad
\kappa_3=\la \sigma_V^3 \ra = {2 \lambda_3 VT^2}\, \xi^6\,;
\\
  &\kappa_4=\la \sigma_V^4 \ra_c \equiv \la \sigma_V^4 \ra - 3 \la \sigma_V^2 \ra^2
= {6VT^3}\, [\,2(\lambda_3\xis)^{2} - \lambda_4\,]\, \xi^8\,,   
  \end{split}
\end{equation}
where in the last line the subscript `$c$' means `cumulant' or `connected'.
Diagrammatically this calculation is represented by tree graphs in
Fig.~\ref{fig:tree-graphs}, where each propagator of the $\sigma$
field (wavy line) contributes a factor $\xi^2$.

Going beyond the tree-level (saddle-point) approximation, which is
necessary near the critical point due to the importance of
fluctuations, requires resummation of infinitely many graphs of all
loop orders, which is what the renormalization group achieves. The result
is that the couplings ``run'', i.e., the dependence of the
non-Gaussian cumulants $\kappa_3$ and $\kappa_4$ on the correlation
length $\xi$ appears not only explicitly, but also via the
coefficients $\lambda_3$ and $\lambda_4$ which, similarly to
$m_\sigma^2$, depend on $\xi$.  Near the critical point this
dependence follows from the invariance of the Landau-Ginzburg free
energy $\Omega$ under rescaling $x\to x/b$, $\xi\to \xi/b$,
$\sigma\to b^{1/2}\sigma$.  One can check that such rescaling leaves
the gradient term in Eq.~(\ref{eq:Omega-sigma})
invariant.\footnote{Fluctuations slightly modify the power $1/2$ into
  $1/2-\eta/4$, but for our purposes it is sufficient to approximate
  $\eta\approx 0$ as we already noted at the end of previous
  subsection.} The invariance of the term $\int d^3\bm x U(\sigma)$ 
requires $m_\sigma^2\sim\xi^{-2}$, $\lambda_3\sim \xi^{-3/2}$ and
$\lambda_4\sim\xi^{-1}$. The scaling invariance is one of the most
important properties of the critical phenomena and is a consequence of
the fact that in this regime the correlation length $\xi$ is much
larger than any other microscopic scale, and thus becomes the only
relevant length scale in the system. Therefore
\begin{align}
  \label{eq:kappa3-sigma}
&  \kappa_3 = 2\tilde\lambda_3 V T^{3/2}\xi^{9/2}\,,\\
\label{eq:kappa4-sigma}
&  \kappa_4 = 6(2\tilde\lambda_3^2-\tilde\lambda_4) V T^{2}\xi^{7}\,,
\end{align}
where we introduced dimensionless couplings $\tilde\lambda_3$ and
$\tilde\lambda_4$ via $\lambda_3 = \tilde \lambda_3 T (T\xi)^{-3/2}$
and $\lambda_4=\tilde\lambda_4 (T\xi)^{-1}$. These rescaled couplings are scale-invariant
and universal.\footnote{Up to very slow ``running''
due to small $\eta$ which, as we already noted, we can neglect.}
Eqs.~\eqref{eq:kappa3-sigma} and~\eqref{eq:kappa4-sigma} demonstrate that the dependence on the correlation length is much stronger in higher-order cumulants, which measure the shape of $\mathcal P[\sigma]$, compared to the second order cumulant in Eq.~\eqref{eq:kappa2-xi}, which measures the width of $\mathcal P[\sigma]$. Furthermore, the higher-order cumulants in Eqs.~\eqref{eq:kappa3-sigma} and~\eqref{eq:kappa4-sigma}  may change sign, depending on $T-T_c$ and $h$ \cite{Stephanov:2011pb}.

    {This is a good place to emphasize that the results such as the ones shown in Eqs.~\eqref{eq:kappa2-xi},~\eqref{eq:kappa3-sigma} and~\eqref{eq:kappa4-sigma} are {\em universal} in the sense that the scaling with certain powers of $\xi$ is independent of the microscopic details of the theory. This universality is the key to our ability to search for the QCD critical point even despite having only incomplete knowledge of the QCD equation of state and the microscopic dynamics. 
}

\subsection{The cumulants and the equation of state in parametric representation}
\label{sec:param-repr-equat}

Cumulants of fluctuations can be determined directly from the equation
of state, i.e., the free energy $F(M)$ in Eq.~(\ref{eq:PF}). Indeed,
identifying
\begin{equation}
  \label{eq:M=sigmaV}
 \sigma_V = N\delta M \,,
\end{equation}
we can write, using Eq.~(\ref{eq:F''}),
\begin{equation}
  \label{eq:kappa_2=chi}
  \kappa_2 = \langle\sigma_V^2\rangle = NT\left(\frac{\partial
      \langle M\rangle}{\partial h}\right)_T \,.
\end{equation}
In order to generalize this to higher-order cumulants we will introduce the
Gibbs free energy as a Legendre transform of $F(M)$ with respect to $M$:
\begin{equation}
  \label{eq:GvsM}
  G(h) \equiv \min_{M}(F(M)-hM)
\end{equation}
Since $\langle M\rangle = -(\partial G/\partial h)_T$ we can rewrite 
Eq.~(\ref{eq:kappa_2=chi}) as $\kappa_2 = -NT(\partial^2 G/\partial
h^2)_T$. This relation can be easily generalized to cumulants of all orders:
\begin{equation}
  \label{eq:kappa-n-eos}
\kappa_{k}  = -NT \left(\frac{\partial^k G}{\partial h^k}\right)_T 
= NT \left(\frac{\partial^{k-1} \langle M\rangle}{\partial h^{k-1}}\right)_T .
\end{equation}

We can thus directly determine the dependence of $\kappa_n$'s on $T$
and $h$ from the equation of
state $\langle M\rangle (h)$, or from $G(h)$,  without knowing the values of $\xi$ and $\lambda_n$'s 
at given $T$ and $h$.

%\subsection{Parametric representation of the equation of state}

Even for the mean-field free energy in Eq.~(\ref{eq:LG}) the explicit
expression for $G(h)$ of Eq.~(\ref{eq:GvsM}) or  $\langle M\rangle (h)$ looks
cumbersome.  A tangible expression of the
equation of state in this case can be written in the form
$h=h(\langle M\rangle)$ where:
\begin{equation}
  \label{eq:hmeos}
  h(M) \equiv \frac{\partial F(M)}{\partial M}
  = a(T) M + b(T) M^3\quad\mbox{(mean-field equation of state)}
\end{equation}
This equation of state is singular at $T=T_c$ and $h=0$, since the
susceptibility $(\partial M/\partial h)_T = 1/a$ diverges because
$a(T_c)=0$. The dominant part of this singularity can be described by
approximating $a(T)$ and $b(T)$ by the leading term of their Taylor
expansion around $T=T_c$:
\begin{equation}
  \label{eq:hmr}
  h = a' r M + b M^3\quad\mbox{(mean-field equation of state)}
\end{equation}
where $a'=da/dT|_{T=T_c}$ and $b=b(T_c)$ and we introduced a ``reduced
temperature'' $r=T-T_c$. The most notable feature of
the equation of state is invariance under rescaling
\begin{equation}
  \label{eq:scalingMrh}
  r\to\lambda r,\quad M\to \lambda^\beta M,\quad h\to\lambda^{\beta\delta}h,
\end{equation}
with specific values of $\beta=1/2$ and $\delta=3$, known as
mean-field critical exponents. The scaling is observed near Curie
critical points in ferromagnets and near liquid-gas critical
points. It is characterized by the universal values of $\beta$ and
$\delta$ for all systems in the Ising model universality class, which
includes uniaxial ferromagnets, liquid-gas and binary fluid
transitions.\footnote{This is a ubiquitous critical universality
  class because it occurs in systems with a one-component (singlet)
  order parameter. For example, the Heisenberg ferromagnet
  universality class requires an exact O$(3)$ symmetry and order
  parameter which is a triplet (a vector).} However, the
universal values of the critical exponents observed
($\beta\approx 1/3$ and $\delta\approx 5$) deviate significantly from
the mean-field values. The role of fluctuations which are responsible
for these deviations from the mean-field values can be understood
within the renormalization group approach to critical phenomena which
is a subject of many classic textbooks (see, e.g.,
\cite{Amit:1984ms,ZinnJustin:2002ru}).

Thus the mean-field equation of state, while providing a simple and
intuitive description of the phase transition, is not adequate for
describing the leading singular behavior of the critical equation of
state {\em quantitatively}. A quantitative description can be
achieved by using the parametric representation
\cite{Schofield:1969zza,ZinnJustin:2002ru}, where, instead of using the variables $h$,
$r$ and $M$ directly, as in Eq.~(\ref{eq:hmeos}), one introduces two
auxiliary variables $R$ and $\theta$, roughly parameterizing the
``distance'' from the critical point and the ``angle'' relative to the
crossover direction (i.e., $h=0$ for $T>T_c$) respectively.
% The equation
% of state is, of course, singular at $T=T_c$ and $h=0$, but the form of
% the singularity is controlled by scaling exponents such as $\beta$ and
% $\delta$ which are known numerically and a universal scaling function
% which can be approximated by a polynomial.
 In terms of $R$ and $\theta$ the variables of the Ising model $M$, $h$ and
$r$ %  \begin{equation}\label{eq:r=T-Tc}
% r\equiv(T-T_c)/T_c
% \end{equation}
are given by
\begin{figure}
  \centering
  \includegraphics[height=20em]{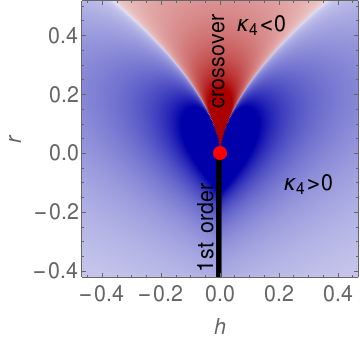}
  \caption{Quartic cumulant $\kappa_4$ of the Ising model magnetization near the critical point~\cite{Stephanov:2011pb}. The cumulant is negative (red) in the sector around the crossover bounded by $ht^{-\beta\delta}=\pm{\rm const}$ lines (white). These lines correspond to $\theta\approx\pm0.32$. }
  \label{fig:k4}
\end{figure}

\begin{equation}
  \label{eq:Rtheta}
M= R^\beta\theta,\qquad r = R(1-\theta^2), \quad h = R^{\beta\delta} H(\theta).
\end{equation}
The correct scaling Eq.~(\ref{eq:hmr}) is built into this
representation and corresponds to $R\to\lambda R$ with $\theta$ being
scale invariant.

 All the
information about the equation of state $M(r,h)$ is in the values of critical
exponents $\beta$ and $\delta$ and the universal scaling function $H(\theta)$
which can be calculated order by order in the expansion around
dimension $d=4$, i.e., $\varepsilon$ expansion where $\varepsilon = 4-d$. To
an approximation sufficient for our purposes % (and correct to order
% $\mathcal O (\varepsilon^2)$)
it is given by
\begin{equation}
  \label{eq:h(theta)}
  H(\theta) = \theta(3-2\theta^2).
\end{equation}

Note that when scaling exponents are assigned their mean-field values
$\beta=1/2$ and $\delta =3$ the mean-field equation of state
Eq.~(\ref{eq:hmeos}) emerges from the parametric representation given
by Eqs.~(\ref{eq:Rtheta}) and~(\ref{eq:h(theta)}), with parameters
$a'=3$ and $b=1$. At $d=4$ (i.e., $\varepsilon=0$) the exponents
$\beta$ and $\delta$ are given by their mean-field values. At order $\varepsilon$ the
exponents differ from mean-field values~\footnote{More precisely,
  $\beta=1/2-\varepsilon/6 + \mathcal O(\varepsilon^2)$ while
  $\beta\delta = 3/2 + \mathcal O(\varepsilon^2)$.}, while Eq.~(\ref{eq:h(theta)})
is correct to order $\varepsilon^2$ \cite{ZinnJustin:2002ru}. %  In three dimensions, for our
% purposes, it will be sufficient to approximate the exact values of
% exponents as $\beta\approx 1/3$ and $\delta \approx 5$.

One can see from Eq.~(\ref{eq:Rtheta}) that parameter $R$ is the
measure of the distance from the critical point $r=h=0$. The
correlation length diverges as $\xi\sim R^{-\nu}$ when
$R\to0$. The mean-field value of the exponent $\nu$ is $1/2$,
as follows from Eq.~(\ref{eq:corr-length}) and $m_\sigma^2\sim
r$. However, the actual value of the exponent $\nu$ (in $d=3$) is
significantly different: $\nu\approx 2/3$. One can check that this
value (together with $\beta\approx 1/3$ and $\delta\approx 5$) is
consistent, via Eq.~(\ref{eq:kappa-n-eos}), with the scaling of the cumulants in
Eqs.~(\ref{eq:kappa3-sigma}) and ~(\ref{eq:kappa4-sigma}).

The parameter $\theta$ is a measure of the direction in the
$r$ vs $h$ plane (or more precisely, $r^{\beta\delta}$ vs $h$), with $\theta=0$ along the crossover line, $\theta=1$
along $r=0$, and
$\theta=\sqrt{3/2}$ along the coexistence (first-order transition)
line.
The sign of the cumulants depends on the value of the scale-invariant
parameter $\theta$.
As an example, the corresponding dependence of $\kappa_4$ on $r$ and $h$ is shown in
Fig.~\ref{fig:k4}. The curves along which $\kappa_4$ changes sign
correspond to $\theta\approx \pm0.32$.
The negative sign of $\kappa_4$ along the crossover line is the
reflection of the fact that the distribution is ``flatter'' than
Gaussian when one approaches the critical point, as seen in Fig.~\ref{fig:3P}.

\subsection{Mapping to QCD equation of state}
\label{sec:4:map_eos}

The QCD equation of state with a critical point
is an essential ingredient for hydrodynamic simulations of the
heavy-ion collision fireball evolution at collision energies relevant
to the beam-energy scan program. 
Since first-principle lattice simulations, hindered by the sign problem, cannot yet produce the needed equation of state at nonzero $\mu_B$, the following approach has been proposed in Ref.~\cite{Parotto:2018pwx}. The available information from lattice (at $\mu_B=0$) can be combined with the known universal behavior of the equation of state (pressure vs $T$ and $\mu_B$) near the critical point. Of course, this information is not sufficient to completely fix the equation of state. However, it allows to construct a family of equations of state which depends on a relatively small number of parameters (e.g., position of the critical point) while satisfying known constrains. 
Comparisons of hydrodynamic simulations with such a flexible parametric equation of state and
experimental data could help detect and identify the
signatures of the QCD critical point.

\begin{figure}
    \centering
    \includegraphics[height=16em]{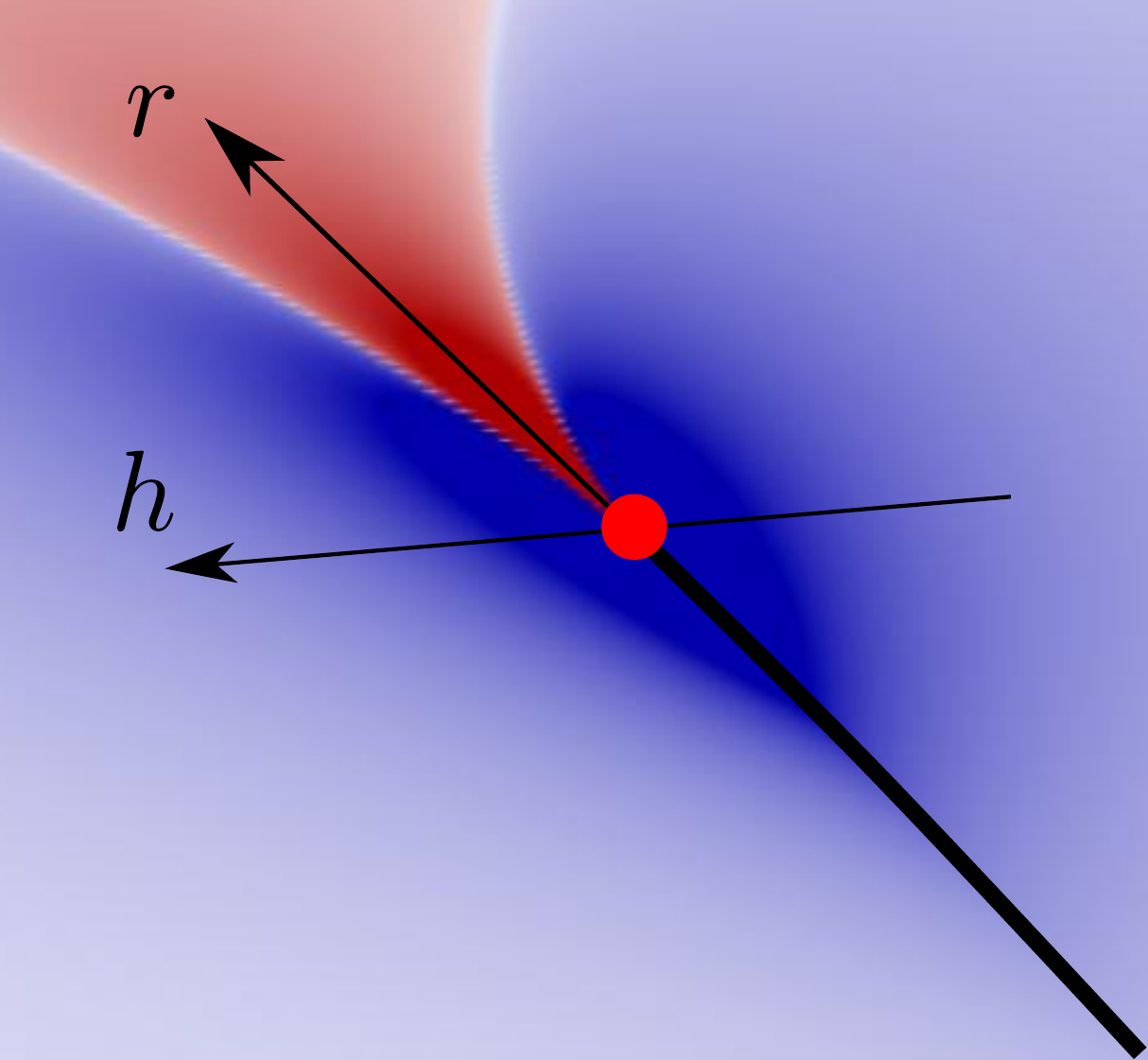}
    \caption{An illustration of the mapping of the Ising model equation of state onto QCD phase diagram given by Eq. \eqref{eq:rh-Tmu}. The quartic cumulant of an order parameter (e.g., baryon density) is plotted for comparison with Fig.~\ref{fig:k4}.}
    \label{fig:rh-map}
\end{figure}

In order to use the universality of the critical phenomena to describe
the leading singularity of the equation of state at the critical point
we need to map the coordinates $T$ and $\mu_B$ on the QCD phase
diagram to the coordinates $r$ and $h$ on the phase diagram of the Ising
model (see Fig.~\ref{fig:rh-map}).~\footnote{ In this section $T$ denotes the temperature  in QCD (or,
  more generally, the temperature in a liquid-gas system under
  consideration), while $r$ is the reduced temperature of the Ising
  model $r=T_{\rm Ising}-T_{c,\rm Ising}$, as before. } The mapping is linear, i.e.,
\begin{equation}
  \label{eq:rh-Tmu}
  r(T,\mu)= r_T (T-T_c) + r_\mu (\mu-\mu_c); \quad
h(T,\mu) = h_T (T-T_c) + h_\mu (\mu-\mu_c). 
\end{equation}
Nonlinearities in the mapping have no effect on the {\em leading} singularity of the equation
of state.
The coefficients $r_T$, etc., determine  directions of
the $r$ and $h$ axes in the $T$ vs $\mu$ plane shown in Fig.~\ref{fig:rh-map}. In particular,
the slope of the coexistence line  at the critical point
is given by setting $h=0$ in Eq.~(\ref{eq:rh-Tmu}):
$(dT/d\mu)_{h=0}=-h_\mu/h_T$. The slope of the $h$-axis ($r=0$) is
non-universal and corresponds to the leading asymmetric correction to
scaling \cite{Rehr:1973zz,Nicoll:1981zz}. Taking the mixing in Eq.~(\ref{eq:rh-Tmu}) into
account one can write the QCD pressure as:
\begin{equation}
  \label{eq:p-rev-scaling}
  p(T,\mu) =  -G (r(T,\mu),h(T,\mu)) + p_{bg}(T,\mu)\,,
\end{equation}
where $ G(r,h)$ is the leading singular contribution to Gibbs free
energy of the Ising model.     {The relation between the QCD pressure (or the pressure in any system in the vicinity of the critical point in the same universality class -- that of the Ising model) in Eq.~\eqref{eq:p-rev-scaling} is the consequence of the universality of critical phenomena which follows from renormalization group theory. This relation allows us to make certain predictions about the behavior of QCD near the critical point despite having only incomplete knowledge about the QCD equation of state, thus enabling an experimental search for this phenomenon.}

As we discussed in the previous section an
explicit expression for $G$ is cumbersome, but for all practical
purposes it can be found from Eq.~(\ref{eq:GvsM}) with $F$ given by
integrating $h=(\partial F/\partial M)_r$ 
using the equation of state $h(M)$ given by the parametric
representation Eqs.~(\ref{eq:Rtheta}) and (\ref{eq:h(theta)}).
The remaining part of the pressure, $p_{bg}(T,\mu)$, is the `background'
which contains less singular (corrections to
scaling) and non-singular contributions.

The parameters $r_T$, $r_\mu$, $h_T$ and $h_\mu$  are nonuniversal. If the information about
the equation of state is available, e.g., from a lattice calculation
or beam energy scan experiment, one can attempt to determine these
parameters by fitting the parametric equation of state such as given
in Eq.~(\ref{eq:p-rev-scaling}). The parameters which need to
be fitted are the location $T_c$ and $\mu_c$ of the critical point,
the slope $-h_\mu/h_T$ of the coexistence line ($h=0$),
and the slope $-r_\mu/r_T$ of the $r=0$ line at the critical
point. To set the scale, one needs to supply also $r_T$ and
$h_\mu$. The overall scale of the singular part is not an additional independent parameter
because of the
scaling property of the leading singularity of the equation of state:
\begin{equation}
  \label{eq:G-scale}
   G(\lambda r, \lambda^{\beta\delta} h) = \lambda^{\beta(\delta+1)}  G(r,h)
\end{equation}
for an arbitrary $\lambda$.

The background part of the pressure, $p_{bg}(T,\mu)$, can be chosen to smoothly match the
equation of state at $\mu=0$ known from the lattice \cite{Parotto:2018pwx}. In
this form the equation of state can incorporate the information
reliably known from lattice QCD calculations as well as the correct
leading singular behavior at the critical point, while being flexible
enough to accommodate a critical point in the range of $T$ and $\mu$
accessible by heavy-ion collisions.

% The $\mu\to-\mu$
% charge conjugation symmetry can be implemented by replacing $p(T,\mu)$
% with $p(T,\mu)+p(T,-\mu)$

\begin{figure}[ht]
\begin{tabular}{llll}
(a) $  \chi_1 = n $  & (b) $  \chi_2 = {\partial n}/{\partial\mu} $ 
& (c) $ \chi_3 = {\partial \chi_2}/{\partial\mu}  $ 
& (d)  $ \chi_4 = {\partial \chi_3}/{\partial\mu}  $  \\ 
  \includegraphics[width=10em]{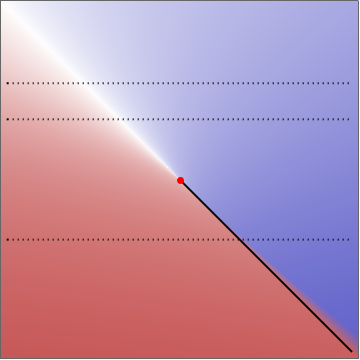}&
  \includegraphics[width=10em]{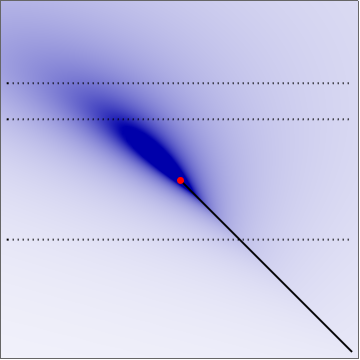}&
  \includegraphics[width=10em]{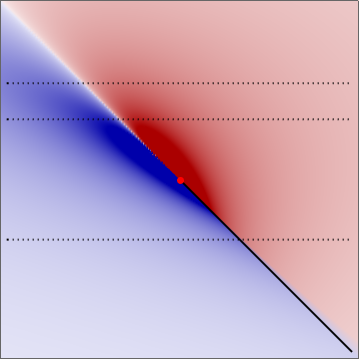}&
  \includegraphics[width=10em]{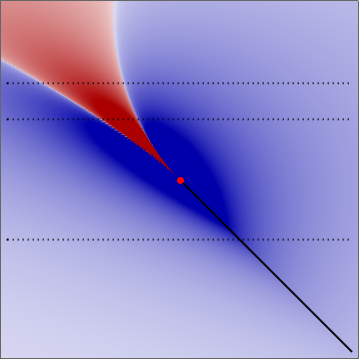}\\
  \includegraphics[width=10em]{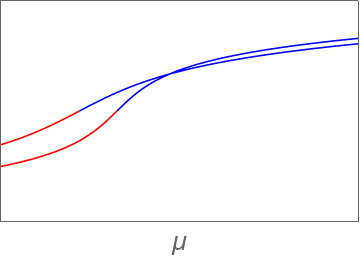}&
  \includegraphics[width=10em]{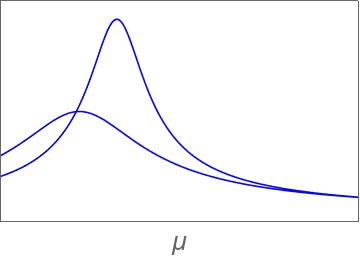}&
  \includegraphics[width=10em]{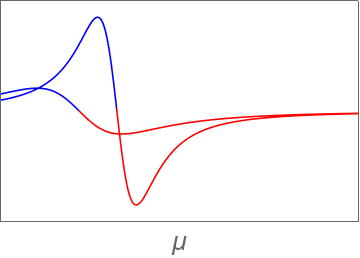}&
  \includegraphics[width=10em]{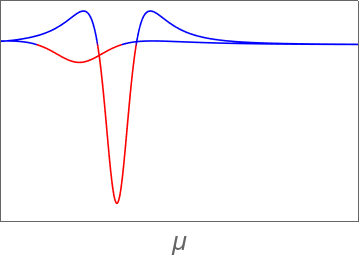}\\
  \includegraphics[width=10em]{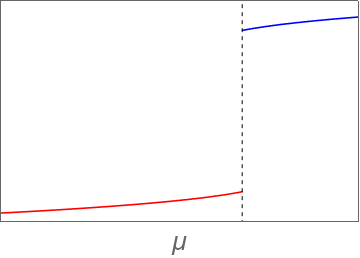}&
  \includegraphics[width=10em]{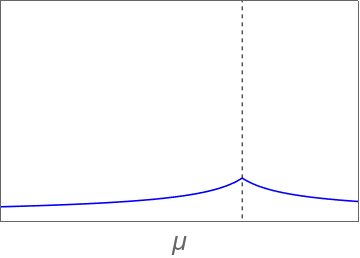}&
  \includegraphics[width=10em]{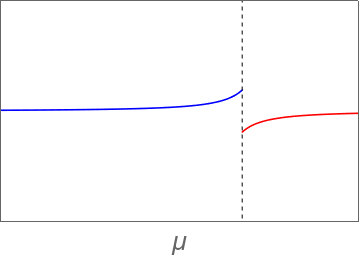}&
  \includegraphics[width=10em]{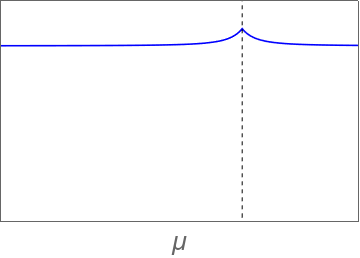}
\end{tabular}
\caption{First three derivatives (susceptibilities) of the baryon
  density $n$ with respect to $\mu$ as a function of $\mu$ along three
  constant $T$ lines (horizontal dotted lines in the first row of
  plots -- the density plots of the susceptibilities vs $T$ and
  $\mu$).  Two temperatures are above (second row) and another
  temperature is below (third row) the critical point. Red denotes
  region of negative and blue -- of positive value of the
  susceptibilities. Only the critical contribution to $n$ and its
  derivatives dictated by the universality near the critical point is
  shown. The vertical range for two graphs of the same quantity
  $\chi_n$ (i.e., in the same {\em column}) are the same, but is
  different across the columns.}
\label{fig:t-scans}
\end{figure}

% \subsection{Signatures of critical behavior in experimentally observable fluctuations and correlations}
% \label{sec:signatures-exp-observable}

% Heavy ion collisions do not directly measure the values of the order
% parameter field, unlike the magnetization in a ferromagnet, or the values
% of thermodynamic quantities, such as the specific heat, which allows to
% determine the location of the critical points in condensed matter
% systems. However, as we just discussed, the cumulants of the net-baryon number distribution, 
% which have a well defined singular behavior at the critical point are accessible to experiment. 

% In addition to being a signature for the critical point the baryon number cumulants are also
% sensitive to the cross-over transition. Why this is is the case we will illustrate next 

\subsection{Baryon number cumulants near the critical point}
\label{sec4:baryon:cumulants}

In order to understand better the equation of state near the QCD
critical point described by Eq.~(\ref{eq:p-rev-scaling}) it is helpful
to study the behavior of baryon number cumulants. Due to the
relationship between the pressure and the partition function of QCD
\begin{align}
  e^{V p(T,\mu)/T} =Z &= \sum_{\textrm{states}\,\, i}
                        \left \langle i \left | e^{-\frac1T\left(
  \hat{H} + \mu \hat{N}\right)} \right| i \right\rangle\,,
\end{align}
the cumulants of the fluctuations of the baryon charge $N$ are
related to the derivatives of the pressure:\footnote{In the context of lattice calculations the susceptibilities are often defined as dimensionless quantities, i.e., $\chi_k^{\rm lattice} = \partial^k (p/T^4)/ \partial (\mu/T)^k$.}
\begin{equation}
  \label{eq:ncp}
  \chi_k  =  
\left( \frac{\partial^k  p}{\partial\mu^k} \right)_T
  =  \left( \frac{\partial^{k-1}     n }{\partial\mu^{k-1}} \right)_T
  = \frac1{V T^{k-1}}\,\langle (\delta N)^k\rangle_c
  ,
\end{equation}
where
\begin{equation}
n\equiv \frac{\langle N\rangle}V 
= \left(\frac{\partial p}{\partial \mu}\right)_T.\label{eq:n=N/V}
\end{equation}

Fig.~\ref{fig:t-scans} shows a close-up of the QCD phase diagram in
Fig. \ref{fig:pd} near the critical point. Using the universal equation of state
given by the mapping in Eqs.~(\ref{eq:p-rev-scaling}) and
(\ref{eq:rh-Tmu}) (where we choose $r_T=0$ for simplicity), we
illustrate the behavior of susceptibilities $\chi_k$. It is instructive
to follow $\chi_k$ along lines of fixed $T$. Three such lines are
shown in Fig.~\ref{fig:t-scans} (top row): two isothermal lines
traverse the crossover region above the critical point and the behavior
of $\chi_k$ along these lines is shown in the second row and one
isothermal line traverses the first-order coexistence line with the
corresponding $\chi_k$ shown in the third row.

% Now, suppose we move across the phase-diagram, Fig.\ref{fig:pd}, along a horizontal
% line, i.e. at constant temperature from small to large chemical potential. Then as we cross the
% pseudo-critical line the baryon density will rapidly increase. This is illustrated in more detail in
% Fig.~\ref{fig:t-scans} where we show the region around the critical point. The top panel in the left columns (a) shows the
% density as both a contour plot in the $T-\mu$ plane. Also shown in black is the first order phase co-existence line
% which ends in the critical point (in red) and continues as a crossover region indicated by the white
% area. The three dotted lines are iso-therms which either traverse the cross over region (the top
% two) or cross  the first order co-existence line (bottom). In the panels below we show how the
% density changes along these lines.
As we traverse the crossover region  (panel (a) in Fig.~\ref{fig:t-scans}) the density increases continuously
with a steeper slope for the case where the isothermal line is closer
to the critical point. When we cross the first order line, the baryon
density, $n$, jumps, as expected.  The baryon number cumulants, or
susceptibilities, $\chi_k$, being derivatives of the density (see
Eq.~\eqref{eq:ncp}), will be sensitive to the proximity of the
critical point in the crossover region as the change of the density $n$ becomes
steeper.  This is illustrated in the panels (b) through (d) in
Fig. ~\ref{fig:t-scans}, where we show the second to fourth order
susceptibilities. We see that, not surprisingly, the steeper
increase in the density when traversing the pseudo-critical region
closer to the critical point is reflected in larger values of the
cumulants. This difference gets more pronounced the higher the
order of the susceptibility or cumulant. Furthermore, the sign changes
of the various cumulants shown in the contour plots can be easily
understood as simply changes in the slope (for $\chi_2$), curvature (for $\chi_{3}$) and higher
derivatives of the density $n$ in the first column. Finally, when crossing the
first-order line (third row) we find that away from the critical line
the cumulants are only modestly changed. On the critical line, of
course, they are undefined due to a discontinuity.~\footnote{The
  absence of visible discontinuity in even cumulants in Fig.~\ref{fig:t-scans}
is a consequence of our simplification $r_T=0$.}

This simple example qualitatively explains what happens near the
critical point as discussed in section~\ref{sec:high-moments-cumul}: The
higher the order of the cumulant the stronger is its dependence on the
correlation length. As we get closer to the critical point, where
correlation length diverges, the transition gets sharper and the
cumulants also diverge at the critical point.

As we have seen, the high-order cumulants show nontrivial dependence on $T$ and $\mu$ in the crossover region. This observation suggests that  the measurement of net-baryon cumulants may also provide an avenue to establish the existence of a cross-over
transition at $\mu_{B}=0$, as predicted by lattice QCD \cite{Aoki:2006we}. As discussed in
\cite{Schmidt:2010xm,Cheng:2008zh,Friman:2011pf,Borsanyi:2018grb} in the context of model as well as lattice QCD
calculations, a cross-over transition results in negative sixth and eighth order cumulants at the
freezeout temperature,
$\cum{6}/\cum{2}<0$ and $\cum{8}/\cum{2}<0$. Therefore, the measurement of these cumulant ratios could
provide experimental evidence that the systems created in high energy heavy ion
collisions freeze out close to the cross-over transition.

\subsection{Fluctuation cumulants in heavy-ion collisions}

The baryon number cumulants, or susceptibilities, are not directly
measurable in heavy-ion collision experiments which detect charge
particles, leaving neutrons out of the acceptance. However, the
fluctuations near the critical point affect fluctuations of charged
particles as well as the neutral ones because the coupling of the
critical mode is isospin blind. Thus cumulants of the fluctuations of
proton number (or net proton number) show a similar pattern near the
critical point. In Section~\ref{sec:crit-pt-correl-momentum} we shall describe how to relate the
critical mode fluctuations with the observable fluctuations of the
particle multiplicities.

\begin{figure}[t]
  \centering 
  \includegraphics[height=16em]{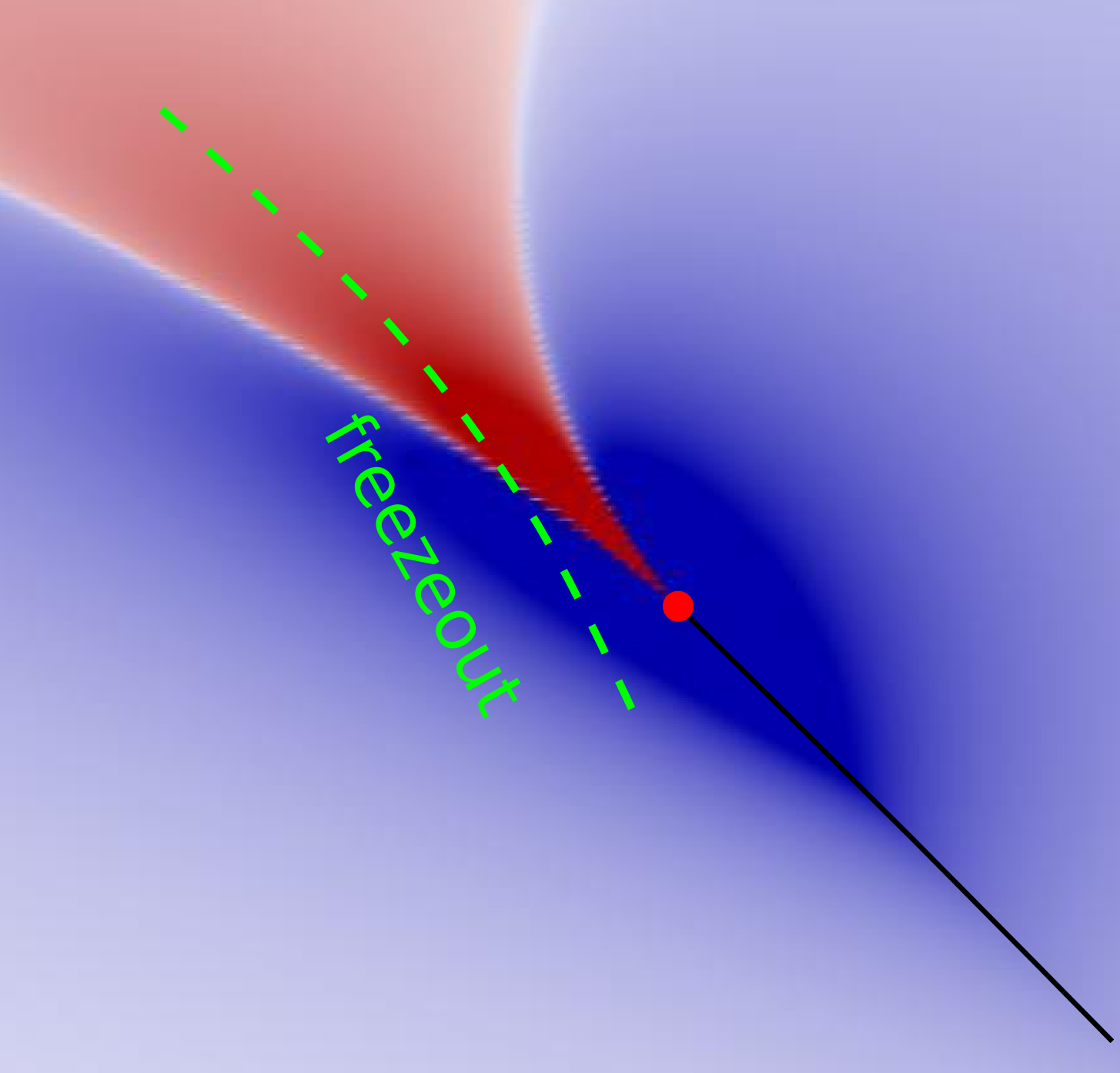}
\caption{Density plot of the quartic cumulant
  of the order parameter obtained by mapping of the Ising equation of
  state onto the QCD equation of state near the critical point. The
  freezeout point moves along the dashed green line as $\sqrt{s_{NN}}$ is
  varied during the beam energy scan.}
\label{fig:scenario-color4}
\end{figure}

The experiments also do not scan the phase diagram along fixed $T$
lines as in Fig.~\ref{fig:t-scans}. The scanning parameter, such as
$\sqrt{s_{NN}}$, affects both $T$ and $\mu$ of the freezeout. A typical
freezeout trajectory along which $T$ and $\mu$ are varied is shown in
Fig.~\ref{fig:scenario-color4} superimposed on the density plot of the
quartic cumulant of a critical order parameter, such as, e.g., baryon
density.  The position of the freezeout point on the curve depends on
the  collision energy $\sqrt{s_{NN}}$ and can be determined
experimentally using the chemical freezeout systematics extracted from
the measured particle yields~\cite{Andronic:2005yp}. The systematics provides an
estimate of $T$ and $\mu$ of the system at freezeout at given
$\sqrt{s_{NN}}$.

\begin{figure}
  \centering
  \includegraphics[height=16em]{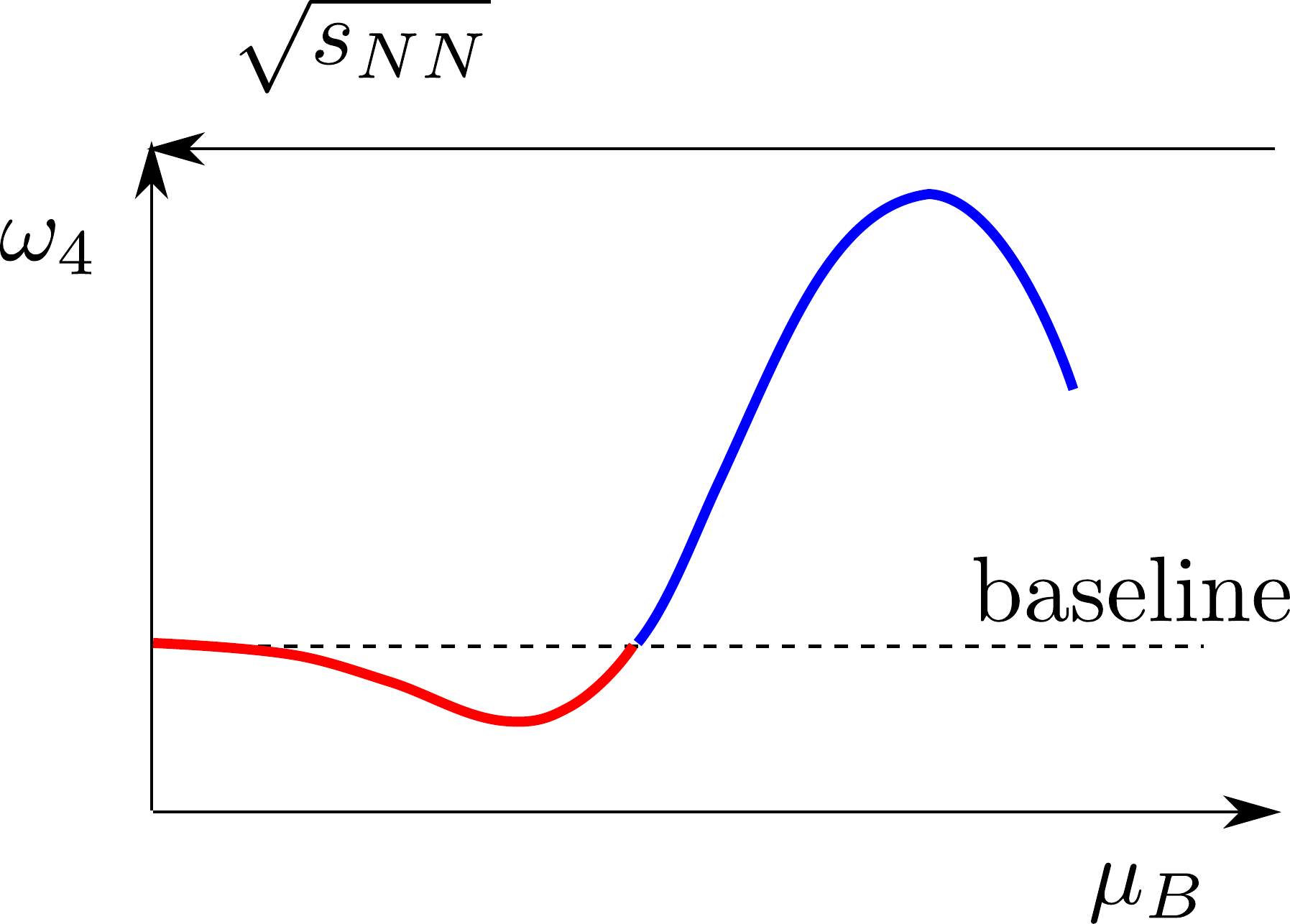}
  \caption{Normalized quartic cumulant of proton multiplicity
    $\omega_4=\kappa_4[N]/N$ as a function of collision energy 
    $\sqrt{s_{NN}}$ or, equivalently, the chemical potential $\mu_B$ along the
    freezeout curve on Fig.~\ref{fig:scenario-color4}~\cite{Stephanov:2011pb,Stephanov:2011zz}.}
  \label{fig:omega-sqrts-mu}
\end{figure}

As we shall see below in Section~\ref{sec:crit-pt-correl-momentum} the critical
contribution to the cumulants of the fluctuations of the observed
particle multiplicities are proportional to the cumulants of the
fluctuations of the critical order parameter~\cite{Stephanov:1999zu,Stephanov:2008qz,Stephanov:2011pb,Athanasiou:2010kw}. Thus the deviation of
the particle multiplicity cumulant from its noncritical baseline will
depend on collision energy (or on $\mu_B$ along the freezeout curve)
as shown qualitatively  in Figure~\ref{fig:omega-sqrts-mu} .

\subsection{Finite time dynamics and limitations on the correlation length in heavy-ion collisions}
\label{sec:limitations-on-xi}

Phase transitions and associated thermodynamic singularities are
properties of {\em static} systems in the limit of {\rm infinite} volume. In a
finite system the phase transition singularities are smeared out. The
maximum correlation length achievable on a finite system is limited by
the system size. In the case of heavy-ion collision this limit is on
the order of 10~fm.

However, a more stringent constraint comes from the fact that the
system does not spend enough {\em time} in the critical region for the
correlation length to build up to its equilibrium value~\cite{Stephanov:1999zu,Berdnikov:1999ph}. The amount
of time that it takes to equilibrate diverges near the critical point
\begin{equation}
  \label{eq:tau-xi}
  \tau \sim \xi^z\,,
\end{equation}
where $z\approx 3$ is the dynamic critical exponent \cite{Son:2004iv}.
Thus, for a system evolving through the critical region with a
characteristic time scale $\tau$ the maximal achievable correlation
length will be proportional to $\tau^{1/z}$~\cite{Stephanov:1999zu,Berdnikov:1999ph}. The
typical evolution time is on the order of $7-10$ fm. Estimating the
coefficient of proportionality requires dynamical modeling. The
simplest models of this type  predict the
maximum correlation length on the order of $2-3$ fm~\cite{Berdnikov:1999ph} (see also Ref.~\cite{Akamatsu:2018vjr} for a more recent discussion of the physics involved in Eq.\eqref{eq:tau-xi}).

    {We note that the maximum correlation length due to critical slowing down limits the applicability of finite size scaling arguments \cite{Palhares:2009tf,Fraga:2011hi,Lacey:2014wqa}.
The finite-size scaling describes the behavior of systems so close to criticality that $\xi$ is as large as the system itself. Such arguments do not apply to systems created in heavy-ion collisions because finite-time dynamics limits $\xi$ to be several times smaller ($2-3$ fm) than the size of the system ($7-10$ fm). }

A more comprehensive description of the effect of space-time evolution
on the fluctuation signatures of the critical point requires embedding
the fluctuations into hydrodynamic evolution code and is work in
progress. % This includes introducing thermal fluctuations into
% relativistic hydrodynamics \cite{}. 
The physics of the critical
slowing down and of critical fluctuations are essentially the same
physics, described by the evolution of equal-time correlators of the slowest
mode (entropy per baryon) towards their equilibrium value. Such an
extension of hydrodynamics (called ``Hydro+'') incorporating additional critically
slow degrees of freedom (correlators of the slow mode) has been introduced in
Ref.~\cite{Stephanov:2017ghc}.     { The first application of this approach to heavy-ion collisions, albeit in a very simplified setup, has been recently reported in Ref.~\cite{Rajagopal:2019xwg}.} Generalization of this approach to non-Gaussian cumulants is still a challenge, but is expected to describe interesting ``memory" effects, as in, e.g., Ref.\cite{Asakawa:2009aj,Mukherjee:2015swa,Mukherjee:2016kyu}.

In order to describe the effects of the critical fluctuations on the
observable particle fluctuations and correlations, below we shall
consider a simplified model scenario~\cite{Stephanov:1999zu,Stephanov:2011pb,Ling:2015yau}, where we parameterize the
effects of the evolution near the critical point by the value of the
correlation length $\xi$, which determines the cumulants of the
fluctuations via scaling relations such as
Eqs.~(\ref{eq:kappa2-xi}),~(\ref{eq:kappa3-sigma})
and~(\ref{eq:kappa4-sigma}).

The fact that the correlation length $\xi$ may not reach large values
makes higher-order cumulants important signals of the criticality in heavy-ion collisions because their dependence on $\xi$ is much
stronger than that of the quadratic cumulant.

%%%%%%%%%%%%%%%%%%%%%% Misha stuff

\subsection{Experimentally observable fluctuations and correlations}
\label{sec4:exp:observable}

Heavy ion collision experiments do not directly measure the values of
the order parameter field or its fluctuations. What quantities should
then serve as signatures for the QCD critical point?

Fluctuations in heavy-ion collisions are studied by collecting data on
event-by-event basis and looking at event-by-event fluctuations. The data
for each event is a set of particle momenta and other quantum numbers
such as charge, baryon number, spin, etc. Let us denote the set of
variables characterizing a given particle by $A$, such that
\begin{equation}
  \label{eq:A=p-etc}
  A = \{\bm p_A, q_A, s_A, \ldots \}\,.
\end{equation}
We can think of $A$ being a point in a generalized phase space,
which includes also (discrete) variables such as charge $q_A$, spin $s_A$, etc.\ and the
data set for each event as a collection of such points.

What experiments cannot measure is the position of a particle at the
time of freezeout. The critical fluctuations we discussed are related
to {\em spatial} correlations (see, e.g., Eq.~(\ref{eq:MM-mean-field})) at
that time. Let us consider how these spatial correlation translate into
the correlations between momenta of the particles. 

Does the correlation length $\xi$ determine the range of correlations
in momentum space? The answer to a first approximation is `no'~\cite{Ling:2015yau}.  The range of
correlations is set by the thermal distribution of particles within
the correlated spatial volume. That range is of order of the average
momentum difference of the particles and is determined by temperature
(e.g., for nonrelativstic particles of mass $m$, it is given by $\Delta p\sim
\sqrt{mT}$). 
In particular, for protons, this corresponds to a typical range of
rapidities of order unity. In principle,
spatial correlations can be translated into longitudinal momentum
correlations by strong longitudinal expansion, as in Bjorken
flow. However, for typical correlation length $\xi\sim 0.5-1$ fm, or even near the critical point $\xi\sim 2-3$ fm, and freezout time of order $\tau\sim 10$ fm, the corresponding Bjorken rapidity correlation range $\Delta\eta\approx\xi/\tau\sim 0.05-0.3$ is significantly smaller than the  range, $\Delta y_{\rm corr}=\mathcal O(1)$, due to thermal momentum spread.

Though the range of momentum correlations is not sensitive to $\xi$,
their {\em magnitude} is directly related to the correlation length
$\xi$. Indeed, the number of particles correlated with any given one grows,
as these are the particles which come from the region of space which
grows with $\xi$. Below we shall estimate that effect using a simple
model.

First, however, we shall define the quantities in terms of which these
momentum correlations can be described. At freezeout the particles
move freely and the state of the system can be described by the phase
space distribution function $f_{A}(x)$, where we denoted by $A$ the
set of momentum and other (conserved) quantum numbers characterizing
the particle, as in Eq.~(\ref{eq:A=p-etc}). The space-time coordinates
$x$ of particles at freezeout are not directly measurable.  The
quantity related to $f_{A}(x)$ but directly measurable in experiment is a
single-particle {\em momentum} distribution function, which is simply
the integral of $f_{A}(x)$ over the fireball at freezeout (over
the space-like freezeout
hypersurface):
\begin{equation}
  \label{eq:n-f}
  n_A = \int d^3x f_{A}(x)\,.
\end{equation}
The event average of this distribution is commonly used in
experimental analysis
\begin{equation}
  \label{eq:rho1-n}
  \rho_1(A) \equiv \frac{dN}{d^3\bm p_A} 
= \langle n_A\rangle\,. %= \int d\bm x_A \langle f_{\tilde A} \rangle
\end{equation}
The distribution function $f$ fluctuates on
event-by-event basis. These fluctuations can be described in terms of the
phase-space  event-averaged product $\langle \delta f_{ A}
\delta f_{ B}\rangle$, where $\delta f = f -\langle f\rangle$.
In the absence of correlations this quantity is only nonzero when
$A=B$, in which case it equals
$\langle (\delta f)^2\rangle=\langle f\rangle$ according to the Poisson distribution.
%(1\pm\langle f\rangle)
% where the sign corresponds to Bose or Fermi statistics.

 Again,
experiments can observe only the momentum space correlations given by
the integral
\begin{equation}
  \label{eq:nn-ff}
  \langle \delta n_A\delta n_B\rangle
= \int d^3x_A \int d^3x_B\, \langle \delta f_{A}(x_A) \delta f_{B}(x_B) \rangle\,.
\end{equation}
The quantity often used in experiments to analyze
two-particle correlations is a
two-particle momentum space density:
\begin{equation}
  \label{eq:rho2}
  \rho_2( A, B) = \frac{d^2N}{d^3\bm p_Ad^3\bm p_B}\,,
\end{equation}
which counts the (event averaged) number of pairs of particles in an
element of 2-particle momentum space $d^3\bm p_Ad^3\bm p_B$. Without
correlations $\rho_2$ factorizes: $\rho_2(A,B)\to\rho_1(A)\rho_1(B)$. Thus
correlations are characterized by 
\begin{equation}
  \label{eq:c2}
  C_2(A,B) = \rho_2(A,B) - \rho_1(A)\rho_1(B)
\end{equation}
This quantity is related to $\langle\delta n_A\delta n_B\rangle$ by
\begin{equation}
  \label{eq:c2-n}
  C_2(A,B) = \langle\delta n_A\delta n_B\rangle 
  - \delta_{AB}\langle n_A\rangle\,. % (1\pm\langle n_A\rangle),
\end{equation}
where $\delta_{AB}=\delta_{q_A,q_B}\delta^3\left(\frac{\bm p_A-\bm
    p_B}{2\pi}\right)$ and the last term makes sure that we do not
count a particle in the
same momentum cell twice as if it was a pair.

It is easy to express fluctuations of inclusive quantities, such as,
e.g., total charge in terms of the correlators we defined:
\begin{equation}
  \label{eq:Q-fluct}
  Q = \int_A q_A n_A,\quad 
\langle(\delta Q)^2\rangle = \int_A\int_B\,q_A q_B
\langle\delta n_A\delta n_B\rangle
\end{equation}
Alternatively, in terms of the correlation function $C_2$ \cite{Bialas:1999tv}:
\begin{equation}
  \label{eq:Q-c2}
  \langle(\delta Q)^2\rangle = \int_A q_A^2 \rho_1(A) 
  +\int_A\int_B\, q_A q_B C_2(A,B) 
.  
\end{equation}
Here and in what follows we use a shorthand notation for the integral
over the momentum space within the detector's acceptance window as
well as particle quantum numbers (cf. Eq.~(\ref{eq:A=p-etc})):
\begin{equation}
  \label{eq:intA}
  \int_A \equiv \underbrace{\int \left(\frac{d\bm
        p}{2\pi}\right)^3 \sum_{q_A,s_A,\ldots}}_{\mbox{within acceptance}}
\end{equation}
Acceptance cuts also apply to discrete quantum numbers. For example,
if some particles are not detected, such as  electrically neutral ones,
they are not to be included in the sum in Eq.~(\ref{eq:intA}). The
relevant example is the case when $Q$ is the baryon number: neutrons
carry nonzero baryon number $q_A=1$ in this case, but should not be included in
$\int_A$ if they are not detected.

\subsection{Translating critical point correlations into observable
  momentum correlations}
\label{sec:crit-pt-correl-momentum}

Clearly, the knowledge of the momentum space correlator
(\ref{eq:nn-ff}) or, equivalently, $C_2$, would allow for quantitative
predictions of fluctuation and correlation measurements. There are
many sources contributing to $C_2$: memory of initial fluctuations,
flow-induced correlations, charge conservation, hadron interactions
    (resonance decays)  and jets, to name just a few \cite{Schuster:2009jv,Bzdak:2012an,Steinheimer:2016cir,Hippert:2017xoj}. We will not
attempt to cover all of them here, but some of the effects will be discussed in Section~\ref{sec:ab:star-data-discussion} . Our focus is on determining the
effect of fluctuations associated with the {\em critical point} on this
experimentally measurable correlator. In other words, we want to know how
fluctuations and {\em spatial} correlations of the critical mode
$\sigma$ affect fluctuations described by {\em momentum}-space
correlator $\langle\delta n_A\delta n_B\rangle$. The main point to
keep in mind is that critical point effects are {\em localized} to the
critical region of the phase diagram, i.e., a relatively small
interval of beam energies $\sqrt{s_{NN}}$, compared to other contributions
unrelated to critical point, which typically persist across the whole
range of $\sqrt{s_{NN}}$ and/or vary monotonously with $\sqrt{s_{NN}}$. This is the rationale put forward in Ref.~\cite{Stephanov:1998dy} for the beam energy scan.

For the basic
estimate and to illustrate the physics at work we shall consider the
idealization of thermal equilibrium, which will allow us to apply the
results of the earlier subsections for the fluctuations of $\sigma$.
Another idealization we shall use exploits the fact that the typical relevant size of
the fireball (transverse radius or Bjorken proper time at freezeout)
$R\sim 7-10$ fm is large compared even with the maximum correlation
length expected in heavy-ion collisions near the critical point. Under
assumption $R\gg\xi$ we can consider the spatial correlations as
almost local compared to the smoothly varying (in position space)
distribution functions $\langle f\rangle$. We can thus describe the
spatial correlator of $\sigma$ as a delta-function normalized by
matching the space integral of the correlator (see
Eqs.~(\ref{eq:MM-mean-field}),~(\ref{eq:kappa2-sV-ss})
and~(\ref{eq:kappa2-xi})):
\begin{equation}
  \label{eq:ss-delta}
  \langle\sigma(\bm x)\sigma(\bm y)\rangle\to T\xi^2\delta^3(\bm x -
  \bm y)\,.
\end{equation}
This can be easily generalized to 3-point or 4-point 
correlation functions~\cite{Ling:2015yau}.

A simple model \cite{Stephanov:2008qz,Stephanov:2011pb} to describe the
correlations between particles induced by the fluctuations of the
$\sigma$ field considers the particles in question in equilibrium
in the presence of a slowly varying~\footnote{This requires the momenta of the
  particles to be large compared to $1/\xi$, which is a reasonable
  approximation near the critical point} background of the field
$\sigma$. We should expect the properties of the particles,
i.e., the energy-momentum (dispersion) relation $E_A(\p;\sigma)$, to
depend on~$\sigma$. Then the equilibrium distribution $f_A$ of the
particles depends on $\sigma$ and, as a result, will fluctuate, contributing to the fluctuation of the distribution function in addition to the trivial statistical fluctuations in a free particle gas:
\begin{equation}
  \label{eq:deltaf}
  \delta f_{A}(\bm x) = (\delta f_A(\bm x))_{\rm free } +
  \frac{\partial f_A}{\partial \sigma}\delta\sigma(\bm x)\,.
%  -\frac{\chi_A}{\gamma_A}g_A\sigma(\bm x)\,,
\end{equation}

For example, we can assume, for simplicity (and the present lack of better knowledge of $E_A$ dependence on $\sigma$),  that the mass $m_A$ of the particles
depends on the local value of $\sigma$, which is the usual assumption in the sigma-model.    {\footnote{    {This approach is inspired by the text-book sigma-model where the $\sigma$-field is an order parameter of the chiral symmetry breaking and in which the masses of hadrons depend on the expectation value of $\sigma$. Near the QCD critical point the field which we denote by $\sigma$ must be understood as a certain combination of scalar quatities such as energy density, baryon density and chiral condensate.}}}  
In this case
\begin{equation}
  \label{eq:chiA}
  \frac{\partial f_A}{\partial \sigma}  %\chi_A\equiv\frac{\partial f_{A}^{\rm eq}}{\partial\mu}
=- \frac{g_A}{T\gamma_A} f_{A}(1\pm f_{A})
\end{equation}
(plus/minus for bosons/fermions),
where we defined a coupling $g_A\equiv dm_A(\sigma)/d\sigma$ and denoted $\gamma_A=(dE_A/dm_A)^{-1}=E_A/m_A$ -
the relativistic gamma-factor for a particle with momentum $\bm p_A$.

The coupling $g_A$ plays an important role in determining the
magnitude of the critical fluctuation effects. The existing rough
estimates of its value for various particle species are based on an
assumption that $\sigma$ couples with strength proportional to the
particle mass, as in the sigma-model. If this assumption is correct
the strength of the critical point signatures in cumulants of the
protons should be much larger than in the similar cumulants for pions~\cite{Athanasiou:2010kw}.

The results obtained in this simple model can be reproduced in a more rigorous diagrammatic approach~\cite{Stephanov:2001zj}. The same results can also be obtained in a dynamical model of the evolution of the particle gas coupled to a scalar field $\sigma$, which can also describe effects of finite-time evolution~\cite{Stephanov:2009ra}.

The fluctuations of $\sigma$ via Eqs.~(\ref{eq:deltaf}) and~\eqref{eq:ss-delta} give rise to an additional contribution
to the correlator (at freezeout): 
\begin{equation}
  \label{eq:ff-sigma}
  \langle \delta f_A(\bm x_A) \delta f_B(\bm x_B) \rangle
= 
f_A\delta_{AB}\delta^3(\bm x_A - \bm x_B)
+T\xi^2\, \frac{\partial f_A}{\partial \sigma} \frac{\partial f_B}{\partial \sigma}%\frac{g_A\chi_A}{\gamma_A}\frac{g_B\chi_B}{\gamma_B}
\delta^3(\bm x_A - \bm x_B)\,.
\end{equation}
This correlator is local in coordinate space (on distance scales much
larger than $\xi$) but is nonlocal in momentum space. This nonlocality
is important for understanding the acceptance dependence of the
fluctuation measures.

Using equation~(\ref{eq:nn-ff}) we find
\begin{equation}
  \label{eq:k2-sigma}
  \langle \delta n_A\delta n_B\rangle
= \langle n_A \rangle \delta_{AB}  
+ \int d^3x T\xi^2\,
\frac{\partial f_A}{\partial \sigma} \frac{\partial f_B}{\partial \sigma}\,.
%\frac{g_A\chi_A(x)}{\gamma_A}\frac{g_B\chi_B(x)}{\gamma_B}
\end{equation}
where the first term is the trivial statistical Poisson contribution (see also Eq.\eqref{eq:c2-n}). 

\begin{figure}
    \centering
    \includegraphics[height=12em]{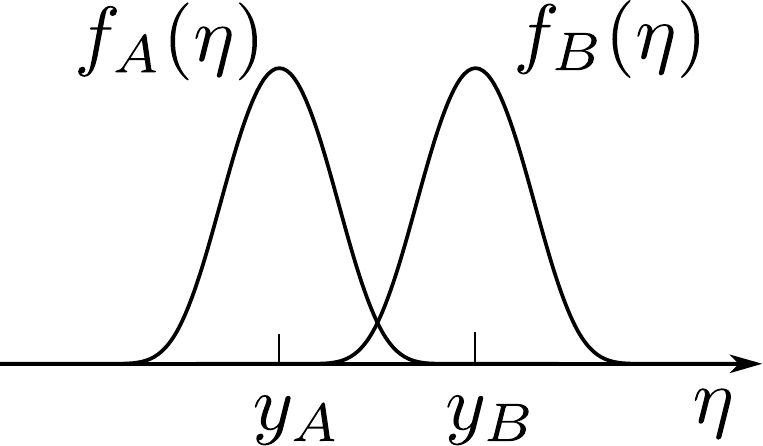}
    \caption{In the Bjorken longitudinal expansion scenario the probability distribution $f_A(\eta)$ for a particle with momentum rapidity $y_A$ to be at Bjorken coordinate rapidity $\eta$ is centered around $\eta=y_A$. The width of $f_A(\eta)$ is given by the thermal rapidity spread of the particles. The product $f_Af_B$ which dominates momentum dependence  in the integrand in Eq.~\eqref{eq:k2-sigma} gives the number of pairs $AB$ which come from the {\em same} locally correlated cell which  quickly vanishes if $|y_A-y_B|$ exceeds the thermal rapidity spread (typically of order unity).}
    \label{fig:ypyk}
\end{figure}

Following Refs.~\cite{Stephanov:2001zj,Ling:2015yau}, let us note that the dominant dependence on momentum in $\partial f_A/\partial\sigma$, comes from factor $f_A$ (see, for example, Eq.\eqref{eq:chiA}, taking into account that typically $f_A\ll1$). Correspondingly, the dominant contribution to the integrand in Eq.~\eqref{eq:k2-sigma} comes from the factor  $f_A(x)f_B(x)$. The product $f_A(x)f_B(x)$   is the number of pairs of particles with momenta $A$ and $B$ from the same cell (of size $\xi$) located at $x$. In the boost-invariant Bjorken expansion scenario the local rest frame of the fireball moves with rapidity equal to its Bjorken coordinate $\eta$. Thus, most of the particles with momentum rapidity $y_A$ come from the cell at $\eta=y_A$ (see Fig.~\ref{fig:ypyk}). The spread of the $f_A(\eta)$ around rapidity $y_A$ is of order the typical thermal rapidity $|\eta-y_A|\lesssim 1$.
Therefore, the product $f_Af_B$ is strongly suppressed if $|y_A-y_B|$ exceeds typical thermal rapidity: a pair of particles with larger rapidity separation could not come from the same spatially correlated cell.  Thus the range of the correlations in the rapidity space,
$\Delta y_{\rm corr}$, is set by thermal rapidity (typically of order
1) of the particles (and not by the correlation length $\xi$, which at
freezeout corresponds to very short interval of Bjorken rapidity
$\Delta\eta \sim \xi/\tau\sim 0.1$) \cite{Ling:2015yau}.

Using Eq.~(\ref{eq:Q-fluct}) one can also obtain the corresponding
cumulant of a given charge $Q$: 
\begin{equation}
  \label{eq:k2-x}
  \kappa_2[Q]\equiv\la (\delta Q)^2\ra
  = \int_A\int_B\,q_A q_B
\langle\delta n_A\delta n_B\rangle
=\int_A q_A^2 \langle n_A \rangle
+\int \!d^3x \ {T}\xi^2
 \left(\int_A q_A
 \frac{\partial f_A}{\partial \sigma}  %\frac{g_A\chi_A}{\gamma_A}
 \right)^2\,.
\end{equation}
The value of the cumulant depends on the experimental acceptance (the
momentum space region over which $\int_A$ and $\int_B$ are
evaluated). This dependence is directly related to the range of the correlator $\langle\delta n_A\delta n_B\rangle$. In the Bjorken boost-invariant expansion scenario the
first term on the RHS of Eq.~(\ref{eq:k2-x}), i.e., the Poisson term,
is simply proportional to the volume of integration,
$\int dy_A = \Delta y_{\rm acc}$, since the integrand is rapidity
independent. The rapidity dependence of the second, correlation term
in Eq.~(\ref{eq:k2-x}) is different in two regimes: small and large
acceptance. For small acceptance, $\Delta y_{\rm acc}\ll\Delta y_{\rm
  corr}\sim 1$, each rapidity integral $\int_A$ is proportional to
$\Delta y_{\rm acc}$ and the correlation term is proportional to
$\Delta y_{\rm acc}^2$. In the opposite regime\footnote{In practice,
  this regime is limited by the total rapidity width of the Bjorken
  plateau, which depends on the collision energy.}, $\Delta y_{\rm
  acc}\gg\Delta y_{\rm corr}$, the double integral $\int_A\int_B$ has
support only for $|y_A-y_B|\lesssim \Delta y_{\rm corr}$ (the diagonal
strip in $y_A-y_B$ space), and thus the
cumulant scales as $\Delta y_{\rm acc}$~\cite{Ling:2015yau}.

It should be kept in mind that $\int_A$, in addition to momentum
integration, also contains summation over other quantum numbers of the
particles, including $q_A$ (see Eq.~(\ref{eq:intA})). Different
particles can carry different quanta $q_A$ of the given charge
$Q$. For example, if $Q$ is net baryon charge -- protons and
antiprotons have $q_A=\pm1$. If their average numbers and couplings
to the sigma mode, i.e., $\partial f_A/\partial\sigma$, are close to each other (e.g., at large
$\sqrt{s_{NN}}$, when baryon/antibaryon asymmetry is small), the positive
and negative contributions to $\int_A q_A (\partial f_A/\partial\sigma)$ will
tend to cancel each other. Therefore, the critical point
contribution to cumulants of positive-definite quantities (e.g.,
number of protons in acceptance) are typically larger than to the cumulants of such quantities as net-charge or
net-baryon number~\cite{Hatta:2003wn,Athanasiou:2010kw}.

\subsection{Higher-order cumulants}
This analysis can be generalized to higher-order cumulants \cite{Stephanov:2008qz,Ling:2015yau}. For
example,
\begin{equation}
  \label{eq:k3-Q}
  \kappa_3[Q]\equiv\la (\delta Q)^3\ra
=\int_A q_A^3 \langle n_A \rangle
+\int \!d^3x \ 2\tilde\lambda_3{T}^{3/2}\xi^{9/2}
 \left(\int_A q_A
 \frac{\partial f_A}{\partial \sigma} 
 %\frac{g_A\chi_A}{\gamma_A}
 \right)^3
\end{equation}
\begin{equation}
  \label{eq:k4-Q}
  \kappa_4[Q]\equiv\la (\delta Q)^4\ra_c
=\int_A q_A^4 \langle n_A \rangle
+\int \!d^3x \ 6(2\tilde\lambda_3^2-\tilde\lambda_4){T}^{2}\xi^{7}
 \left(\int_A q_A
 \frac{\partial f_A}{\partial \sigma} 
 %\frac{g_A\chi_A}{\gamma_A}
 \right)^4
\end{equation}
One can also generalize this analysis to correlations between different charges and species \cite{Athanasiou:2010kw}.

\begin{figure}[h]
  \centering
  \includegraphics[width=12em]{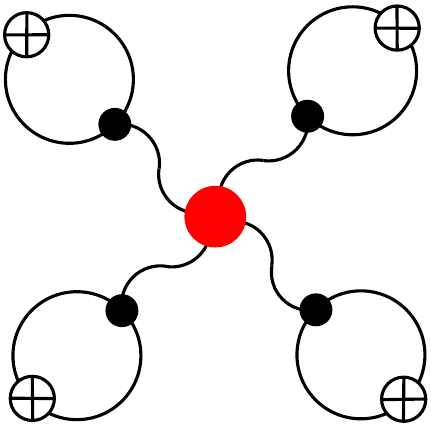}
  \caption{The 4-point correlation due to the fluctuations of the critical
    field in Eq.~\eqref{eq:k4-Q}. The red point in the center represents the cumulant of the
    critical field, $\la(\delta\sigma_V)^4\ra_c$, given by Eq.~\eqref{eq:kappa4-sigma}. The black points
  correspond to $\partial/\partial\sigma$ insertions, or
  couplings $g_A$ in Eq.~\eqref{eq:chiA}, and the crossed circles -- to insertions of~$q_A$. The loops with the insertions represent the four particles
  which are being correlated, and whose momenta are being integrated over in Eq.~\eqref{eq:k4-Q} ~\cite{Stephanov:2001zj,Stephanov:2008qz}.}
  \label{fig:4x}
\end{figure}

Comparing to Eqs.~(\ref{eq:sigma-moments}) one can see that the
magnitude of the critical contribution is proportional to the
cumulants of the order parameter $\sigma$. I.e., critical fluctuations
of the order parameter are ``imprinted'' on the fluctuations of the
observed particles. One can view the critical contribution diagrammatically as shown in
Fig.\ref{fig:4x}, which also illustrates that it contributes to the
irreducible correlation function.
As already noted, the cumulants contain non-critical contributions, such as the trivial statistical contribution or Poisson term. Some of these non-critical contributions can be removed by
subtracting the trivial statistical contribution. 
Even better, by subtracting contributions from all lower order correlations, one can
construct a {\em factorial} cumulant which represents irreducible
correlations \cite{Ling:2015yau}, as we shall discuss in Section~\ref{sec:ab:star-data-discussion}.

Similarly to the quadratic cumulant in Eq.~(\ref{eq:k2-x}), the
critical contribution to the higher order cumulants for small $\Delta y_{\rm
  acc}\ll \Delta y_{\rm corr}$ scale as $(\Delta y_{\rm acc})^k$, where
$k$ is the order of the cumulant $\kappa_k$ and scale linearly with
$\Delta y_{\rm acc}$ in the (idealized) limit of  $\Delta y_{\rm
  acc}\gg \Delta y_{\rm corr}$.
Since in realistic heavy-ion collisions $\Delta y_{\rm acc}\lesssim
\Delta y_{\rm corr}\sim 1$, one should be able to observe the
beginning of the transition from the fast $\Delta y_{\rm acc}^k$ rise
to a slower $\Delta y_{\rm acc}^1$ rise. To which extent this is possible with the presently available data will be discussed in Section~\ref{sec7:correlation_rapidity}. It is often convenient to
consider the ratio of cumulants such as $\kappa_k/\kappa_2$, or
$\omega_k = \kappa_k/N$. In this
case the $\Delta y_{\rm acc}^{k-1}$ growth would begin to saturate at
a constant value at larger $\Delta y_{\rm acc}$. This behavior can be used
to determine the rapidity range of an observed correlation. {Since the rapidity range is different for the critical correlations ($\Delta y_{\rm corr}\sim 1$) and for, e.g., initial state correlations ($\Delta y_{\rm corr}\gg 1$), the range of correlations could be used to separate critical point signatures from certain  background effects. }

From the experimental perspective it is important to note~\cite{Stephanov:2008qz} that, even though the $\xis$ dependence of
$\kappa_4$ is stronger than that of $\kappa_3$ or $\kappa_2$, its
measurement involves subtraction of two contributions, such as
$\la(\delta N)^4\ra-3\la(\delta N)^2\ra^2$, each of which is order $N$
times larger than their difference ($\mathcal O(N^2)$ vs
$\mathcal O(N)$), which typically dilutes the signal-to-noise ratio in
experimental measurement. This issue becomes more problematic as the
order of the cumulant grows (e.g.,
$\langle (\delta N)^6\rangle_c=\mathcal O(N)$ is obtained by
cancellation of contributions of order $\mathcal O(N^3)$.) Therefore,
even though the higher-order cumulants are even more sensitive to the
divergence of the correlation length and equations such as
(\ref{eq:k4-Q}) can be generalized to higher-order cumulants, the
measurement of such cumulants would become impractical at too high orders.

Our analysis in this section was focused on integral measures, such as
fluctuation cumulants in Eq.~(\ref{eq:k2-x}). It is clear, however,
that more information is contained in the correlator, such as in
Eq.~(\ref{eq:k2-sigma}). Some of this information, such as the range of
the correlations, can be extracted by studying acceptance dependence,
but some information is lost during the integration.  The correlator in
Eq.~(\ref{eq:k2-sigma}), and similar higher-point correlators
$\langle \delta n_A\delta n_B\delta n_C \ldots\rangle$ are, of course,
directly measurable, albeit with more difficulty.

It is important to note again that other sources may and do contribute to
fluctuation measures: remnants of initial fluctuations, flow,
jets -- to name just a few obvious contributors. 
A quantitative study of these effects may be necessary 
to unambiguously identify the critical point signal.
 This serves to emphasize that the energy scan of the QCD phase
diagram is needed to separate such background contributions from the
genuine critical point effect, the latter being {\em non-monotonous}
function of the initial collision energy  $\sqrt{s_{NN}}$ \cite{Stephanov:1998dy,Stephanov:1999zu} (or accepted particle rapidity \cite{Brewer:2018abr}) as the critical
point is approached and then passed. The fact that non-Gaussian cumulants
have stronger dependence on $\xis$ than, e.g., quadratic moments,
makes those higher-order cumulants more sensitive signatures of the critical
point~\cite{Stephanov:2008qz}.

%%% Local Variables:
%%% mode: latex
%%% TeX-master: "BES_Main_current"
%%% End:

%-------------Section 5-----------------------------------
%
%=====================================================================================
%=====================================================================================

\newpage 

\section{Theory and Phenomenology of Anomalous Chiral Transport }
\label{sec5}

\subsection{The many-body physics of the chiral anomaly}
\label{subsec_5_1}

Macroscopic properties of matter and symmetries of the microscopic dynamics are often in deep and direct correspondence. For example, if a  physical theory is invariant under space and time translation, then energy and momentum must be conserved in any individual interaction process. Correspondingly, this mandates that the many-body system of this theory must also conserve energy momentum at the macroscopic level. The hydrodynamic equations for the energy-momentum tensor   are the direct consequences of these conservation laws.  Similarly the phase invariance of an underlying theory leads to the conserved charge, which in turn gives rise to the current conservation equation in hydrodynamics.  
These observations raise an interesting question with regards to the chiral anomaly discussed in Section 2.  In a way, the axial symmetry is like a ``semi-symmetry'',  in that it is classically preserved but quantum mechanically broken. One naturally wonders: how would the microscopic chiral anomaly manifest itself in the macroscopic properties of chiral matter, i.e  a many-body system containing chiral fermions?  In the past decade, significant progress has been achieved in understanding this question, strongly driven by motivations and efforts from multiple branches of physics, notably from heavy ion collisions as well as from topological semimetals (see recent reviews in e.g. \cite{Kharzeev:2015znc,Kharzeev:2015kna,Liao:2014ava,Landsteiner:2016led}).    The short answer is that the anomaly will induce a number of novel anomalous chiral transport effects, such as the Chiral Magnetic Effect (CME), Chiral Vortical Effect (CVE) and Chiral Magnetic Wave (CMW). The many-body theoretical frameworks for the quantitative description of these effects are also under active investigation. In the following, we briefly discuss  these developments. 

\subsubsection{Chiral Magnetic and Vortical Effects}

Let us consider a certain type of chiral materials which could be, e.g., the quark-gluon plasma (QGP) with (approximately) chiral quarks or the Dirac/Weyl semimetals with chiral quasi-particles. The macroscopic transport properties, which typically involve collective motion (i.e. momentum correlations) of underlying particles, become markedly different from normal materials because of the nontrivial interplay between the particles' momentum and spin due to their chirality and the anomaly. In particular, certain transport processes that are forbidden in normal environment become possible (and necessary) in such chiral materials. As a typical example of a normal transport process, let us consider electric conductivity in usual materials:  an electric current is generated in response to an applied electric field. In those materials, no current could be generated by applying a constant magnetic field. But in the chiral materials discussed here, an electric current can be generated in response to an applied magnetic field. This is the so-called Chiral Magnetic Effect (CME)~\cite{Kharzeev:2007jp,Fukushima:2008xe}, the most widely studied example among a class of anomalous transport effects in chiral materials. Specifically the CME predicts the following relation: 
\begin{eqnarray} \label{eq:jl:cme}
\mathbf{J}_e =(q e)\mathbf{J} =  \sigma_5 \mathbf{B} \quad ,  \quad \sigma_5 =  C_A \mu_5 (qe)^2 \quad , \quad \mathbf{J} = C_A \mu_5 (qe) \mathbf{B}
\end{eqnarray}
where $\mathbf{J}_e$ is the electric current, $\mathbf{J}$  is the underlying fermion vector current 
\footnote{Note that  the vector current associated with a given type of fermion field  is $J^\mu=\bar \psi \gamma^\mu \psi$, while the electric current or baryonic current associated with it can be constructed from this vector current based on the fermion's electric or baryonic charge accordingly, e.g. $J^\mu_e=(qe)J^\mu$.} 
and $\mathbf{B}$ is the magnetic field. 
The coefficient $\sigma_5$ represents  the chiral magnetic conductivity for each species of Dirac fermions with electric charge $qe$. It is  expressed in terms of  the   anomaly coefficient $C_A=1/(2\pi^2)$ and the chiral chemical potential $\mu_5$ which quantifies the imbalance in the densities of right-handed (RH) and left-handed (LH) fermions (see Eq.~(\ref{eq:jl:PsiRL})). For QCD with $N_c$ colors and $N_f$ light quarks of charges $Q_f e$, one needs to sum over the quark colors and flavors, so that $(qe)^2  \to \left( N_c \sum_{f=1,...,N_f}  Q_f^2 e^2 \right )$.

%\vspace{-0.15in}
\begin{figure}[!hbt] 
\begin{center}
\includegraphics[height=5cm,width=12.5cm]{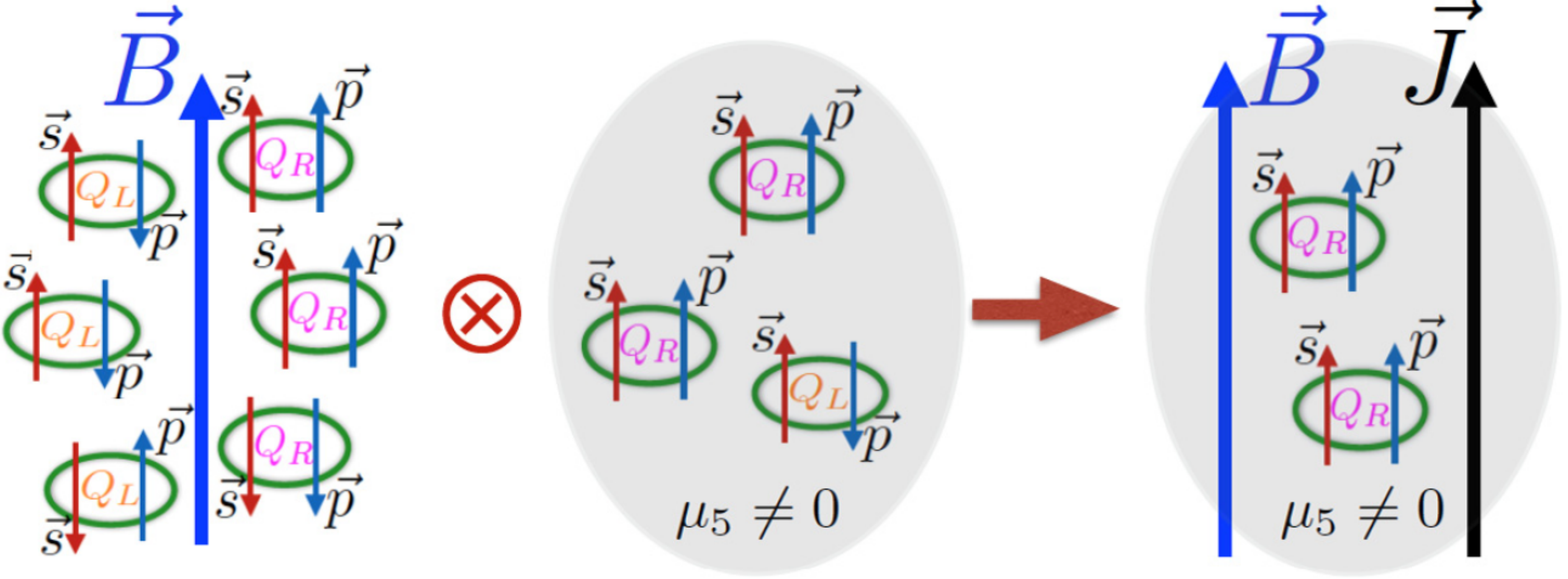} %\hspace{0.6in}
%\includegraphics[height=4.5cm]{cme_triangle.pdf}
%\vspace{-0.1in}
\caption{An   illustration of the Chiral Magnetic Effect (adapted from~\cite{Kharzeev:2015znc}). See text for details.} 
 \label{fig:jl:f00}
%\vspace{-0.15in}
\end{center}
\end{figure}

In order to provide an intuitive understanding of the CME, let us consider a system of chiral fermions with positive charge, as illustrated in Fig.~\ref{fig:jl:f00}. The magnetic field aligns the fermions' spin preferentially along its direction, i.e. $\la \mathbf{s} \ra \propto (qe)\mathbf{B}$. At the same time the momentum and spin of a chiral fermion is correlated so that the momenta of RH particles are preferentially parallel and those of LH particles anti-parallel to the direction of the magnetic field, respectively. Therefore, if the system has a chirality imbalance, say we have more RH than LH particles ($\mu_5>0$), the net momentum (and spin) would be parallel to the magnetic field, $\la \mathbf{p} \ra \propto \mu_5 \la \mathbf{s} \ra$. Since the particles all carry a positive charge, this would result in an electric vector current along the magnetic field, $\mathbf{J} \propto \la \mathbf{p} \ra \propto \mu_5 (qe) \mathbf{B}$.

  {As one can see from this intuitive illustration of CME, an essential
ingredient for the CME transport current to occur, is the correlation
between spin and momentum for a massless chiral fermion of specific RH
or LH chirality. A finite fermion mass, either found in the Lagrangian
or dynamically generated by spontaneous symmetry breaking, as in
the case of the QCD vacuum, would spoil this correlation. More
precisely, in a massive fermion theory 
the chirality imbalance will relax to zero at a rate
proportional to mass squared (see, e.g., Refs.~\cite{Grabowska:2014efa,Manuel:2015zpa}). In order for chirality to be maintained, its relaxation rate needs to be considerably smaller than the thermal kinetic relaxation rate, which would require the mass to be small compared with the relevant matter scale like temperature. 
Therefore, while the effect of a small current quark mass on CME is negligible, the effect of large constituent mass %of order $T$ 
due to chiral symmetry breaking is not. For this reason, the
chiral symmetry restoration is important for observation of the CME.
}

It is instructive to contrast this phenomenon with the usual conductivity driven by an electric field $\mathbf{J}_e = \sigma_e \mathbf{E} $. Contrary to the electric conductivity $\sigma_{e}$ which depends on the properties of the material such as mean free path etc, the chiral magnetic conductivity  $\sigma_{5}$ is ``universal'' in that it only depends on the chiral anomaly coefficient $C_A$, $\mu_5$ and particle charge, and is {\em not affected} by the interaction among the particles~\cite{Fukushima:2008xe}. 
This reflects the   fact that  CME is  essentially the macroscopic manifestation of the fundamental quantum anomaly in a many-body setting.  
Furthermore, the chiral magnetic conductivity, $\sigma_5$,  is ${\hat{\mathcal P}}$-odd and ${\hat{\mathcal C}}{\hat{\mathcal P}}$-odd, due to the ${\hat{\mathcal P}}$ and ${\hat{\mathcal C}}{\hat{\mathcal P}}$  transformation properties of the chiral chemical potential $\mu_5$. This explains why the CME may  occur only in chiral matter with macroscopic chirality (i.e. nonzero $\mu_5$). 
Finally the $\sigma_5$  is  time-reversal even~\cite{Kharzeev:2011ds} which implies the non-dissipative nature of the CME transport current.  In contrast, the electric conductivity $\sigma_e$  is   time-reversal odd, thus involving dissipative transport.  

Analogous to the anomaly relations of Eq.~\eqref{eq:jl:ARL}, the  CME-current may be written in terms of  the RH/LH   sectors:  
\begin{eqnarray} \label{eq:jl:cme2}
\mathbf{J}_R = + \frac{C_A}{2} \mu_R (qe)\mathbf{B} \quad ,  \quad  \mathbf{J}_L = - \frac{C_A}{2} \mu_L (qe) \mathbf{B}.
\end{eqnarray} 
Here $\mu_{R/L}$ are the chemical potentials for RH/LH fermions, respectively, which satisfy $\mu_R-\mu_L=2\mu_5$. 
 The above RH/LH formulation also allows one to recognize  the so-called Chiral Separation Effect (CSE)~\cite{Son:2004tq,Metlitski:2005pr} in a straightforward way. By noting that  $\mathbf{J}_5=\mathbf{J}_R-\mathbf{J}_L$, one obtains from (\ref{eq:jl:cme2}): 
\begin{eqnarray} \label{eq:jl:cse}
 \mathbf{J}_5 = C_A \mu_V (qe) \mathbf{B}
\end{eqnarray} 
where $\mu_V=(\mu_R+\mu_L)/2$. It predicts the generation of an axial current along the external magnetic field when there is a nonzero vector charge density (i.e. with its corresponding chemical potential $\mu_V\neq 0$). 

In addition to a magnetic field $\mathbf{B}$, the rotation of the fluid could also induce anomalous transport in chiral matter. This can be understood by the analogy between magnetic field and fluid rotation, such as the similarity  between Lorentz force and Coriolis force in classical mechanics and that between magnetic flux and orbital angular momentum in quantum mechanics. 
The linear motion of a fluid is described by a velocity field $\mathbf{v}$ in analogy to the vector field $\mathbf{A}$ in electromagnetism.  The rotational motion of a fluid, on the other hand, is described by a vorticity field $\bm{\omega} = \frac{1}{2} \nabla\times \mathbf{v}$, which plays the role similar to magnetic field $\mathbf{B}$. In the covariant formulation of relativistic fluid dynamics, one has correspondingly the velocity vector $u_\mu$ and the vorticity tensor $\omega_{\mu\nu}=\frac{1}{2}(\partial_\nu u_\mu - \partial_\mu u_\nu)$. The counterpart of the CME in chiral matter under fluid rotation, is the so-called Chiral Vortical Effect (CVE)~\cite{Kharzeev:2007tn,Son:2009tf,Landsteiner:2011cp,Hou:2012xg}. The CVE predicts the generation of the  underlying fermion vector current  in response to  the fluid rotation: 
\begin{eqnarray} \label{eq:jl:cve}
\mathbf{J} = C_A \, \mu_5  \, (2\mu) \, \bm{\omega} \,\, .   
\end{eqnarray}
In comparison to the CME in Eq.~(\ref{eq:jl:cme}), the  anomaly coefficient $C_A$ and  the $\mu_5$ are common to both, while the vorticity $ \bm{\omega}$ replaces the magnetic field $\mathbf{B}$ as the driving ``force''. 
Finally, let us mention in passing that there are additional transport effects such as the Chiral Electric Separation Effect (CESE)~\cite{Huang:2013iia,Jiang:2014ura}, characterizing the generation of axial current in chiral matter in response to an external  electric field. More detailed discussions may be found in Refs. \cite{Kharzeev:2015znc,Liao:2014ava,Hattori:2016emy}.

\subsubsection{Chiral collective excitations}

Collective excitations are important ``degrees of freedom'' in many-body systems. They often emerge in the long-time, large-distance behavior of a system and play key roles in its transport properties. A prime example is the sound wave, a gapless excitation arising from the coupled evolution between the fluctuations of pressure and energy density. 

A system of chiral fermions in the presence of an external magnetic field  $\mathbf{B}$ exhibits new collective excitations as a result of the aforementioned anomalous transport effects. Physically this is not hard to imagine: Consider such a system with a neutral background (i.e. without any net vector or axial density for the underlying fermions) and suppose an axial density fluctuation occurs somewhere. This will generate a vector current via the CME, which transports vector charges, and thus charge fluctuations, from the initial location along the direction of the magnetic field $\mathbf{B}$. These charge fluctuations, in turn trigger an axial current via the CSE, which transports axial charge fluctuations further away along the magnetic field direction, which again induce further vector charge fluctuation via the CME and so on. Consequently, as a result of the anomalous transport in the presence of a magnetic field, the fluctuations of vector and axial charges get entangled and propagate along the $\mathbf{B}$ field direction.

In order to describe this phenomenon mathematically, it is convenient to use the chiral basis i.e. to examine the fluctuations in the RH and LH sectors. We consider an equilibrated neutral background chiral matter under the presence of a constant magnetic field $\mathbf{B}$ and for simplicity assume no electric field. In this case, the small charge fluctuations of the RH and LH sectors satisfy the following continuity equations as well as the CME relations:
\begin{eqnarray} 
\partial_t (\delta J^0_\chi )+ \nabla\cdot (\delta \mathbf{J}_\chi) = 0  \quad , \quad   (\delta \mathbf{J}_\chi)  = \frac{\chi \, C_A}{2} \, (\delta \mu_\chi) (qe) \mathbf{B} 
\end{eqnarray} 
where the label $\chi=+1$ for RH sector and $\chi=-1$ for LH sector.
Furthermore the fluctuations of charge density and its corresponding chemical potential can be connected via the susceptibilities, i.e. $(\delta J^0_\chi ) = c_\chi \, (\delta \mu_\chi )$ with $c_\chi$ the thermal susceptibility evaluated from the background matter. Combining all these together, we arrive at the following equation for the chiral density fluctuations: 
\begin{eqnarray}
\partial_t (\delta J^0_\chi )+ \frac{\chi C_A (qe)}{2 c_\chi } (\mathbf{B}\cdot \nabla)  (\delta J^0_\chi) = 0 \quad \to 
\quad 
\left[ \partial_t  + \chi \, v_B \, \hat{\mathbf{B}}\cdot \nabla \right] (\delta J^0_\chi ) = 0
\end{eqnarray} 
which takes the form of a wave equation. Here $v_B=\frac{C_A(qe) |\mathbf{B}|}{2c_\chi}$ and $\hat{\mathbf{B}}=\mathbf{B}/|\mathbf{B}|$ is the unit vector along the magnetic field direction.  
Decomposing  the density fluctuation $(\delta J^0_\chi )$ into its Fourier modes  $(\delta J^0_\chi ) \sim e^{-(i\omega t - \mathbf{k}\cdot \mathbf{x} )}$ with frequency $\omega$ and wave-vector $k \hat{\mathbf{B}}$, we find the dispersion relation 
\begin{eqnarray}
\omega - \chi \,  v_B \, k = 0 
\end{eqnarray}
where the $v_B$  plays the role of the wave-speed. This is the Chiral Magnetic Wave (CMW), a gapless collective excitation in chiral matter~\cite{Kharzeev:2010gd,Burnier:2011bf}. It has two independent modes (with $\chi=\pm 1$ respectively): the RH wave that propagates in parallel to the magnetic field direction $\hat{\mathbf{B}}$ while the LH wave propagates in the opposite direction to $\hat{\mathbf{B}}$. 
Just like sound waves that can transport energy and momentum in the usual medium, the CMW can transport vector and axial charges   in chiral matter. 
Different from  the CME for which a chirality imbalance (i.e. nonzero axial density) is necessary, the existence of CMW does not require any particular background density and can be simply triggered by density fluctuations in chiral matter as long as there is a magnetic field.

The above analysis can be further generalized into the situation of collective modes on top of any non-neutral background matter under the presence of both electric and magnetic fields~\cite{Huang:2013iia}. Furthermore it is natural to expect that collective excitations similar to the CMW should be present also under fluid rotation. Indeed, chiral collective modes for RH and LH density fluctuations, referred to as the Chiral Vortical Wave (CVW), were identified in~\cite{Jiang:2015cva} and further analyzed in~\cite{Chernodub:2015gxa,Abbasi:2016rds,Hattori:2017usa}.

\subsubsection{Fluid dynamics with chiral anomaly}
\label{subsec_5_1_fluid}

Fluid dynamics is a theoretical framework that provides a universal effective description of the long wavelength behavior of many-body systems~\cite{landau1959fluid}, and it is successfully applied to describe a large variety of physical systems, ranging from the smallest fluid droplet (e.g. the QGP on a femto scale) to the largest fluid (e.g. the cosmic evolution).  In the large wave length limit the system may be considered to be  close to local thermal equilibrium and its behavior is dictated by conservation laws such as energy-momentum and charge conservation. If a system is in strict local equilibrium, it is governed  by {\em ideal fluid dynamics}.  Deviations from local thermal equilibrium can be systematically incorporated via an order-by-order ``gradient expansion'' which leads to {\em viscous fluid dynamics}.   

In order to illustrate the gradient expansion let us consider a charged current $J^\mu$ (for a single degree of freedom) in  fluid dynamics. For such a conserved current, the fluid dynamics equation is simply the continuity equation, $\partial_\mu J^\mu =0$ where the current is expressed in terms of fluid dynamical variables such as velocity field $u^\mu(x)$, charge density field $n(x)$, temperature field $T(x)$, etc. Such an expression for $J^\mu$ is the so-called constituent relation.  For ideal fluid dynamics (0th order in gradient expansion), one has $J^{\mu, (0)} = n u^\mu$ which simply says that the charges move along with the local fluid flow.  However a fluid cell cannot be totally isolated from its surroundings and the behavior of one cell will be affected by neighboring cells. Such a ``gradient'' in the fluid leads to the viscous transport currents. Diffusion is a prime example, which simply says that if the charge density is higher at one location compared to other, nearby locations, then the charges will develop a diffusion current and spread out from the dense spot to the dilute spots. This is a typical dissipative process that increases the total entropy of the system and tends to bring the system toward global thermal equilibrium. Turning on an external electric field has the equivalent effect of creating a gradient in the charge chemical potential along the electric field direction.  As it turns out, the second law of thermodynamics (i.e. entropy must not decrease with time) completely fixes the possible form of such leading-order viscous corrections, denoted by  $J^{\mu, (1)} $ and known as Navier-Stokes current, as follows: 
%.  Thus  the usual 1st-order fluid dynamics for $J^\mu$ is summarized as: 
\begin{eqnarray} \label{eq:jl:Jfluid}
&& \partial_\mu J^\mu =0 \,\, , \,\, J^\mu = J^{\mu, (0)} +   J^{\mu, (1)}_{viscous} + ... \,\, \\ 
\label{eq:jl:Jfluid2}
&& J^{\mu, (0)} = n u^\mu \,\,  , \,\, 
 J^{\mu, (1)}_{viscous}   =  \frac{\sigma}{2} T \Delta^{\mu\nu}   \partial_\nu \left(\frac{\mu}{T}\right) +  \frac{\sigma}{2} q E^\mu
\end{eqnarray}
In the $J_{viscous}^{\mu, (1)} $,  the first term is the diffusion current and the second term is the electric conducting current. Here, T is temperature, $\mu$ is the  chemical potential corresponding to the charge density $n$ in thermal equilibrium, $q$ is the electric charge, and the projection operator $\Delta^{\mu\nu}=(g^{\mu \nu} - u^\mu u^\nu)$ with the usual Minkowski metric tensor $g^{\mu \nu}=Diag(+,-,-,-)$. The electric conductivity,  $\sigma$, is a diffusion coefficient which is a characteristic transport property of a given material.

We now turn to the discussion of a chiral fluid (i.e. a fluid containing chiral fermions), with a chiral current $J^\mu_\chi$ (where $\chi=\pm 1$ is a chirality label) that could be from either the right-handed (RH, with $\chi=+1$) sector or the left-handed (LH, with $\chi=-1$) sector. 
Such a chiral current is associated with the  underlying  ``semi-symmetry'' and subject to   the chiral anomaly. As a consequence the continuity equation needs to be modified in order to account  for the anomaly relation Eq.~\eqref{eq:jl:ARL}.  Furthermore, the CME and CVE discussed before would hint at additional anomalous transport currents that are absent in usual fluid dynamics. As found  in  \cite{Son:2009tf}, the second law of thermodynamics is still able to fix completely the first order gradient expansion terms of the chiral current and in fact mandates   the presence of anomalous transport currents in the form of  CME and CVE. Such fluid dynamics which incorporates the chiral anomaly, up to the first order gradient expansion, can be summarized as follows: 
\begin{eqnarray} \label{eq:jl:Cfluid}
&& \partial_\mu J_\chi^\mu = \chi \, \frac{C_A}{2} E_\mu B^\mu  \,\, , \,\, J_\chi^\mu = J^{\mu, (0)} +   J^{\mu, (1)}_{viscous} + J^{\mu, (1)}_{anomalous} + ... \,\, \\ 
%&& J^{\mu, (0)} = n u^\mu \,\,  , \,\, 
% J^{\mu, (1)}_{viscous}   =  \frac{\sigma}{2} T \Delta^{\mu\nu}   \partial_\nu \left(\frac{\mu}{T}\right) +  \frac{\sigma}{2} q E^\mu \,\, \\
 \label{eq:jl:Cfluid2}
 &&  J^{\mu, (1)}_{anomalous}   = \xi_B  \, B^\mu + \xi_\omega \, \omega^\mu
\end{eqnarray}
Here $E^\mu=F^{\mu\nu} u_\nu$ and $B^\mu = \frac{1}{2} \epsilon^{\mu \nu \alpha \beta} u_\nu F_{\alpha \beta}$ are external electromagnetic fields in the fluid local rest frame. The covariant fluid vorticity is given by $\omega^\mu=\frac{1}{2} \epsilon^{\mu \nu \alpha \beta} u_\nu \partial_\alpha u_\beta$. While $0^{th}$ and $1^{st}$ order currents, $J^{\mu, (0)}$ and $J^{\mu, (1)}_{viscous}$, remain the same as in usual fluid dynamics of Eq. (\ref{eq:jl:Jfluid2}), we find  two new terms in  $J^{\mu, (1)}_{anomalous}$ which correspond precisely to the CME and CVE respectively.  The two coefficients $\xi_B$ and $\xi_\omega$ are entirely fixed by chiral anomaly and thermodynamics, and (in appropriate frame) reproduce exactly the coefficients in Eqs. (\ref{eq:jl:cme}) and (\ref{eq:jl:cve}). Different from the viscous currents, these anomalous currents are universal for various chiral fluids. Therefore, this new or extended type of fluid dynamics  incorporates macroscopic quantum transport currents arising from microscopic anomaly and  distinguishes the ``left'' from the ``right''. It also provides a useful framework for the quantitative modeling of anomalous chiral transport in heavy ion collisions~\cite{Jiang:2016wve,Shi:2017cpu,Yin:2015fca,Hirono:2014oda,Hongo:2013cqa,Yee:2013cya}, to be discussed later in Sec.~7.

\subsubsection{Chiral kinetic theory}
\label{subsec_5_1_kinetic}

While the preceding discussions address the macroscopic effects of the anomaly for chiral matter in or near thermal equilibrium, there remains the question about the behavior of such system in an out-of-equilibrium situation. 
%Given the emergence of anomalous chiral transport in the fluid dynamical context, it is quite obvious that similar phenomena must manifest themselves in a certain way even in a far-from-equilibrium setting.  
The natural framework to address this situation is  kinetic theory based on transport equations for the phase space distribution function of such a system. Different from classical kinetic theory, a proper description of the chiral fermions must account for intrinsic quantum and relativistic effects. During the last several years, such a kinetic chiral kinetic theory has been developed using a variety of approaches, see e.g.~\cite{Son:2012wh,Son:2012zy,Stephanov:2012ki,Chen:2014cla,Chen:2015gta,Chen:2012ca,Gao:2012ix,Hidaka:2016yjf,Hidaka:2017auj,Mueller:2017lzw,Huang:2018wdl}. Also, several phenomenological attempts to study out-of-equilibrium anomalous chiral transport have been proposed~\cite{Mace:2016shq,Ebihara:2017suq,Sun:2016nig,Huang:2017tsq}. In the following we will outline the key elements of the chiral kinetic theory.

Let us start by considering the classical kinetic theory, which describes the time evolution of  the classical phase space distribution   $f^{(c)}(t,{\mathbf x},{\mathbf p})$ for a given type of particles. Various physical quantities of interest (e.g. physical density and current, energy-momentum tensor, etc) may be constructed by properly integrating the distribution. The corresponding transport equation (under the presence of external electromagnetic fields $\mathbf{E}$ and  $\mathbf{B}$) reads~\cite{pitaevskii2012physical}: 
\begin{eqnarray} \label{eq:jl:ckinetic}
&& \Bigg\{ \partial_{t} + \dot{\mathbf x} \cdot \vec{\triangledown}_{\mathbf{x}} + \dot{\mathbf p}\cdot \vec{\triangledown}_{\mathbf{p}}  \Bigg \}  f^{(c)}(t,\mathbf{x},\mathbf{p}) = \mathcal{C}[f^{(c)}]  \,\, , \\
\label{eq:jl:ceom} 
&& \dot{\mathbf{x}} =   \mathbf{v}  =\vec{\triangledown}_{\mathbf{p}} \, E_{\mathbf{p}} \,\, , \,\, 
 \dot{\mathbf{p}} =  q \left ( \mathbf{E} + \mathbf{v}\times \mathbf{B} \right) \,\, .
\end{eqnarray}
In the above, $q$ is the charge of the particle and $E_{\mathbf{p}}$ is the single particle energy for given momentum.  Notably, the second line (\ref{eq:jl:ceom}) gives the classical equation of motion: individual particles simply follow classical trajectories in between sequential collisions with other particles. The  $\mathcal{C}[f^{(c)}]$ schematically represents the collision term, and its precise form would depend on the details of scattering processes under consideration (see e.g. \cite{pitaevskii2012physical}).

Let us next consider the kinetic description of chiral fermions. In this case, the spin degrees of freedom cannot be ignored and thus quantum mechanical effects have to be accounted for. For example, a spin-$\frac{1}{2}$ charged fermion will have a nonzero magnetic moment which interacts with an external magnetic field, in addition to the classical Coulomb and Lorentz forces. More importantly,  a fermion with fixed chirality has its spin and momentum states ``locked up'', for example a RH fermion has its  spin and momentum always aligned. Consequently a change of its momentum (e.g. when the particle moves in an external magnetic field) will necessarily be accompanied by the appropriate change in its spin direction and vice versa. To properly treat the spin degrees of freedom and thus chiral fermions in kinetic theory one therefore has to adopt a quantum transport theory such as the Wigner function formalism discussed in ~\cite{Vasak:1987um,Zhuang:1995pd}. It turns out that a semi-classical treatment which takes into account quantum effects up to the first order in $\hbar$  appears to be sufficient for describing the anomaly-induced transport~\cite{Son:2012wh,Son:2012zy,Stephanov:2012ki}.   In such a chiral kinetic theory (CKT), the corresponding transport equation for the semi-classical phase space distribution $f^{(q)}$  can be written as: 
\begin{eqnarray} \label{eq:jl:qkinetic}
&& \Bigg\{ \partial_{t} + {\mathbf{G}}_{\mathbf{x}} \cdot\triangledown_{\mathbf{x}} + {\mathbf{G}}_{\mathbf{p}} \cdot\triangledown_{\mathbf{p}}  \Bigg \}  f^{(q)}(t,\mathbf{x},\mathbf{p}) = \mathcal{C}[f^{(q)}]  \,\, , \\
\label{eq:jl:qeom} 
&&  {\mathbf{G}}_{\mathbf{x}} = \frac{1}{\sqrt{G}} \left [ \widetilde{\mathbf{v}}+\hbar q(\widetilde{\mathbf{v}}\cdot\mathbf{b}_{\chi})\mathbf{B}+\hbar q\widetilde{\mathbf{E}}\times\mathbf{b}_{\chi} \right] \,\, , \,\, 
 {\mathbf{G}}_{\mathbf{p}} = \frac{ q}{\sqrt{G}} \left[\widetilde{\mathbf{E}}+\widetilde{\mathbf{v}}\times\mathbf{B}+\hbar q(\widetilde{\mathbf{E}}\cdot\mathbf{B})\mathbf{b}_{\chi} \right] \,\, .
\end{eqnarray}
with the  Jacobian factor $\sqrt{G}=\left(1+\hbar q\mathbf{b}_{\chi}\cdot\mathbf{B} \right)$ and   
$\widetilde{\mathbf{E}} =  \mathbf{E}-\frac{1}{q}\triangledown_{\mathbf{x}}E_{\mathbf{p}}$. The single particle energy is given by $E_{\mathbf{p}}=|\mathbf{p}|\left( 1-\hbar q  \mathbf{B}\cdot \bf{b}_{\chi}\right)$ and the corresponding velocity is $\widetilde{\mathbf{v}}=\frac{\partial E_{\mathbf{p}}}{\partial\mathbf{p}}=\hat{\mathbf{p}}\left(1+2\hbar q\mathbf{B}\cdot\mathbf{b}_{\chi} \right)-\hbar q b_{\chi}\mathbf{B}$ with $\hat{\mathbf{p}}=\mathbf{p}/|\mathbf{p}|$. Here the factor $\mathbf{b}_{\chi}=\chi\frac{\mathbf{p}}{2|\mathbf{p}|^{3}}$ is the  Berry curvature, with $\chi=\pm 1$ for RH/LH fermions respectively.  The collision term also needs to be carefully modified in an appropriate way  (see, e.g., Ref. \cite{Chen:2015gta}).

In comparison with the classical kinetic theory, in the chiral kinetic theory the order-$\hat{O}(\hbar)$ level quantum effects appear in a number of places. First of all, the single particle energy $E_{\mathbf{p}}$ receives a shift by an order-$\hat{O}(\hbar)$ level magnetization energy which accounts for the interaction between the particle's magnetic moment and the magnetic field. This also results in a quantum correction to the  corresponding particle velocity $\widetilde{\mathbf{v}}$. Further quantum corrections at the same order in  ${\mathbf{G}}_{\mathbf{x}}$ and ${\mathbf{G}}_{\mathbf{p}}$ arise from the so-called Berry phase, which originates from the adiabatic progression of the particle spin state in order to ``catch up'' with the change in momentum. Additionally, the phase space integration measure  gets modified by the factor $\sqrt{G}$.  Such a CKT formalism has been shown to correctly reproduce the chiral anomaly relation as well as the CME and CVE transport currents in or near the thermal equilibrium~\cite{Son:2012wh,Son:2012zy,Stephanov:2012ki}. Also, it allows the study of such anomalous transport  even  far from equilibrium, see e.g. recent examples in~\cite{Kharzeev:2016sut,Sun:2016nig,Huang:2017tsq}.  
Currently the exploration of the transport theory for chiral fermions is under intensive development, aiming to fully develop its conceptual foundation as well as various applications.

\subsection{Anomalous chiral transport in  heavy ion collisions}
\label{subsec_5_2}

As discussed above, the anomalous chiral transport effects (such as the CME, CVE and CMW) stem from highly nontrivial interplay between chirality, topology and anomaly, which all concern fundamental aspects of QCD. A conclusive  observation of them in heavy ion collisions would be of paramount significance,   providing the first tantalizing experimental verification for the high-temperature restoration of the spontaneously broken chiral symmetry of QCD as well as a direct manifestation of the elusive gluonic topological fluctuations. In the following let us discuss the necessary conditions for these effects to occur in heavy ion collisions. 

As previously discussed, both CME and CVE would  occur only under the presence of macroscopic chirality, i.e. requiring a nonzero chiral chemical potential $\mu_5$. But how does such a situation occur in the quark-gluon plasma created by heavy ion collisions? The answer lies in the gluonic topological fluctuations. These fluctuations are strong due to nonpertubative gauge coupling and they induce the chirality imbalance of the light quarks via Eq. (\ref{eq:jl:N5Qw}), therefore creating a sizable nonzero $\mu_5$ (or more precisely 
$\sqrt{\mu_5^{\, 2}}$) in the quark-gluon plasma on an event-by-event basis.  
The CMW, on the other hand, does not require a nonzero  chirality imbalance to begin with.   Given the initial conditions, the propagation of CMW   leads to specific charge distributions at the end,  to be discussed later.

%\vspace{-0.15in}
\begin{figure}[!hbt] 
\begin{center}
\includegraphics[height=5cm]{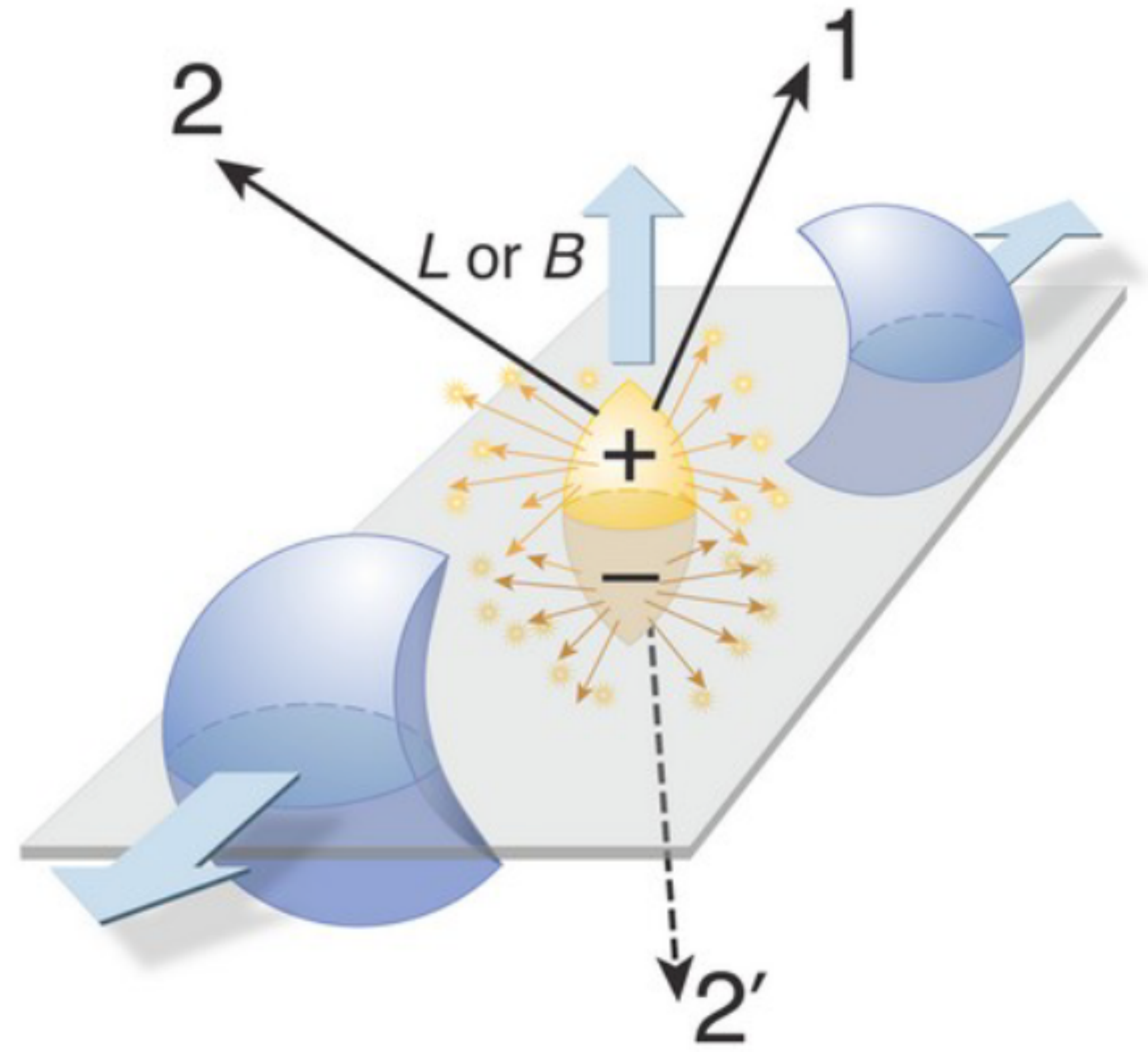} \hspace{0.5in}
\includegraphics[height=5cm]{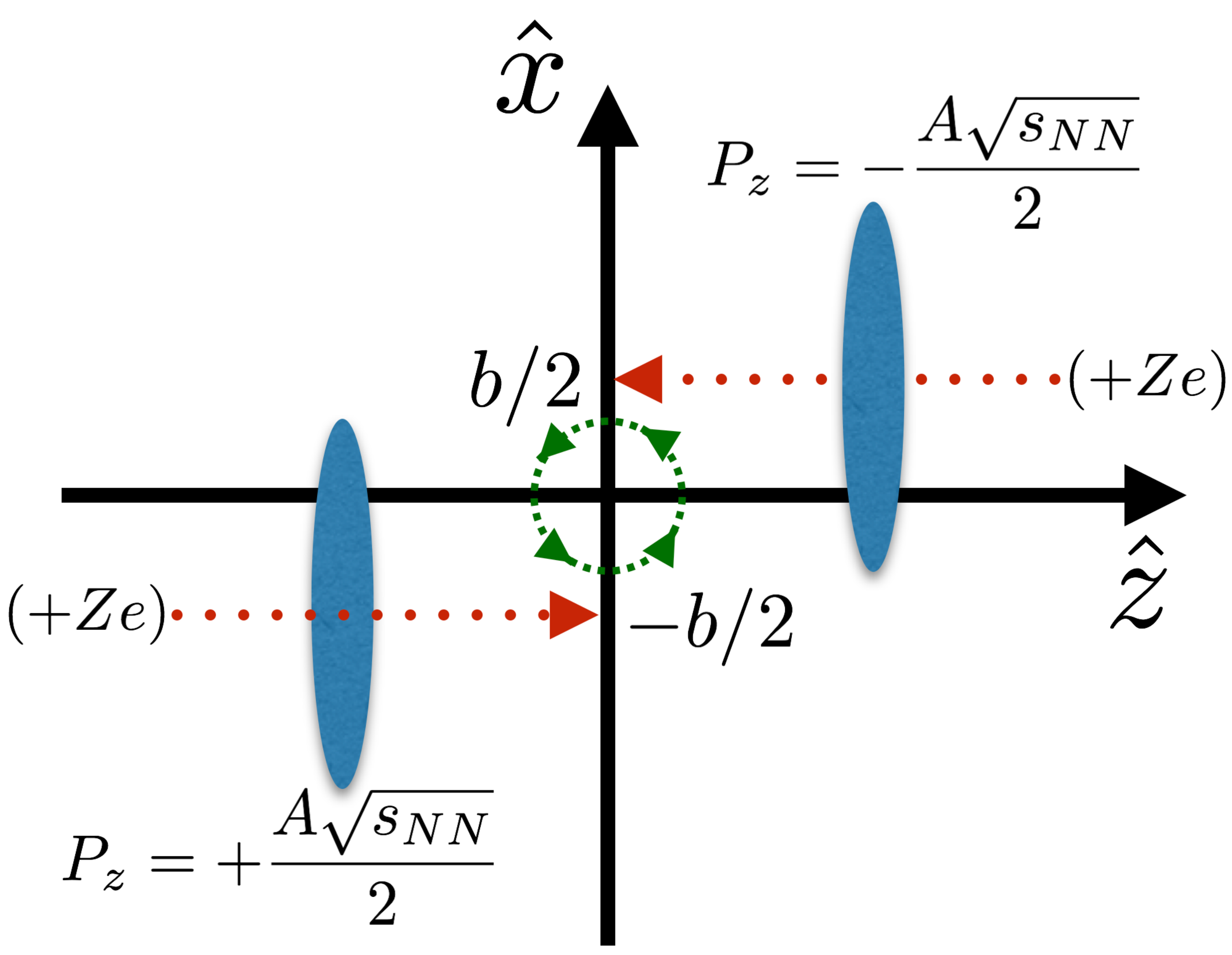}  
%\vspace{-0.1in}
\caption{Illustrations of the large angular momentum and strong magnetic field as well as possible CME-induced particle correlations  in a non-central heavy ion collision (see text for details). Left panel is adapted from \cite{Abelev:2009ac,APS_Physics_CME} (image credit: APS/Carin Cain/B. I. Abelev et al. (STAR Collaboration), Phys. Rev. Lett. (2009)).}  
 \label{fig:jl:f3}
%\vspace{-0.15in}
\end{center}
\end{figure}

Furthermore the necessary ``driving forces'' for these effects, namely the  fluid vorticity  and magnetic field, are available in  a typical off-central collision, as illustrated in Fig.~\ref{fig:jl:f3} (left). In the reaction plane spanned by the impact parameter direction $\hat{x}$ and beam direction $\hat{z}$ the two incident nuclei have their center-of-mass separated by impact parameter $b$, see Fig.~\ref{fig:jl:f3} (right). They both carry a longitudinal momentum $P_z=\pm A\sqrt{s_{NN}}/2$ (with opposite direction) where $A$ is the number of nucleons in each nucleus and $\sqrt{s_{NN}}$ is the center-of-mass energy per nucleon pair of the collision. 
Clearly such a colliding system carries a  considerable orbital angular momentum,  $L=bA\sqrt{s_{NN}}/2$.   For example, a collision of AuAu at a center of mass energy of $\sqrt{s_{NN}}=200\gev$ with $b=8$ fm has $L\simeq 1.6\times 10^5 \hbar$. For comparison,  the highest spin states of atomic nuclei (which are typically heavy ones with a spatial volume similar to the fireball in a heavy ion collision) have their total angular momentum values only at the order of $10^{1\sim 2} \hbar$~\cite{deVoigt:1983zz,Afanasjev:2013zda}. 
The direction of this angular momentum is perpendicular to the reaction plane $\vec{L}\sim \hat{y}$. Given that each fast-moving nucleus also carries a large positive electric charge (e.g. $Z=79$ for Au), a collision configuration as illustrated in Fig.~\ref{fig:jl:f3} (right) would also imply a strong magnetic field along the same direction as the angular momentum.

%\vspace{-0.15in}
\begin{figure}[!hbt] 
\begin{center}
\includegraphics[height=5cm,width=5cm]{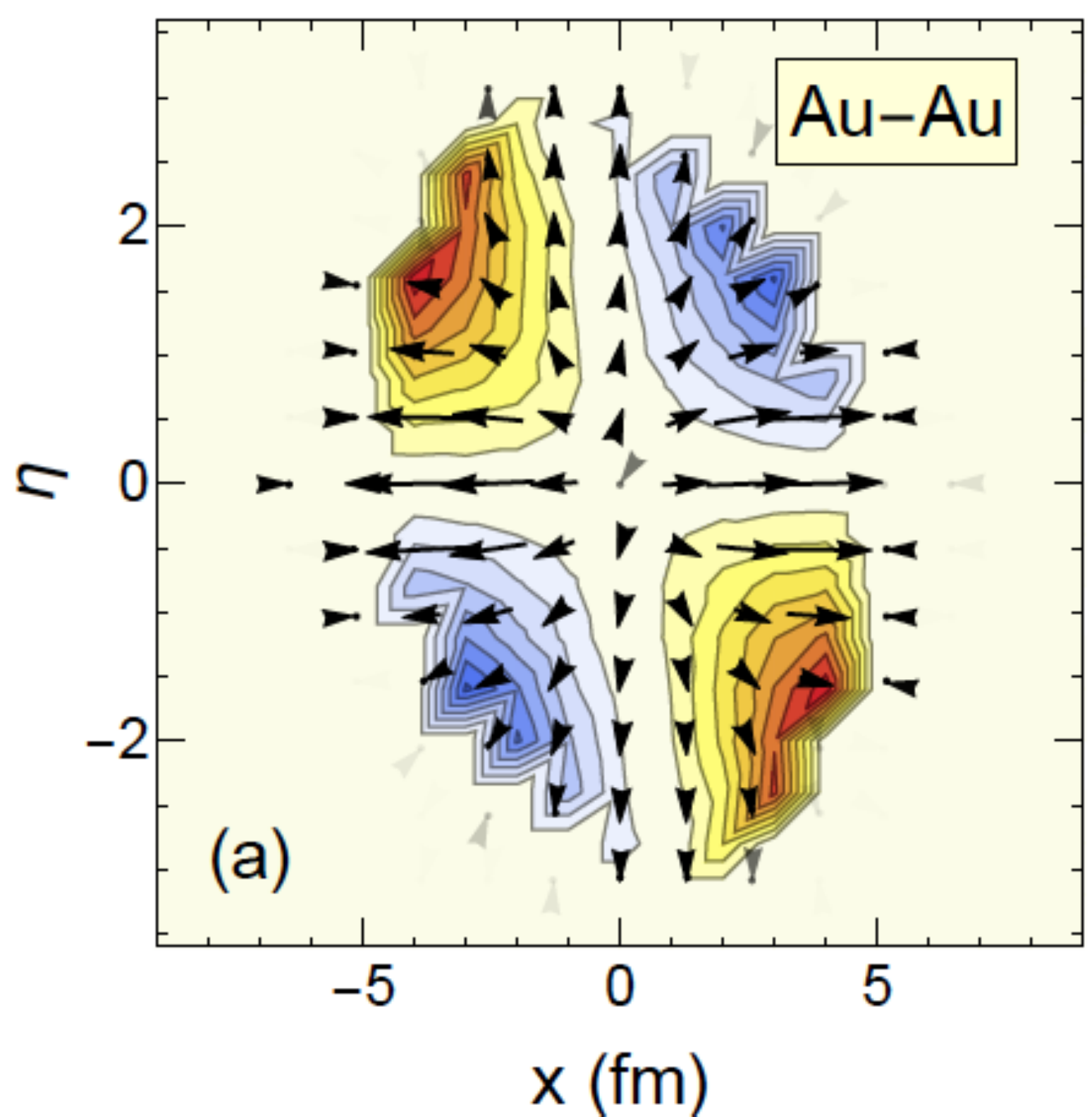} \hspace{0.5in}
\includegraphics[height=5cm,width=5.5cm]{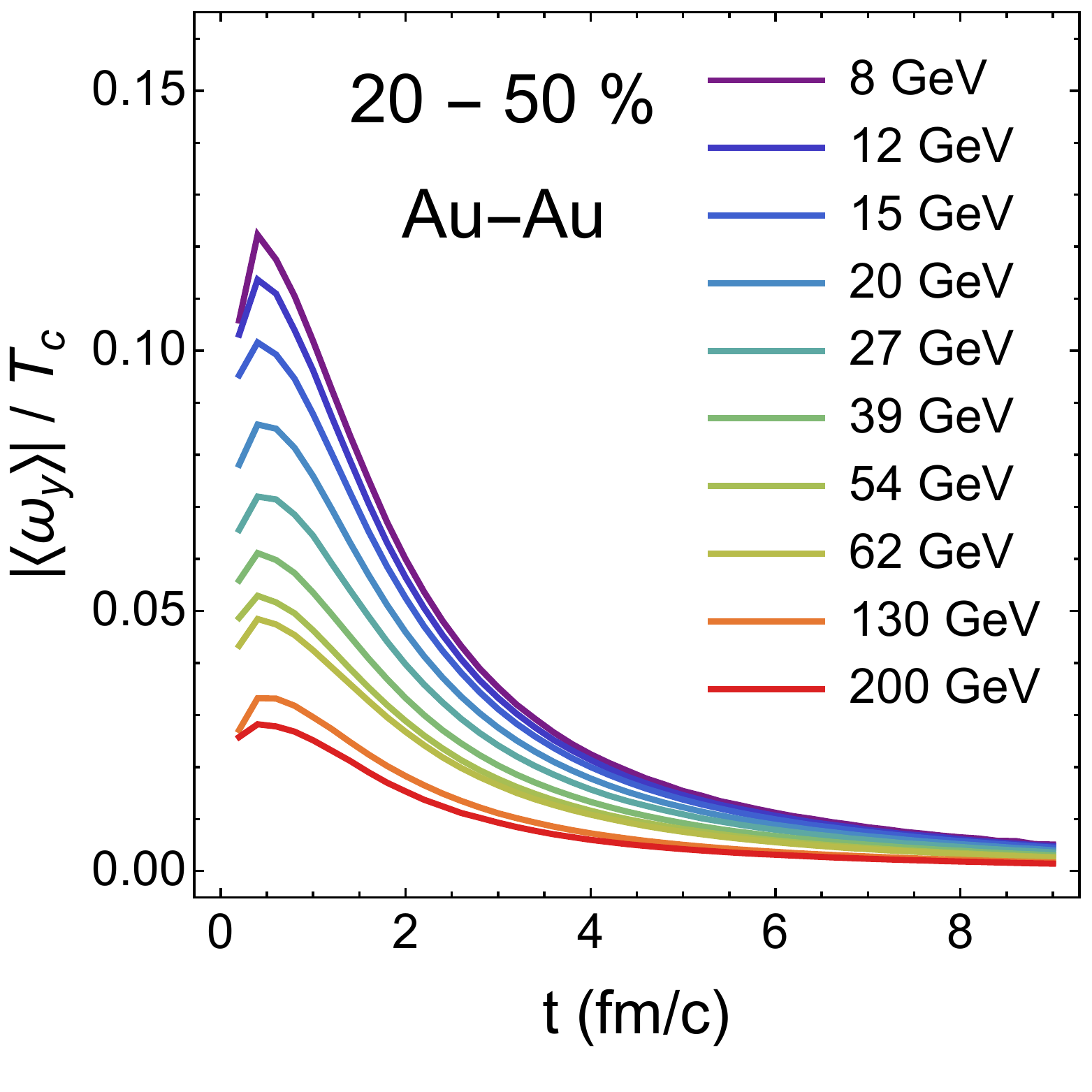}
%\vspace{-0.1in}
\caption{(left) The spatial distribution patterns of the fluid velocity (arrows) and vorticity (color contours)  on the $x-\eta$ plane at $y=0$ (with $\eta$ the spatial rapidity and $x,y$ the transverse coordinates along in/out-of plane directions),   {  for $(20-50)\%$ centrality} Au-Au  collisions at $\sqrt{s_{NN}}=20$ GeV;  (right) Averaged vorticity $|\la \omega _y\ra|$ (scaled by  $T_c=160$ MeV) versus time in Au-Au collisions for a wide range of beam energy, showing a monotonic increase with decreasing $\sqrt{s_{NN}}$. Figures are adapted from~\cite{Jiang:2016woz,Shi:2017wpk}.} 
 \label{fig:jl:f4}
%\vspace{-0.15in}
\end{center}
\end{figure}

The majority of the initial angular momentum is carried away by the spectator nucleons, i.e. those outside the overlapping zone and not actively colliding, see Fig.~\ref{fig:jl:f3} (left). Nevertheless, a fraction of the angular momentum, about $10\sim20\%$~\cite{Jiang:2016woz}, remains to be carried by the hot matter created in the overlapping zone. Despite the subsequent complicated evolution of the hot fireball,  angular momentum is conserved and therefore will be distributed across the system's constituents, affecting both their orbital motion and their individual spins. As a consequence, as the fireball undergoes collective expansion in a fluid dynamical way, rotational orbital motion develops, which is characterized by nonzero fluid vorticity $\bm{\omega}$. Owing to the compressible nature of this fluid, it will develop a nontrivial distribution of local fluid vorticity which has a nonzero average $\la \bm{\omega} \ra$ over the whole fluid  along the out-of-plane direction~\cite{Jiang:2016woz,Shi:2017wpk,Becattini:2015ska,Csernai:2013bqa,Deng:2016gyh,Pang:2016igs}.  
In Fig.~\ref{fig:jl:f4} (left) we show the results from AMPT model~\cite{Jiang:2016woz,Shi:2017wpk} for the fluid velocity and vorticity distribution patterns, with the time evolution of averaged vorticity  shown in Fig.~\ref{fig:jl:f4} (right) for a wide range of collision energies. We note that the averaged vorticity, $\la \bm{\omega}_y \ra$,  increases rapidly with decreasing collision energy, $\sqrt{s_{NN}}$. Thus the energy range explored by the RHIC beam energy scan (BES)  appears to be much preferred for the search of any vorticity-driven effects. Furthermore the nonzero global angular momentum and average vorticity would also imply a preference of their spins to be aligned along the same direction due to the spin-orbital coupling~\cite{Liang:2004ph,Voloshin:2004ha,Becattini:2007sr,Becattini:2013vja,Becattini:2016gvu,Pang:2016igs,Jiang:2016woz,Shi:2017wpk,Li:2017slc}. 
 A robust evidence of such global polarization effect has been reported recently by the STAR Collaboration, via sophisticated measurements of the hyperon spin orientation~\cite{STAR:2017ckg}.

%\vspace{-0.15in}
\begin{figure}[!hbt] 
\begin{center}
\includegraphics[height=4.6cm]{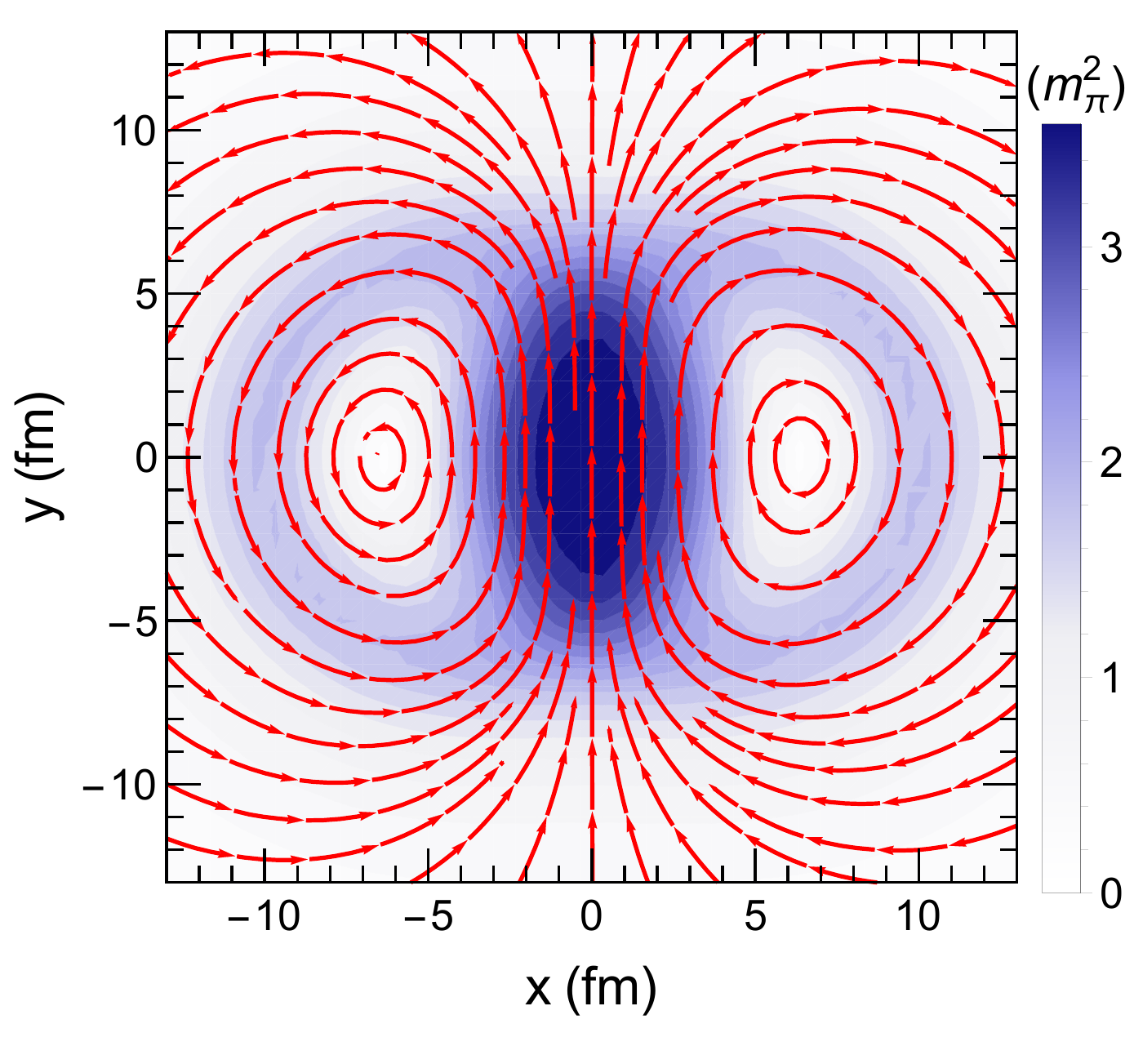}\hspace{0.6cm}
\includegraphics[height=4.6cm,width=5.4cm]{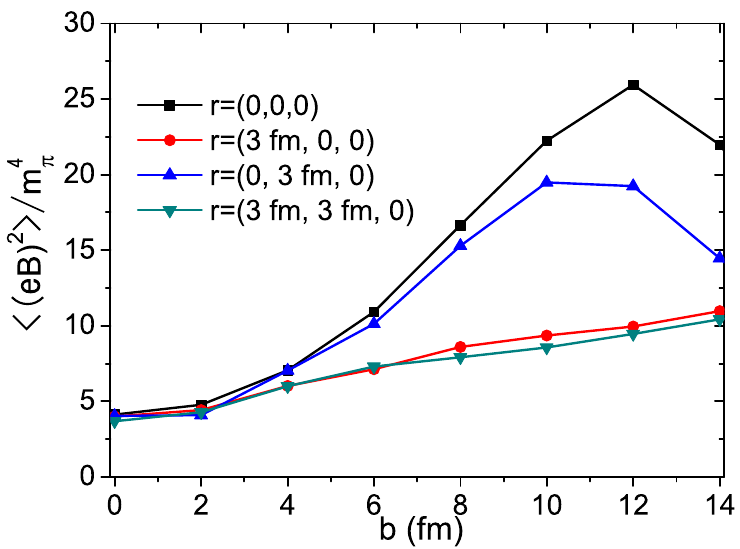}\hspace{0.6cm}
\includegraphics[height=4.4cm]{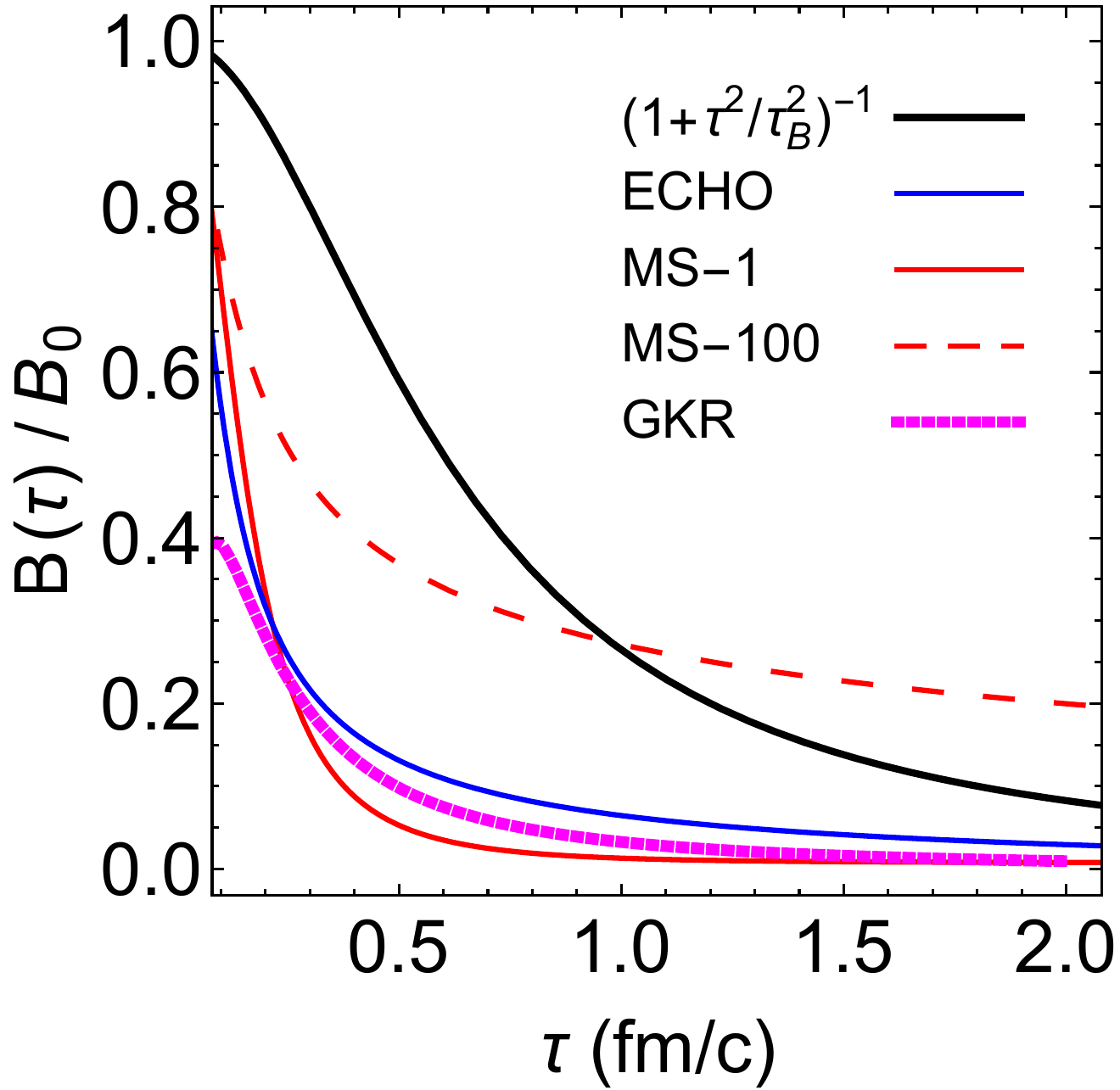}
%\vspace{-0.1in}
\caption{(left) The spatial distribution of the magnetic field orientation (arrow) and strength (color contour), computed for Au-Au collisions at $b=8$ fm and $\sqrt{s_{NN}}=200$ GeV; (middle) The magnetic field strength, measured by $\langle (eB)^2 \rangle /m_\pi^4$, at several points on the transverse plane in the collision zone, from event-by-event simulations; (right) The possible magnetic field time dependence computed in various models, with ECHO from \cite{Inghirami:2016iru}, MS-1 and MS-100 from \cite{McLerran:2013hla}, and GKR from \cite{Gursoy:2014aka}.  The middle panel is adapted from~\cite{Bloczynski:2012en} and the right panel is adapted from~\cite{Huang:2017tsq}.  } 
 \label{fig:jl:f5}
%\vspace{-0.15in}
\end{center}
\end{figure}

We next examine the strong magnetic fields from the two colliding ions which are similar to  two oppositely-running strong electric currents, as depicted in Fig.~\ref{fig:jl:f3} (right). An example of the spatial distribution of the magnetic field orientation and strength, calculated for Au-Au collisions at $b=8$ fm and $\sqrt{s_{NN}}=200$ GeV, is shown in Fig.~\ref{fig:jl:f5} (left).   In the overlapping zone the magnetic fields from each current add and  become very strong, pointing approximately along the angular momentum (i.e. the out-of-plane) direction.  A simple estimate suggests the strength of such field to be $(eB) \sim \frac{Z\alpha_{EM}\, \gamma}{R_A^2}$ where $\alpha_{EM}=1/137$, $R_A\simeq 1.1\, A^{1/3}$ fm is the nuclear radius and $\gamma = \sqrt{s_{NN}}/(2M_N)$ is the Lorentz factor for the moving nucleons of mass $M_N\simeq 0.939$ GeV. For example a AuAu collision at $\sqrt{s_{NN}}=200$ GeV would have  $(eB) \sim 3 m_\pi^2$ or about $10^{17}$ Gauss,  which is significantly stronger than any other known sources of magnetic fields.  
For comparison the magnetic field of a Magnetar, a neutron star which carries extremely strong magnetic field, is estimated to be on the order of $10^{14\sim 15}$ Guass~\cite{Kaspi:2017fwg}. 
Quantitative computations of such magnetic field strength, using event-by-event simulations, have been done at various beam energies and centralities as well as for different colliding systems~\cite{Bzdak:2011yy,Deng:2012pc,Bloczynski:2012en,Bloczynski:2013mca}, with an example shown in Fig.~\ref{fig:jl:f5} (middle)~\cite{Bloczynski:2012en}.   
While many details of the magnetic fields are known at the beginning of a heavy ion collision, its subsequent time evolution is, however, less understood at the moment. The dominant contribution to the initial magnetic fields comes from the spectators which quickly fly away from the fireball.  Thus, naively one would expect $eB$ in the collision zone  to decay away rapidly, on a time scale as short as $\sim 2R_A/\gamma$. However, it has been argued that the partonic medium containing charged quarks/antiquarks could prevent the decrease of the magnetic field through the induction mechanism just as in the Lenz's Law~\cite{Tuchin:2013apa,McLerran:2013hla,Gursoy:2014aka,Inghirami:2016iru}. Fig.~\ref{fig:jl:f5} (right) shows the widespread model results and parameterizations of the magnetic field time dependence,  illustrating the significant uncertainty associated with this quantity. To further constrain the time evolution of the magnetic field is the most important challenge for the quantitative modeling of magnetic field driven effects.

\subsection{Experimental signals of anomalous chiral transport} 
\label{subsec_5_3}

Given that the conditions for anomalous chiral transport effects are in principle realized in the environment created in heavy ion collision, it is of great interest to examine their potential signals in order to verify these phenomena in experiment. Here, we will focus on the magnetic field driven effects, the CME and CMW, for which intensive experimental efforts have been carried out.  

The CME leads to an electric current $\bf{J}_e$ along the magnetic field $\mathbf{B}$ direction which is approximately perpendicular to the reaction plane, as indicated by the grey plane in the left panel of Fig.~\ref{fig:jl:f3} .  This current implies transport of excessive positive charges toward one tip of the fireball (e.g. above the reaction plane) and transport of excessive negative charges toward the other tip of the fireball (e.g. below the reaction plane). As a result, a charge separation across the reaction plane emerges~\cite{Kharzeev:2007jp,Kharzeev:2004ey}, as illustrated in the left panel of Fig.~\ref{fig:jl:f3}. Such a spatial separation of positive/negative charges will convert into a charge dipole in the momentum space distribution of final state hadrons, and  the strong outward or radial expansion of the fireball helps amplify this effect. Taking the configuration shown in Fig.~\ref{fig:jl:f3} (left) as an example, the positive charges above the reaction plane will have their momenta preferably pointing upward while the negative point mostly  downward. This results in a charge asymmetry in the positive/negative hadrons'  distributions in azimuthal angle $\phi$ with respect to the reaction plane (RP): 
\begin{eqnarray} \label{eq:jl:separation}
\frac{dN_{\pm}}{d\phi} \propto  1 \pm a_1 \sin\left( \phi - \Psi_{RP} \right) + ...  
\end{eqnarray}
%Such charge-dependent azimuthal distributions are readily measurable. 
There is however a subtlety: the CME current could be either parallel or antiparallel to the $\mathbf{B}$ field, depending upon the sign of chirality imbalance $\mu_5$ arising from event-wise fluctuations. Consequently, it is equally probable for the electric dipole in momentum space to be aligned or anti-aligned with the magnetic field or angular momentum. Therefore, the coefficient $a_1$ in Eq.~(\ref{eq:jl:separation}) also fluctuates between positive/negative values from event to event and averages to zero, $\langle a_1 \rangle \propto \langle \mu_5 \rangle  =  0$. This  simply reflects the fact that the parity symmetry is respected by QCD and is not violated in heavy ion collisions.   What can be measured, however, is the variance of this charge separation coefficient, $\sqrt{\langle a_1^2 \rangle } \propto \sqrt{ \langle \mu_5^2 \rangle }  \neq  0$. 
To measure the variance of $a_{1}$, the following azimuthal correlator $\gamma$ for a pair of charged hadrons with either same-sign (SS) or opposite-sign (OS) was proposed~\cite{Voloshin:2004vk}: 
\begin{eqnarray} \label{eq:jl:gamma}
\gamma^{\alpha \beta} = {\big \langle} \cos\left(\phi^\alpha+\phi^\beta - 2\Psi_2 \right) {\big \rangle}  
\end{eqnarray} 
where $\Psi_2$ is the 2nd harmonic event plane as an experimental proxy for the reaction plane. 
The correlation of pairs with same electric charge, $\gamma^{SS}$, has $\{ \alpha \beta \} \to \{ ++\}$ or $\{ -- \}$ while that for opposite charged pairs,  $\gamma^{OS}$, has $\{ \alpha \beta \} \to \{ +- \}$ or $\{ -+ \}$. A CME-induced charge dipole distribution in (\ref{eq:jl:separation}) will contribute to the above correlation: $\gamma^{SS}_{CME} \to - \langle a_1^2 \rangle $ and 
$\gamma^{OS}_{CME} \to + \langle a_1^2 \rangle $. 
To maximize the signal and reduce the backgrounds, one can further examine the difference between the correlation of same and opposite charged pairs, $\left( \gamma^{OS}-\gamma^{SS}\right) \sim 2 \langle a_1^2 \rangle $. Taking such difference would also make the  charge-independent backgrounds  cancel out.  Extensive measurements of the $\gamma$ correlations were done for a variety of colliding systems (AuAu, CuCu, UU, PbPb, pPb, dAu) by STAR,  ALICE and CMS over a wide span of beam energies~\cite{Abelev:2009ac,Abelev:2009ad,Abelev:2012pa,Khachatryan:2016got,Sirunyan:2017quh}.

The $\gamma$ correlator by itself, however, does not unambiguously determine if a pair is moving in parallel or anti-parallel to the reaction plane. For example: a pair of particles co-moving along the out-of-plane direction (which would be the case for CME-induced same-sign pairs) would have $\gamma<0$, but a pair of particles moving back-to-back along the in-plane direction would also have $\gamma<0$; likewise a pair of particles moving back-to-back along the out-of-plane direction (which would be the case for CME-induced opposite-sign pairs) and a pair of particles co-moving along the in-plane direction would both have $\gamma>0$. This ambiguity can be resolved by considering in addition the following correlator~\cite{Bzdak:2009fc,Bzdak:2012ia}
\begin{eqnarray}\label{eq:jl:delta}
\delta^{\alpha \beta} = {\big \langle} \cos\left(\phi^\alpha - \phi^\beta  \right) {\big \rangle} \,\, .
\end{eqnarray}  
In this case, a pair of particles co-moving along the out-of-plane direction  would have $\gamma<0$ and $\delta>0$, while a pair of particles moving back-to-back along the in-plane direction would also have $\gamma<0$ and $\delta<0$; likewise a pair of particles moving back-to-back along the out-of-plane direction would have $\gamma>0$ and $\delta<0$,  while a pair of particles co-moving along the in-plane direction would   have $\gamma>0$ and $\delta>0$. 
 
 A careful analysis of $\gamma$ and $\delta$ together indeed pointed to a dominance of background contributions~\cite{Bzdak:2009fc,Bzdak:2012ia}. Sources of those non-CME backgrounds include e.g. transverse momentum conservation, local charge conservation, resonance decay and clusters, etc~\cite{Bzdak:2009fc,Liao:2010nv,Bzdak:2010fd,Wang:2009kd,Pratt:2010gy,Schlichting:2010na,Pratt:2010zn,Chatterjee:2014sea,Tribedy:2017hwn}.   
Note that  the $\gamma$ correlator measures a difference between the in-plane  and out-of-plane projected azimuthal correlations. As a result, the background contributions to $\gamma$ are controlled by  the elliptic anisotropy coefficient $v_2$ that quantifies the difference between the in-plane versus out-of-plane  collective expansion of the bulk matter.  The substantial backgrounds pose a significant challenge for the experimental search of CME (see e.g. recent reviews~\cite{Kharzeev:2015znc,Bzdak:2012ia}). In order to remove or at least suppress backgrounds associated with the elliptic anisotropy, a two-component decomposition strategy was proposed~\cite{Bzdak:2012ia}: 
\begin{eqnarray} \label{eq:jl:FH}
\gamma^{\alpha \beta} = \kappa \, v_2 \, F^{\alpha \beta}  - H^{\alpha \beta} \quad , \quad  \delta^{\alpha \beta} =  F^{\alpha \beta}  +  H^{\alpha \beta}
\end{eqnarray} 
where the $H$  represents the CME-like out-of-plane correlations while the $F$ represents the various  background correlations whose contributions to $\gamma$ are modulated by the elliptic anisotropy $v_2$ as well as a kinematic factor $\kappa$. 
While one may expect $\kappa$ to be of the order unity, the precise value of $\kappa$ would sensitively depend on detector acceptance and analysis kinematic cut, as illustrated by the example of transverse momentum conservation in~\cite{Bzdak:2010fd}.  
By measuring $\gamma$, $\delta$ and $v_2$, one may extract the $H$-correlation under certain plausible assumption of the $\kappa$ factor. Analysis based on this method was successfully carried out by STAR~\cite{Adamczyk:2014mzf}. Clearly  the $\gamma$ or $H$ correlations are not perfect   due to strong background contamination, and a firm conclusion would require efforts to accurately calculate the signal, to  get those backgrounds under full control, and to develop additional observables sensitive to the CME.   { We defer a detailed treatment of these issues to subsections 6.4 and 7.2 where the current status of measurements, their interpretations as well as pertinent background contributions will be discussed. }

Another interesting experimental signal is related to the Chiral Magnetic Wave (CMW). As previously discussed, small fluctuations of vector and axial (or equivalently the RH and LH) charge densities propagate as CMW under the presence of a magnetic field, just like sound waves. In particular the RH wave propagates along the $\mathbf{B}$ field direction while the LH wave propagates in the opposite direction. If the fireball created in heavy ion collisions has a nonzero electric charge density (which is an equal mix of RH and LH) in its initial condition, then a simple calculation in terms of CMW solutions suggests that the RH wave moves toward one tip of the fireball along the out-of-plane direction while the LH wave moves toward the opposite tip. 
As a result, the nonzero initial (say, positive) electric charges  will  be transported toward the two tips along the azimuthal direction $(\phi-\Psi_{RP})=\pm \frac{\pi}{2}$. Due to charge conservation, this also leaves  less positive charges around the ``equator'' region of the fireball along the azimuthal direction $(\phi-\Psi_{RP})=0$ or $ \pi$. Such a spatial  distribution pattern of the electric charge on the transverse plane has a nonzero quadruple moment with respect to the reaction plane, as first predicted and estimated in~\cite{Burnier:2011bf}.  
%The QGP fireball in heavy ion collisions is not an infinitely large medium, and the waves stop at the boundary of the cold hadronic corona surrounding the QGP. 
%A realistic estimate predicts that a quadruple pattern of the electric charge density distribution will emerge in the fireball, with excessive charges of one sign on both tips of the fireball while excessive charges of the opposite sign on the equator of the fireball~\cite{Burnier:2011bf}.   
Again,  the strong radial flow will ``blow'' all the charges outward and correlate their momenta with positions. As a consequence, the spatial quadruple pattern of the electric charge distribution will translate into a quadruple pattern in the final state charged hadrons' azimuthal distributions with respect to the reaction plane. That is, there will be  a little more in-plane charged hadrons of one sign while a little more out-of-plane charged hadrons of the opposite sign. Such a pattern implies a splitting in the elliptic flow coefficient $v_2$ for the positive/negative hadrons. The magnitude of this effect is proportional to  the CMW-induced quadruple moment which in turn is determined by the amount of initial electric charge density. If the initial density is positive (negative), more positive (negative) charges will be transported by CMW to the out-of-plane tips of the fireball, leading to a  smaller $v_2$ for the positive (negative) charged hadrons as compared with hadrons of the opposite sign. This initial density   mainly comes from random ``stopping'' of the initial charges carried by the colliding nuclei. One could use the observed charge asymmetry $A_{ch}=\frac{N_+ - N_-}{N_+ + N_-}$ of an event (that is, the difference in the measured numbers, $N_{\pm}$ of positive/negative charged hadrons in this event) as a proxy  which is correlated with  the initial nonzero charge density of an event.  
It was predicted~\cite{Burnier:2011bf} that a specific splitting of the  elliptic flow for positively and negatively charged pions should be observed: 
\begin{eqnarray} \label{eq:jl:v2Ach}
\Delta v_2 = v_2^{\pi-} - v_2^{\pi+} \simeq  r A_{ch} + \Delta^{base}
\end{eqnarray}
In other words, the CMW predicts  a linear dependence of the splitting on the overall charge asymmetry $A_{ch}$, with the slope parameter $r$ computed from the CMW. The  intercept $\Delta^{base}$ is independent of $A_{ch}$ and unrelated to CMW. This prediction was  experimentally verified later by the STAR collaboration at RHIC~\cite{Adamczyk:2015eqo}. Similar measurements were carried out by ALICE and CMS  at the LHC as well~\cite{Adam:2015vje,Sirunyan:2017tax}. Certain potential background contributions were proposed and studied, such as the local charge conservation~\cite{Bzdak:2013yla} and viscous transport~\cite{Hatta:2015hca}. At the moment  the CMW  appears to be the viable 
  interpretation  for the measured flow splitting signal at RHIC energies.

\subsection{The beam energy dependence of anomalous chiral transport}
\label{subsec_5_4}

In this last subsection, we discuss how the anomalous chiral transport may vary with the collision energy, which is an important question for the BES experimental program. The answer relies upon a number of key ingredients, to be discussed below. 

First, the necessary condition for anomalous transport to occur is a sufficiently hot environment such that  a substantial amount, in terms of both spatial volume and lifetime, of quark-gluon plasma with restored chiral symmetry is created~\footnote{  {As discussed in subsection 5.1.1., a vacuum condensate would spoil chirality of quarks and the restoration of chiral symmetry in high temperature phase is a prerequisite for CME. }}.  One would then expect that with increasing collision energy this condition is satisfied better and better.   {Another relevant factor is the topological transition rate which controls the initial chirality imbalance for CME and which in general increases with the relevant medium scale, be it the pre-thermal saturation scale for the early stage or the temperature scale in the quark-gluon plasma. In either case, such a rate is expected to increase with the collision energy.}
\footnote{{At present there is not yet a reliable way of precisely determining such a topological transition rate, which remains a major theoretical uncertainty. In fact, the main early motivation for studying CME was actually to help manifest and extract the topological transition rate.}}
Thus it is natural to  expect a ``threshold energy'' below which a QGP with sufficient topological transitions would ceases to exist in the fireball.  A number of measurements from BES-I (such as the constituent quark scaling of elliptic flow, jet energy loss, directed flow, net proton fluctuations, light cluster production, etc) appear to hint at a qualitative change in the observed properties of the created bulk matter occurring around the beam energy range of $\sqrt{s_{NN}}\simeq (10\sim 20)$ GeV, as we shall discuss in Sec.~6 in more details.  
Therefore, one should expect any signal from anomalous chiral transport to decrease towards the low enough beam energy regime and to eventually turn off when the collision energy drops below the QGP production threshold.

The other necessary condition is the presence of the ``driving forces'', i.e. the fluid vorticity $\bm{\omega}$ and the magnetic field $\mathbf{B}$. As already shown in Fig.~\ref{fig:jl:f4}, the vorticity $\bm{\omega}$ decreases with increasing beam energy and becomes negligible at the high energy end.  The situation for the magnetic field is more complicated, due to an important difference between angular momentum and magnetic field: the former is conserved during the system evolution while the latter is strongly dependent on time. The peak strength of $\mathbf{B}$ scales with the nucleon Lorentz factor $\gamma \propto \sqrt{s_{NN}}$ and thus increases with increasing beam energy. On the other hand, without any medium feedback the time duration of this strong initial magnetic field scales inversely with $\gamma$  and thus decreases rapidly with increasing beam energy. This time scale can be estimated as $\tau_B \sim \frac{2R}{\gamma}\sim \frac{4R M_N}{\sqrt{s_{NN}}}$ with $R$ the nuclear radius (e.g. $R\sim 7$ fm for Au or Pb) and $M_N\simeq 0.939$ GeV the nucleon mass. Another important time scale is the formation time of the quarks after the initial collisions. Estimates of this timescale $\tau_f$ based on the glasma picture for the early stage in heavy ion collisions would suggest that 
$\tau_f \sim \frac{1}{Q_s} $, where $Q_s$ is the so-called saturation scale~\cite{Gelis:2010nm,Gelis:2012ri}.  $Q_s$ is expected to scale with beam energy in a specific way, $ {Q_s} \sim {Q_0} \left ( \frac{\sqrt{s_{NN}}}{E_0}  \right )^{\frac{0.3}{2}}$ with $Q_0$ the saturation scale at a reference energy scale $E_0$. We may use the RHIC values $Q_0\simeq 1.5$ GeV and $E_0\simeq 200$ GeV for a quick order-of-magnitude estimate~\cite{Gelis:2010nm,Gelis:2012ri}. Let us now compare the two time scales $\tau_B$ versus $\tau_f$, with the former decreasing rapidly with beam energy $\sqrt{s_{NN}}$ while the latter only decreasing mildly. Note that it  is the quarks that are needed for both the anomalous chiral transport and for the medium induction mechanism which has the potential of prolonging lifetime of the magnetic field. 
One may therefore expect that at certain high enough beam energy where $\tau_B\ll \tau_f$,  the initial magnetic field will exist only  like an extremely short pulse before the formation of any quark or antiquark medium and would not cause any anomalous transport. 
That is, the signals of anomalous chiral transport effects may be expected to eventually  {\em disappear at the high  beam energy end}. 
Let us give a concrete estimate for beam energies relevant to RHIC and the LHC. 
At RHIC top energy $\sqrt{s_{NN}}=200$ GeV, one has $\tau_B\sim 0.13$ fm/c and $\tau_f\sim 0.13$ fm/c, with the two scales  comparable $\tau_B\sim \tau_f$. At LHC energy e.g. $\sqrt{s_{NN}}=5020$ GeV,   $\tau_B \sim 0.005$ fm/c significantly decreases from RHIC while the $\tau_f\sim 0.08$ fm/c only slightly decreases, resulting in a situation where $\tau_B \ll \tau_f$.

To conclude this brief discussion, it is quite plausible that potential signals from anomalous chiral transport effects would have nontrivial  dependence on the collision beam energy, possibly  disappearing in collisions  both at very low energy end (e.g. below $\sim \hat{O}(10)\, \rm GeV$) and at very high energy end (e.g. beyond $\sim \hat{O}(1)\, \rm TeV$). From the estimate above, it appears 
quite likely that the  optimal  beam energy window   may be  in the  range of $\hat{O}(10\sim 100)\, \rm GeV$.  
Due to this non-monotonic trend, the beam energy scan program at RHIC provides the unique opportunity to look for the signals from anomalous chiral transport.  In fact, such a pattern of  beam energy dependence could in itself be considered as a characteristic signature  for the search of such effects. 

\newpage 

%=====================================================================================
%=====================================================================================
%

%%% Local Variables:
%%% mode: latex
%%% TeX-master: "BES_Main_current"
%%% End:

%-------------Section 6-----------------------------------
%latest in BES_current
%=====================================================================================
%=====================================================================================
\section{The Beam Energy Scan Program at RHIC}
\label{sec6}

\subsection{Introduction}
\label{sec:exp_intro}

Soon after the observation that the matter created at the top RHIC energy collision behaves like a
highly opaque, nearly perfect fluid \cite{Back:2004je,Arsene:2004fa,Adcox:2004mh,Adams:2005dq},
subsequently dubbed as the strongly coupled Quark Gluon Plasma (sQGP), a natural question arose: how
would such properties change as one increases or lowers the collision energy?  One way to look at
this question is from the perspective of QCD phase diagram as shown in
Fig.~\ref{fig:qcdpd}.  
Generally speaking, an increase/decrease of the collision energy would create
systems with higher/lower temperature and less/more net-baryon density.  If nearly equilibrated
QCD matter in the cross-over region at nearly vanishing baryon number density is being created at
RHIC, the expectation would be that at even higher energies, available at the LHC, the created
system would be in the same region and its basic thermodynamic properties should not change.
This expectation has meanwhile been confirmed by measurements at the LHC \cite{Aamodt:2010pa,Abelev:2013vea,ATLAS:2011ah,Chatrchyan:2012ta,Muller:2012zq}. 
Lowering the beam
energy, and thus probing the baryon-rich region of the phase diagram, one would expect qualitative
changes. Based on many models predictions, it is commonly expected that at sufficiently high baryon
density the crossover turns into a first-order phase transition with a critical point marking the
position in the phase diagram where the crossover ends and the first order transition begins.
If the critical point and/or first order transition happen to be within reach of low energy
collisions, one would expect certain significant changes in the properties of created matter. If, on
the other hand, there is no true phase transition or the transition is at too high a value of the
baryon density, one would at least expect the disappearance of the  hot  quark-gluon plasma at
sufficiently low collision energy.  In either case, it is plausible and tempting to expect
qualitative changes in many observables as one lowers the collision energy. It were these
considerations which gave rise to the RHIC beam energy scan (BES) program, which started its first
phase in 2010.

The first phase of the beam energy scan program (BES-I) at RHIC has been completed with great
success. The collision energies and event statistics for the BES-I as well as the planned second
phase (BES-II) are listed in Table~\ref{tab:exp:bes}.  The program covers the beam energies of
$\sqrt{s_{NN}}= 200, 62.4, 54.4, 39, 27, 19.6, 14.5, 11.5, 7.7\, \gev $ 
\footnote{   {By switching to fixed-target mode, the RHIC BES program has  been extended further toward even lower collision energies: see caption of Table.~\ref{tab:exp:bes}.}}
corresponding to a range of
chemical potentials $25 \lesssim \mu_B \lesssim 420 \mev$. \footnote{As we shall discuss below, at a given
  energy, especially at the lower beam energy, the value of the chemical potential varies with
  collision centrality. These values were extracted for the most central (0-5\%) Au+Au
  collisions.} Comprehensive experimental measurements of observables have been carried out by both
the PHENIX and STAR collaborations, 
aiming to characterize the system's bulk properties and its
collective expansion as well as to search for evidences that are sensitive to a possible   phase
transition and/or critical point as well as the aforementioned anomalous chiral transport effects.

While in the following we concentrate on the results from the RHIC Beam Energy Scan, we should mention that there are other experiments addressing the same physics. There are on-ongoing efforts at SPS energies by the NA61/SHINE Collaboration and its predecessor NA49. Their main focus is the search for possible structures in the QCD phase diagram with various fluctuation and correlation observables for various energies and colliding systems \cite{Gazdzicki:2015ska}. In particular, NA49 and NA61/SHINE obtained results on the scaled variance \cite{Aduszkiewicz:2015jna,Mackowiak-Pawlowska:2014ipa,Alt:2007jq,Alt:2006jr}, strongly intensive quantities \cite{Andronov:2018bln,Aduszkiewicz:2015jna,Anticic:2015fla,Anticic:2008aa,Alt:2004ir,Anticic:2003fd,Appelshauser:1999ft}, intermittency \cite{Davis:2017mzd,Anticic:2012xb,Anticic:2009pe}, and many others \cite{Prokhorova:2018tcl,Anticic:2012cf,Anticic:2011am,Alt:2008ab,Afanasev:2000fu}. Also, the HADES experiment at GSI Darmstadt, which measures at very low collision energies, has some capability to address fluctuation observables.
In addition, there are several planned and approved experiments,
namely: the Compressed Matter Experiment (CBM) at the future FAIR facility in Darmstadt, Germany, the
Multi Purpose Detector (MPD) experiment at the NICA facility in Dubna, Russia, the  CSR-External
target Experiment (CEE) at HIAF in Hushou,  China. Furthermore, there is the possibility to carry
out an experiment at J-PARC in Tokai, Japan. 

Let us now turn to reviewing the major experimental results from BES-I, 
 with an emphasis on the beam energy dependence of the various observables.

\begin{table}[htb]
\begin{center}
  \begin{tabular}{|c|c||c|c|c||c|c|c|} 
  \hline 
 \multirow{2}{*}{BES-II} & STAR                  & \multirow{2}{*}{BES-I} & STAR                  & PHENIX             &$\sqrt{s_{NN}}$ & $\mu_B$ & $T_{ch}$ \\ 
                                     & Events (10$^6$)  &                                    & Events (10$^6$) &Events (10$^6$)  &  (GeV)              & (MeV)      & (MeV) \\ \hline \hline
                  &                                   & 2010  &         238                     &    1681                        &200                               &  25                     & 166           \\ \hline
                  &                                   & 2010  &          45                      &    474                          &62.4                              &  73                     & 165           \\ \hline
                  &                                   & 2017  &        1200                    &                                    &54.4                              &  92                     & 165           \\ \hline
                  &                                   & 2010  &          86                      &    154                          &39                                 & 112                    & 164           \\ \hline
                  &                                   & 2011  &          32                      &     21                           &27                                  & 156                   & 162           \\ \hline
    2019 &            580                 & 2011  &          15                      &     6                             &19.6                              & 206                     & 160           \\ \hline
%     2021 &            ???                 &   &                               &                                    & 16.7                              & ???                    & ???           \\ \hline
    2019 &            320                 & 2014  &          13                      &                                    &14.5                              & 264                    & 156           \\ \hline
    2020-21 &            230                 & 2010  &          7                      &                                    &11.5                              & 315                    & 152           \\ \hline
    2020-21 &            160                 & 2008  &          0.3                     &                                    &9.2                                & 355                    & 140           \\ \hline
    2020-21 &            100                & 2010  &          3                        &    2                              &7.7                                & 420                    & 139           \\ \hline
  \end{tabular}
\caption{RHIC beam energy scan schedule (as originally approved by BNL management),  event statistics, collision energies as well as the corresponding values 
of the chemical freeze-out temperature and baryonic chemical potential. These values are for the most central Au+Au collisions at RHIC. 
   {The 2019 BES-II runs not shown in this Table include a short RHIC  electron-cooling test run (3M events at $\sqrt{s_{NN}}=7.7\rm GeV$) as well as a series of fixed-target mode collisions (50M events at $7.7\rm GeV$, 50M events at $3.9\rm GeV$, 200M events at $3.2\rm GeV$ and 300M events at $3.0\rm GeV$). In addition to the originally planned energies, a new request for collisions with  $\sqrt{s_{NN}}$ at about $17\rm GeV$ (possibly taking 250 million events in year 2021) has also been put forward. }
}
\label{tab:exp:bes}
\end{center}
\end{table}

%--===================================================================================
\subsection{Bulk properties and collectivity}
In order to characterize the system created in a heavy ion reaction one typically measures the
spectra for various hadrons from which one then obtains the hadron yields and extracts information about
possible collective expansion, also referred to as flow. Further information about the size of the
system and its lifetime is deduced from interferometry, or HBT measurements (see \cite{Lisa:2005dd}
for a review).  In this subsection we briefly review the collision energy dependence of these
measurements as obtained from the first phase of the RHIC beam energy scan (BES-I). 

\subsubsection{Particle spectra and ratios}
\label{sec:exp:spectra}

\begin{figure}[hbt] 
\begin{center}
\includegraphics[scale=0.39]{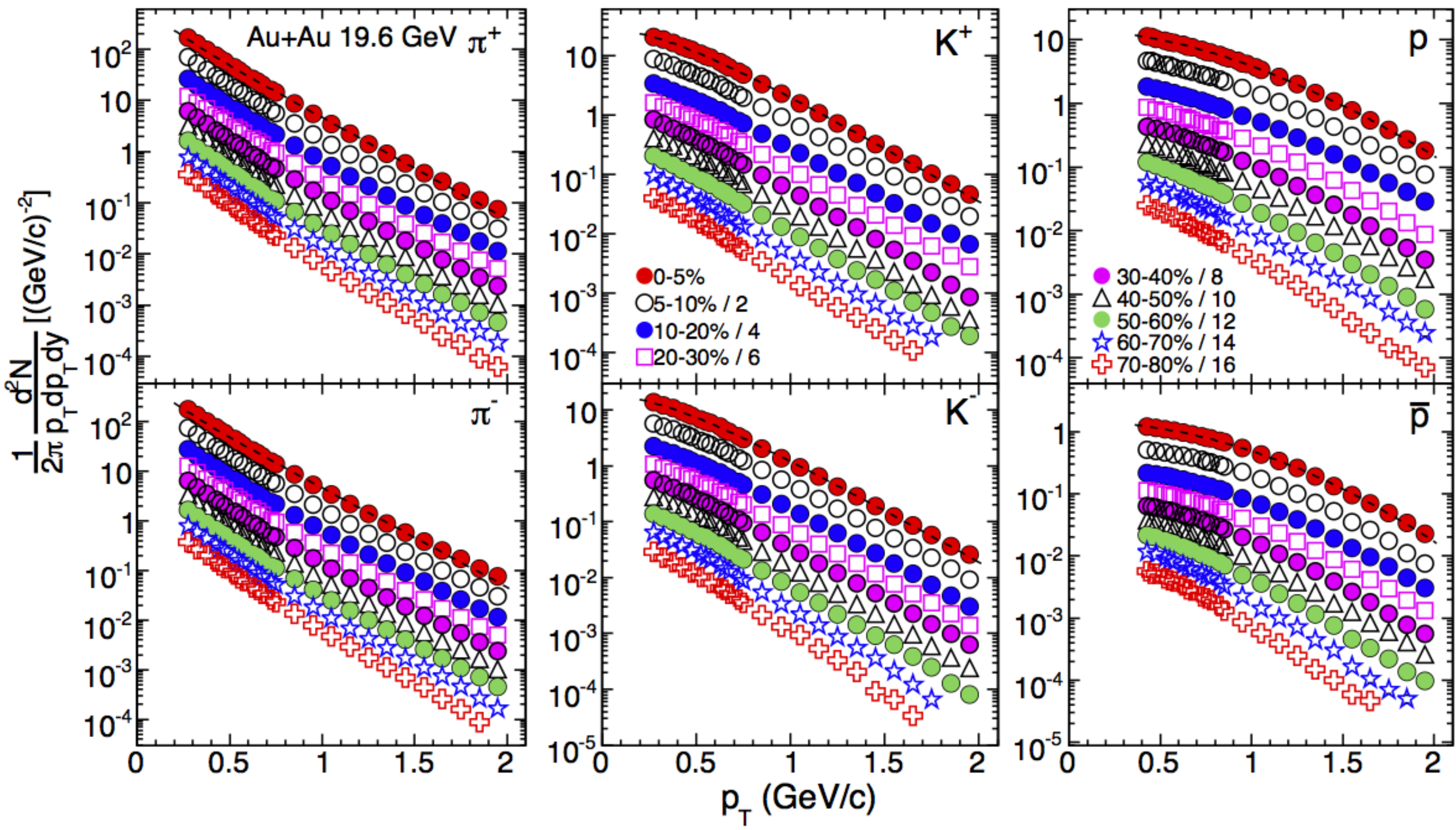}
\caption{
The centrality dependence of the hadron transverse momentum spectra from $19.6$ GeV Au+Au collisions. 
From top to bottom in each plot are different centrality bins: 0-5\%, 5-10\%, 10-20\% and 20-30\%, ..., 70-80\%, respectively. 
For the most central data, the dashed-lines are the results of the blast-wave fits. These plots are adapted  from 
Ref. \cite{Adamczyk:2017iwn} where the full sets of the RHIC BES-I spectra data are listed. 
}
 \label{fig:exp:f0}
\end{center}
\end{figure}

As part of BES-I, both the STAR and PHENIX experiments have measured  the collision
energy as well as centrality dependence of the transverse momentum spectra 
\cite{Adler:2004hv,Adler:2003cb,Adcox:2002au,Adams:2003xp,Adamczyk:2017iwn}
for the following hadron
species: $\pi^{\pm}$, $K^{\pm}$, $p$, $\bar{p}$, $\phi$, $\Lambda$, $\bar{\Lambda}$, $\Xi^{\pm}$,
$\Omega^{\pm}$ at the center of mass energies of 7.7, 11.5, 14.5, 19.6, 27, and 39~GeV. As an
example, the results of the acceptance and efficiency corrected spectra for $\pi^{\pm}$, $K^{\pm}$,
$p$, and $\bar{p}$ taken at $\sqrt{s_{NN}} = 19.6 \gev$, are shown in Fig.\ref{fig:exp:f0}
\cite{Adamczyk:2017iwn}.  These spectra are measured at mid-rapidity, $|y| \le 0.1$, 
and over a
transverse momentum range of $0.2 \le p_T \le 2 \gev/\rm c$. The hadron yields at mid-rapidity are then
obtained by extrapolating the measured $p_{T}$ spectra to $p_{T}=0$ and integrating them. In the case of
 the STAR measurements, this extrapolation gives rise to systematic uncertainties, which are of the
order of 9-11\%, 8-10\% and 11-13\% for pions, kaons and protons, respectively. Being a collider
detector, for a given type of hadron, the acceptance of the STAR detector does not change as a
function of collision energy. Therefore the systematic errors do not change much in these BES energies
as compared with fixed target experiments.
Given the hadron spectra and yields one can then proceed to extract some basic properties of the created system.

\begin{figure}[!hbt] 
\begin{center}
\includegraphics[width=0.9\textwidth]{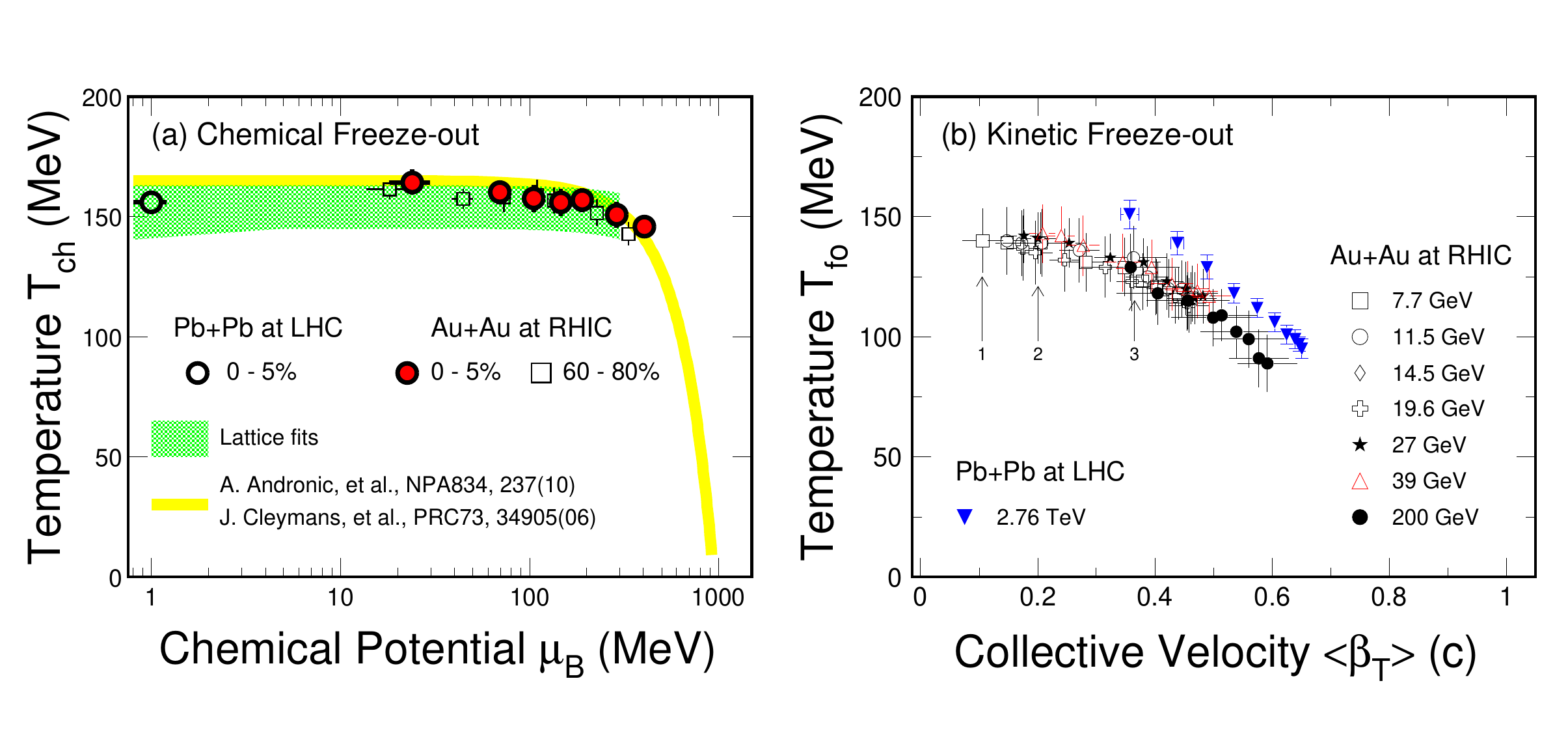}
\caption{
(Left-plot) Experimental results of the temperature versus baryonic chemical potential, both extracted at chemical freeze-out, for a variety of collision energies  from the RHIC BES-I \cite{Adamczyk:2017iwn}. Red-circles and black-squares represent results 
from the top 5\% and 60-80\% Au+Au collisions, respectively. Also shown is the result for Pb+Pb
collisions at the LHC \cite{Andronic:2017pug,Citron:2018lsq}. The hatched green-band represents the Lattice 
results for the region of the cross-over transition \cite{Bazavov:2011nk,Kaczmarek:2011zz}. The yellow-line shows the empirical thermal fits 
results \cite{Andronic:2009jd,Cleymans:2005xv}. (Right-plot) Blast-wave fit results of kinetic 
freeze-out temperature T$_{fo}$ and collectivity velocity $\langle \beta_T \rangle$. The error bars indicate 1-sigma values in the 2-D contour plot. The labels 1,2,3 in the right plot indicate the most peripheral collisions of  beam energy 7.7 GeV (1), 200 GeV (2) and 2.76 TeV (3) respectively. 
}
 \label{fig:exp:f1}
 \vspace{-0.15in}
\end{center}
\end{figure}

First, one observes that the shape of the spectra depends on the mass of the
particles. This can be understood as the result of collective radial expansion of the system. As
first discussed in \cite{Siemens:1978pb}, such a collective expansion leads to a larger apparent
temperature similar to the well known blue shift for photons. In addition, for massive particles the spectrum
flattens at low transverse momentum, and this flattening is more prominent for particles with  higher mass.  In order to test whether such a collective expansion is the plausible cause  for the difference in
the spectral shape of the various hadrons, as well as to extract the collective flow velocity and intrinsic 
temperature, one employs the so-called blast-wave model \cite{Schnedermann:1993ws}. This model
assumes a Hubble type radial expansion with a velocity profile, $\beta(r) = \langle \beta_{T} \rangle  \left(\frac{r}{R}\right)^{\alpha}$,
where $R$ is the radius of the source, 
$\langle \beta_{T} \rangle$ is the collective flow velocity at the surface,  
and $\alpha$ is some
parameter which is of the order of unity. The blast-wave model works remarkably
well in describing spectra of produced particles in heavy ion collisions at all energies.  
In Fig. \ref{fig:exp:f0}, we show the fit for the most central collisions as dashed lines. Given the
success of the blast-wave model, one can extract the flow velocity $\langle \beta_{T} \rangle$ and the temperature, $T_{fo}$, of the source at the moment of the so-called kinetic freeze-out when the particles cease to scatter with each other~\cite{Sinyukov:1998ex}.  The kinetic freeze-out temperature and collective velocity parameters, $T_{fo}$
and $\langle \beta_{T} \rangle$,  extracted from the blast-wave fits to the BES-I hadron spectra, are plotted in the right-panel of
Fig. \ref{fig:exp:f1}. 
For comparison, the results from 200~GeV Au+Au collisions at RHIC and 2.76~TeV
Pb+Pb collisions at LHC are also shown~\cite{Abelev:2013vea}.
The general trend seen in these data
indicates that with decreasing beam energy the system freezes out at a higher temperature with
smaller collective velocity. This may be understood as the consequence of a lower initial energy
density and thus lower initial pressure of the system as well as a shorter fireball lifetime at
lower collision energy. Similarly the smaller systems produced in the more peripheral collisions appear to have a higher freeze-out temperature and weaker collective expansion.  The labels 1,2,3 in the  plot indicate the most peripheral collisions of  beam energy 7.7 GeV (1), 200GeV (2) and 2.76TeV (3) respectively, showing a stronger collectivity for the most peripheral collisions at   higher collision energy.  

The blast-wave analysis of the spectra provides information about the kinetic freeze out, i.e
the moment the system seizes to interact entirely.  The particle yields, on the other hand,  allow us to deduce the
conditions when the system freezes out chemically, i.e. when inelastic, or particle number changing
processes, such as, e.g., $4\pi \leftrightarrow 2\pi$,\footnote{We note that resonant (quasi) elastic
  processes, such as $n+\pi\rightarrow \Delta \rightarrow n+\pi$ do not change the number of final
  state nucleons or pions.} are no longer effective. To extract the
conditions, i.e. the temperature $T_{ch}$ and baryonic chemical potential $\mu_{B}$ at the
chemical freeze-out, one compares the measured particle yield with the expectation of  a hadron resonance gas
(HRG) at certain given $T_{ch}$ and $\mu_{B}$. The HRG model is simply an ideal gas of all known
hadronic resonances, with  their abundances  given by their thermal weights $N \sim \int d^{3}p \,n(p)$, where $n(p)$ is the
appropriate thermal Fermi-Dirac or Bose-Einstein distribution functions for a given hadron resonance. The final state particles are then
determined by taking into account all the decays of the unstable resonances (for
details see e.g. \cite{Andronic:2014zha, Becattini:2009fv}). This approach, often referred as the thermal model for particle yields, also works remarkably well. 
In the left panel of Fig. \ref{fig:exp:f1} we show the extracted values for $T_{ch}$ and
$\mu_{B}$ for the systems created in BES-I.   
The filled-circles and open-squares are from central (top 5\%) and 
peripheral (60-80\%) Au+Au collisions, respectively. The yellow-line represents the systematics of 
hadron resonance gas (HRG) fit of chemical freeze-out \cite{Andronic:2009jd,Cleymans:2005xv}
over a whole range of collision energies ranging from $2\gev \lesssim \sqrt{s_{NN}} \lesssim 200\gev$. 
The green-band represents the region of the cross-over transition as determined from lattice 
QCD \cite{Bazavov:2011nk,Kaczmarek:2011zz}. As one can see, 
the available collision energies of  BES-I cover a wide range of baryonic chemical potential, $20\mev \lesssim
\mu_B \lesssim 420\mev$, with the lowest energy corresponding to the highest value of $\mu_B$. While at a given collision energy the temperature $T_{ch}$ hardly changes with
centrality, we observe a clear centrality dependence of the chemical potential, namely the more
central collision tends to have larger chemical potential. The collision energy as well as the centrality dependence of the
baryon number chemical potential can be easily understood. At lower energies and  more central
collisions, more of the incoming nucleons get stopped and thus increase the net-baryon number at mid-rapidity. 
We also observe, that at low baryon number chemical potential, $\mu_B \lesssim 300\mev$, the chemical
freeze temperature, $T_{ch}$, stays almost constant at a value close to the crossover transition temperature 
determined by lattice QCD. 
Above  $\mu_B \gtrsim 400 \mev$, on the other hand,  $T_{ch}$ starts to
decrease rather rapidly according to the global fit (yellow line).

% Assuming that the thermal equilibrium is reached in the heavy-ion collisions at RHIC, the chemical freeze-out 
% parameters temperature and baryonic chemical potential can be readily extracted and are shown in 
% Fig.\ref{fig:exp:f1} (left-plot). 

%Ref [4]: LHC REF B. Abelev, et al. (ALICE Collaboration), Phys. Rev. C88, 044910(2013)
%Ref [5]: Yu.M. Sinyukov S.V. Akkelin and N. Xu, Phys. Rev. C59, 3437(1999).

%\vspace{-0.15in}
\begin{figure}[!hbt] 
\begin{center}
\includegraphics[scale=0.4]{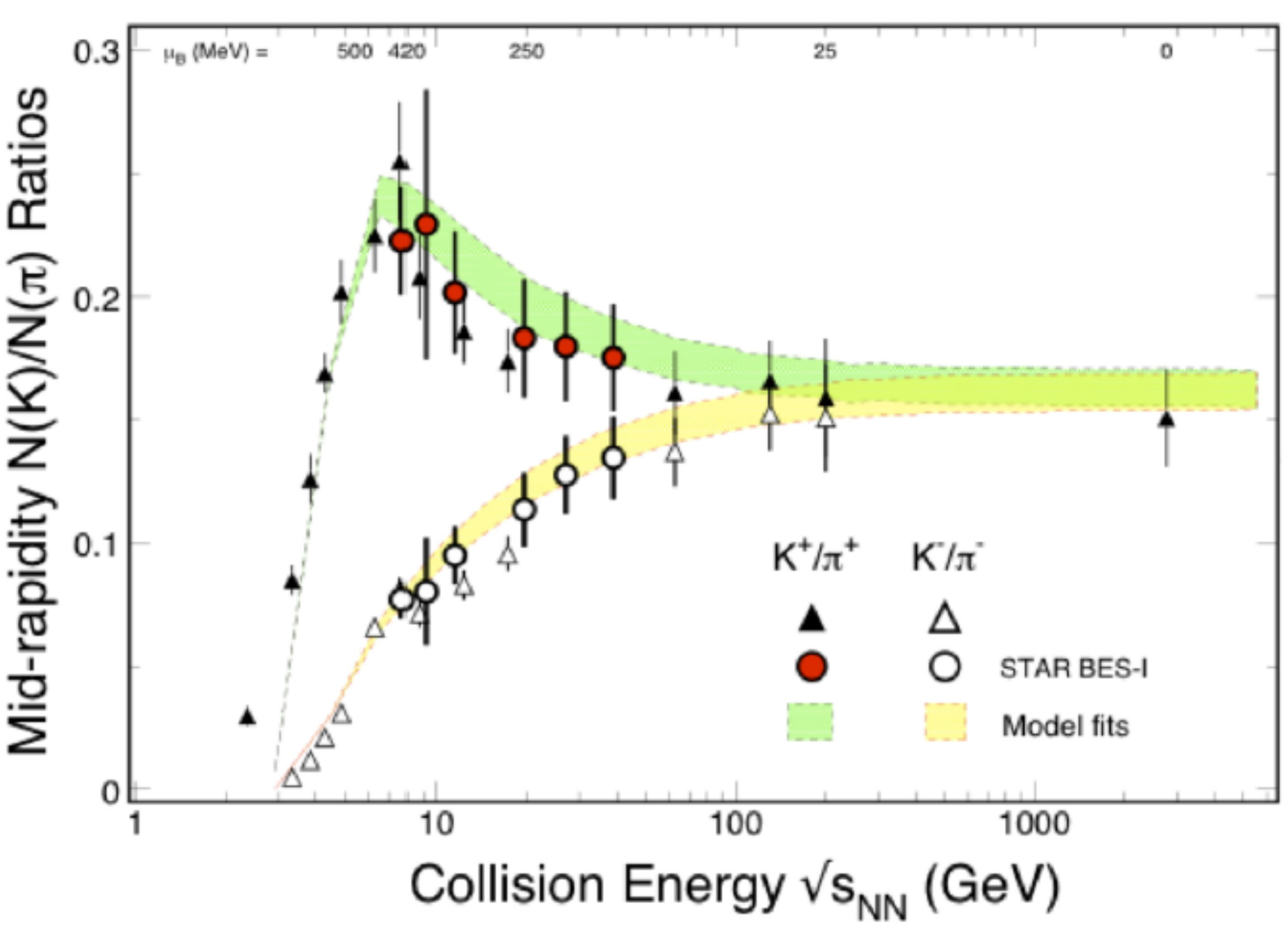}
%\vspace{-0.1in}
\caption{
Energy dependence of the ratios of K$^+$/$\pi^+$ (filled-symbols) and K$^-$/$\pi^-$ (open-symbols) 
\cite{Adamczyk:2017iwn}. All  results are from the top few-10\% central heavy-ion collisions 
\cite{Ahle:1998jc,Ahle:1999uy,Ahle:1999va, Ahle:2000wq,Afanasiev:2002mx,Abelev:2013vea}. 
The errors from the RHIC BES-I are the quadratic sum of statistical and systematic uncertainties 
where the latter dominates \cite{Adamczyk:2017iwn}. The corresponding values of the chemical 
potential for the most central heavy-ion collisions are listed at the top of the plot. 
The results of the model fits are shown as the hatched-bands.
}
 \label{fig:exp:f2}
%\vspace{-0.15in}
\end{center}
\end{figure}

The observed higher baryon density at lower collision energy originates from the stronger baryon
stopping. The effect of such baryon stopping is prominently seen in the energy dependence of the
kaon to pion ratio, which is shown in Fig. \ref{fig:exp:f2}. The BES-I results from the STAR experiment
are plotted as circles in the figure. While the $K^{-}/\pi^{-}$ ratio (open symbols) shows a smooth
energy dependence, the $K^{+}/\pi^{+}$ ratio (filled symbols) exhibits a maximum around
$\sqrt{s_{NN}}\sim 8\gev$. This behavior is well reproduced by the HRG-model \cite{Ahle:1999uy}
as seen by the yellow and green shaded areas. We note that the $\Lambda/\pi^{-}$ ratio (not shown here)
also has a maximum at the same energy which is equally well reproduced by the HRG model
\cite{Andronic:2008gu}. The occurrence of this maximum can be understood by realizing that the
threshold energy for associate production, $N + N \rightarrow N + \Lambda + K^+$ is
$E^{associate}_{threshold} = 1.58 \gev$ while that for kaon pair production is
$E^{pair}_{threshold} = 2.5 \gev$.  Therefore at low energy, associate production dominates and
gives rise to the steep increase of $K^{+}$ production compared to $K^{-}$ production, which are
mostly produced via pair-production.  With increasing collision energy very few baryons are stopped
at mid-rapidity. Consequently, the contribution to the kaon yield from associate production drops so
that it is dominated by kaon pair production.  Therefore, the maximum of the $K^{+}/\pi^{+}$ ratio indicates the
collision energy where the maximum baryon density at chemical freeze out is reached
\cite{Randrup:2006nr,Andronic:2008gu}. For completeness we note that other model studies
\cite{Rafelski:2008av,Tomasik:2006qs,Nayak:2005tz} have claimed that the peak might be an
indication of the QCD phase transition. 
Indeed the existence of such a maximum was first proposed in Ref.~\cite{Gazdzicki:1998vd} based on arguments concerning the onset of de-confinement. 
  In Refs. \cite{Cassing:2015owa,Nara:2016phs},
the authors attempted to relate the peak in the ratio of $K^+/\pi^+$ to the restoration of
chiral symmetry, which however relied on a rather complicated scheme in explaining why the chiral
symmetry restoration does not affect $K^-/\pi^-$. At the moment, it appears that the simple HRG
model provides a very successful and perhaps the most natural and ``economic'' explanation for the
energy dependence of these ratios which serves as a useful indicator of the baryon-rich environment
at low collision energy.

To summarize this subsubsection, the single particle observables such as spectra and hadron yields already characterize the bulk
properties of the system very well. They provide information about the chemical and kinetic freeze-out conditions as well as the collective radial flow and baryon stopping. In order to gain further insight about bulk expansion dynamics, one needs to look at correlations,
the most prominent of which are the azimuthal asymmetries often refereed to as directed, elliptic
and higher order flow~\cite{Ackermann:2000tr, Adler:2001nb, Adler:2003kt, Adare:2014bga, Afanasiev:2007tv, Adare:2011tg, Adare:2014kci}.

\subsubsection{Collective flow and azimuthal asymmetries }
\label{sec:exp:flow}
The system created in off-center heavy ion collision initially will not be azimuthally symmetric
in configuration space, but rather has  roughly an elliptic shape with the long axis pointing
perpendicular to the reaction plane. As a result, in the hydrodynamics framework, the pressure gradient and thus the expansion will
be stronger in the in-plane direction. As a consequence, one expects an (elliptic) azimuthal
asymmetry in the final state momentum distribution \cite{Ollitrault:1992bk}. The measurement of this
asymmetry, therefore, provides information about the pressure evolution of the system which in turn
constrains the equation of state as well as the various transport coefficients such as the
shear viscosity \cite{Alver:2010dn,Schenke:2010rr}.  Given the azimuthal direction of the reaction
plane, $\Psi_{R}$, the momentum distribution in the transverse direction may be generally written as
a Fourier series in the azimuthal angle $\phi$
\begin{align}
  \frac{d^{2}N}{p_{T} dp_{T}d\phi} = \frac{1}{2\pi}\frac{dN}{p_{T} dp_{T}}\left( 1 + \sum_{n=1}^{\infty} 2
  v_{n}(p_{T})\cos\left(n (\phi-\Psi_{R}) \right) \right) 
  \label{eq:sec6:azimuthal}
\end{align}
where the Fourier coefficients, $v_{n}(p_{T})$, characterize the azimuthal distribution for a given
transverse momentum bin. Integrating the above expression over the transverse momentum and dividing by
the number of particles, $N$, one arrives at the so-called integrated $v_{n}= (1/N) \int dp_{T} \frac{dN}{dp_T}
v_{n}(p_{T})$. Due to the weighting with the transverse momentum distribution, the integrated
$v_{n}$ are also sensitive to the aforementioned radial flow, or blue shift.
Since for each event the orientation of the reaction plane $\Psi_{R}$ in the laboratory is
different, the determination of the Fourier coefficients requires the measurement of at least a certain two-particle correlation. And indeed the $v_{n}$ can be extracted from two particle and higher order correlation
functions, for example
\begin{align}
  v_{n}^{2}{\{2\}} = \ave{\cos\left(n (\phi_{1}-\phi_{2}) \right)},
  \label{eq:sec6:vn_correlation}
\end{align}
which also holds for each momentum bin separately.  Here, the argument $\{2\}$ denotes that we use a
two-particle correlation function. However, the above formula holds only under the assumption that no
contributions from other intrinsic particle correlations are present. However, in reality hadronic decays give
rise to azimuthal two particle correlations, so that a two-particle measurement,
Eq.~\eqref{eq:sec6:vn_correlation}, will receive substantial corrections. In addition event-by-event
fluctuations of the initial shape of the fireball introduce additional subtleties in determining the flow coefficients. Therefore, as
discussed in detail in \cite{Borghini:2000sa,Adare:2011tg}, the extraction of the $v_{n}$ from higher order
correlations, if possible, is preferred.  However, for the discussion at hand, where we are more
concerned with the energy dependence of the various $v_{n}$, the $v_{n}^{2}{\{2\}}$ are sufficient
as the contamination for hadron decays are likely similar for all energies, and therefore, will not
be responsible for a possible non-monotonic behavior.  Let us now proceed to the energy dependence
of the first two harmonics, $v_{1}$ and $v_{2}$.

\begin{figure}[!hbt] 
\begin{center}
\includegraphics[scale=0.4]{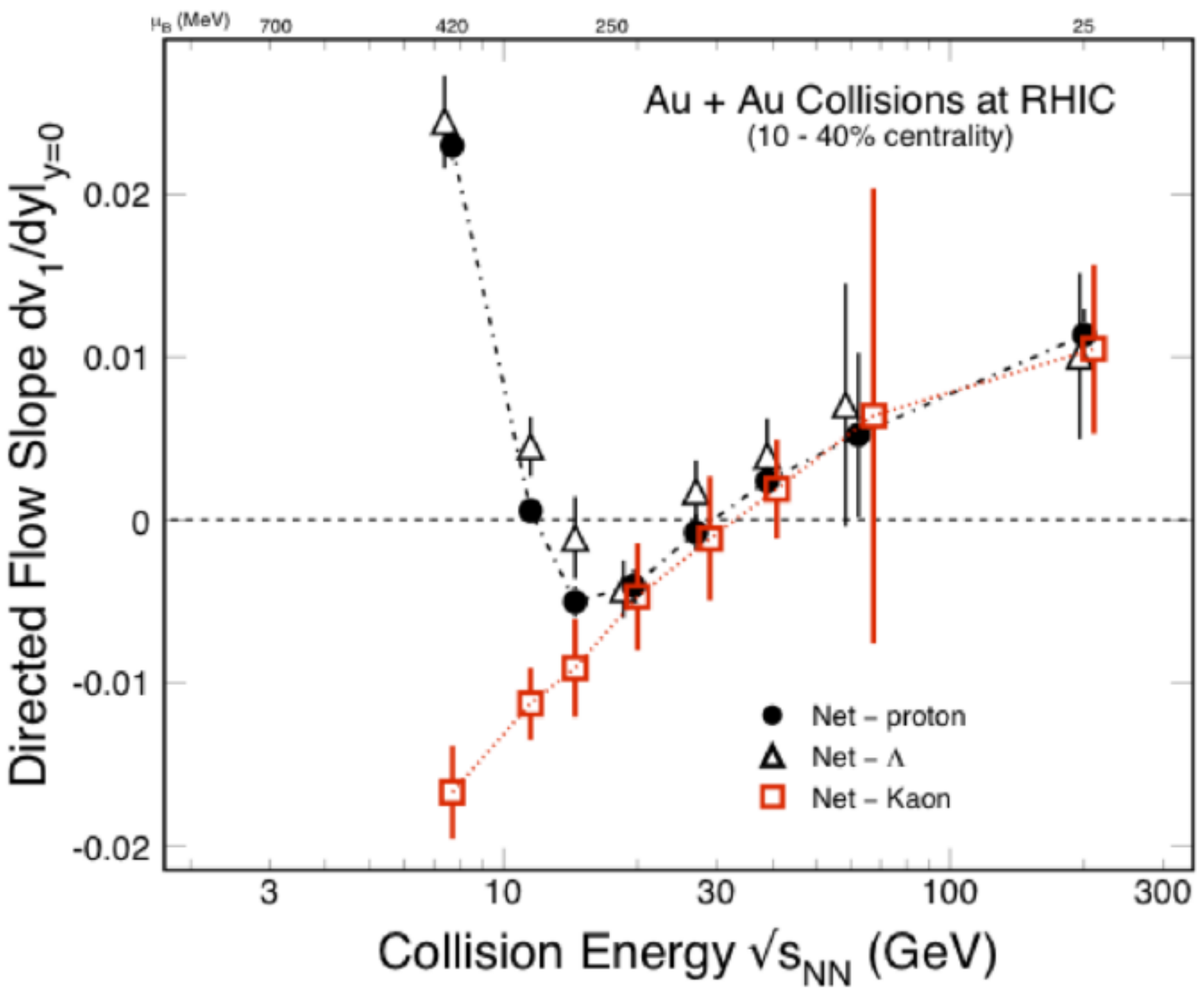} 
\caption{Collision energy dependence of the $v_1$ slope parameters for mid-rapidity net-particle of proton (filed-circles), 
Lambdas (open-triangles) and Kaons (open-squares), from 10-40\% Au+Au collisions 
at RHIC \cite{Adamczyk:2014ipa,Adamczyk:2017nxg}.}
 \label{fig:exp:f13}
\end{center}
\end{figure}

\paragraph{Directed flow}
\ \\
The first harmonic, $v_1$, measures a dipole deformation of the azimuthal momentum
distribution. Consider an off-central collision of a symmetric system such as Au+Au. If there is
repulsion, forward going particle will be pushed to one side (along the impact parameter direction)
and backward going particle will be pushed to the other side. Consequently $v_{1}$ will be similar
in magnitude but of opposite signs for particles with positive and negative center of mass
rapidities. Therefore, in order to measure the strength of this ``bounce off'' one considers the slope
of $v_{1}$ at mid-rapidity, which is often referred to as directed flow. It has been proposed
\cite{Nara:2016phs,Stoecker:2004qu} that the directed flow is sensitive to the equation of state
and its softest point, i.e. the point where the speed of sound is minimum. It was predicted that the
softest point, should reveal itself as a minimum in the excitation function of the net-proton
directed flow, where its value is predicted to be negative.  In Fig. \ref{fig:exp:f13} we show the
result of a systematic study by the STAR collaboration on the energy dependence of the directed flow
of net-protons as well as strange hadrons including Kaons and $\Lambda$ for Au+Au collisions at 7.7,
11.5, 14.5, 19.6, 27, 39, 62.4 and 200 GeV at RHIC \cite{Adamczyk:2014ipa,Adamczyk:2017nxg}.
There is a clear minimum in both the net-proton and net-lambda directed flow whereas the kaons show a monotonic decrease with collision energy.  
However, so far none of the available dynamical
calculations involving a first order phase transition are able to even get a qualitative agreement
with the STAR data \cite{Steinheimer:2014pfa, Nara:2016hbg,Hillmann:2018nmd}. Also, the slope of
 $v_{1}$ at mid-rapidity, i.e. the directed flow,   for protons but also other baryons, is sensitive
to the stopping of the incoming nuclei \cite{Bertsch:1988xu}. This may be the reason why the
produced particles such as kaons do not show a minimum while the net protons do. \footnote{Indeed,
  when looking at protons and antiprotons separately, one finds that the energy dependence of the
  antiprotons is quite similar to all produced particles, whereas the protons exhibit a clear minimum
  at $\sqrt{s_{NN}}\simeq 15 \gev$.} In addition, at lower energies the repulsive nuclear force among
nucleons is expected to eventually result in a positive directed flow \cite{Isse:2005nk}.  And
indeed one way to obtain a directed flow consistent with the data is by including a repulsive
interaction for the nucleons in the hadronic phase \cite{Hillmann:2018nmd}.  Whatever the detailed
mechanism is, the different behavior for kaons and baryons clearly indicates that the bulk expansion
dynamics in such low energy region is much more complex than that of a single fluid being pushed out
as it appears to be the case in high energy collisions. 
   {The present status certainly calls for much better theoretical calculations, either hydro-based or hydro-transport-hybrid, for collisions in the BES energy range. Such calculations need to carefully model the initial conditions of hydrodynamics by fully accounting for the baryon stopping dynamics and for the pre-hydro baryon transport. Substantial progress has been made and developments continue addressing such needs~\cite{Shen:2017bsr,Denicol:2018wdp,Du:2019obx,Du:2018mpf,McLerran:2018avb,Li:2018ini,Batyuk:2016qmb,Bialas:2016epd}.}

\paragraph{Elliptic and higher order flow}
\ \\
The second harmonic, $v_{2}$ corresponds to a quadrupole deformation of the azimuthal distribution
and is expected to be symmetric around mid-rapidity (for symmetric collisions). As already mentioned, for off-central collisions
the difference in the pressure gradients for in-plane and out-of-plane directions are expected to give rise to such a
quadrupole deformation in momentum space. And indeed  a finite and initially unexpected large
$v_{2}$ has been observed. This, together with the observed strong jet-quenching \cite{Adcox:2001jp, Adler:2002tq}
have been the major steps towards the discovery of the so-called strongly coupled quark-gluon plasma.   Furthermore it was found that the $v_{2}$ for baryons and mesons behaves
such as if the baryons and mesons are produced via coalescence of constituents quarks, 
for example $1/2 v^{\pi}_{2}(p_{T}/2) \simeq 1/3 v_{2}^{proton}(p_{T}/3)$. This systematic pattern, often
referred to as number of constituent quark (NCQ) scaling, works remarkable well for essentially all
observed hadrons at top RHIC and LHC energies.
With decreasing energy and increasing baryon number chemical potentials one
  observes an increasing  difference in the elliptic flow of particles and anti-particles. However,  the number of quark scaling still holds among particles and antiparticles separately~\cite{Adamczyk:2013gv,Adamczyk:2015fum}.   

\begin{figure}[!hbt] 
\begin{center}
\includegraphics[scale=0.36]{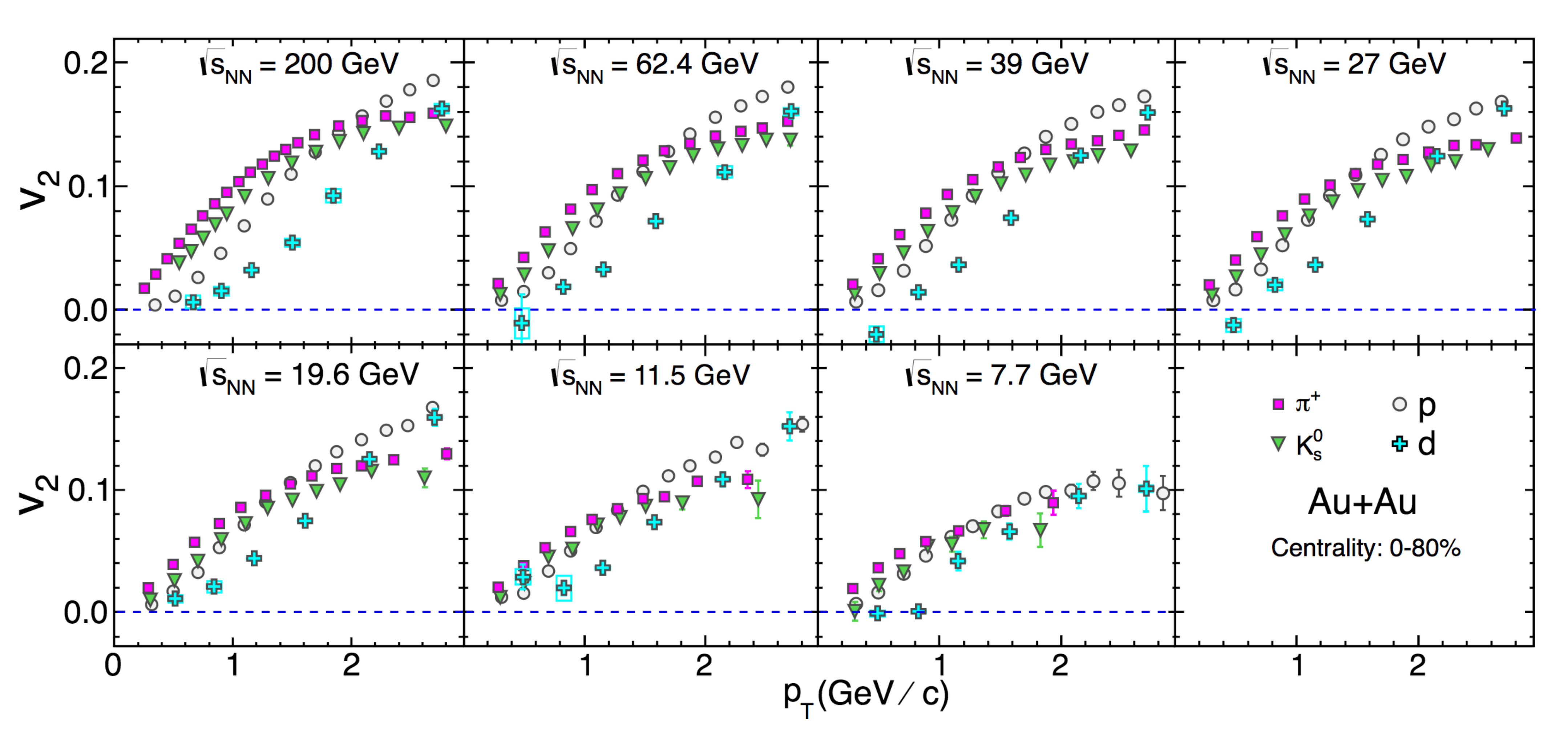}
\caption{Beam energy dependence of   identified particle $v_2$ as a function of $p_T$ for pion, $K^0$-short, 
proton and deuteron are shown in 7 panels from 200 GeV down to 7.7 GeV in $\sqrt{s_{NN}}$. The gradual reduction 
of $v_2$ magnitude is seen from 200 to 19.6 GeV in $\sqrt{s_{NN}}$ which is followed by additional stronger 
reduction from 19.6 to 7.7 GeV. Increased splitting of $v_2$ between different particles is observed 
especially around 1-2 GeV/c in $p_T$. Figure adapted from Ref. \cite{Adamczyk:2016gfs}} 
 \label{fig:exp:f4}
\end{center}
\end{figure}

\begin{figure}[!hbt] 
\begin{center}
\includegraphics[scale=0.45]{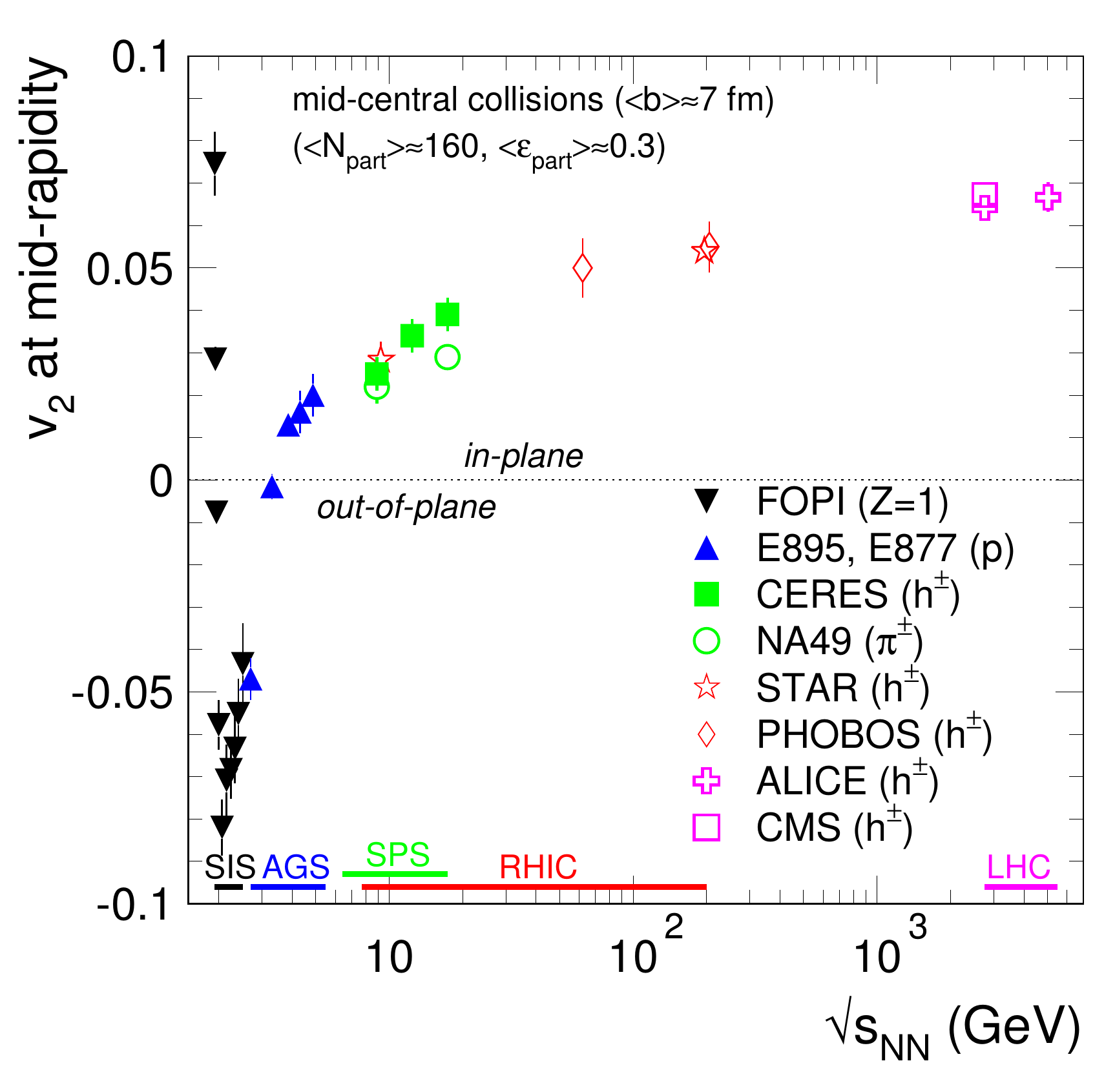}
\caption{The integrated elliptic anisotropy coefficient $v_2$ is shown as a function of beam energy. This plot is an updated version from \cite{Andronic:2014zha}.} 
\label{fig:exp:f6}
\end{center}
\end{figure}

In Fig. \ref{fig:exp:f4} we present the second harmonic, $v_{2}(p_{T})$, for pions, $K^0_{S}$, protons,
and deuterons for collision energies ranging from 7.7 GeV to 200 GeV at 0-80\% centrality
\cite{Adamczyk:2016gfs}. The dependence on particle mass of the observed $v_2$ at relatively small
$p_T$ has been well reproduced by various hydrodynamical models as well as simple blast-wave model
calculations, and can be understood as an effect of the previously discussed radial flow. We further
observe that the magnitude of $v_2(p_{T})$ increases rather mildly with increasing collision energy and
seems to level off for collision energies above $\sqrt{s_{NN}} \gtrsim 20\gev$, contrary to the
directed flow, $v_{1}$, for net-protons (see Fig. \ref{fig:exp:f13}), the increase of $v_2(p_{T})$ is
monotonic as a function of beam energy. This can also be seen in the beam energy dependence of the
integrated $v_2$ for all charged particles, shown in Fig.~\ref{fig:exp:f6}.\footnote{The change of
  sign of the integrated $v_{2}$ at $\sqrt{s_{NN}}\sim 4\gev$ is understood to come from the shielding of
  the in-plane transverse expansion by the spectators. The increase of $v_{2}$ at even lower
  energies, on the other hand, reflects the energy dependence of the nuclear force which turns
  attractive at low energies.}  We note that the integrated $v_{2}$ receives also contributions from
the radial flow. Therefore, we do not observe the leveling off in the integrated $v_{2}$ since the
radial expansion velocity steadily increases with beam energy, see Fig.~\ref{fig:exp:f1}. We  finally
note, that higher order flow coefficients such as  the triangular flow $v_3$ have also been measured over a wide
  span of beam energies~\cite{Adamczyk:2016exq}, and  the general trends of these higher order flow coefficients with changing
  beam energy resemble that of the elliptic flow.

\begin{figure}[!hbt] 
\begin{center}
\includegraphics[width=0.95 \textwidth]{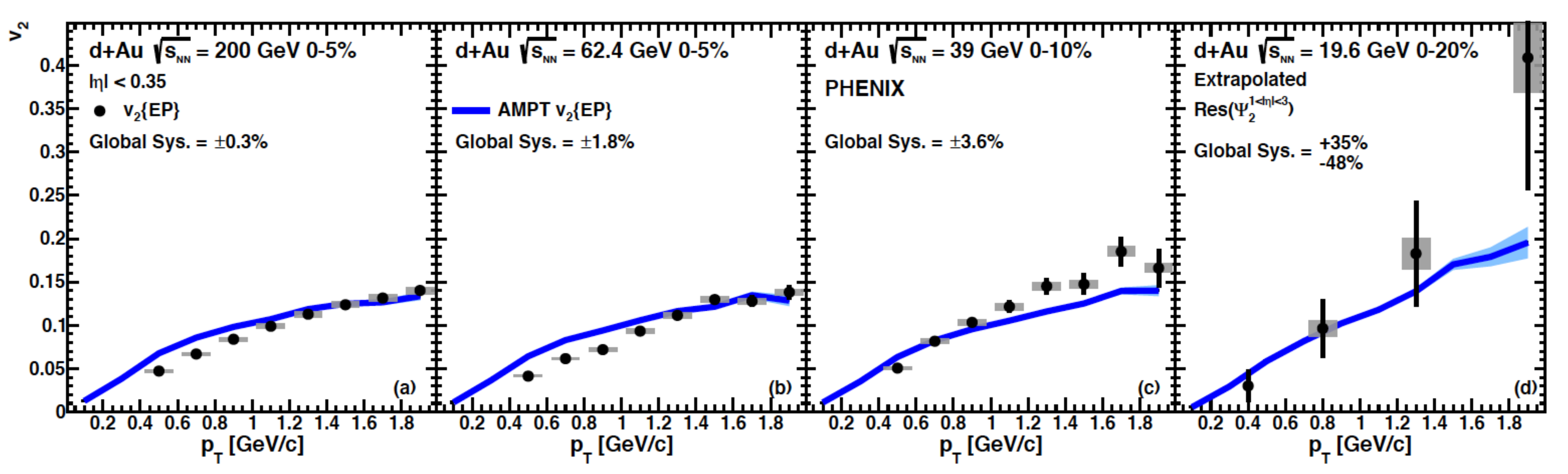} 
\caption{Beam energy dependence of the elliptic flow $v_2(p_{T})$ for d+Au
  collisions at 200, 62.4, 39, and 19.6 GeV 
for central collisions (high multiplicity events) 0-5\% \cite{Aidala:2017pup}.} 
 \label{fig:exp:f7}
\end{center}
\end{figure}

\paragraph{Flow in small systems}
\ \\
Recently it has been observed that flow-like multi-particle correlations also exist in small
systems, such as the ones created in high energy p+p, d+Au and p+Pb collisions. While it is still being debated
if the origin of these correlations is hydrodynamic evolution as it is in nucleus-nucleus
collisions \cite{Bozek:2011if, Schenke:2014gaa, Romatschke:2015gxa}, it is interesting to see 
if these flow-like correlations disappear at lower collision 
energies. Such a measurement has been recently carried out by the PHENIX collaboration
\cite{Aidala:2017pup} and in Fig. \ref{fig:exp:f7}  we show their results for the beam energy
dependence of $v_{2}(p_{T})$ for deuteron-gold reactions for collision energies ranging from $\sqrt{s_{NN}}=19.6\gev$ to
$\sqrt{s_{NN}}=200\gev$. Similar to the $v_{2}$ in heavy-ion collisions shown in Fig.~\ref{fig:exp:f4}, the
elliptic flow hardly changes over this range of collision energies. 

To summarize this subsubsection, the measured collective flow coefficients demonstrate interesting
patterns with changing collision energy. The elliptic and higher order flows appear to be mainly
driven by anisotropic pressure gradients arising from the initial geometric shape which does not change much with
beam energy and which occurs for both large and small colliding systems. The directed flow for baryons, on the
other hand, shows highly nontrivial, non-monotonic collision energy dependence, with a minimum
structure around $15 \gev$. The excitation function for directed flow of  mesons, however, is monotonic. In addition a difference in the
particle and antiparticle elliptic flow also becomes important in the low energy region.  These
observations suggest the presence of additional interaction dynamics in the bulk collectivity that
pertains to the baryon-rich matter created at low collision energy and that is absent at high
energy.  Aside from an 
enhanced role of baryon stopping, this may provide a first hint that one probes the region in the
phase diagram where the equation of state changes rapidly with the control parameters, as one would expect in
the vicinity of a phase transition.

\subsubsection{HBT and femtoscopic correlations}
\label{sec:exp:hbt}

At very low relative momentum,  correlations between a pair of identical particles are sensitive to the quantum
interference resulting from the symmetry or anti-symmetry of the two-particle
wave-function for bosons and fermions, respectively. Analogous to the well known Hanbury-Brown Twiss
interferometry measurements of the sizes of stellar objects \cite{HanburyBrown:1956bqd}, one can utilize the quantum
interference to determine the size of the systems produced in heavy ion collisions. These
measurements are often referred to as HBT or femtoscopic correlation analysis, and are mostly done
with pion, kaon and proton pairs. These measurements require appropriate corrections for the Coulomb
interaction as well (in case of protons) correction for the s-wave scattering amplitude, see, e.g., the review~\cite{Lisa:2005dd} for details. 

To extract the size of a colliding system, one examines the two-particle correlation function 
\begin{align}
  C_{2}\left( \vec{q},\vec{P} \right) = 
     \frac{\frac{d^{2}N}{d^{3}p_{1}d^{3}p_{2}}}{\frac{dN}{d^{3}p_{1}}\frac{dN}{d^{3}p_{2}}}
  \label{eq:sec6:hbt_corrleation}
\end{align}
for a given total pair momentum $\vec{P}=(\vec{p_{1}}+\vec{p_{2}})$ as a function of the relative
momentum $\vec{q}=(\vec{p_{1}}-\vec{p_{2}})$ and parameterizes it in the following form (in case of pions)
\begin{align}
  C(\vec{q}) &= N\left[  \lambda\left( 1+G(\vec{q}) \right) 
  F_{c} +\left( 1-\lambda \right) 
  \right] \non
  G(\vec{q}) &\simeq \exp\left( -R_{side}^{2} q_{side}^{2} - R_{out}^{2}q_{out}^{2} - R_{long}^{2}q_{long}^{2}\right) 
  \label{eq:sec6:hbt_source}
\end{align}
Here, $N$ is an overall normalization, $\lambda$ represents the correlation strength and $F_{c}$ is
the Coulomb correction factor \cite{Sinyukov:1998fc}. In the function $G(\vec{q})$, we have decomposed the relative momentum
$\vec{q}$ into three components:  ${q}_{long}$ which points along the beam axis, and the transverse part
$\vec{q}_{T}$ is further split into the component ${q}_{out}$ which points along the transverse
pair-momentum, $\vec{P}_{T}$, and the component perpendicular to it, ${q}_{side}$. While the above
formula is an approximation -- corrections are discussed in detail, e.g., in Ref. \cite{Lisa:2005dd} -- it
captures the relevant physics we wish to discuss here. In the above formula, one assumes  that the
source in configuration space is a three dimensional Gaussian.  The parameters $R_{long}$,
$R_{out}$, and $R_{side}$ then represent the {\em spatial} size along the beam axis, $R_{long}$,
and the transverse sizes along the pair momentum, $R_{long}$, and  perpendicular to it,
$R_{side}$. 

\begin{figure}[!hbt] 
\begin{center}
  \includegraphics[width=0.7\textwidth]{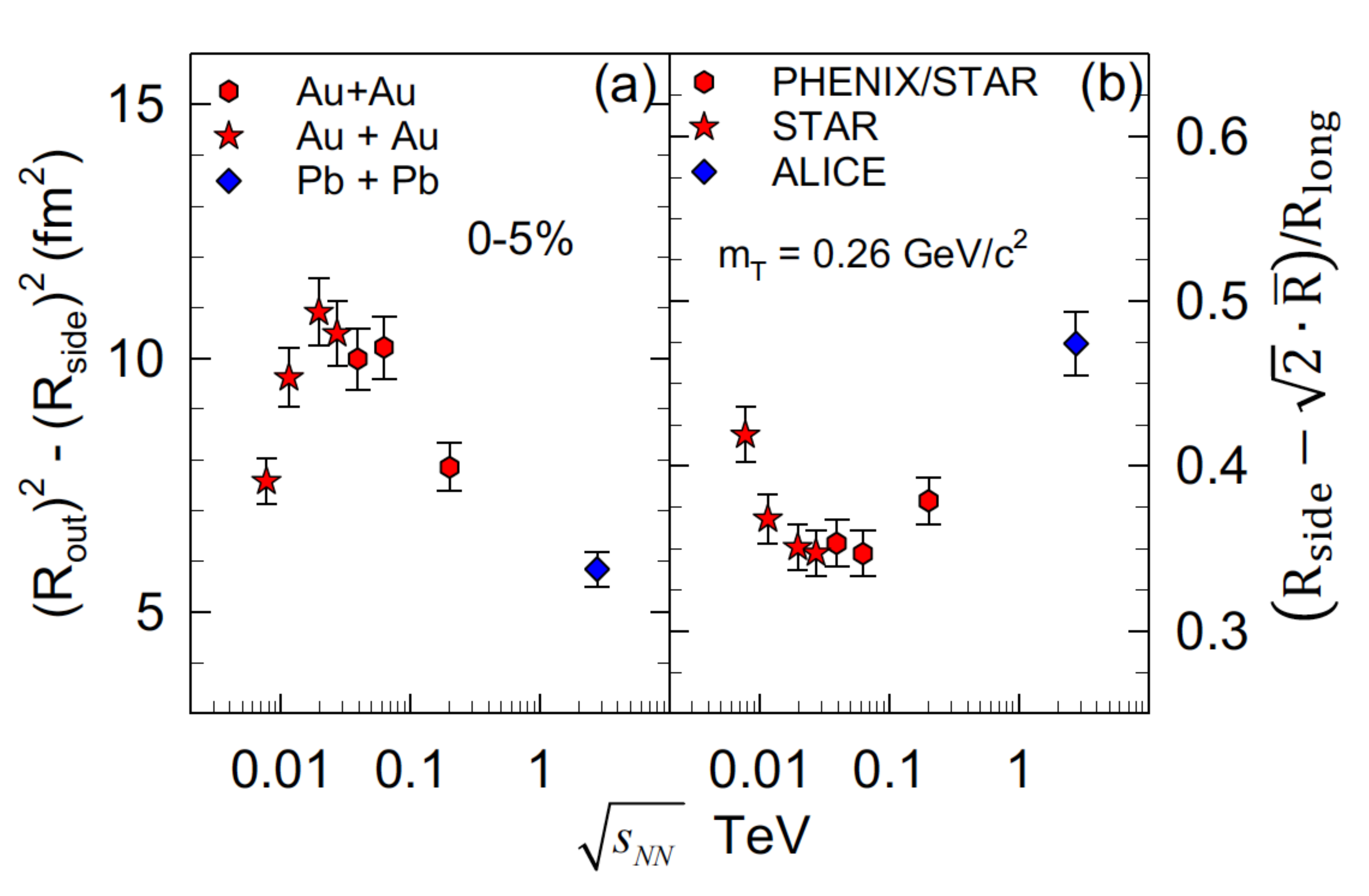}
\caption{Beam energy dependence of HBT-radii: Estimator of the emission time $\Delta \tau^{2}$,
  $R_{out}^{2}-R_{side}^{2}$ (left) and estimator of the expansion velocity $v_{expand}$, 
  $(R_{side}-\sqrt{2}\bar{R})/R_{long}$ (right). Figures are adapted from \cite{Lacey:2014wqa,Adare:2014qvs}.}
 \label{fig:exp:f8}
\end{center}
\end{figure}

Suppose we have a system, where particles are slowly emitted from the surface in the transverse
direction. In this case the apparent size in the ``out'' direction appears larger than that in the ``side''
direction, since the first emitted particle has already traveled a long distance before the last emitted
particle emerges. In other words, the difference between $R_{out}$ and $R_{side}$ is sensitive
to the emission duration, $\Delta \tau^{2} \sim (R_{out}^{2}-R_{side}^{2})$
\cite{Pratt:1984su,Pratt:1986cc,Bertsch:1989vn}. Such a slow emission could, for example happen, if
the speed of sound is very small due to a first order phase transition or a critical point
\cite{Hung:1994eq,Rischke:1996em}. Therefore, a maximum in $R_{out}^{2}-R_{side}^{2}$ as a
function of the collision energy could
indicate the energy where the system gets closest to the ``softest point'' in the QCD equation of
state. This is actually seen in the experimental data shown in the left panel of Fig. \ref{fig:exp:f8}
\cite{Adare:2014qvs}. In addition to the emission duration $\Delta \tau$ one can also attempt to
estimate the expansion velocity from the HBT-correlation measurements. For a system which expands
longitudinally in a nearly boost invariant fashion \cite{Bjorken:1982qr} the longitudinal size of a
system $R_{long}$ provides a measure for the total lifetime of the system, $\tau_{life}\sim
R_{long}$. The spatial expansion in the transverse direction, on the other hand, can be estimated by
difference between  the final side-ward radius as determined by $R_{side}$ and the initial size of the fireball  
$R_{init}=\sqrt{2} \bar{R}$, where $\bar{R}=\frac{1}{\sqrt{1/\sigma_{x}^{2}+1/\sigma_{y}^{2}}}$ with the
Gaussian widths $\sigma_{x}$, $\sigma_{y}$ of the fireball density distribution determined by a Glauber model calculation. Therefore,
the combination $(R_{side}-\sqrt{2}\bar{R})/R_{long}\sim \Delta R/\tau_{life}\simeq v_{expand} $
provides an  estimator for the expansion velocity, $v_{expand}$. The beam energy dependence of this quantity is shown in the
right panel of Fig. \ref{fig:exp:f8}. We observe a minimum at about the same energy where the
emission time exhibits a maximum. This combined behavior is rather intriguing and may be a hint
about the softest point of the equation of state. However, a note of caution is in order
here. First, the blast-wave analysis discussed above (see Fig.~\ref{fig:exp:f1}) shows no minimum
of the expansion velocity. Second,  both
quantities shown in Fig. \ref{fig:exp:f8} change as much when going from the LHC energy (blue points) to the top RHIC energy, as they change from
the top RHIC energy to the maximum/minimum at around $\sqrt{s_{NN}}\simeq 20\gev$. However, for both the LHC and top RHIC energy collisions
one probes the equation of state at approximately $\mu_{B}\simeq 0$. Therefore, in addition to a
possible effect of the equation of state there is obviously some other  mechanism which changes these
two observables in between top RHIC energy and LHC energy. Unless this other mechanism has been identified and subtracted, it is too premature to
conclude that the softest point of the equation of state has been found. \newline

To summarize this subsection on bulk properties and collectivity, we find that the kinetic and chemical freeze-out
temperatures as well as the radial and elliptic flow  exhibit a rather smooth
dependence on the collision energy. However, some observables such as the $K/\pi$ ratio, directed
flow as well as the emission time extracted from HBT show a clear non-monotonic behavior, with
maxima/minima  occurring in a  collision energy range of $10 \gev \lesssim \sqrt{s_{NN}} \lesssim
20\gev$, corresponding to a baryon-rich region. To which extend these maxima/minima are related to the equation of state or are more a
reflection of the change from baryon to meson dominance of the system cannot be easily concluded
based on flow and global observables alone. To make further progress one needs to study
observables that are more sensitive to a possible phase transition, such as fluctuations, which we will turn to next.

%--===================================================================================
%--===================================================================================
%Nu
\subsection{Criticality}

\subsubsection{Experimental data on proton, anti-proton, net-proton and net-charge high order cumulants}
\label{sec6:exp:cumulants}

%\vspace{-0.15in}
\begin{figure}[!hbt] 
\begin{center}
\includegraphics[scale=0.15]{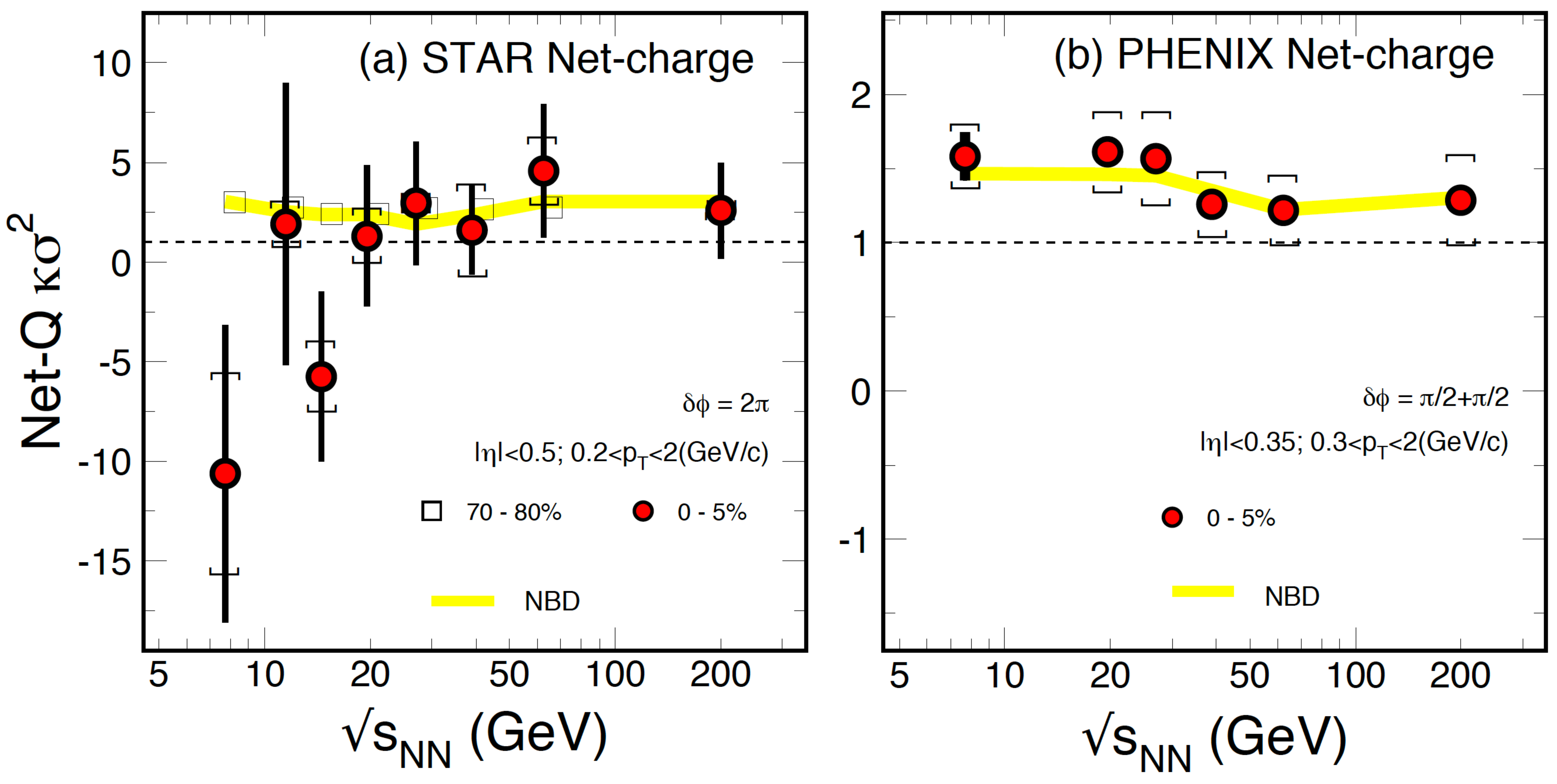} 
%\vspace{-0.1in}
\caption{Energy dependence of the fourth-order cumulant ratio $\cum{4}/\cum{2}=\kappa\sigma^2$ for net-charge (a) from STAR \cite{Adamczyk:2014fia} 
and (b) from PHENIX \cite{Mitchell:2015bfp,Garg:2015anr,Adare:2015aqk} experiments, respectively. For STAR data, results of both top 5\% central and 70-80\% peripheral 
Au+Au collisions are shown while for PHENIX data, only the top 5\% central Au+Au collisions  are shown. The expected values from the 
negative binomial distribution are shown as the yellow-band. The event statistics from both
experiments are listed in Table~\ref{tab:exp:bes}. } 
 \label{fig:exp:f15a}
%\vspace{-0.15in}
\end{center}
\end{figure}

%ref: --=======================
%(1) \cite{Mitchell:2015bfp}
%J. Mitchell, (PHENIX Collaboration),  CPOD2014 proceedings, PoS CPOD2014(2015) 075
%"Phenix global observables and fluctuations from the RHIC beam energy scan"
%(2) \cite{Garg:2015anr}
%P. Garg, (PHENIX Collaboration),  QM2015 proceedings, Nucl. Phys. A956, 369(2016).
%"PHENIX results on fluctuations and Bose-Einstein correlations in Au+Au collisions from the RHIC beam energy scan"
%(3) \cite{Adare:2015aqk}
%A. Adare et al., (PHENIX Collaboration), Phys. Rev. C93R, 011901(2016)
%"Measurement of higher cumulants of net-charge multiplicity distributions in Au+Au collisions at  $\sqrt{s_{NN}}=7.7\sim 200 \rm GeV$"
%(4) \cite{Adamczyk:2014fia}
%L. Adamczyk, et al. (STAR Collaboration), Phys. Rev. Lett. 113, 92301(2014)
% "Beam energy dependence of moments of the net-charge multiplicity distributions in Au+Au collisions at RHIC "
%ref: --=======================

In the previous section, Sec.~\ref{sec4}, we demonstrated that fluctuations, including fluctuations of conserved charges, are expected to be  
sensitive observables for the existence of a potential QCD critical point and/or first order phase transition. We further discussed that the correlation length, which would be infinite in a static infinitely large thermal system, will be limited to $\xi\simeq 2-3 \rm \,fm$, because the systems created in heavy-ion collisions have a finite size and a finite lifetime. 
Therefore, it is advantageous to study cumulants and correlations of higher order, since they scale with the higher powers of the
correlations length (see Sec.~\ref{sec:high-moments-cumul} and
\cite{Stephanov:2008qz,Gupta:2011wh,Luo:2017faz}). 
Meanwhile all collider heavy-ion experiments ALICE, PHENIX and STAR have carried out the measurement of cumulants of various orders and for several particle species~\cite{Behera:2018wqk,Adamczyk:2013dal, Adamczyk:2017wsl, Adare:2015aqk}. 
 In the context of the RHIC BES-I, cumulants of  net-proton, net-Kaon and
net-charge multiplicity distributions have been measured as part of the search for the QCD critical region \cite{Adamczyk:2013dal, Adamczyk:2017wsl, Adare:2015aqk}.
In the following we will discuss selected results from these measurements with an emphasis on
net-protons, as they are believed to couple most strongly to the fluctuating $\sigma$ field near the critical point, as we discussed in Sec.~\ref{sec:crit-pt-correl-momentum}.

Let us start with the net-electric charge distribution. In Figure~\ref{fig:exp:f15a} we show the
beam energy dependence of the cumulant ratio $\kappa\sigma^2 \equiv \kappa_{4}^{Q}/\kappa_{2}^{Q}$ for the net-charge distribution.\footnote{   {See the Appendix for discussions on quantities like kurtosis, skewness etc. and their relations to the cumulants.}}
Panel (a) shows the results obtained by the STAR collaboration for two centralities (0-5\%, filled-circles,
and 70-80\% open-squares) at collision energies $\sqrt{s_{NN}}=$ 7.7, 11.5, 14.5, 19.6, 27, 39, 62.4 and 200
GeV~\cite{Adamczyk:2014fia}. The results from the PHENIX experiment, shown in panel (b), are for the top 5\%
central Au+Au collisions at energies $\sqrt{s_{NN}}=$ 7.7, 19.6, 27, 39, 62.4 and 200
GeV~\cite{Adare:2015aqk}. Within statistical and systematic uncertainties, both data sets show
no significant features as a function of the collision energies over the 
range of  $7.7 \gev \le \sqrt{s_{NN}} \le 200 \gev$, corresponding to a range of the  baryon-chemical
potential $420\mev \gtrsim \mu_B \gtrsim 20\mev$. We note
that the statistical errors for the STAR measurement are significantly larger than those for the
PHENIX data. The reason for this difference is essentially the larger acceptance of the STAR
detector. It can be shown \cite{Luo:2014rea} that the statistical error for the measurement for the cumulant ratio
$\kappa \sigma^{2}$ scales like\footnote{While this scaling is true in general, there are distributions such as
  the bi-modal one discussed in Section~\ref{sec:6:bimodal} which are remarkably statistics friendly.} $error(\kappa\sigma^2) \propto
\sigma^2/\left( \sqrt{N} \epsilon^{4}\right)$, where $N$ is the number of events, $\epsilon$ the detection efficiency, and $\sigma$ is the width of the {\em observed} net-charge
distribution.
Even though PHENIX recorded less events at $7.7\gev $ and $19.6\gev$ than STAR, see Table~\ref{tab:exp:bes},
the width of the {\em observed } net-charge distribution, which, to a good
approximation is given by the total number of {\em observed } charged particles, $\sigma\sim
N_{ch}$, is significantly larger for STAR. This is simply a result of the larger acceptance in the
STAR detector. To be specific, the
STAR events were collected with full azimuthal acceptance within $|\eta|<0.5$ and
$0.2 \le p_T \le 2.0\gev/ \rm c$ while the PHENIX events were taken  with its two central arm
spectrometers with $|\eta|<0.35$ and $0.3 \le p_T \le 2.0 \gev/\rm c$, and each of the two arms cover
$\pi/2$ radians in azimuth.

As mentioned in Section~\ref{sec:crit-pt-correl-momentum}, the coupling between the fluctuating sigma field with baryons is expected to be stronger than that for pions, which are the most abundant charged particles. Therefore, the rather featureless beam energy dependence  of the cumulant ratio
$\kappa\sigma^{2}=\cum{4}/\cum{2}$ \footnote{In the following we will use the notation for the
  cumulant ratios $\cum{3}/\cum{2}=S\sigma$ and $\cum{4}/\cum{2}=\kappa\sigma^{2}$
  interchangeably.}
for the net-charge distribution may very well reflect this weaker coupling. Since
one expects a stronger signal in the baryon sector, for the rest of this Section we will focus our
discussions on the energy dependence of the proton and anti-proton distributions.

%\vspace{-0.15in}
\begin{figure}[!hbt] 
\begin{center}
\includegraphics[scale=0.65]{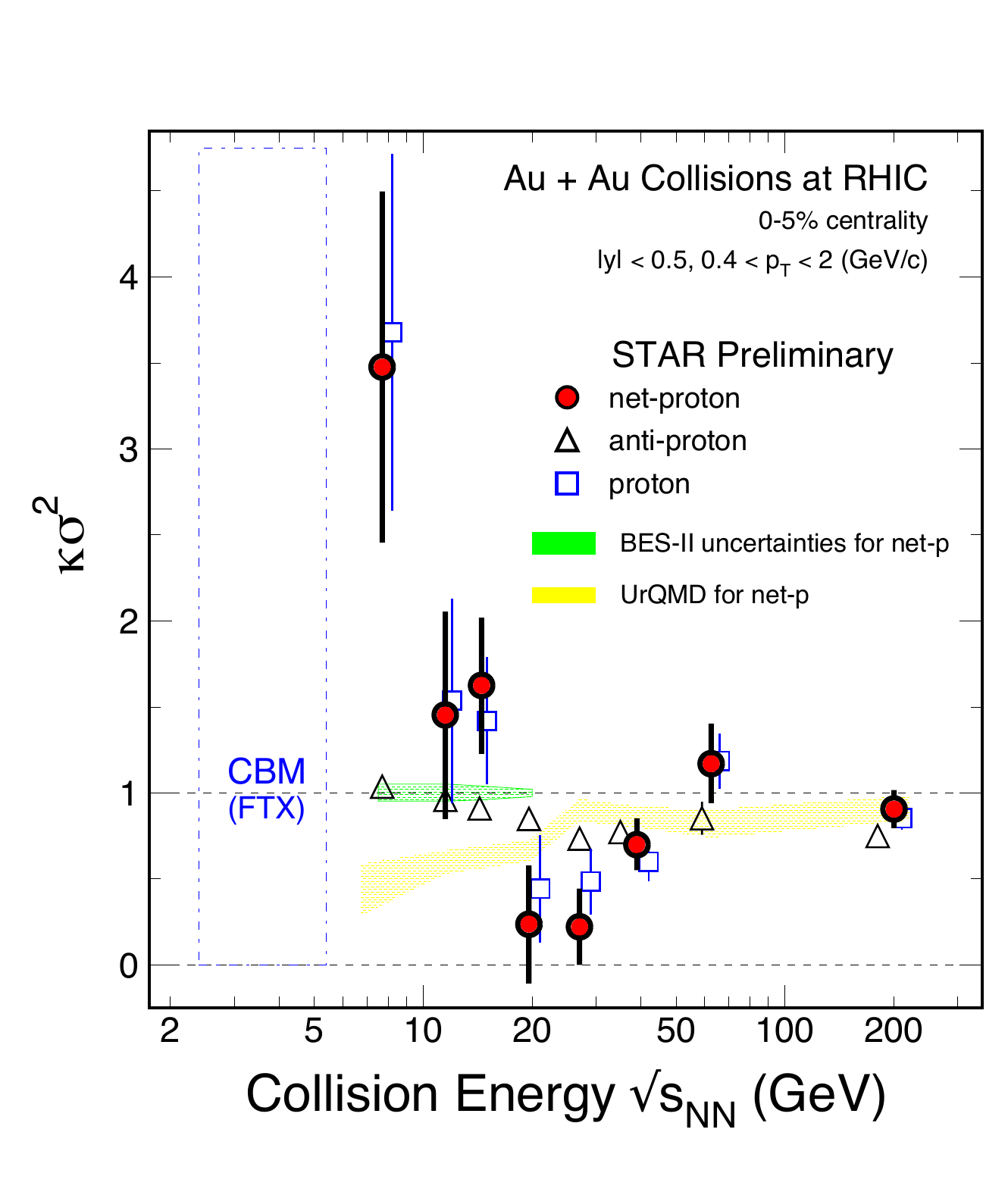} 
%\vspace{-0.1in}
\caption{STAR preliminary results \cite{Luo:2015ewa} on the energy dependence of the
  fourth-order cumulant ratio 
$\cum{4}/\cum{2}=\kappa\sigma^2$ of net-protons (filled-circles), anti-protons (open-triangles) and protons (open-squares) 
from the top 5\% central Au+Au collisions. Those data were taken from the first phase of the RHIC BES-I 
and in the kinematic region of rapidity $|y|\le0.5$ and transverse momentum $0.4 \gev/c \le p_T
\le 2.0 \gev/c$. 
The yellow-band represents the results from UrQMD model calculations. The green-band shows the estimated 
statistical errors for the net-proton $\kappa\sigma^2$ from RHIC BES-II, in the collision energy window of 
$\sqrt{s_{NN}}=7.7$ to 20~GeV. Future fixed-target experiment at FAIR will cover the energy region of 
$\sqrt{s_{NN}}=2.5$ to 5.5~GeV. 
{Part of the data has been 
submitted for publication by the STAR Collaboration in \cite{Adam:2020unf}. }
} 
 \label{fig:exp:f15}
%\vspace{-0.15in}
\end{center}
\end{figure}

In Fig.~\ref{fig:exp:f15} we present the preliminary results from the STAR collaboration  for the
cumulant ratio  $\cum{4}/\cum{2}=\kappa\sigma^2$ of net-protons (filled-circles), 
anti-protons (open-triangles) and protons (open-squares) for the 5\% most central Au+Au
collisions. We also show, as yellow band, the results obtained with the  UrQMD event generator. We 
see that with decreasing collision energy, and thus increasing net-baryon density, the proton
and net-proton $\kappa\sigma^2$ decreases until about 
$\sqrt{s_{NN}} \sim 20 \gev$, followed by a sharp increase above unity for collision energies below
10~GeV. As discussed in more detail in Section~\ref{sec4}, such a non-monotonic behavior was
predicted in \cite{Stephanov:2011pb} for a scenario where the freezeout line traverses the critical region, as shown in Figs.~\ref{fig:scenario-color4} and~\ref{fig:omega-sqrts-mu}.  
The UrQMD prediction, on the other hand, exhibits a continuous monotonic 
decrease, which is mostly due to  baryon-number conservation \cite{Schuster:2009jv}(see
Section~\ref{sec7:baryon_conserve}).
As we shall discuss in
Section~\ref{sec7:fact_cumulants}, a more detailed analysis of all cumulants shows that the
observed non-monotonic behavior could result from the interplay of negative two-particle correlations, which are likely the result of baryon number
conservation, and rapidly increasing four-particle correlations, which then dominate the signal at the
lowest measured energy of $\sqrt{s_{NN}} =7.7\gev$.

%\vspace{-0.15in}
\begin{figure}[!hbt] 
\begin{center}
\includegraphics[scale=0.7]{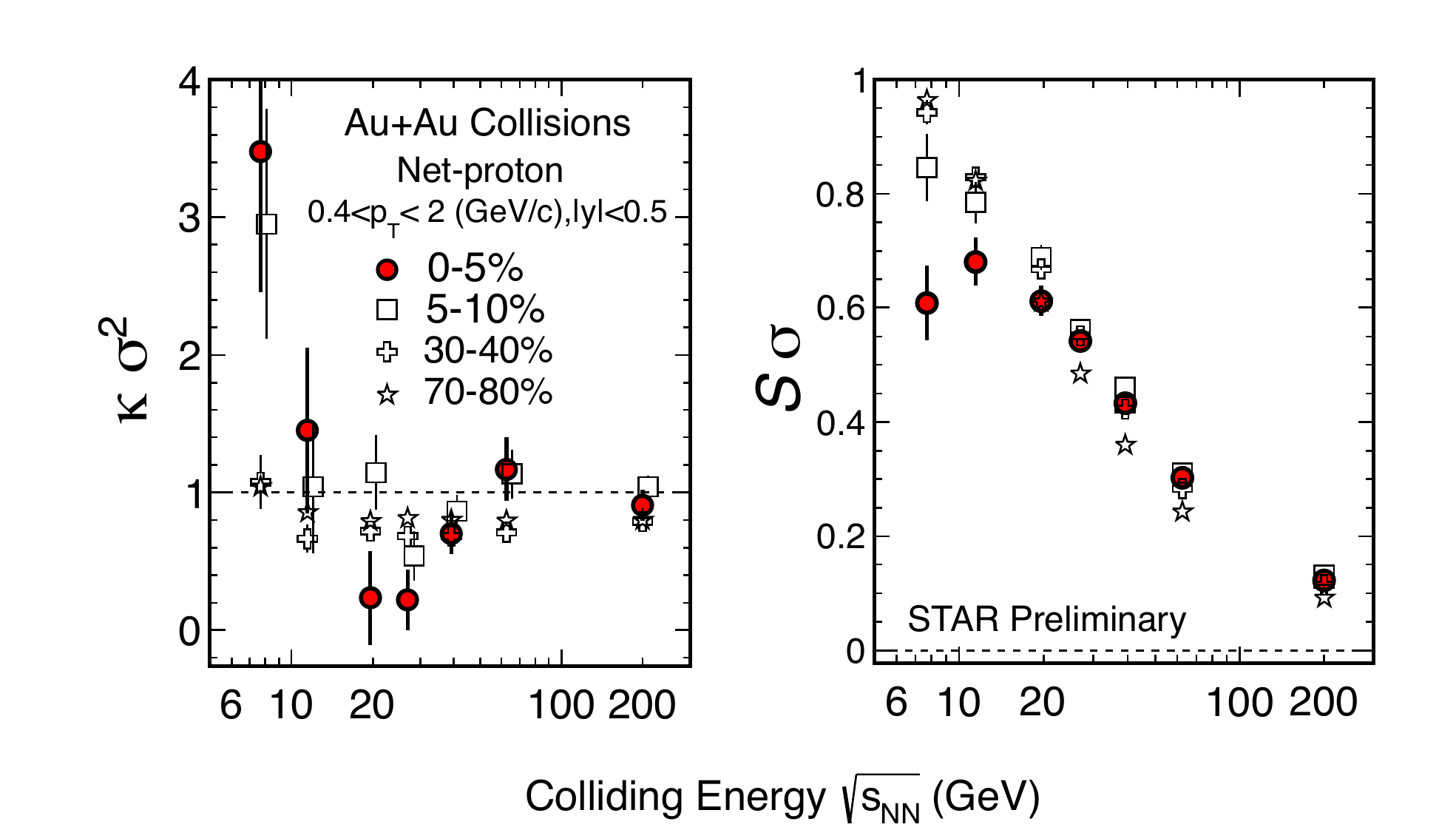} 
%\vspace{-0.1in}
\caption{Cumulant ratios $\cum{4}/\cum{2}=\kappa  \sigma^2$ (left panel) and
  $\cum{3}/\cum{2}=S\sigma$ (right panel) for net-protons in Au+Au collision as a 
function of beam energy for different centralities (STAR preliminary data \cite{Luo:2015ewa}). } 
 \label{fig:nu:exp:f16}
%\vspace{-0.15in}
\end{center}
\end{figure}

The non-monotonic behavior of the cumulant ratio $\cum{4}/\cum{2}=\kappa\sigma^2 $ is most pronounced
for  the most central collisions and gradually disappears as the collisions become less central. At
the most peripheral centrality, 70-80\%,  $\kappa\sigma^2$ for net-protons is flat in the measured energy range 
$\sqrt{s_{NN}} = 7.7 -  200\gev$. This can be seen in Fig.~\ref{fig:nu:exp:f16} where in the left
panel we show the STAR preliminary data \cite{Luo:2015ewa} for $\cum{4}/\cum{2}=\kappa\sigma^2 $ for several centralities. The right panel of Fig.~\ref{fig:nu:exp:f16}
shows the third order cumulant ratio, $\cum{3}/\cum{2} \equiv  S\sigma$, for the same
centralities. Here the situation is different. While a  non-monotonic 
behavior is seen in the most central collisions (filled-circles) all data from other centrality bins show a 
smooth increase as the energy decreases. In addition for the lowest two energies we see a reduction
of $S\sigma$ with increasing centrality while for $\kappa\sigma^2$  we see an enhancement. As we
shall discuss in Section~\ref{sec7:fact_cumulants}, this decrease can be attributed to
negative three particle correlations.
At present the  statistical uncertainties for $\kappa\sigma^2$  are rather sizable, especially for
the collision energies below 20 GeV. This should be remedied in the planed  second phase of the RHIC
beam energy scan at RHIC, as we shall discuss in Section~\ref{sec8:future}.
%\vspace{-0.15in}
\begin{figure}[!hbt] 
\begin{center}
\includegraphics[scale=0.6]{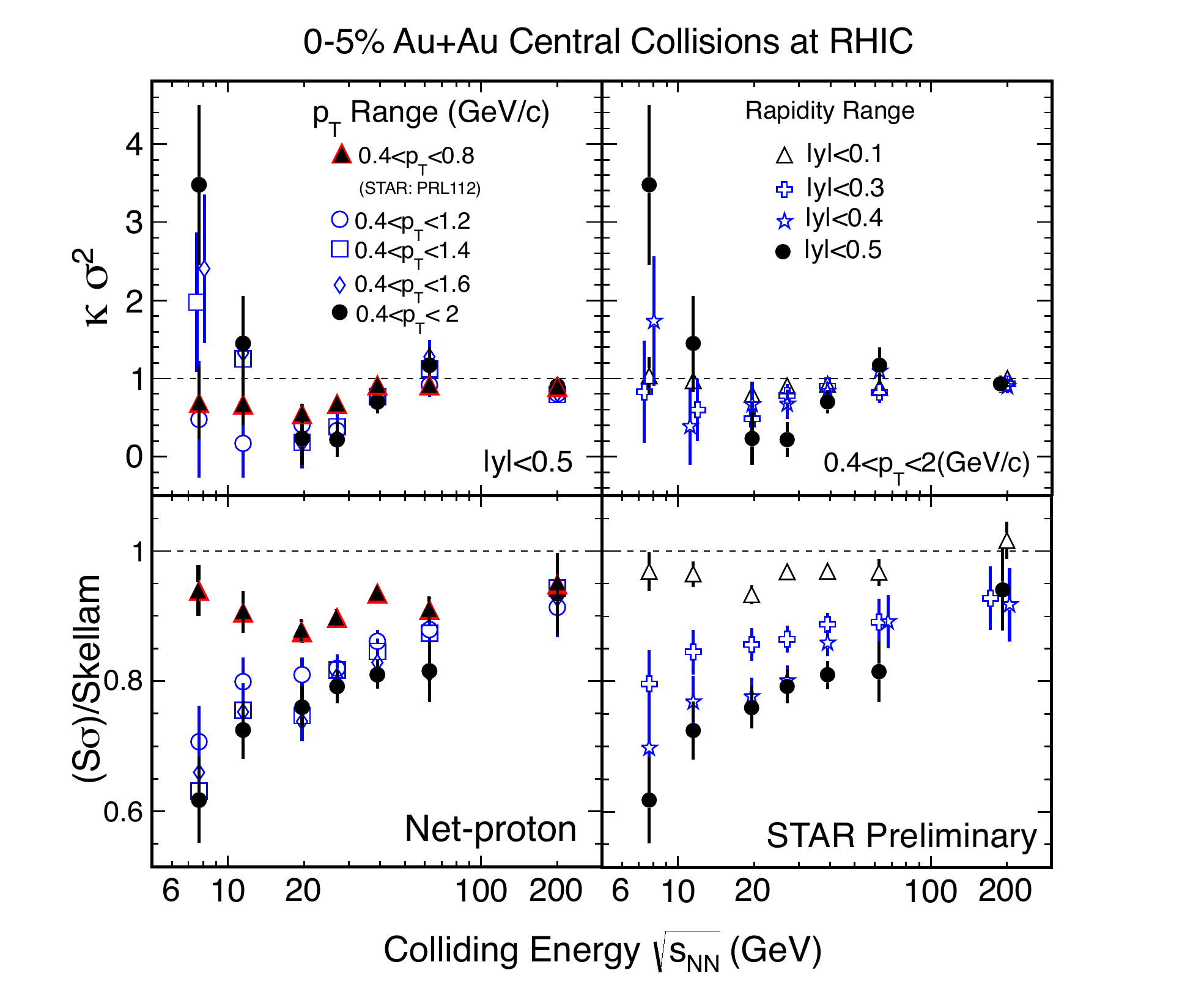} 
%\vspace{-0.1in}
\caption{The beam energy dependence of the cumulant ratio $\cum{4}/\cum{2}=\kappa \sigma^2$ (top panels) and $S\sigma$/Skellam (lower panels) for net-protons in Au+Au collisions 
(STAR preliminary data \cite{Luo:2015ewa}). The left panels illustrate the effects of transverse momentum selection while the right panels indicate 
the effects of rapidity cuts. Dotted horizontal lines are expectations from Poisson/Skellam distributions.} 
 \label{fig:exp:f17}
%\vspace{-0.15in}
\end{center}
\end{figure}

Note that when $\sqrt{s_{NN}} \lesssim 15 \gev $ all data points for $\kappa \sigma^{2}$ are above unity, for which a natural possible interpretation could be clustering of protons due to certain attractive interactions among them \cite{Shuryak:2018lgd}. However, all known transport models fail to reproduce the observed enhancement of $\kappa_4/\kappa_2$ in the high baryon density region $\mu_B \gtrsim 300 \mev$ even when ``attractive interaction'' was turned on~\cite{He:2016uei}. A recent work \cite{Shuryak:2018lgd}
suggested an enhanced attraction in the vicinity of the critical point. Using molecular dynamics the authors showed that this enhanced attraction lead to cluster formation, which in turn was able to  reproduce the observed enhancement of $\kappa4/\kappa_2$. However, this approach failed to reproduce the observed suppression of the third-order cumulant ratio, $\kappa_3/\kappa_2$, which is a general problem of cluster formation models, as we shall discuss in more detail in Sec.~\ref{sec7:fact_cumulants}. 
As discussed earlier in
this paper, Section~\ref{sec:exp:flow}, a similar sharp increase 
of net-proton directed-flow slope parameter ($dv_1/dy|_{y=0}$) has also been observed in the same energy region 
\cite{Adamczyk:2014ipa, Adamczyk:2017nxg}, see Fig. \ref{fig:exp:f13}. An increase in the directed
flow, however, would rather suggest repulsive 
interactions. Also, as we shall discuss in Section~\ref{sec7:fact_cumulants}, the third order
factorial cumulant is negative, which would also naively suggest repulsion instead of attraction.
Obviously, a simple picture of nucleon interactions does not suffice to explain the observed trends
and more refined calculations are needed in order to understand the nature of the energy dependence of the observed correlation pattern. 

\begin{figure}[!hbt] 
\begin{center}
\includegraphics[scale=0.35]{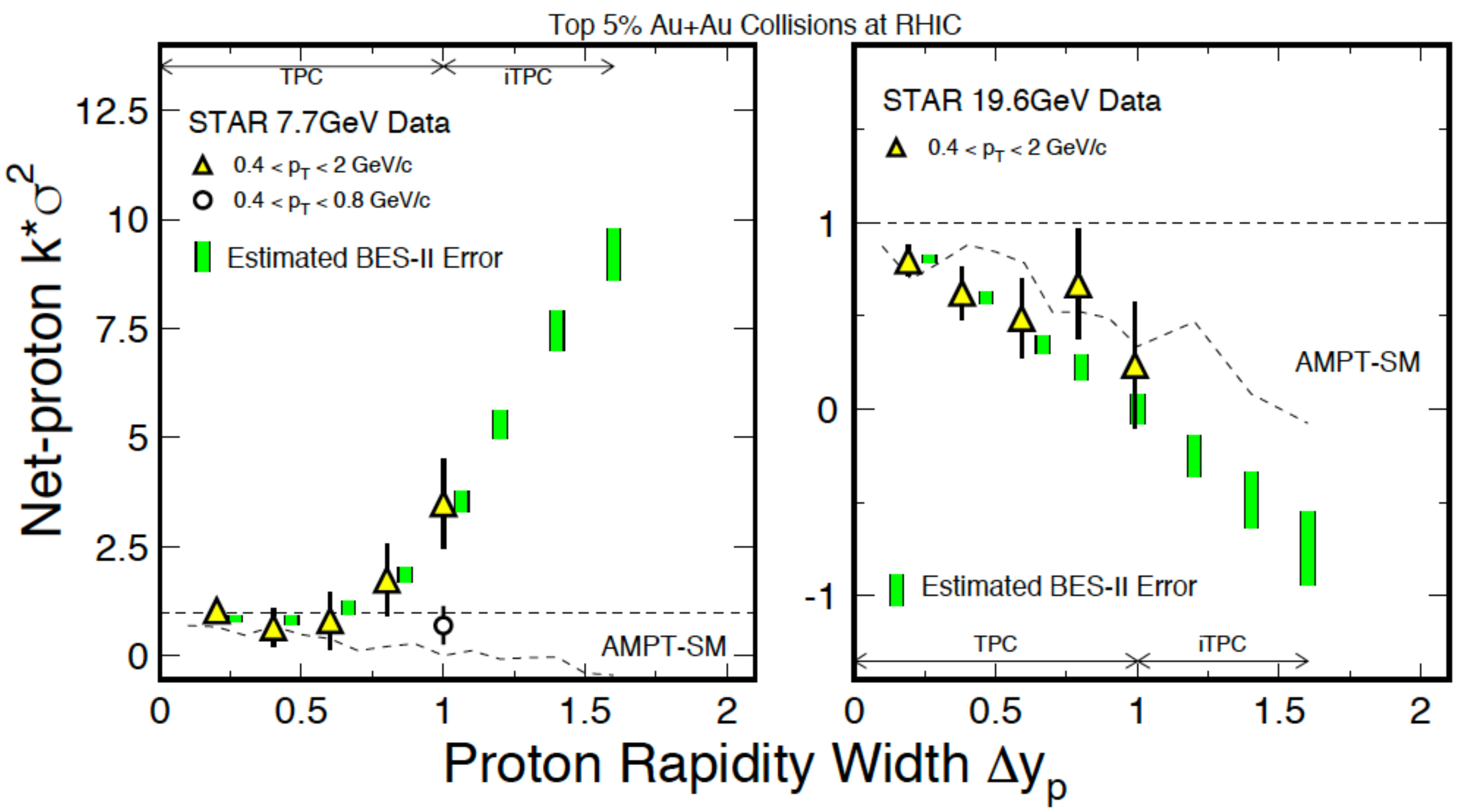} 
\caption{Rapidity width dependence (that is, protons are measured in $|y|<\Delta y_{p}/2$) of the net-proton fourth-order cumulant ratio $\cum{4}/\cum{2}=\kappa \sigma^2$ from the top 5\% Au+Au collisions at 
$\sqrt{s_{NN}}=7.7$ GeV (left panel) and $\sqrt{s_{NN}}=19.6$  GeV. The preliminary STAR data, from $|y|<0.5$ and $0.4 <p_T < 2.0$ GeV/c, 
are shown as triangles and the green-bars are the estimated error bars from the second phase beam energy scan program (BES-II). 
The long-dashed lines are the results from AMPT (with string-melting) simulations. The extended rapidity coverage by STAR's iTPC 
upgrade are shown in both plots.} 
 \label{fig:nu:exp:f18}
\end{center}
\end{figure}

Let us next turn to the dependence of the cumulant ratios $\kappa\sigma^{2}$ and $S\sigma$ on the size
of the rapidity and transverse momentum space window. This will provide insight on the correlation
range in momentum space. As discussed in Section~\ref{sec:crit-pt-correl-momentum} and in Ref.~\cite{Ling:2015yau},  one expects
that the momentum correlations due to critical fluctuations should be similar to that of the thermal system, which translate into
a rapidity correlation length of $\Delta y_{\rm corr} \simeq 1$. In Fig.~\ref{fig:exp:f17} we present the
preliminary STAR results for various transverse momentum windows (left-plots) and rapidity windows (right-plots) for net-proton $\kappa \sigma^2$ (top panels). In the bottom
panels we show the same dependence for the third order cumulant ratio, $S\sigma$/Skellam, which is
normalized by the expectation for an uncorrelated system which follows a Skellam distribution.
As indicated in the Figure the data are for  the top 5\% Au+Au collisions. In case of the
fourth order cumulant ratio $\kappa \sigma^2$, the maximum non-monotonicity is observed for the largest possible
phase-space, i.e.  $0.4 \le p_T \le 2$ GeV/c and $|y| \le 0.5$.  When the  phase space is reduced,
either by reducing transverse momentum window or the rapidity coverage, the observed
non-monoticity is reduced and the cumulant ratio $\kappa \sigma^{2}$  move closer to the expected
value for an uncorrelated system, where $\kappa \sigma^{2} =1$. A similar
trend is also seen in case of the normalized third order cumulant ratio, $S\sigma$/Skellam: the maximum
deviation from unity is seen for the maximum phase-space available to  the STAR experiment.  Once the
phase-space is reduced, the cumulant ratio  again approaches the expectation for an uncorrelated
system, $S\sigma/\rm Skellam = 1$. As we shall discuss in more
detail in Section~\ref{sec7:correlation_rapidity}, these data are consistent with 
long-range 
correlations in both transverse momentum and rapidity. However, we should keep in mind that the maximum rapidity window is rather small so that the present data are consistent
with the expectation that the correlation length is $\Delta y_{\rm corr} \simeq 1$, the aforementioned
expectation of a thermal rapidity correlation length. Clearly, the presently
available acceptance of the STAR detector is  too small to determine the momentum space
correlation range, which would be a very important piece of information in general and in
particular with regards to possible criticality. If the momentum correlation range is indeed
thermal, as suggested in \cite{Ling:2015yau}, then one should eventually see a leveling off of the
cumulant ratio with increasing size of the rapidity window. In the future, as part of its
upgrade program, the STAR experiment will extend the rapidity coverage for protons significantly.
The upgrade extends the proton rapidity coverage from $|y|<0.5$ to $|y|<0.8$, as shown in Figure~\ref{fig:nu:exp:f18}. Also shown are the expected statistical errors from BES-II for both 7.7~GeV 
and 19.6~GeV central collisions  as green-bars, while the expected statistics from
BES-II are listed in Table~\ref{tab:exp:bes}. Also shown in Fig.~\ref{fig:nu:exp:f18} as dashed lines are results obtained with the AMPT event generator. In this case the cumulant ratios decrease with increasing rapidity window
as one would expect from baryon number conservation. 

%\vspace{-0.15in}

\subsubsection{ Efficiency corrections}
\label{sec6:efficiency}

Let us finish this subsection with some remarks concerning detection efficiencies and detector effects. Before comparing to theoretical expectations, all measured quantities will have to be corrected for experimental efficiencies 
including (i) detecting efficiencies, (ii) identification probabilities and (iii) and fiducial
coverages. Although the last correcting factor involves a constant number for
a given measurement, both (i) and (ii) involve the stochastic nature of the detector performance
under specific circumstances. Furthermore, they are all dependent on the momentum
of the particle under study. For the proton and anti-proton measurements discussed in this section,
this implies that one must understand all of the aspects involving proton (and anti-proton) measurements at RHIC including effects of run-to-run fluctuating luminosity, pile-up, event-multiplicity, 
particle momenta and so on. For single spectra measurements, the most relevant  are the efficiency corrections. In that case, the averaged values of the 
efficiency are often extracted, either as a function of either transverse momentum or rapidity or
event-multiplicity or all of them, from the careful simulations of a given colliding system.
However, for the multi-particle measurements, this procedure becomes much
more involved as we shall briefly explain below. For more details see, e.g., Ref.~\cite{Bzdak:2016qdc}. 
 
In general, if there are $N$ produced particles, say protons, the
detector will, due to finite detection efficiencies register $n$ particles, with $n\leq
N$. However, the fraction of particles  detected
fluctuates from event to event. Therefore, we may detect the same number of protons for events with
different number of produced protons, and, vice versa, we may detect different numbers of protons for
events which have the same number of protons. To formalize this situation, let us denote by $P(N)$ the multiplicity distribution of the produced
particles and by $p(n)$ the multiplicity
distribution of the observed (detected) particles. The relation between the
two is then given by 
\begin{equation}
p(n)=\sum_{N=n}^{\infty }B(n,N)P(N),  \label{eq:ab:B-effi}
\end{equation}%
where $B(n,N)$ is the probability to observe $n$ particles given
$N$ produced particles. In general $B(n,N)$ may depend on phase space, particle  multiplicity etc,
which we ignore here for simplicity. 
The aforementioned  (single particle) detection efficiency,
$\epsilon$, is given by the ratio of the number of detected particles, $\ave{n}=\sum_{n}n p(n)$,     
over the average number of produced particles, $\ave{N}=\sum_{N}N P(N)$,
\begin{align}
  \epsilon \equiv \frac{\ave{n}}{\ave{N}}= \frac{\sum_{n=0}^{\infty}\sum_{N=n}^{\infty } n B(n,N)P(N)}{\sum_{N}N P(N)} .
\end{align}
In practice, the detection probability, $\epsilon$, for single particles is, to first  approximation
independent from the presence of other particles. In this case the $B(n,N)$ is given by a binomial
distribution with the Bernoulli probability equal to the detection efficiency $\epsilon$
\begin{equation}
  B(n,N )=\frac{N!}{n!(N-n)!}\epsilon ^{n}(1-\epsilon )^{N-n}.
  \label{eq:sec63:binomial_dist}
\end{equation}%
In this case it is easy to convince oneself that the single particle efficiency is indeed
$\epsilon$. Also, in this case the efficiency to detect pairs
\begin{align}
  \epsilon_{2} \equiv \frac{\ave{n(n-1)}}{\ave{N(N-1)}}= \epsilon^{2}
\end{align}
is simply the product of two single particle efficiencies. And more generally one can show that
the factorial moments for the produced protons, $F_{i}$, are given by
\begin{equation}
F_{i}=\left\langle \frac{N!}{(N-i)!}\right\rangle =\frac{1}{\epsilon ^{i}}%
\left\langle \frac{n!}{(n-i)!}\right\rangle ,  \label{eq:ab:Fi-effi}
\end{equation}
where $\langle \frac{n!}{(n-i)!}\rangle $ is a factorial moment for the
measured (detected) particles.  Given the above relation between the factorial moments of produced
and detected particles, Eq.~\eqref{eq:ab:Fi-effi}, the efficiency corrections to the cumulants
then simply involve expressing the cumulants in terms of factorial moments. We note that the
preliminary results for the cumulant measurements shown in this section are all based on efficiency
corrections based on the assumption that $B(n,N)$ follows a binomial distribution,
Eq.~\eqref{eq:sec63:binomial_dist} \cite{Adamczyk:2013dal,Thader:2016gpa}.

As already mentioned, in reality, the situation is more complicated. For example, $\epsilon $
usually depends on the phase-space, $y,$ $p_{t}$ and $\varphi $. Also in case of net-protons one has
to deal we two particle species, protons and anti-protons. In this case more refined
methods are needed as discussed  e.g. in Refs.
\cite{Bzdak:2012ab,Bzdak:2013pha,Luo:2014rea,Kitazawa:2016awu,Luo:2017faz,Nonaka:2017kko}.
It is also
possible that $B(n,N)$ in Eq. (\ref{eq:ab:B-effi}) is not given by a binomial
distribution which seriously complicates the experimental analysis,  see Ref. 
\cite{Bzdak:2016qdc}. If $B(n,N)$ is binomial then we decide for each particle independently if it is
detected or not. This procedure obviously does not introduce any (artificial) correlations.
If $B(n,N)$ is non-binomial, on the other hand, the detection
mechanism is not independent for each particle and  artificial
correlations are introduced into the system by the detector. In this case one needs to invert the
matrix $B(n,N)$ so that one can extract the distribution of produced particles, $P(N)$ from that of the
measured ones, $p(n)$, 
\begin{align}
P(N) = \sum_{n=0}^{\infty}B^{-1}\left( n,N \right) p(n). 
\label{eq:sec63:unfold}
\end{align}
This procedure, which is commonly referred to as unfolding, is rather
complicated \cite{DAgostini:1994fjx,Schmitt:2012kp,Bzdak:2016qdc}. A simple
example is the situation where $\epsilon $ depends on the number of produced
particles, $\epsilon =\epsilon (N)$. Usually the more produced particles the
smaller efficiency and this is exactly the case for the STAR detector, see for example
Fig.~1 in Ref. \cite{Luo:2015ewa}. In Ref. \cite{Bzdak:2016qdc} a simple
model was considered where the efficiency linearly depends on the number of
produced particles%
\begin{equation}
\epsilon (N)=\epsilon _{0}+\epsilon ^{\prime }(N-\left\langle N\right\rangle
),
\end{equation}%
where $\epsilon _{0}=\sum_{N}P(N)\epsilon (N)$ is the average efficiency. It
was found that even a very small value of $\epsilon ^{\prime }\sim -0.0005$
can generate quite large effects. The problem of non-binomial efficiency and its effect on the measured
cumulants are currently under study  by the STAR Collaboration.

\subsection{Chirality}
\label{subsec_6_chirality}

In non-central heavy ion collisions, the positively charged and fast moving nuclei  not only bring in tremendous amount of angular momenta  but also produce a very strong magnetic field at  mid-rapidity, as already discussed at length  in Sec.~\ref{subsec_5_2}. 
In a typical collision at RHIC, the angular momentum reached could be as large as  $10^4\, \hbar$~\cite{Jiang:2016woz} and the strength of the magnetic field is on the order of  10$^{17}$ Gauss~\cite{Kharzeev:2007jp}, both being  aligned approximately along the direction perpendicular to the event plane. In analogy to the quantum hall effect in condensed matter physics where a strong magnetic field plays the crucial role, the angular momentum and magnetic field   provide the necessary  extreme conditions to induce novel quantum phenomena, such as the chiral magnetic and chiral vortical effect (see Sec.~\ref{sec5}). These effect, if seen in experiment,  in turn reveal   nontrivial properties of the hot and dense matter created in heavy ion collisions such as chiral symmetry restoration.

%\vspace{-0.15in}
\begin{figure}[!hbt] 
\begin{center}
\includegraphics[scale=0.45]{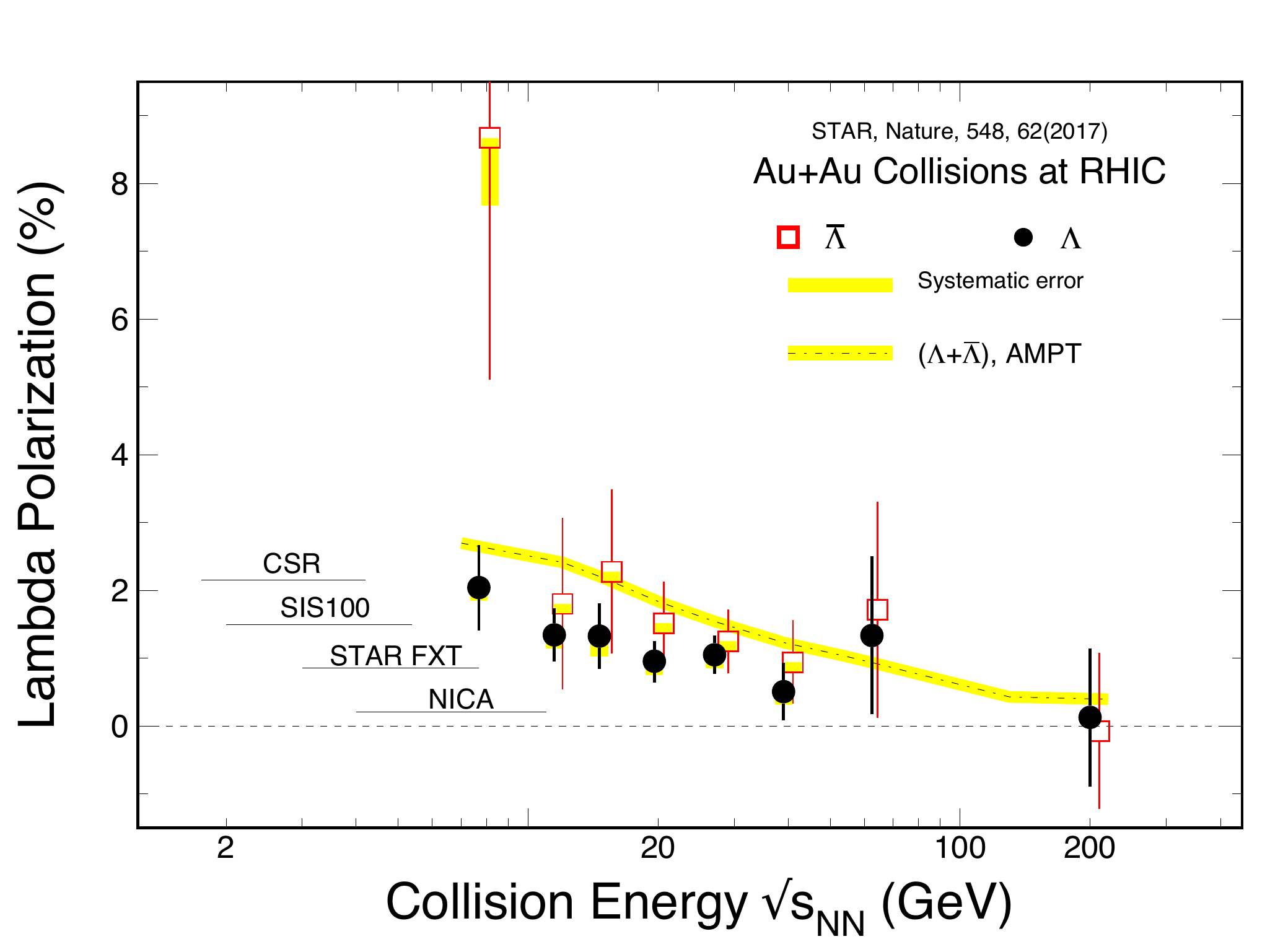} 
%\vspace{-0.1in}
\caption{The collision energy dependence of Lambda (filled-circles) and anti-Lambda (open-squares) polarizations from the non-central (20-50\%) Au+Au collisions at RHIC \cite{STAR:2017ckg}. The dot-dashed-lines represent the AMPT model calculations for the average polarization of combined $\Lambda$ and $\overline{\Lambda}$ hyperons \cite{Jiang:2016woz,Shi:2017wpk,Li:2017slc}. } 
 \label{fig:exp:f11}
%\vspace{-0.15in}
\end{center}
\end{figure}

The measurement of hyperon spin polarization with respect to the global event plane by the STAR collaboration during phase-I of BES has been the first experimental observation connected with the angular momentum of a  high-energy nuclear collision.  The results for  Au+Au collisions at RHIC are shown in Fig. \ref{fig:exp:f11}, where the observable $P$ quantifies the net polarization of hyperons, i.e. their level of preferential spin orientation  along the direction of global angular momentum, which in turn is determined from the directed flow of forward- and backward-traveling fragments via the beam-beam counters~\cite{STAR:2017ckg}.    
Both $\Lambda$ and $\overline{\Lambda}$ hyperons show a nonzero global  polarization at a  level of a few percent. Model estimates indicate the vorticity of the system to be as high as $(9\pm1)\times 10^{21}\, sec^{-1}$ \cite{STAR:2017ckg,Becattini:2016gvu}.
As an example, the result from the AMPT transport model is shown as yellow line in Fig. \ref{fig:exp:f11}. The estimated magnitude of vorticity of approximately $10^{21}\, sec^{-1}$ is many orders of magnitude larger than the previously known maximum vorticity of about $10^{7} \, sec^{-1}$ in superfluids~\cite{STAR:2017ckg}. The hot QCD fluid created in non-central heavy ion collisions, therefore, represents the ``most vortical fluid'' so far produced in the laboratory. 
We further observe that the global polarization signal strongly increases with decreasing collision energy, in line with the arguments given in Sec.~\ref{subsec_5_2}. Obviously, the observation of the global $\Lambda$ polarization provides strong evidence that the fireball created in these collisions carries angular momentum. 
 Since  the global polarization becomes stronger with decreasing collision energy,
 future experiments with the STAR fixed-target program, and at NICA, FAIR and CSR are well positioned to study this new phenomenon.

While the angular momentum with its associated vorticity is necessary for the spin polarization and possible existence of the Chiral Vortical Effect (CVE), the presence of a magnetic field is required for the Chiral Magnetic Effect (CME) and Chiral Magnetic Wave (CMW) in such collisions.
As already discussed in Sec.~\ref{subsec_5_2} (see, e.g., Fig.~\ref{fig:jl:f3}), an observation of nonzero angular momentum and vorticity would be an indirect indication that the initial strong  magnetic field may persist for a while in the system created as a consequence of the Faraday effect \cite{Tuchin:2013apa}. 
What is more, the systematic difference in the polarization between lambda (filled-circles) and anti-lambda (open-squares) in Fig.~\ref{fig:exp:f11}, with the latter having a larger polarization than the former,  may have already provided a direct hint at the effect of the magnetic field. Clearly a magnetic field would affect the polarization of particles and anti-particles with opposite sign~\cite{Becattini:2016gvu,Muller:2018ibh,Ye:2018jwq,Guo:2019mgh}. However, as shown in \cite{Csernai:2018yok}, a vortical net-baryon current, may lead to a similar splitting.
In addition, recent STAR measurements  on the dielectron production at very low transverse momentum and low mass regime at RHIC~\cite{Adam:2018tdm},
the  difference in the directed flow $v_1$ for charmed-hadron $D^0$ and $\overline{D^0}$~\cite{Singha:2018cdj} at RHIC as well as the large $v_2$ of $J/\Psi$ at high transverse momentum at LHC~\cite{Guo:2015nsa}  might provide independent evidence for the   strong magnetic fields that may persist long enough to affect observables in such collisions. 
We now turn to the measurements related to the anomalous chiral effects driven by the magnetic field.

%\vspace{-0.15in}
\begin{figure}[!hbt] 
\begin{center}
\includegraphics[scale=0.75]{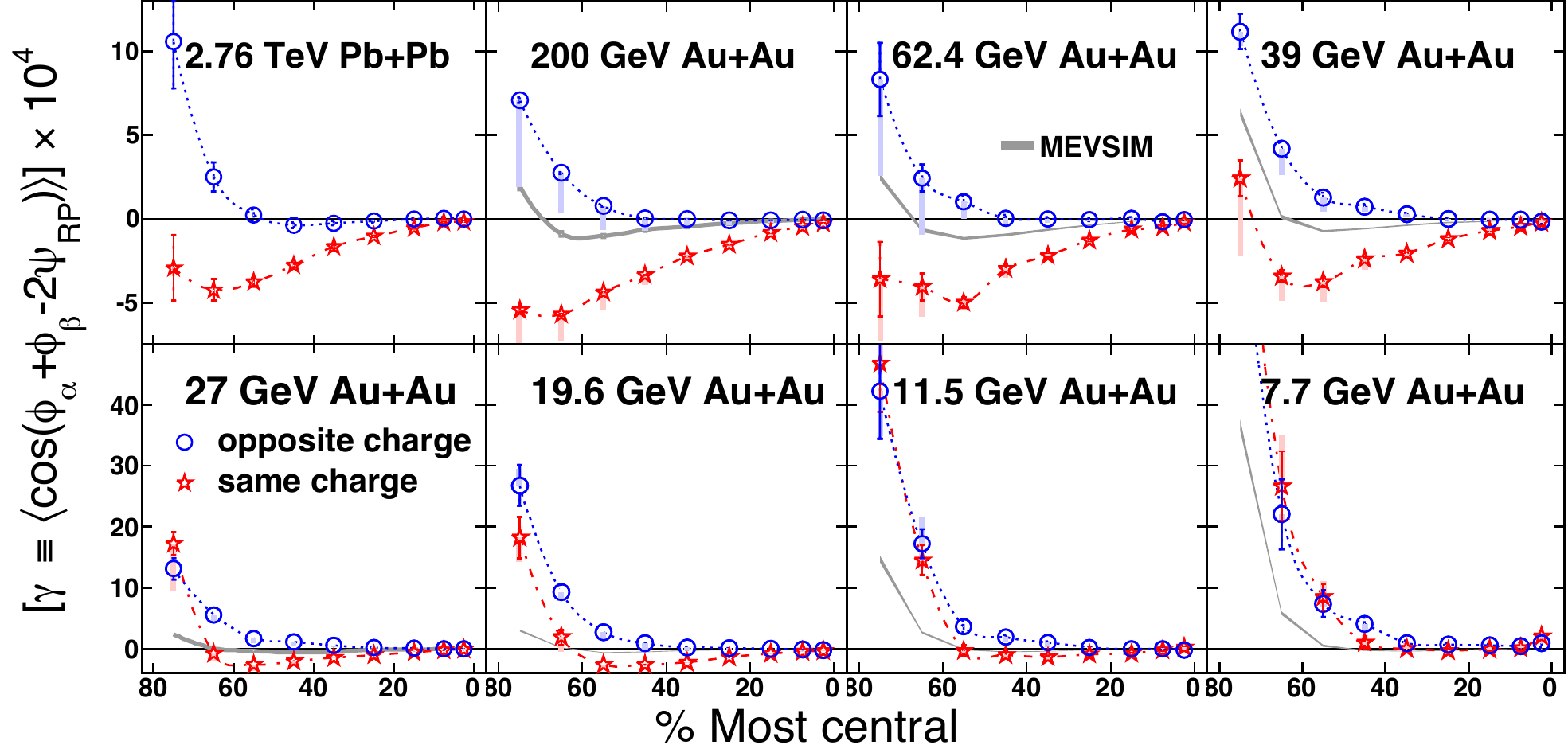} 
%\vspace{-0.1in}
\caption{The $\gamma$ correlator as a function of centrality for Au+Au collisions at beam energy from $7.7 \gev$ to $200 \gev$~\cite{Adamczyk:2014mzf} and for Pb+Pb collisions at beam energy $2760 \gev$~\cite{Abelev:2012pa}. Figure from \cite{Kharzeev:2015znc}.} 
 \label{fig:exp:f12n}
%\vspace{-0.15in}
\end{center}
\end{figure}

As discussed in Sec.~\ref{subsec_5_3}, one observable consequence of the CME is the electric charge separation along the direction of the external magnetic field. Such a charge separation can be measured through the charge-dependent azimuthal correlator $\gamma^{OS}$ and $\gamma^{SS}$, which are defined in Eq.~\eqref{eq:jl:gamma}. These observables have been measured at both RHIC and the LHC over a wide span of collision energies~\cite{Abelev:2009ac,Abelev:2009ad,Abelev:2012pa,Khachatryan:2016got,Sirunyan:2017quh} and are shown in Fig.~\ref{fig:exp:f12n}. For collision energies of $\sqrt{s_{NN}} \gtrsim 19.6 \gev$  we observe a significant  charge asymmetry in the correlators, that is, a difference between $\gamma^{OS}$ and $\gamma^{SS}$. This difference disappears for the lowest two energies, $7.7\gev$ and $11.5\gev$.
Also, as discussed in Sec.~\ref{subsec_5_3}, one expects for the CME that $\gamma^{OS}>\gamma^{SS}$ and that this difference increases toward more peripheral collisions. This trend can be seen in the data.

\begin{figure}[!hbt] 
\begin{center}
\includegraphics[scale=0.3]{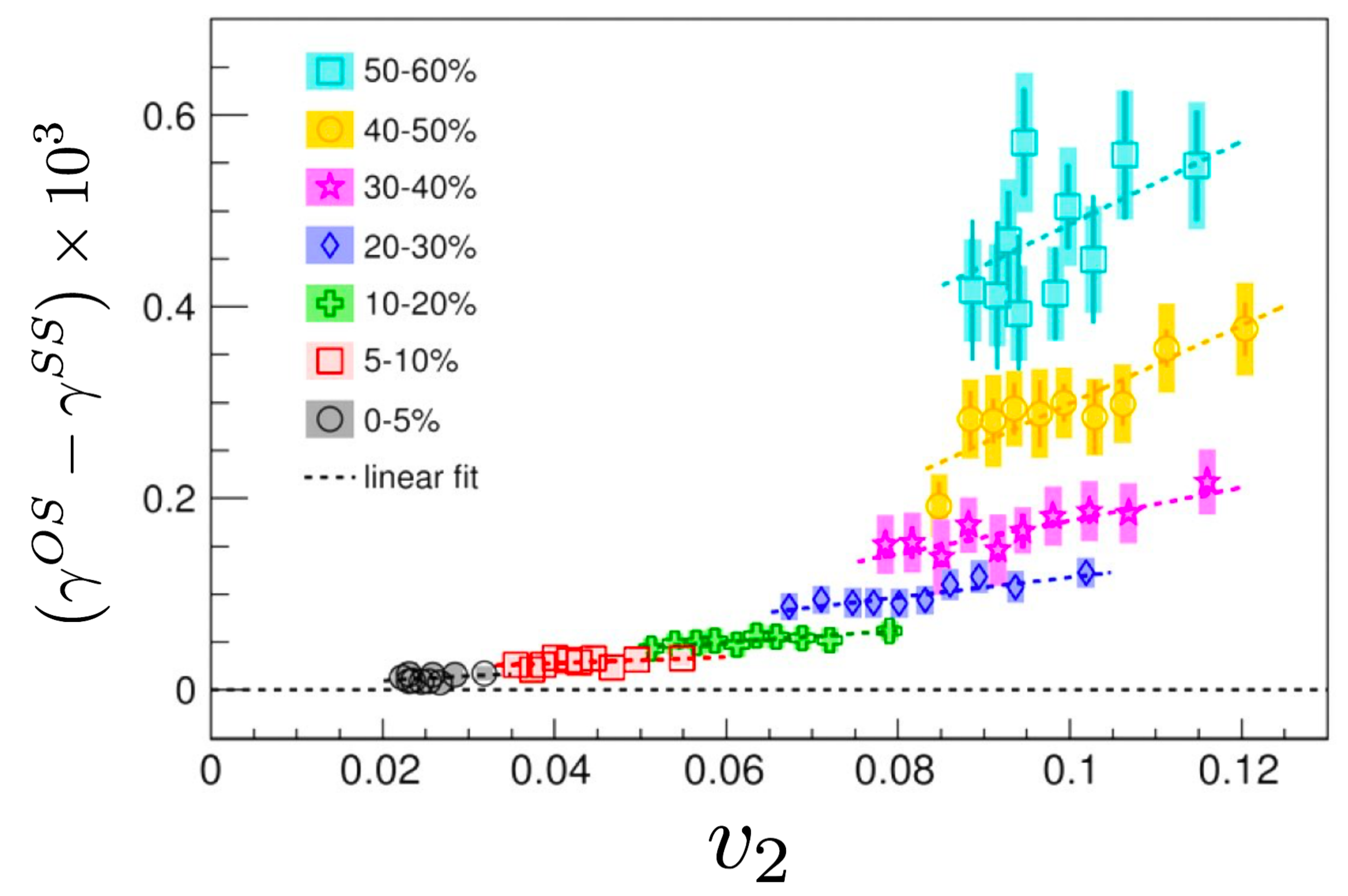} 
%\vspace{-0.1in}
\caption{The correlation of charge separation $(\gamma^{OS} - \gamma^{SS})$ as a function of $v_2$ for a variety of centrality in $\sqrt{s_{NN}}=2760$ GeV  Pb+Pb  collisions. This plot is adapted from the publication of ALICE collaboration~\cite{Acharya:2017fau}.} 
 \label{fig:exp:f12a}
%\vspace{-0.15in}
\end{center}
\end{figure}

However, as discussed in detail in Sec.~\ref{subsec_5_3}, at this point we should remind the reader that the interpretation of the $\gamma$-correlator is complicated by strong background contributions. 
These backgrounds arise from certain genuine two-particle correlations coupled with the anisotropic collective expansion, and their contributions to $\gamma$ are thus (roughly) linearly dependent on the elliptic coefficient $v_2$, see Eq.~(\ref{eq:jl:FH}). One possible way to study and potentially remove this flow induced background is by means of the so-called event-shape engineering analysis~\cite{Bzdak:2011np,Acharya:2017fau}. Within a given centrality (where the magnetic field is expected to approximately stay the same), one could further divide the events into subgroups according to their measured $v_2$ and then examine how the $\gamma$-correlator varies with respect to $v_2$. An example of such an analysis for Pb+Pb collisions at $2.76 \rm\, TeV$ by ALICE is shown in Fig.~\ref{fig:exp:f12a}~\cite{Acharya:2017fau}. There, in a given centrality class we see a more or less linear dependence  of $\gamma$ on $v_2$. An extrapolation of this trend toward $v_2=0$ indicates a possibly nonzero intercept that could be the CME signal. Combing results from the $10-50\%$ centrality range,  ALICE concludes  that the fraction of CME signal   could be  about $8\sim10\%$ of the measured $\gamma$ correlators but this is still subject to significant  systematic and statistical uncertainties.

%Fig.~\ref{fig:exp:f12a} shows the preliminary results of correlation of charge separation $\Delta \gamma = \gamma^{opposite-sign}-\gamma^{opposite-sign}$ as a function of $v_2{2}$ from central collisions of top 0-20\% Au+Au at $\sqrt{s_{NN}} = 200$ GeV and top 0-10\% U+U at  $\sqrt{s_{NN}} = 193$ GeV cite{Tribedy:2017hwn}. In both collisions, $\Delta \gamma$ is proportional to the event anisotropy $v_2$ but the values of $\Delta \gamma$ become zero at finite value of $v_2$.

%\vspace{-0.15in}

%\vspace{-0.15in}
\begin{figure}[!hbt] 
\begin{center}
\includegraphics[scale=0.24]{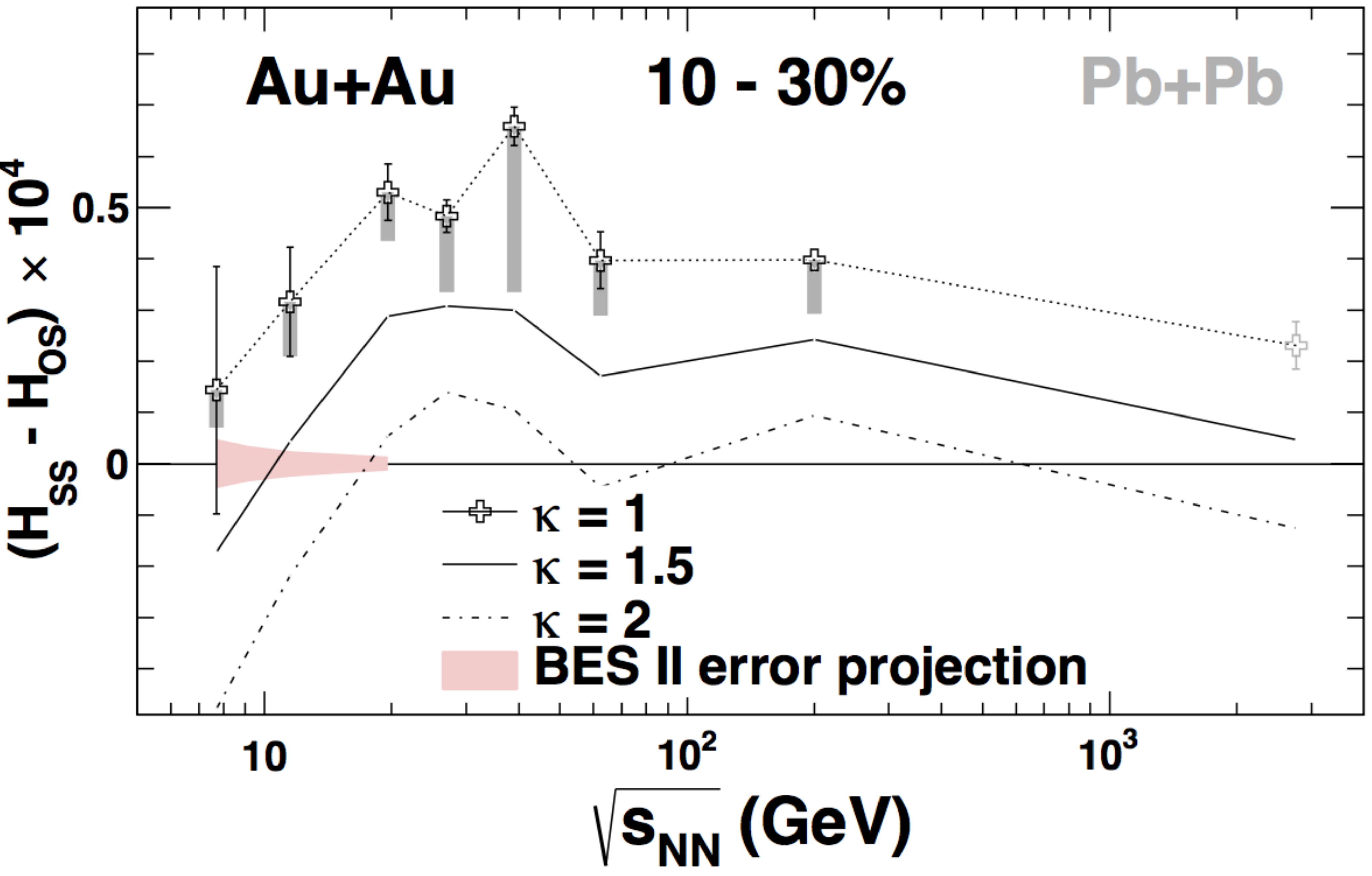} 
%\vspace{-0.1in}
\caption{Collision energy dependence of the charge separation for mid-central (10-30\%) Au+Au collisions from STAR Collaboration~\cite{Adamczyk:2014mzf}. 
The default values (dotted curves) are from H$^{\kappa = 1}$, and the solid (dash-dot) curves are obtained 
with $\kappa$ = 1.5 ($\kappa$ = 2). For comparison the results for Pb+Pb collisions at 2.76 TeV are also 
shown~\cite{Abelev:2012pa}. 
% \cite{CMEPhysRevLett.103.251601,CMEPhysRevC.81.054908,CMEPhysRevLett.113.052302,CMEPhysRevLett.110.012301}.  
The vertical asymmetric bands represent the systematic errors and the colored band 
indicates the statistical errors from the proposed RHIC BES-II program. } 
 \label{fig:exp:f12}
%\vspace{-0.15in}
\end{center}
\end{figure}

Another approach \cite{Bzdak:2012ia}, discussed in detail in Sec.~\ref{subsec_5_3}, is to decompose the $\gamma$-correlator into one part which scales with $v_{2}$, and one which does not (see Eq.~\eqref{eq:jl:FH}).
An analysis along this line  was completed with BES-I data by STAR~\cite{Adamczyk:2014mzf}, with suitable assumptions for the kinematic $\kappa$ factor introduced in Eq. (\ref{eq:jl:FH}) and discussed therein.
Fig.~\ref{fig:exp:f12} shows the STAR results~\cite{Adamczyk:2014mzf} for  the collision energy dependence of the component $H$, which is independent of $v_{2}$ and thus may be interpreted as the signal for charge separation. These results are from the mid-central ($10-30$\%) Au+Au collisions from $7.7 \gev$ to $200 \gev$. For comparison, the ALICE result \cite{Abelev:2012pa} is also shown on the high beam energy end in the figure. One uncertainty involved in such an analysis, as discussed already in Sec.~\ref{subsec_5_3}, is the choice of the $\kappa$ factor for background correlations. As shown in, e.g., Ref. \cite{Bzdak:2010fd,Bzdak:2012ia}, the value of this factor is sensitive to the momentum spectra of charged particles, the differential flow coefficient $v_2(p_T)$, the detector acceptance and kinematic cuts, etc.
The factor  $\kappa$ may be estimated by performing simulations of the bulk evolution (without the presence of CME) with e.g. event generators such AMPT or UrQMD or using hydrodynamics. Such an exercise, taking into account the STAR acceptance as well as kinematic cuts, leads to values of $\kappa$ in the  range of $1\lesssim \kappa \lesssim 2$~\cite{Wang:2017wzm}. 
For the analysis shown in Fig.~\ref{fig:exp:f12}, three lines are presented corresponding to the choice of $\kappa=1, 1.5, 2$ respectively to reflect the influence of this uncertainty.  
As one can see from the Figure, the so-obtained signal shows a nontrivial dependence on the collision energy, reaching a maximum around $39$ GeV while approaching zero when the energy is lower than 11 GeV.  This specific trend appears in qualitative agreement with expectations from CME, as discuss in Sec.~\ref{subsec_5_4}. The apparent decrease of the H-correlator toward the very high energy end deserves a special note. The ALICE collaboration~\cite{Acharya:2017fau} combined a similar two-component decomposition strategy with the event-shape engineering analysis to constrain the level of a possible CME signal at $2.76\rm \,TeV$ in the measured $\gamma$-correlator and the results are consistent with the above H-correlator results. Furthermore, recently the CMS collaboration reported its results on charge dependent azimuthal particle correlations with respect to the second-order event plane  in both p+Pb and Pb+Pb collisions at 5.02 TeV~\cite{Khachatryan:2016got}.  At the same multiplicities, a similar behavior in p+Pb and Pb+Pb collisions is observed. 
A further analysis by CMS carefully examined such correlations in Pb+Pb collisions with respect to both the second and third harmonic flow event planes and attempted to extract the background contributions~\cite{Sirunyan:2017quh}. 
These  analyses put a very stringent upper limit, no more than about $4\%$ for Pb+Pb according to CMS, on any potential CME fraction in the $\gamma$-correlator at 5.02 TeV collisions~\cite{Sirunyan:2017quh}.  
The ALICE and CMS measurements at the LHC, therefore, consistently suggest a gradual decrease of the potential CME signal  with increasing beam energy and its possible disappearance at the very high energy end (e.g. $\sim 5$ TeV and beyond).

%\vspace{-0.15in}
\begin{figure}[!hbt] 
\begin{center}
\includegraphics[scale=0.085]{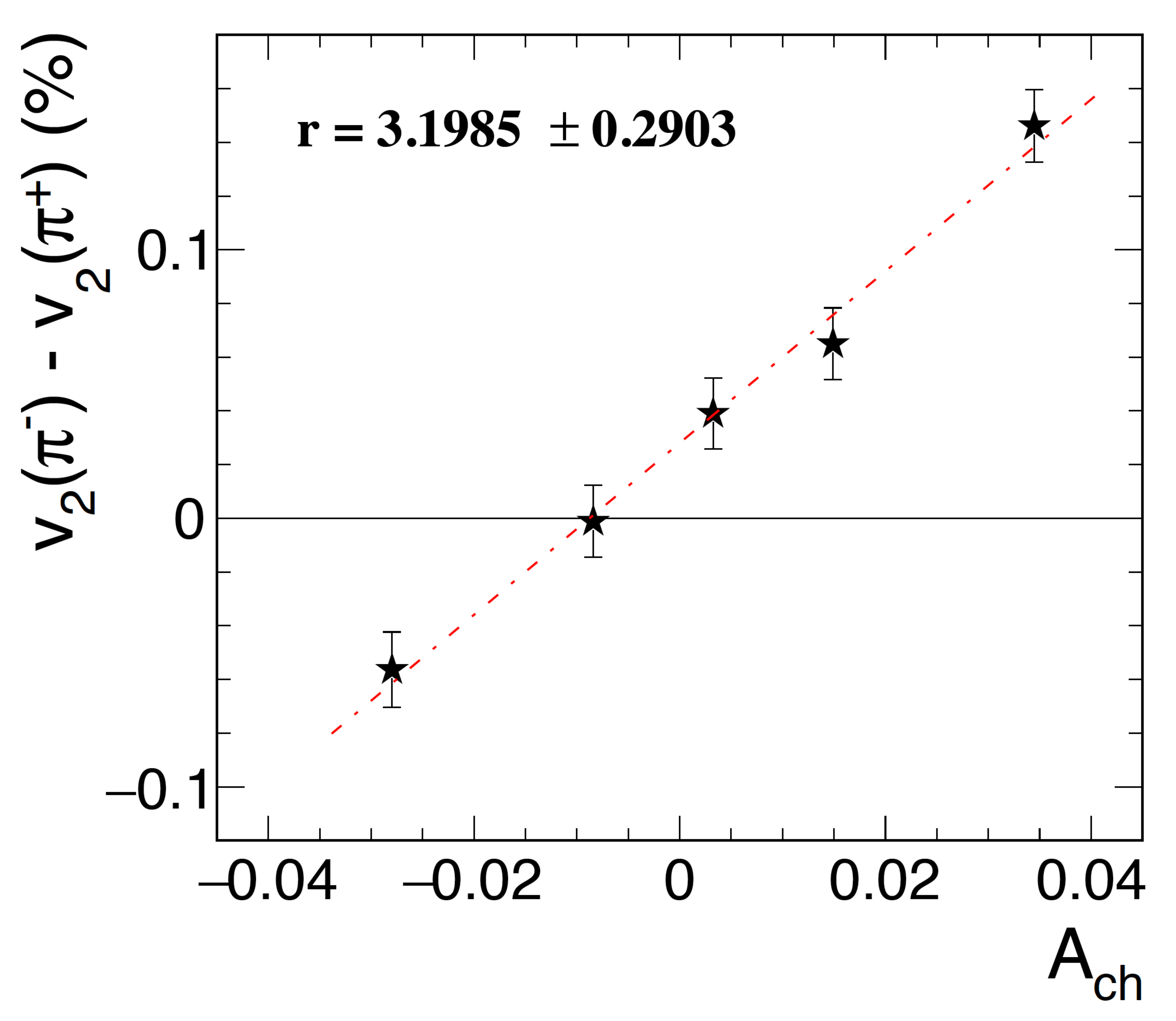} \hspace{0.2in}
\includegraphics[scale=0.28]{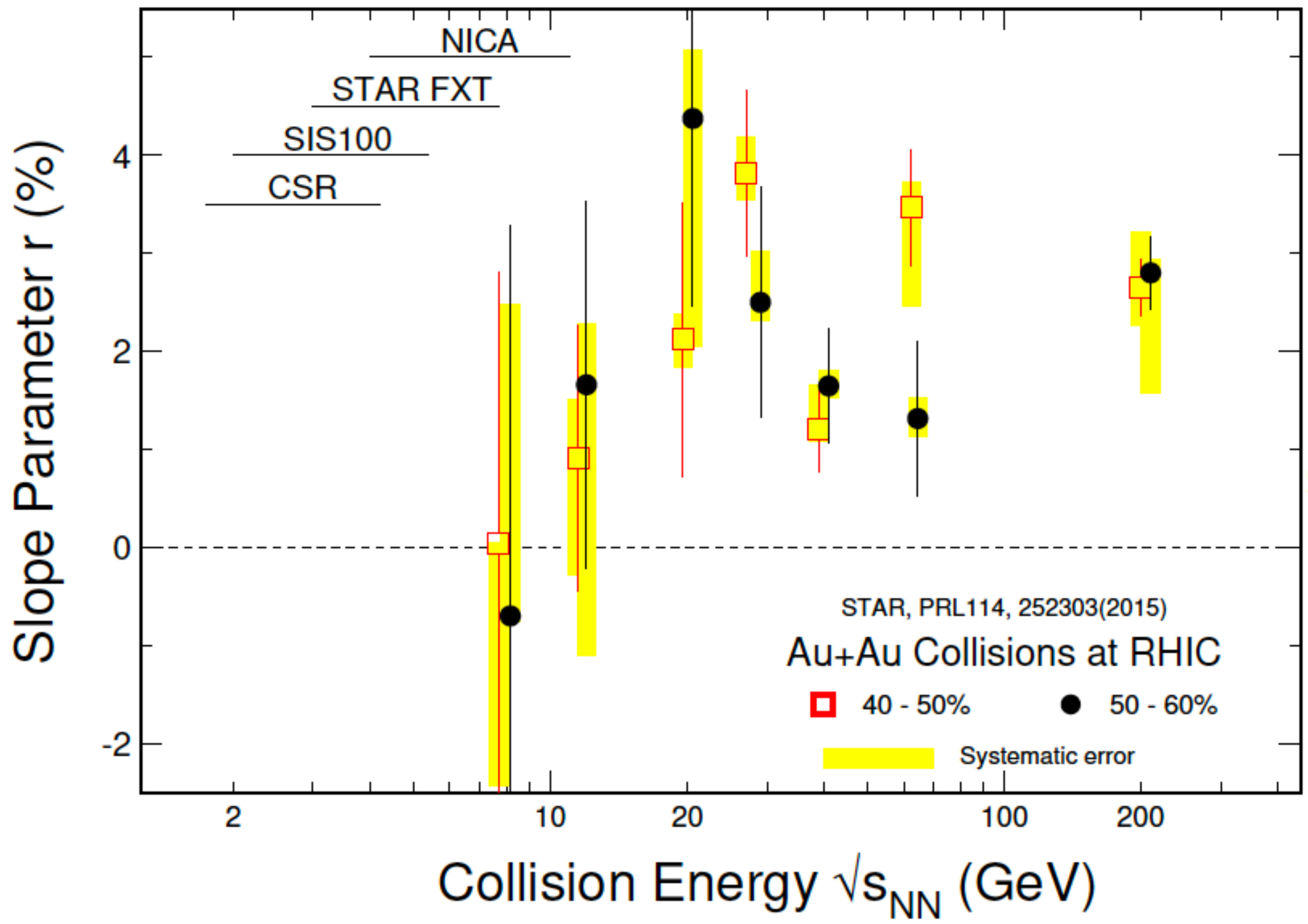} 
%\vspace{-0.1in}
\caption{(left) The difference in $v_2$ between $\pi^-$ and $\pi^+$ as a function of charge asymmetry for 30-40\% central Au+Au collisions at $200$ GeV. (right) Collision energy dependence of the pion charge asymmetry slope parameter $r$ from the Au+Au collisions at RHIC \cite{Adamczyk:2015eqo}. Open-squares and the filled-circles represent the collision centrality of 40-50\% and 50-60\%, respectively.  Both panels are adapted from \cite{Adamczyk:2015eqo}.} 
 \label{fig:exp:f12b}
%\vspace{-0.15in}
\end{center}
\end{figure}

As discussed in Secs.~\ref{subsec_5_1} and \ref{subsec_5_3},  in addition to the CME, the magnetic field can also induce the Chiral Magnetic Wave (CMW).
The CMW predicts a specific splitting between the elliptic flow, $v_2$, of $\pi^-$ and $\pi^+$, as given in Eq.~\eqref{eq:jl:v2Ach} of Sec.~\ref{subsec_5_3}, with a linear dependence on the charge asymmetry $A_{ch}= \frac{N_+ - N_-}{N_+ + N_-}$. Such a feature is indeed confirmed by  experimental data, as shown in Fig.~\ref{fig:exp:f12b} (left). In the CMW scenario, the slope parameter $r$ extracted from this linear dependence is directly related to the CMW-induced quadruple moment of the charge distribution in the QGP. The estimated values for $r$ from CMW calculations are in good agreement with STAR data~\cite{Burnier:2011bf,Adamczyk:2015eqo}.

The STAR measurements  for the collision energy dependence of the  parameter    $r$ are shown in Fig.~\ref{fig:exp:f12b} (right), for Au+Au collisions in two centrality bins, 40-50\% and 50-60\% at RHIC BES energies.   
As one can see in this figure,  at high energies,   $\sqrt{s_{NN}} > 20$ GeV, the slope parameter is clearly finite, $r>0$, consistent  with the CMW expectations. At energies below 15 GeV, on the other hand, $r$ is consistent with  zero within the albeit large error bars. As discussed earlier in  Sec.~\ref{subsec_5_3}, the effect of the CMW requires the existence of QGP with chiral restoration. Therefore, the vanishing trend of the $r$ parameter at low energies could be an indication of broken chiral symmetry in the  hadronic phase dominating the fireball at such energies~\cite{Adamczyk:2015eqo}. 
%The observed positive $r$ in mid-central Au+Au collisions cannot be reproduced by any of the conventional models \cite{CMWPPNP4.255,CMWJPhysG31.S49,CMWPhysLettB735.383}. 
A similar nonzero slope parameter was also measured and  reported by ALICE for Pb+Pb collisions at $2.76$ TeV \cite{Adam:2015vje}. At the even higher energy of $5.02 \rm \, TeV$, the CMS  measurements~\cite{Sirunyan:2017tax}  appear to suggest a negligible CMW signal at such energy,  which is similar to and consistent with the CME measurements by CMS at the same  energy.    
 
Let us briefly summarize this subsection, which  focused on measurements pertinent to the chirality aspect in high energy nuclear collisions. There are strong  experimental  evidences for a nonzero angular momentum 
and vorticity structure in the  fireball created by such collisions.  There also appear compelling  indications for  the existence of very strong magnetic fields that may last for sufficient amount of time during the  fireball evolution to induce observable effects.   
Given such external conditions, one expects the occurrence of anomalous transport like the CME and CMW. An unambiguous observation of these effects would provide,
 for the first time,  the long sought-after  
 experimental evidence for a QGP with restored chiral symmetry as well as the presence of gluonic topological fluctuations. The CME is predicted to induce a charge separation that can be measured through charge dependent azimuthal correlations, while the CMW is predicted to induce a charge quadruple in the fireball that can be measured through charge dependent elliptic flow. These measurements have been done over a wide span of collision energies from RHIC to LHC, with interesting hints of CME and CMW signals that qualitatively agree with theoretical expectations. A conclusion however can not be drawn yet, due to strong background contamination that is hard to separate unambiguously or evaluate precisely. The current  understanding of these measurements and relevant issues will be further discussed in the next section.  However, it is important to note here that the potential signals of both CME and CMW appear to have a non-monotonic dependence on collision energy, vanishing below a certain threshold energy while also disappearing at  very high energies. In light of the discussions on the beam energy dependence of anomalous chiral transport in Sec.~\ref{subsec_5_4}, this nontrivial trend may actually be considered as a characteristic to be expected for CME and CMW signals. It also appears   that  the optimal  window  for the precision search of anomalous chiral transport effects would be in the RHIC BES energy region.  

We end this subsection by mentioning the exciting prospect, brought by the collisions of
isobars~\cite{Skokov:2016yrj}, for disentangling the background contamination and unambiguously
identifying the CME signal. These  experiments were completed in 2018 and substantial amount of data has
been taken, which are being processed and analyzed at the time of this writing. A more detailed
discussion of the isobaric collisions will be presented in Sections~\ref{sec7c} and \ref{sec8}.

%=====================================================================================
%

%%% Local Variables:
%%% mode: latex
%%% TeX-master: "BES_Main_current"
%%% End:

%-------------Section 7-----------------------------------
%
%=====================================================================================
%=====================================================================================
\section{Discussions on BES-I Results}
\label{sec7}
In the following we will be discussing the experimental results obtained during the first phase of the
RHIC beam energy scan, which we have presented in the previous section. We will concentrate on the
fluctuation measurements, which are relevant for the critical point and first-order transition search, and the results for the
correlation functions which are sensitive to anomalous transport.

\subsection{Discussion of the STAR data on proton number cumulants}
\label{sec:ab:star-data-discussion}

In Section~\ref{sec6:exp:cumulants} we have presented the current status with regards to the measurement of
particle number cumulants. Especially the beam energy dependence of the net-proton cumulants show
some rather interesting features for energies below $\sqrt{s_{NN}}\lesssim 19\gev$. 
In this Subsection we present several observations regarding the preliminary
STAR data on the (net)-proton cumulants. We will focus mostly on the lowest energies, $%
\sqrt{s_{NN}}\leq 19$ GeV, where several experimental observables are
characterized by non-monotonic behaviour, including the cumulant ratios of
the proton distribution. We will further discuss effects on the cumulants due to baryon-number
conservation and participant (or volume) fluctuations, which need to be understood in order to
relate the measurements with most model calculations. Finally, since the number of anti-protons is
negligible for energies below $\sqrt{s_{NN}}\lesssim 19\gev$ in the following we will discuss mostly
cumulants of the protons, which simplifies some of the developments.

\begin{figure}[h]
\begin{center}
\includegraphics[scale=0.3]{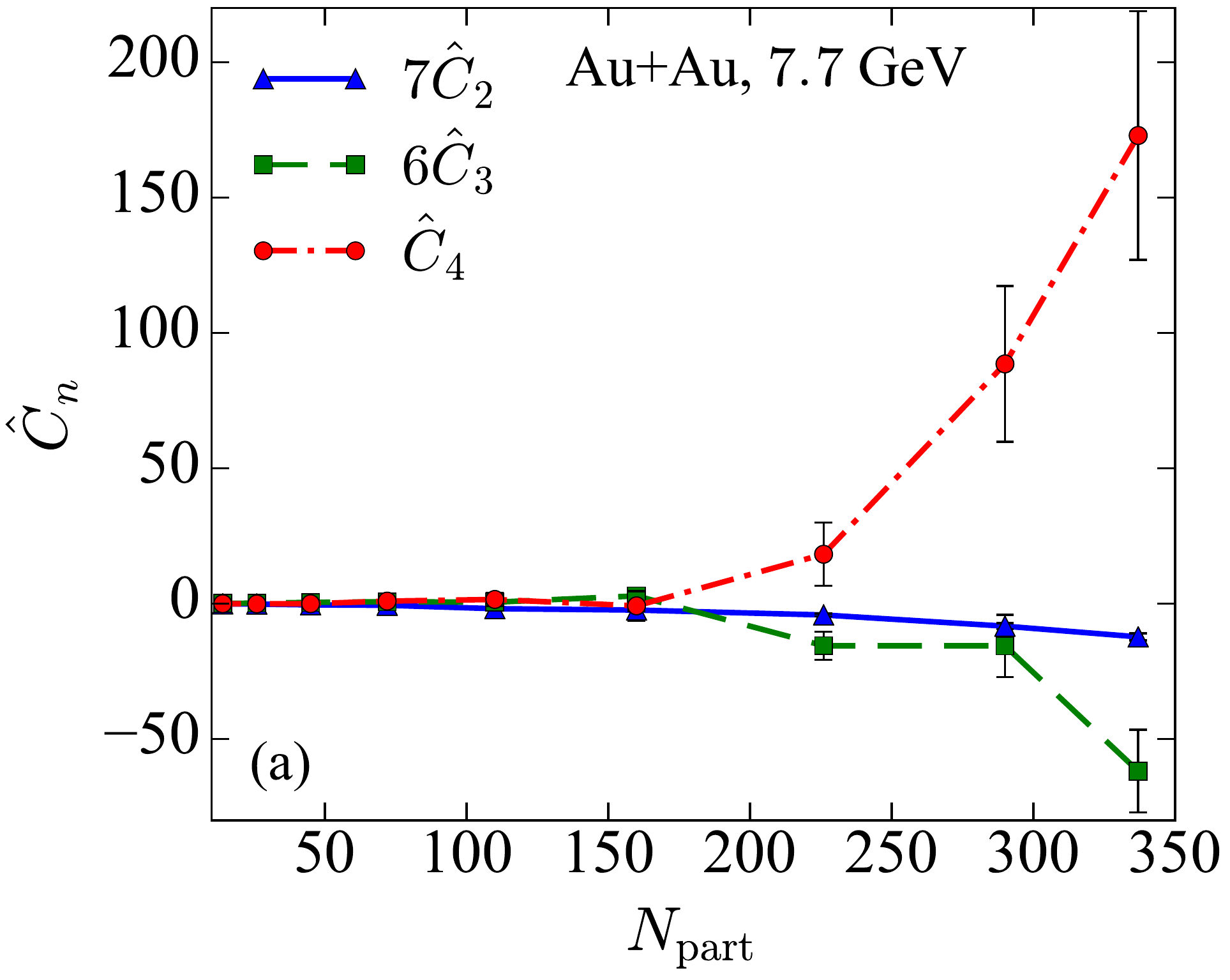} %
\includegraphics[scale=0.3]{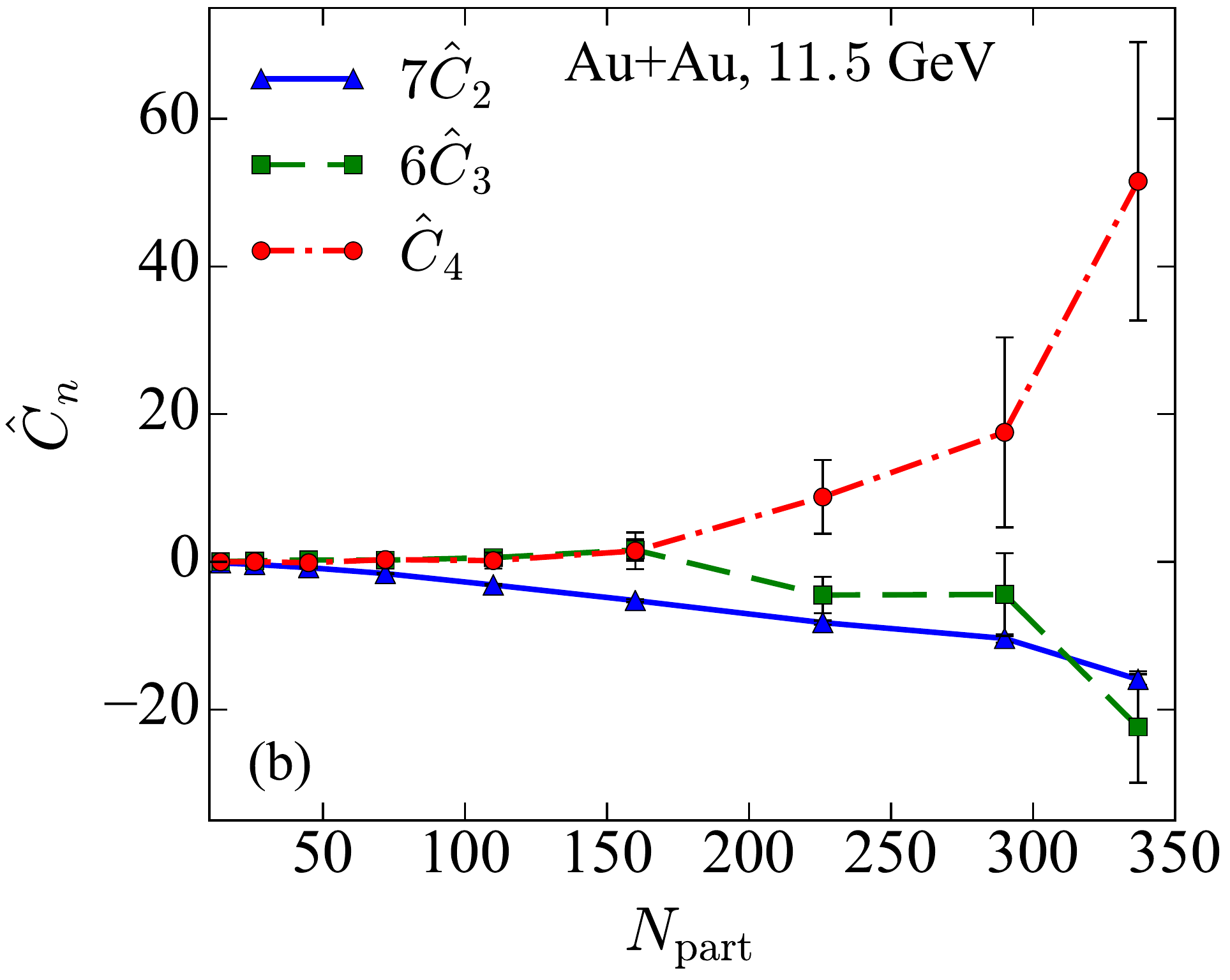} %
\includegraphics[scale=0.3]{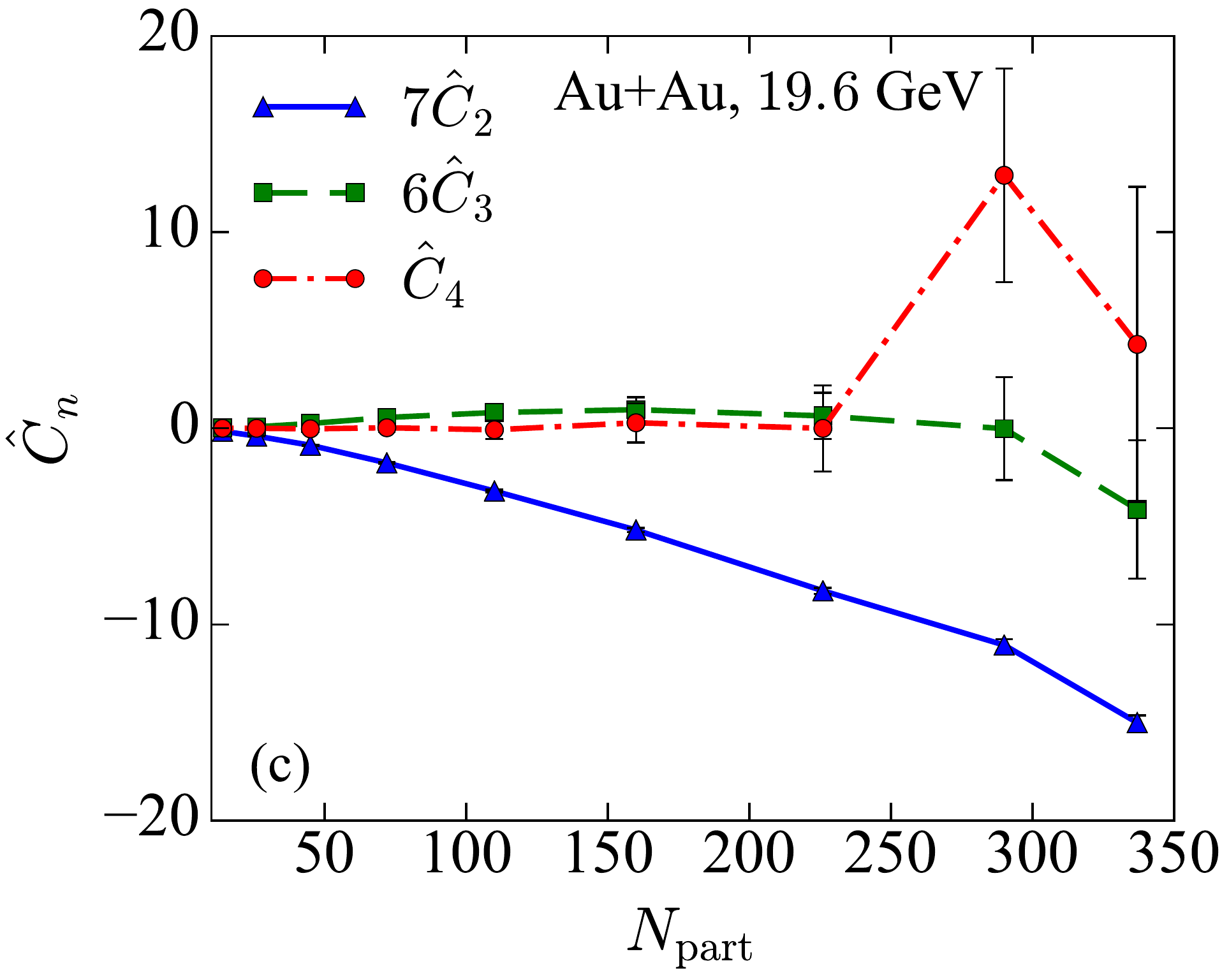}
\end{center}
\par
\vspace{-5mm}
\caption{The factorial cumulants $\hat{C}_{k}$ for $k=2,3,4$ as a function
of the number of wounded nucleons, $N_{\text{part}}$, in Au+Au collisions
for energies (a) $\protect\sqrt{s_{NN}}=7.7$ GeV, (b) $11.5$ GeV and (c) $19.6$
GeV. Results are based on preliminary STAR data \protect\cite{Luo:2015ewa}.
Figure adapted from Ref. \protect\cite{Bzdak:2016sxg}.}
\label{fig:ab:Ckhat}
\end{figure}

\subsubsection{Proton factorial cumulants}
\label{sec7:fact_cumulants}
As already eluded to in Section~\ref{sec4} factorial cumulants provide an
alternative measure of  fluctuations. As discussed in \ref{sec:ab:cumulant-intro}, 
factorial cumulants represent the integrated
irreducible (genuine) correlation functions. Therefore, they are more sensitive to the underlying interactions and have
no contribution from simple finite number statistics, i.e. they vanish in absence of any
correlations. The factorial cumulants of the proton
distribution, $\hat{C}_{n}$,  are given in terms of the regular cumulants, $\kappa_{n}$, by (see Eq.~\eqref{eq:ab:kappa-Chat})
\begin{eqnarray}
\hat{C}_{2}&=& - \ave{N} + \kappa_{2} , 
\non 
\hat{C}_{3}&=& 2\ave{N} - 3\kappa _{2} +\kappa_{3} ,  
\non
\hat{C}_{4}&=& -6\ave{N} + 11\kappa _{2} - 6\kappa_{3} + \kappa_{4} .  
\end{eqnarray}
Based on the above relations and using the preliminary STAR data \cite%
{Luo:2015ewa} measured in the acceptance region $|y|<0.5$ and $0.4<p_{t}<2.0$
GeV, we obtain the results for the factorial cumulants $\hat{C}_{2}$, $\hat{C}_{3}$ 
and $\hat{C}_{4}$ presented in Fig. \ref{fig:ab:Ckhat}.  
To best illustrate the contributions of the various factorial cumulants to the
fourth-order cumulant $\kappa _{4}$ we plot them with the appropriate weights: $7\hat{C}_{2}$, $6\hat{C}_{3}$
and $\hat{C}_{4}$, see Eq. (\ref{eq:ab:kappa-Chat}). For the most central collision points, which
shall be the focus of the subsequent discussion, the average number of protons is
approximately $\left\langle N\right\rangle \approx 39,$ $31$ and $25$ for
energies $7.7$, $11.5$ and $19.6$ GeV, respectively \cite%
{Luo:2015ewa,Luo:2017faz}.

As can be seen in Fig. \ref{fig:ab:Ckhat}, at $7.7$ GeV the fourth order cumulant in
central collisions is clearly dominated by the fourth-order factorial
cumulant and thus there are substantial four-proton correlations present in
the system. We obtain%
\begin{equation}
\text{STAR, }7.7\text{ GeV:\quad }\ave{N}\approx 40, \quad 7\hat{C}_{2}\approx -15,\quad 6\hat{C}%
_{3}\approx -60,\quad \hat{C}_{4}\approx 170.  \label{eq:ab:Ckhat-STAR}
\end{equation}%
As we increase the energy the situation is clearly changing.
At $11.5$ GeV, the $\hat{C}_{4}$ and thus four particle correlations are still sizable. However,
since the contribution of $\hat{C}_{4}$ to the fourth order cumulant, $\kappa_{4}$, is
approximately balanced by these from $\hat{C}_{2}$ and $\hat{C}_{3}$, the presence of the
four-particle correlation is less apparent in the forth order cumulant. This is a nice example of
the usefulness of factorial cumulants, which exhibit the correlation structure more directly than
the regular cumulants, $\kappa_{n}$. 
At $19.6$ GeV, for the most central collisions, 
$\kappa _{4}$ is dominated by the (negative) second order factorial
cumulant, $\hat{C}_{2}$. The results shown in 
Fig. \ref{fig:ab:Ckhat} can be summarized as follows: As we decrease the
energy from $19.6$ GeV to $7.7$ GeV, the relative importance of the
four-order factorial cumulant, and therefore four-proton correlations, is increasing. A more  detailed discussion, including the centrality
dependence, can be found in Ref. \cite{Bzdak:2016sxg}.

To put the above numbers, and in particular the value of $\hat{C}_{4}$ at $%
7.7$ GeV into perspective, let us consider a simple model, already described
in the \ref{sec:ab:cumulant-intro}, see also
\cite{Bzdak:2016jxo}.
Suppose that we have clusters of protons, distributed according to the Poisson distribution,
which always decay into $m$ protons. According to Eq. (\ref%
{eq:ab:Ckhat-clust}), we obtain $\hat{C}_{k}=\left\langle N_{c}\right\rangle 
\frac{m!}{(m-k)!}$. Having four-proton clusters ($m=4$) we obtain $\hat{C}%
_{4}=24\left\langle N_{c}\right\rangle $ and in order to get $\hat{C}_{4}=170$, see Eq. (%
\ref{eq:ab:Ckhat-STAR}), we need to assume $\left\langle N_{c}\right\rangle
\sim 7$. Consequently, on average roughly $28$ of the observed 40 protons should originate from
such clusters. For five-proton clusters ($m=5$) we have  $\hat{C}%
_{4}=120\left\langle N_{c}\right\rangle $ thus $\left\langle
  N_{c}\right\rangle \sim 1.5$. Given these results, it is  clear that
$\hat{C}_{4}\sim 170$ is a rather large number and a
strong source of four-proton correlations is required to explain the
magnitude of the preliminary STAR signal. 
However, the simple cluster  model also predicts all $\hat{C}_{k}$ to be  positive
in obvious disagreement with the preliminary data, where both $%
\hat{C}_{2}$ and $\hat{C}_{3}$ are negative. 
Therefore, it is not straightforward to explain the observed large correlations and their signs with baryonic clusters or any other more conventional means. Could this be the first hint for the critical point and/or first order transition? This question will be addressed to some extent in Section~\ref{sec:6:bimodal}. Before we turn to this and further discussion of the data, however, let us
next discuss corrections due to baryon-number conservation and participant (volume) fluctuations.

\subsubsection{Baryon number conservation}
\label{sec7:baryon_conserve}
Since baryon number is conserved in a heavy ion reaction, the fluctuation of the protons is
different that that obtained, e.g., in a grand canonical ensemble, where most theoretical calculation
are carried out. Therefore, it is essential to determine the corrections due to baryon number
conservation on the various fluctuation measures such as cumulants. 
At low energy, $\sqrt{s_{NN}}\lesssim 20\gev$, where the cumulants show the most interesting features
almost all measured baryons originate from the incoming nuclei. The
baryon number is simply decelerated to the mid-rapidity region.
The simplest model, which does not introduce any new correlations, is to assume that baryons stop
independently in a given rapidity interval. In this case the
distribution of protons is given by the binomial distribution%
\begin{equation}
P(N)=\frac{B!}{N!(B-N)!}p^{N}(1-p)^{B-N},  \label{eq:ab:P-bar-cons}
\end{equation}%
where $p$ is the probability that the initial baryon will end up in our
rapidity bin as a proton. $B$ is the initial number of baryons that can
potentially be observed as protons. In the following we will assume $%
B\approx 400$, being approximately the total baryon number in both colliding
nuclei. Here we assume that all initial baryons are subjected to a maximum
isospin randomization. That is, the initial proton can be stopped either as
a proton or a neutron with equal probability, see Refs. \cite%
{Kitazawa:2011wh,Kitazawa:2012at}.

The generating function reads\footnote{%
For one baryon, the probability to be observed as a proton in a given
interval is $p$ and the probability not to be observed is $1-p$. Thus, the
generating function for a single baryon equals $(1-p)z^{0}+pz^{1}$. Having $%
B $ independent sources, the final generating function is given by a product
of $B$ one-baryon generating functions, which gives Eq. (\ref%
{eq:ab:H-bar-cons}).} 
\begin{equation}
H(z)=\sum_{N}P(N)z^{N}=\left( 1-p+pz\right) ^{B},  \label{eq:ab:H-bar-cons}
\end{equation}%
and $\hat{C}_{1}=\left\langle N\right\rangle =pB$, and%
\begin{equation}
\hat{C}_{2}=-\frac{\left\langle N\right\rangle ^{2}}{B},\quad \hat{C}_{3}=2%
\frac{\left\langle N\right\rangle ^{3}}{B^{2}},\quad \hat{C}_{4}=-6\frac{%
\left\langle N\right\rangle ^{4}}{B^{3}},  \label{eq:ab:Ckhat-bar-cons}
\end{equation}%
where $\left\langle N\right\rangle $ is the average number of measured
protons, which varies with the size of the chosen rapidity interval. Taking $%
B=400$ and for $7.7$ GeV $\left\langle N\right\rangle =40$ we obtain: $%
\hat{C}_{2}=-4$, $\hat{C}_{3}=0.8$ and $\hat{C}_{4}=-0.24$. We observe that $%
7\hat{C}_{2}=-28$ is of the same order of magnitude and roughly a factor of $2$
larger in magnitude than the STAR data. $6\hat{C}_{3}=4.8$ is significantly smaller than in
Eq. (\ref{eq:ab:Ckhat-STAR}) and the baryon conservation driven $\hat{C}%
_{4}=-0.24$ is completely negligible, when compared to STAR $\hat{C}%
_{4}\approx 170$. Finally, we note that the above results for the factorial cumulants due baryon number
conservation together  with Eq.~\eqref{eq:ab:kappa-Chat} give a cumulant ratio of
$\cum{4}/\cum{2} = 0.46$ for $7.7\gev$, consistent with the UrQMD result shown in Fig.~\ref{fig:exp:f15}.

We conclude that at lower energies, the effect of baryon conservation plays
some role in $\hat{C}_{2}$, is an order of magnitude too small for $\hat{C}%
_{3}$, and is almost three orders of magnitude too small for $\hat{C}_{4}$.
This shows a clear advantage of the factorial cumulants when compared to the
cumulants, which are always ''contaminated'' by $\hat{C}_{2}$. A detailed
discussion of the baryon-number  conservation effects, including the presence of
anti-baryons,  can be found in Refs.~\cite{Bzdak:2012an,Braun-Munzinger:2016yjz}.

\subsubsection{Volume fluctuation}
Another source of fluctuations in heavy ion collisions arises from the fact that even with
the tightest centrality cuts the impact parameter and thus the size of the created system
fluctuates. These volume or participant fluctuations \cite{Skokov:2012ds} give rise to
additional non-dynamical fluctuations which need to be understood and, if large,
subtracted from the experimentally determined fluctuation measures. This aspect has been
addressed in, e.g., Ref.~\cite{Bzdak:2016jxo}, where a minimal model, incorporating the
volume fluctuation and baryon conservation at $%
\sqrt{s_{NN}}=7.7$ GeV was discussed. In general,  for a given centrality class,
the fluctuations of the volume of the system
is due to the fluctuations of the number of participating or wounded nucleons,
$N_{\text{part}}$. Following Ref.~~\cite{Bzdak:2016jxo} the distribution of number of 
participating nucleons $P(N_{\text{part}})$ for a given
centrality selection can be determined from a standard Glauber model. In addition, at low energies,
where the production of baryon--anti-baryon pairs is negligible, the number of wounded
nucleons in a given event is, to a good approximation, identical to the total number of
baryons. In absence of any correlations, each wounded nucleon will then end up with a 
probability $p$ as a proton a given rapidity (or acceptance) bin. Consequently, 
the distribution of protons at a 
given $N_{\text{part}}$ is described by the binomial distribution. Note, in this approach we 
assume that each wounded nucleon is decelerated independently. Any deviation from
this assumption would introduce new correlations and we are interested in
the minimal model, where only $N_{\text{part}}$ fluctuations (and baryon
conservation) are present. Therefore, in this model the distribution of protons is
given by 
\begin{equation}
P(N)=\sum_{N_{\text{part}}}P(N_{\text{part}})\frac{N_{\text{part}}!}{%
N!\left( N_{\text{part}}-N\right) !}p^{N}\left( 1-p\right) ^{N_{\text{part}%
}-N},  \label{eq:ab:P-vol-fluct}
\end{equation}%
with the generating function%
\begin{equation}
H(z)=\sum_{N_{\text{part}}}P(N_{\text{part}})\left( 1-p+pz\right) ^{N_{\text{%
part}}}.  \label{eq:ab:H-vol-fluct}
\end{equation}

This situation is very similar to the one discussed above in the context of baryon number
conservation, see Eq.~(\ref{eq:ab:P-bar-cons}), 
except that now in each collision the maximum number of
baryons that can end up in our rapidity bin is not the total baryon-number, $B$ but
the number of wounded nucleons, $N_{\text{part}}$, which fluctuates from event to event.
A straightforward calculation then gives $\hat{C}_{1}=\left\langle
N\right\rangle =p\left\langle N_{\text{part}}\right\rangle ,$ with $%
\left\langle N_{\text{part}}\right\rangle $ being the average number of wounded
nucleons in a given centrality class. $\langle N\rangle$ is the average number of observed protons. The higher order factorial cumulants
read%
\begin{eqnarray}
\hat{C}_{2} &=&\left\langle N\right\rangle ^{2}\left[ -\frac{1}{\left\langle
N_{\text{part}}\right\rangle }+\frac{\langle \left[ N_{\text{part}%
}-\left\langle N_{\text{part}}\right\rangle \right] ^{2}\rangle }{%
\left\langle N_{\text{part}}\right\rangle ^{2}}\right] ,  \notag \\
\hat{C}_{3} &=&\left\langle N\right\rangle ^{3}\left[ \frac{2}{\left\langle
N_{\text{part}}\right\rangle ^{2}}+\frac{\langle \left[ N_{\text{part}%
}-\left\langle N_{\text{part}}\right\rangle \right] ^{3}\rangle -3\langle %
\left[ N_{\text{part}}-\left\langle N_{\text{part}}\right\rangle \right]
^{2}\rangle }{\left\langle N_{\text{part}}\right\rangle ^{3}}\right] ,
\label{eq:ab:Ckhat-vol-fluct}
\end{eqnarray}%
and $\hat{C}_{4}$ ($\hat{C}_{4}/\langle N\rangle ^{4}$ to be more precise)
can be found in Ref.~\cite{Bzdak:2016jxo}.\footnote{%
In Ref. \cite{Bzdak:2016jxo} $\hat{C}_{k}/\langle N\rangle ^{k}$ is denoted
by $c_{k}$.} As seen from Eq. (\ref{eq:ab:Ckhat-vol-fluct}) the fluctuation
of $N_{\text{part}}$ contributes to the factorial cumulants.

%\cut{Having $N_{\text{part}}$ in a given event (coming from the Glauber model) we randomly generate final pions from the Poisson distribution with the mean pion number given by $aN_{\text{part}}(N_{\text{part}}/2)^{0.1}$, where for $7.7$ GeV we take $a=0.75$. In our analysis we closely follow the STAR procedure and apply the tightest centrality cuts, that is, we calculate the factorial cumulants at at given number of produced pions}{not sure we need all this detail here}

\begin{figure}[h]
\begin{center}
\includegraphics[scale=0.32]{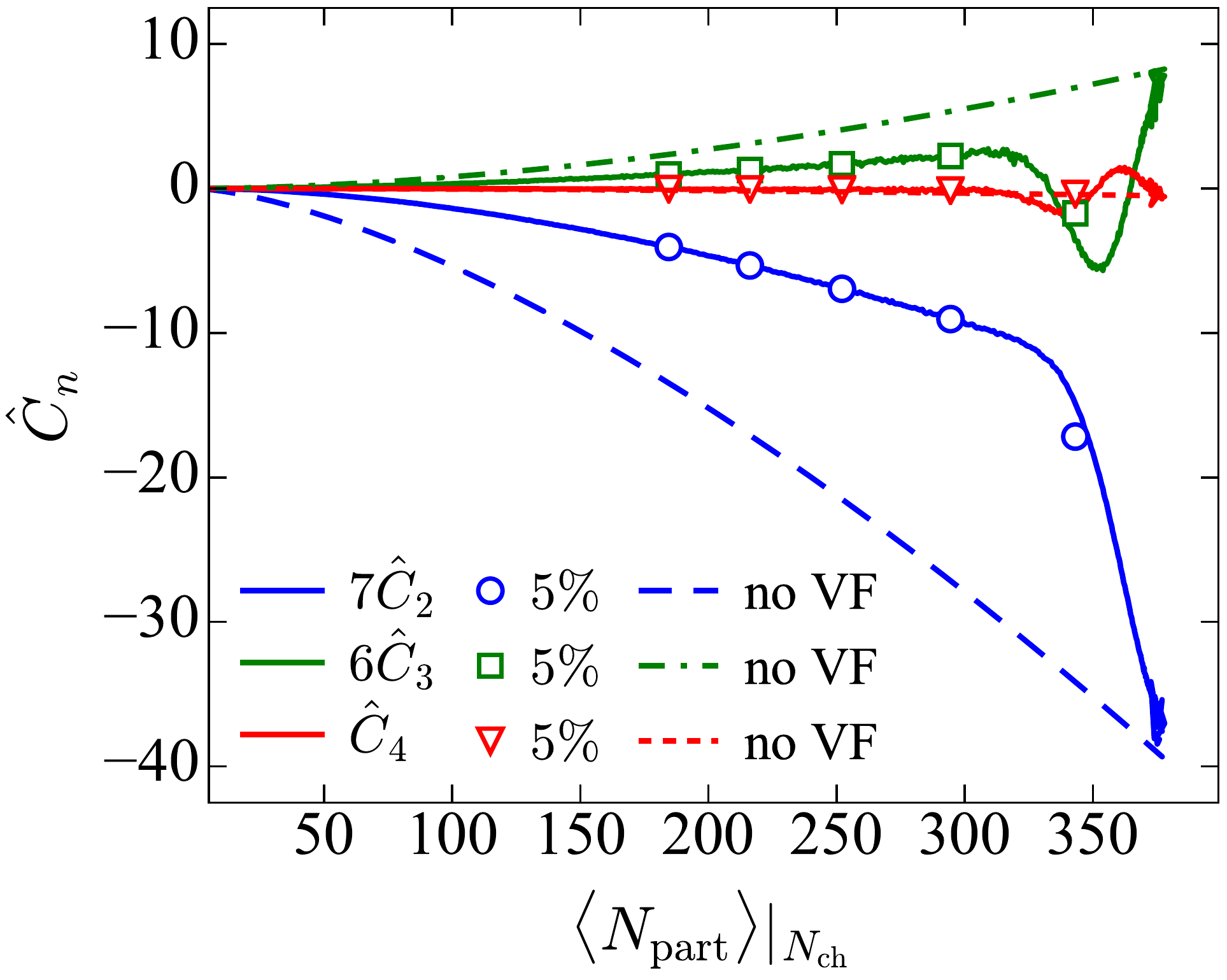} \hspace{0.5cm}
\end{center}
\par
\vspace{-5mm}
\caption{The factorial cumulants $\hat{C}_{k}$ in Au+Au collisions at $%
\protect\sqrt{s_{NN}}=7.7$ GeV with and without $N_{\text{part}}$ fluctuations. $%
\langle N_{\text{part}}\rangle |_{N_{\text{ch}}}$ is the average number of
wounded nucleons, where average is calculated at a given number of produced
pions. See text for details. Figure adapted from Ref. \protect\cite%
{Bzdak:2016jxo}.}
\label{fig:ab:vol-fluc}
\end{figure}

Applying the centrality cuts similar to those used in the STAR measurement
\cite{Luo:2015ewa} (for details see  Ref.~\cite{Bzdak:2016jxo}) one arrives at results  
presented in Fig. \ref{fig:ab:vol-fluc}, where we plot the factorial cumulants as a function
of the average number of wounded nucleons, $\langle N_{\text{part}}\rangle
|_{N_{\text{ch}}}$, and the average is calculated at a given number of
produced pions.
The solid lines represent the full calculations and the dashed
lines represent the results without $N_{\text{part}}$ fluctuations, that is,
only the leading terms in Eq. (\ref{eq:ab:Ckhat-vol-fluct}) are included, $%
\hat{C}_{k}\sim \langle N\rangle ^{k}/\langle N_{\text{part}}\rangle ^{k-1}$.
The open symbols represent the results after averaging $\hat{C}_{k}$ over
bins in centrality $0-5\%$, $5-10\%$, etc. (the first five points are
shown). We see that, in general volume fluctuation introduces some
nontrivial features, however the signal is negligible for $\hat{C}_{4}$ when
compared to the preliminary STAR data, see Eq. (\ref{eq:ab:Ckhat-STAR}). $%
\hat{C}_{3}$ is visibly modified by $N_{\text{part}}$ fluctuations but again
the signal is way to small to explain the STAR data. Interestingly, $\hat{C}%
_{2}$ is strongly modified by volume fluctuation and the signal is modified
(when compared to the case without $N_{\text{part}}$ fluctuations) by
roughly a factor of two.

From this we conclude that at $7.7$ GeV both $\hat{C}_{3}$ and $\hat{C}_{4}$ in
central collisions cannot be explained by a simple volume fluctuation or
baryon conservation.

\subsubsection{Rapidity dependence}
\label{sec7:correlation_rapidity}
Next, we will comment on the acceptance dependence of the cumulants which has been discussed
already in Sections~\ref{sec:crit-pt-correl-momentum} and~\ref{subsec_6_chirality}. STAR has measured cumulants up to the maximum rapidity
interval given by $|y|<0.5$. Therefore, the maximum rapidity distance between
two measured protons is one unit of rapidity.  
We will demonstrate that the preliminary STAR data are
consistent with a constant long-range rapidity correlation within this
interval. Here, for simplicity we assume that the single-particle rapidity
distribution, $dN/dy\equiv \rho(y)$, can be approximated by a constant in $|y|<0.5$, 
$\rho(y) = \rho_{0}=const$
which is not far from reality \cite{Anticic:2010mp}.

Long-range rapidity correlations mean that the normalized
multi-particle correlation function, defined as\footnote{%
In Refs. \cite{Bzdak:2016sxg,Bzdak:2017ltv} this object is denoted by $%
c_{k}(y_{1},...y_{k})$.}%
\begin{equation}
R_{k}(y_{1},...,y_{k})=\frac{C_{k}(y_{1},...,y_{k})}{\rho (y_{1})\cdots \rho
(y_{k})},  \label{eq:ab:Rn-rapi}
\end{equation}%
does not depend on the rapidities of the particles, $y_{i}$,%
\begin{equation}
R_{k}(y_{1},...,y_{k})=R_{k}^{0}=const.
\end{equation}%
Therefore, the factorial cumulants%
\begin{eqnarray}
\hat{C}_{k} &=&\int C_{k}(y_{1},...,y_{k})dy_{1}\cdots dy_{k}=R_{k}^{0}\int
\rho (y_{1})\cdots \rho (y_{k})dy_{1}\cdots dy_{k}  \notag \\
&=&R_{k}^{0}\left\langle N\right\rangle ^{k},
\end{eqnarray}%
scale with $\left\langle N\right\rangle ^{k}$, where in the above equations
the integrations are performed over a rapidity interval $-\Delta y/2<y_{i}<$ 
$\Delta y/2$. Since the single particle rapidity density is, to a good
approximation, constant, $\rho(y) = \rho_{0}$, the average number of particles scales with the size of the
rapidity interval, $\left\langle N\right\rangle = \rho_{0} \Delta y \sim  \Delta y$,
and we obtain\footnote{We should also point out that short-range correlations lead to 
$\hat{C}_{k}\sim \Delta y$ and the cumulant ratios do not depend on $\Delta y$,
see, e.g., Ref. \cite{Bzdak:2016sxg}. For example assuming $%
R_{2}(y_{1},y_{2})\sim \delta (y_{1}-y_{2})$ we have $\hat{C}_{2}\sim \int
dy_{1}dy_{2}\rho (y_{1})\rho (y_{2})\delta (y_{1}-y_{2})$ which gives $%
\hat{C}_{2}\sim $ $\int dy\rho (y)^{2}\sim \Delta y$.} 
\begin{equation}
\hat{C}_{k}\sim \left\langle N\right\rangle ^{k}\sim (\Delta y)^{k}.
\end{equation}
Following Eq. (\ref{eq:ab:kappa-Chat}) we see that the cumulants $\kappa
_{i} $ scale with $\left\langle N\right\rangle $ and $\Delta y$ in a rather
non-trivial way. For example
\begin{equation}
\frac{\kappa _{4}}{\kappa _{2}}=\frac{\left\langle N\right\rangle
+7R_{2}^{0}\left\langle N\right\rangle ^{2}+6R_{3}^{0}\left\langle
N\right\rangle ^{3}+R_{4}^{0}\left\langle N\right\rangle ^{4}}{\left\langle
N\right\rangle +R_{2}^{0}\left\langle N\right\rangle ^{2}}.
\label{eq:sec7:cumulant_ratio_N}
\end{equation}%
We note that as  $\Delta y\rightarrow 0$ ($\left\langle N\right\rangle \rightarrow 0$), the cumulants are dominated by $%
\left\langle N\right\rangle $ and the cumulant ratios go to unity which is the Poissonian
limit, even in the presence of large correlations (given here by $R_{k}^{0}$).

A similar scaling of the factorial cumulants with the particle number, $\hat{C}_{k}\sim \left\langle
N\right\rangle ^{k}$  has been already
obtained when we discussed baryon number conservation and volume fluctuations,
see Eqs. (\ref{eq:ab:Ckhat-bar-cons}) and (\ref{eq:ab:Ckhat-vol-fluct}). This is not
unexpected, since both baryon conservation and volume fluctuation
are, in the simple approach presented here, rapidity independent long-range phenomena.

The maximum value of $\Delta y$ presently available at RHIC is $\Delta y=1$.
Now, we can fit the values of $R_{k}^{0}$ using the data for $\Delta y=1$
and calculate the rapidity dependence of the cumulants. This is discussed
carefully in Ref. \cite{Bzdak:2016sxg} and demonstrated in Fig. \ref%
{fig:ab:k4k2rapi} for (a) $7.7$ GeV and (b) $19.6$ GeV. We see that the
preliminary STAR data are consistent with a constant rapidity
correlation. Of course, since the maximum rapidity interval covered by STAR is $|y|<0.5$, in
  practice this means that the correlation range in rapidity is larger than
  one unit, $\delta y > 1$. 
\begin{figure}[t]
\begin{center}
\includegraphics[scale=0.3]{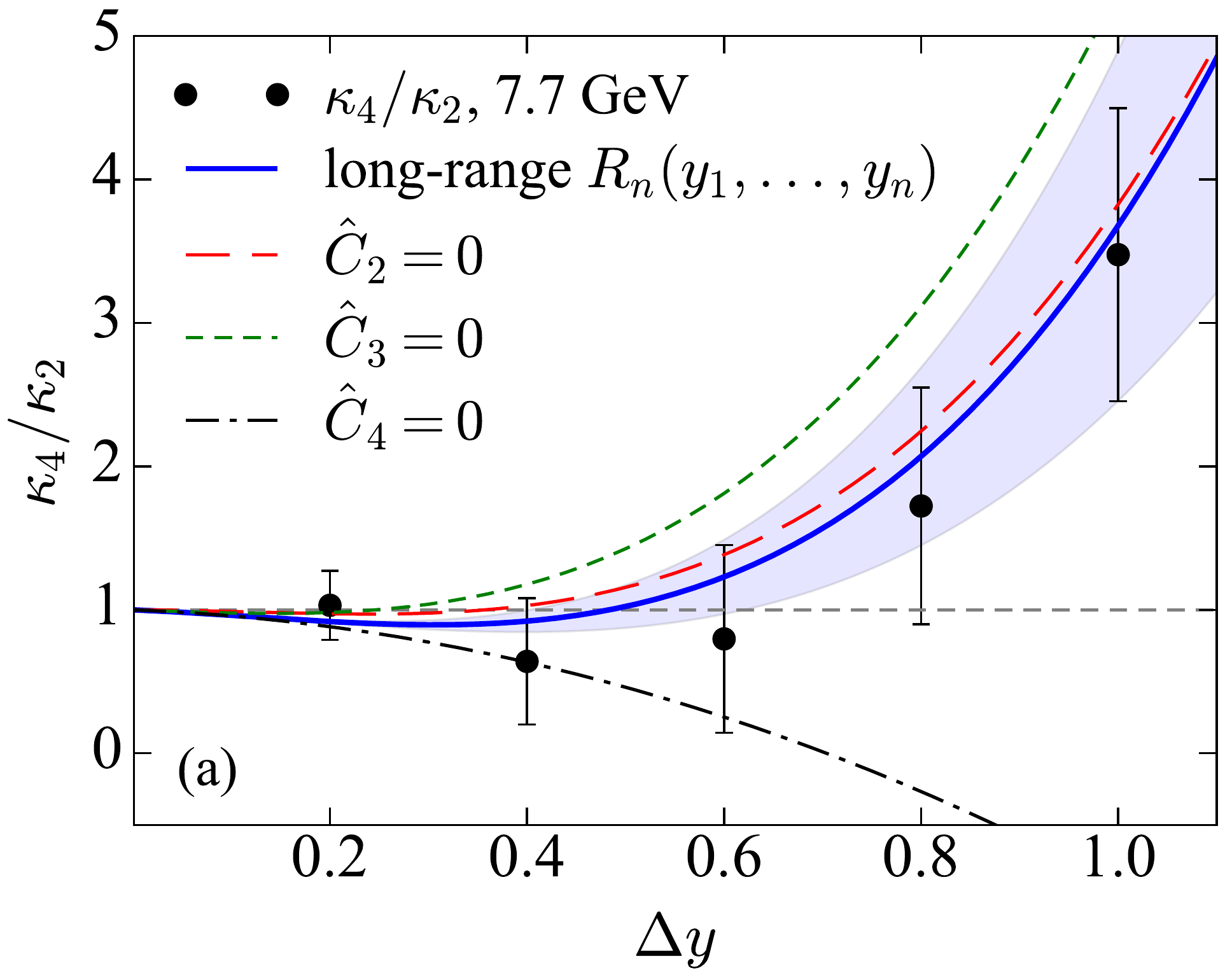} \hspace{1cm} %
\includegraphics[scale=0.3]{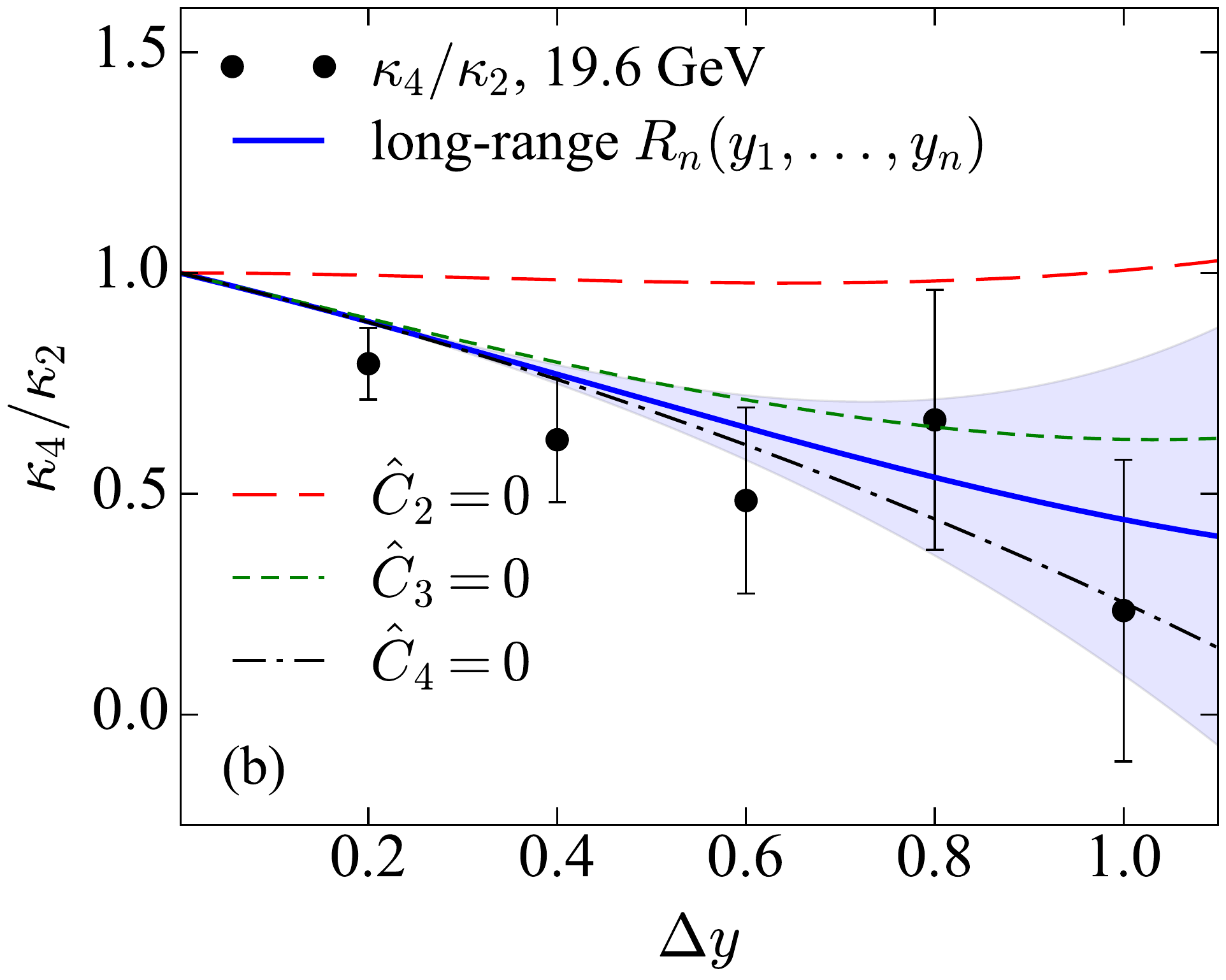}
\end{center}
\par
\vspace{-5mm}
\caption{The rapidity dependence of the cumulant ratio $\protect\kappa _{4}/%
\protect\kappa _{2}$ for (a) 7.7 GeV and (b) 19.6 GeV. The data are
preliminary STAR results \protect\cite{Luo:2015ewa}. Figure adapted from
Ref. \protect\cite{Bzdak:2016sxg}.}
\label{fig:ab:k4k2rapi}
\end{figure}

In Fig. \ref{fig:ab:k4k2rapi} we also demonstrate the importance of various $%
\hat{C}_{n}$ contributions to $\kappa _{4}/\kappa _{2}$. Clearly the signal
at $7.7$ GeV is driven by $\hat{C}_{4}$ and at $19.6$ GeV by $\hat{C}_{2}$,
in agreement with Fig. \ref{fig:ab:Ckhat}.

\begin{figure}[t]
\begin{center}
\includegraphics[scale=0.35]{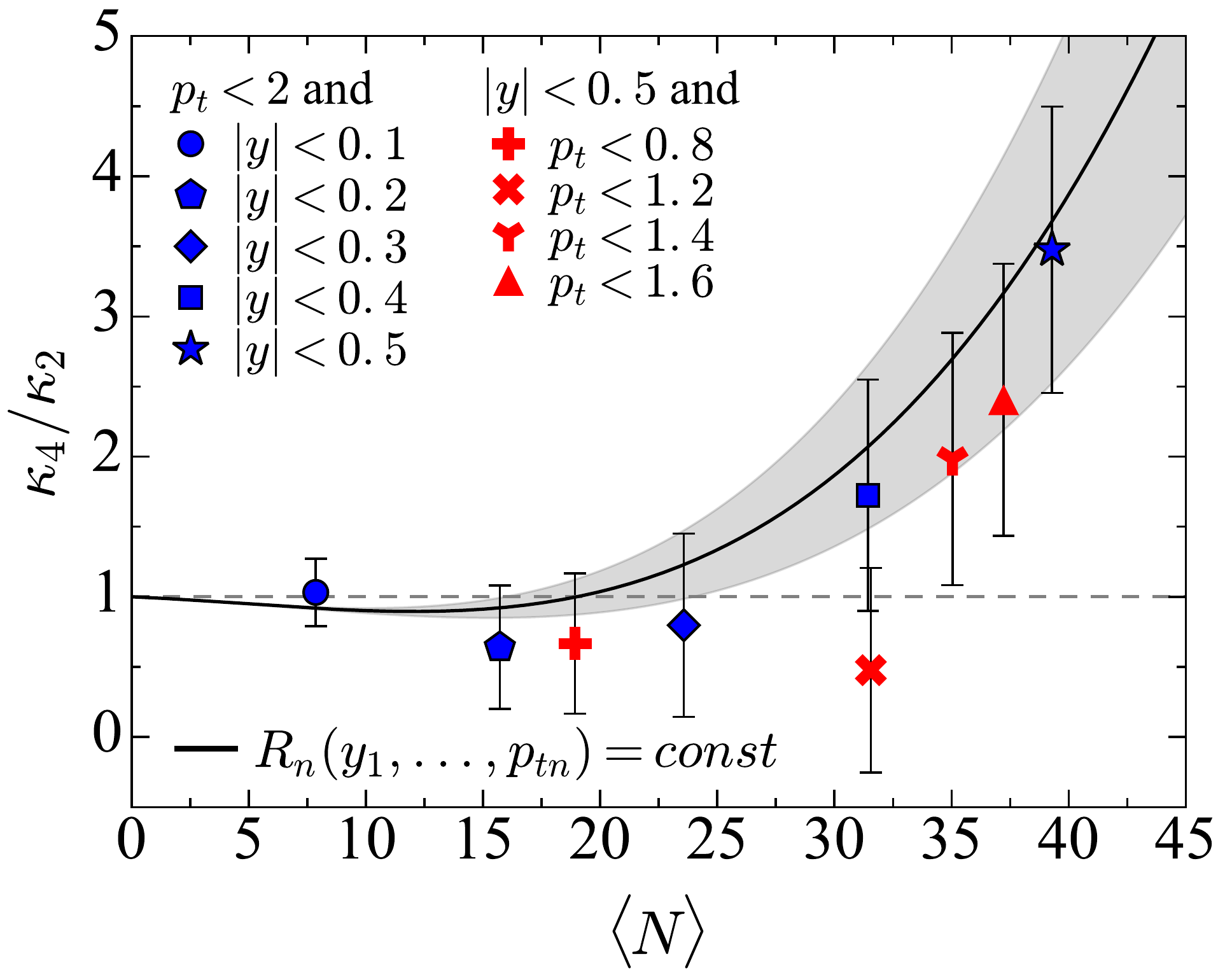}
\end{center}
\par
\vspace{-5mm}
\caption{The cumulant ratio $\protect\kappa _{4}/\protect\kappa _{2}$ for
different intervals in rapidity and transverse momentum as a function of the
average number of observed protons at $\sqrt{s_{NN}}=7.7$ GeV. In this calculation $R_k(y_1,...,y_k;p_{t1},...,p_{tk})=const$. Based on preliminary STAR data %
\protect\cite{Luo:2015ewa}. Figure adapted from Ref. \protect\cite%
{Bzdak:2017ltv}.}
\label{fig:ab:k4k2scaling}
\end{figure}

As explained in Ref. \cite{Bzdak:2017ltv}, similar analysis can be done in
the transverse direction. Here we need to take into account $\rho (p_{t})$
when calculating the dependence of the cumulant ratio on the size of the
transverse momentum interval. Assuming that, within the current acceptance range,%
\begin{equation}
R_{k}(p_{t1},\ldots ,p_{tk})=const,
\end{equation}%
we again arrive at the conclusion that the cumulants and the factorial
cumulants depend on the average number of measured protons $\left\langle
N\right\rangle $ (in a given transverse momentum interval) as demonstrated
in Fig. \ref{fig:ab:k4k2scaling}. 

To summarize, the preliminary STAR data at low energies are consistent with
$R_k(y_1,...,y_k;p_{t1},...,p_{tk})=const$ within the acceptance range of the STAR detector,
$|y|<0.5$ and $0.4 \gev\leq p_{T}\leq 2 \gev$. However, baryon number conservation
\cite{Bzdak:2017ltv} as well as the arguments given in Section~\ref{sec:crit-pt-correl-momentum}
suggest that the correlations will eventually change as one increase the rapidity coverage.
For the upcoming second phase of the RHIC
beam energy scan the rapidity coverage of STAR detector has been increased by a factor of 1.6. Thus, it will be interesting to see if and when the scaling presented in Figs. \ref{fig:ab:k4k2rapi}
and \ref{fig:ab:k4k2scaling} is broken.\footnote{We add that a mild {\it repulsive} rapidity 
dependence, studied in Ref. \cite{Bzdak:2017ltv}, is also consistent with the data at $7.7$ GeV.}

\subsubsection{Bimodal distribution}
\label{sec:6:bimodal}
\begin{figure}[t]
\begin{center}
\includegraphics[scale=0.38]{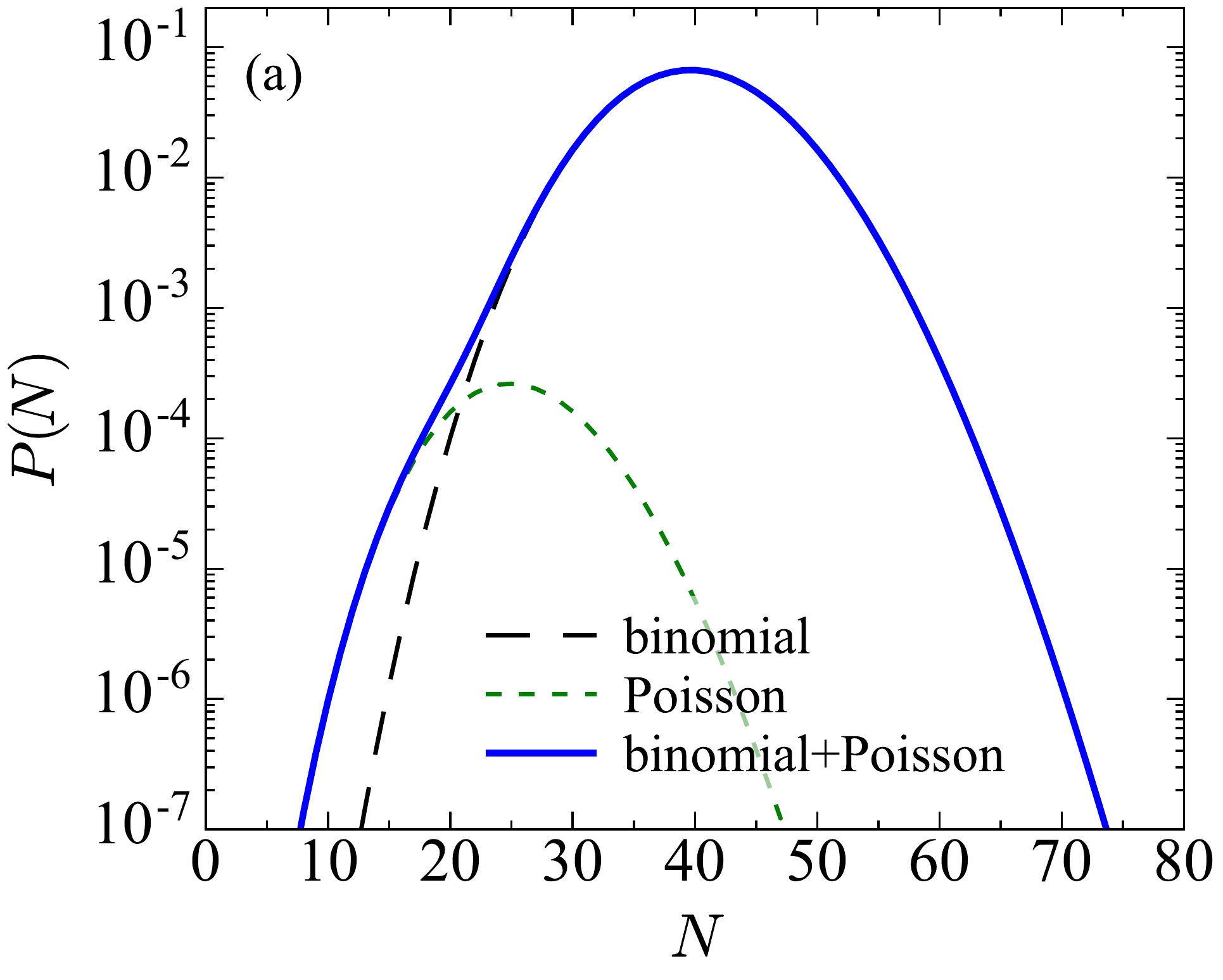}\hspace{10mm} %
\includegraphics[scale=0.38]{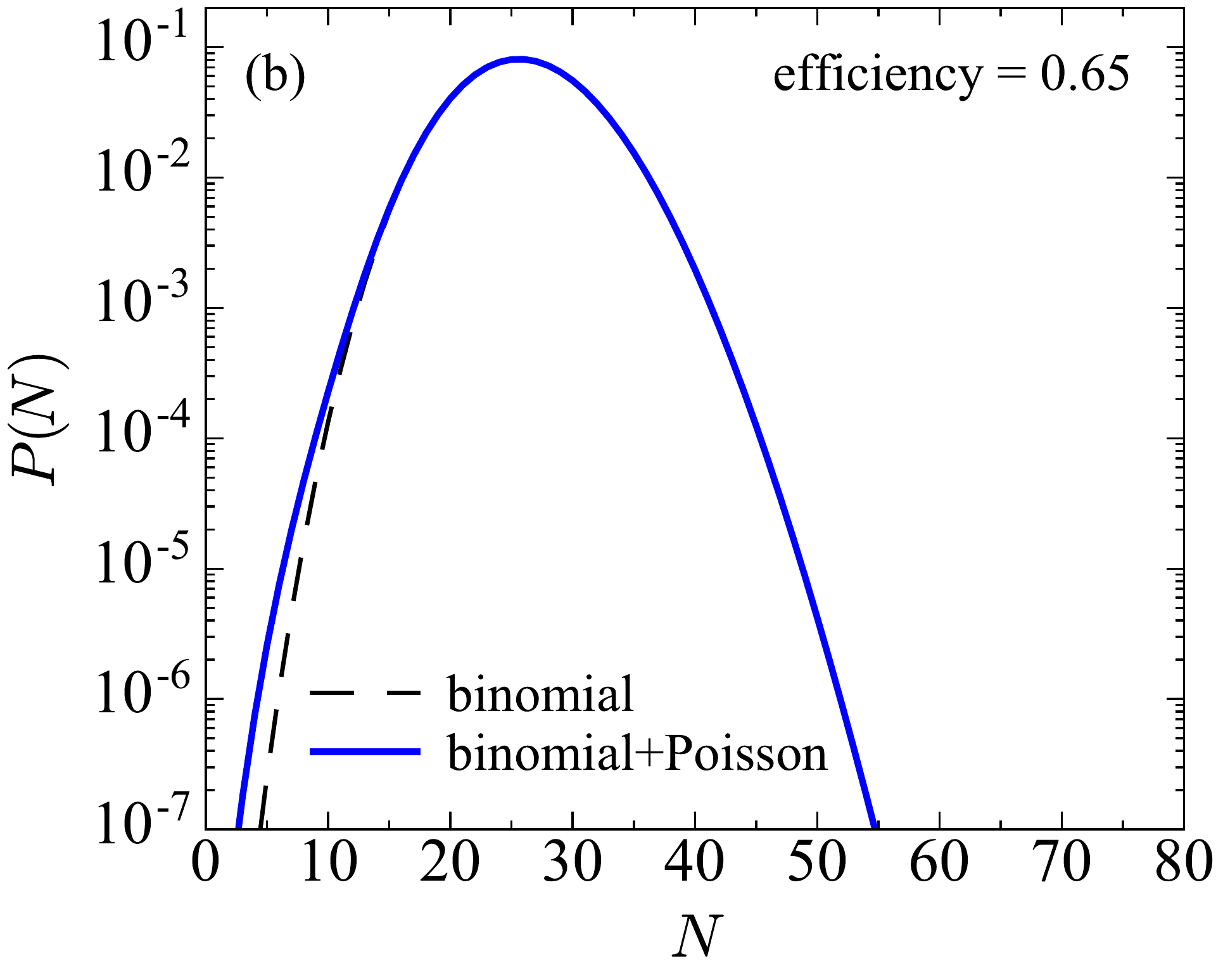}
\end{center}
\par
\vspace{-5mm}
\caption{The proton multiplicity distribution at $\sqrt{s_{NN}}=7.7$ GeV given by Eq. (\ref{eq:two_component}) constructed with (a) efficiency unfolded values for $\langle N \rangle$, $C_3$ and $C_4$ and (b) with efficiency of $0.65$. Figure adapted from Ref. \cite{Bzdak:2018uhv}.}
\label{fig:bimodal_star}
\end{figure}

As already discussed in this Section, the surprisingly large factorial cumulants measured
at $7.7$ GeV cannot be understood with baryon conservation or volume fluctuation. One can
easily generate large values of certain factorial cumulants with, e.g., baryonic clusters
but the signs of the measured $\hat{C}_{n}$ are rather challenging to obtain. Recently in
Ref. \cite{Bzdak:2018uhv} it was demonstrated that the preliminary STAR data at $7.7$ GeV
can be naturally reproduced assuming that the proton multiplicity distribution has a
bimodal form
\begin{equation}
P(N)=(1-\alpha )P_{(a)}(N)+\alpha P_{(b)}(N),
\label{eq:two_component}
\end{equation}
where $P_{(a)}$ and $P_{(b)}$ are the proton multiplicity distributions characterized by
different means $\langle N_{(a)}\rangle$ and $\langle N_{(b)}\rangle$. It is interesting
that one can reproduce the STAR data assuming that both $P_{(a)}$ and $P_{(b)}$ are
characterized by small or even vanishing factorial cumulants. In the above equation
$\alpha$ determines the relative strength of the two distributions. The STAR data (up to
$\hat{C}_{4}$) can be reproduced assuming that $\alpha \approx 0.0033$, $\langle
N_{(a)}\rangle \approx 40$ and $\langle N_{(b)}\rangle \approx 25.3$, as shown in panel
(a) of Fig. \ref{fig:bimodal_star}. Here $P_{(a)}$ and $P_{(b)}$ are taken as binomial
($B=350$ and $p \approx 0.144$) and Poisson distributions, respectively. In panel (b) we
show the same distribution with an imposed efficiency of $0.65$, which is more relevant for the STAR environment. 

\begin{figure}[t]
\begin{center}
\includegraphics[scale=0.35]{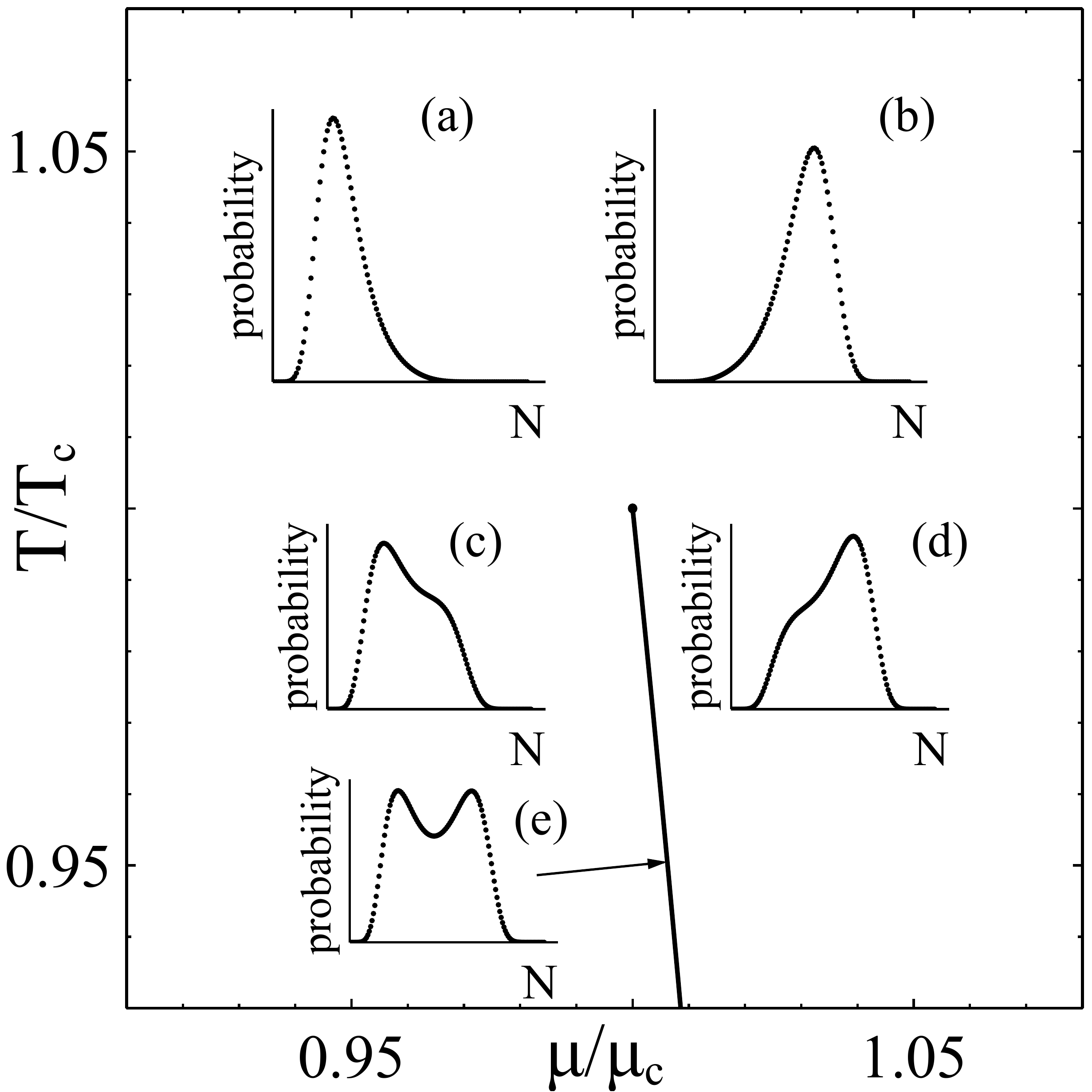}
% width=0.70\textwidth
\end{center}
\par
\vspace{-5mm}
\caption{Multiplicity distributions at various points for the van der Waals model for system at fixed volume: 
$(T/T_c, \mu/\mu_c) = $ (1.02, 0.99) (a), (1.02, 1.004) (b), (0.98, 1.0015) (c), (0.98, 1.004) (d), (0.95, 1.0062) (e).
Figure adapted from Ref. \cite{Bzdak:2018uhv}.}
\label{fig:dima_plot}
\end{figure}

The bimodal distribution given by Eq. (\ref{eq:two_component}) and presented in
Fig. \ref{fig:bimodal_star} arises naturally when the system fluctuates between two distinct phases, as described in Section \ref{sec:fluctuations-role}.  A (small) system close to the
critical point or the first order phase-transition is described by a similar bimodal distribution.
This is demonstrated in Fig. \ref{fig:dima_plot} (see also Fig.~\ref{fig:3P}), where the multiplicity distributions at various points in
the phase diagram for the van der Waals model are presented (see \cite{Bzdak:2018uhv} for details).
If the two modes of the bimodal distribution indeed represent two distinct phases, one would 
expect this to be visible also in other observables, such as spectra, flow etc. As
discussed in \cite{Bzdak:2018uhv}, this can be tested by selecting events with small or large
proton number to enhance either of the two phases. Of course, there is also a more mundane
explanation for the appearance of a bi-modal distribution: A small fraction of events can be, e.g., contaminated by some detector effects, which would
result in (at least) two event classes. This scenario can be carefully studied and
hopefully excluded. 
% In the first one, a
% small fraction of events can be, e.g., contaminated by some detector effects, which would
% result in (at least) two event classes. This scenario can be carefully studied and
% hopefully excluded. In the much more interesting scenario, it is expected that a system
% close to the critical point or the first order phase transition is described by a similar
% bimodal distribution. This is presented in Fig. \ref{fig:dima_plot}, where the multiplicity distributions at various points for the van der Waals model are presented. The details of this calculation can be found in the Appendix B of Ref. \cite{Bzdak:2018uhv}.  

It is imperative to test whether the underlying proton distribution at $7.7$ GeV is indeed
described by the bimodal distribution. As shown in \cite{Bzdak:2018uhv} and detailed 
in \ref{sec:ab:cumulant-intro}, the factorial cumulants 
are approximately given by 
$\hat{C}_{n} \approx -\alpha \langle (N_{(a)} - N_{(b)}) \rangle^{n}$ and thus are quickly
growing with $n$ with alternating sign. Moreover, as found very recently in
Ref. \cite{Bzdak:2018axe}, the bimodal distribution describing the STAR data is statistically
friendly, i.e., can be successfully measured through the factorial cumulants of high orders with a
relatively small number of events. In this case, the factorial cumulants are driven by the bimodal
structure away from the tails and thus can be measured with a limited number of events. This is in
contrast to the Poisson, binomial, negative binomial, etc. distributions, which are statistically
very demanding. In Ref. \cite{Bzdak:2018axe} the prediction up to $\hat{C}_{9}$ (with errors) are
presented based on the current STAR statistics in $0-5\%$ central Au+Au collisions. Confirmation of
these results could very well indicate the observation of the first-order phase transition.

%
%=====================================================================================
%=====================================================================================
\subsection{Discussions on the measurements for anomalous chiral transport}
\label{sec7c}

In this subsection, we discuss the current understanding and existing issues of the various measurements for anomalous chiral transport. We also discuss the theoretical and experimental progress in addressing these issues.

\subsubsection{The issues with background correlations}

A main observable that has been used for the search of the chiral magnetic effect (CME) is the $\gamma$-correlator, Eq.~\eqref{eq:jl:gamma}. As discussed in Sec.~\ref{subsec_5_3}, it is sensitive to the CME  but  also receives background contributions. Thus, the key question is what fraction of the observed signal arises from  the CME signal compared to  those originating from backgrounds. In terms of Eq.~\eqref{eq:jl:FH}, it is a question of quantitatively separating the F and H components. 

\begin{figure*}[hbt!]
\begin{center} 
\includegraphics[scale=0.25]{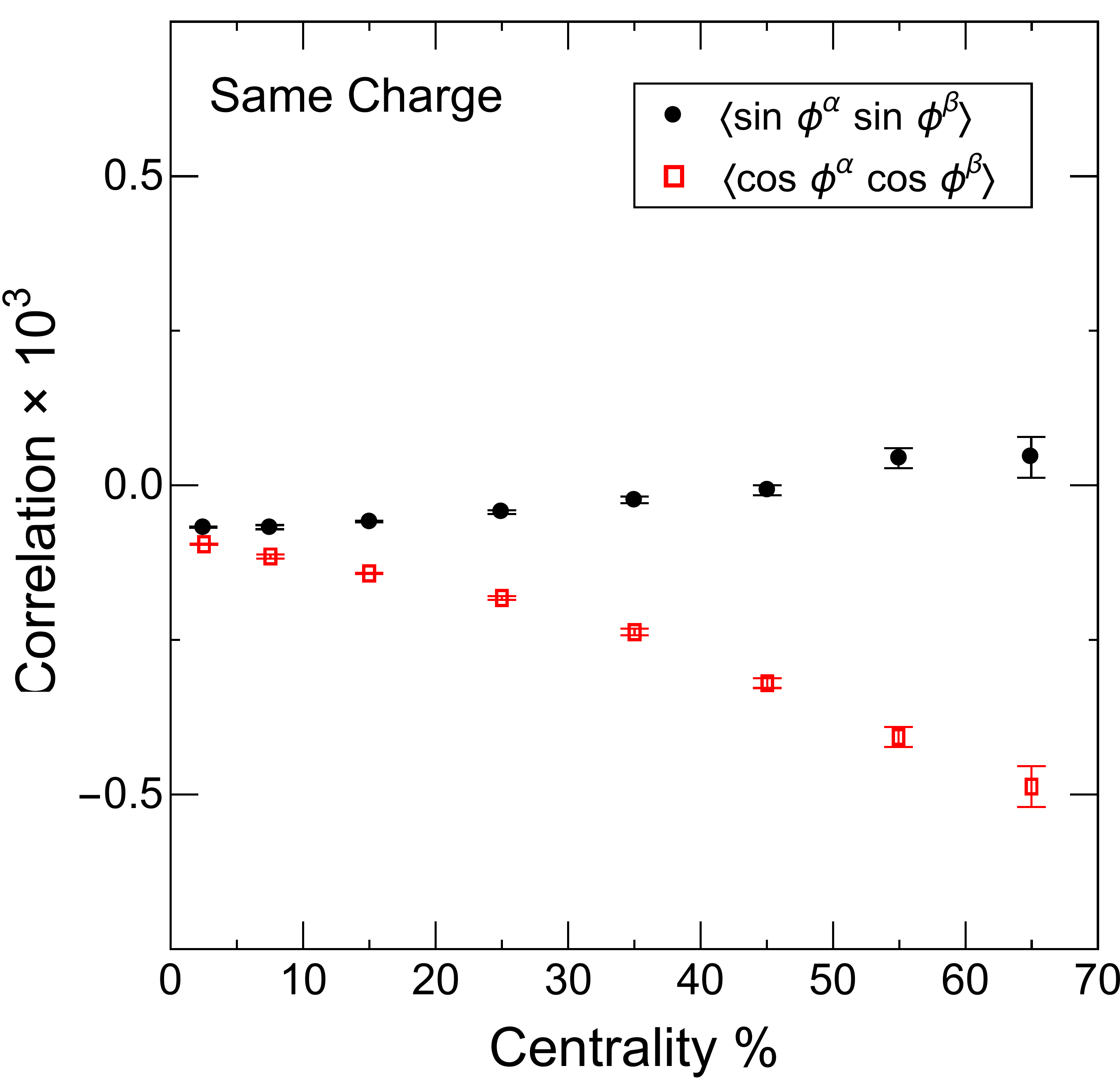} \hspace{0.2in}
\includegraphics[scale=0.25]{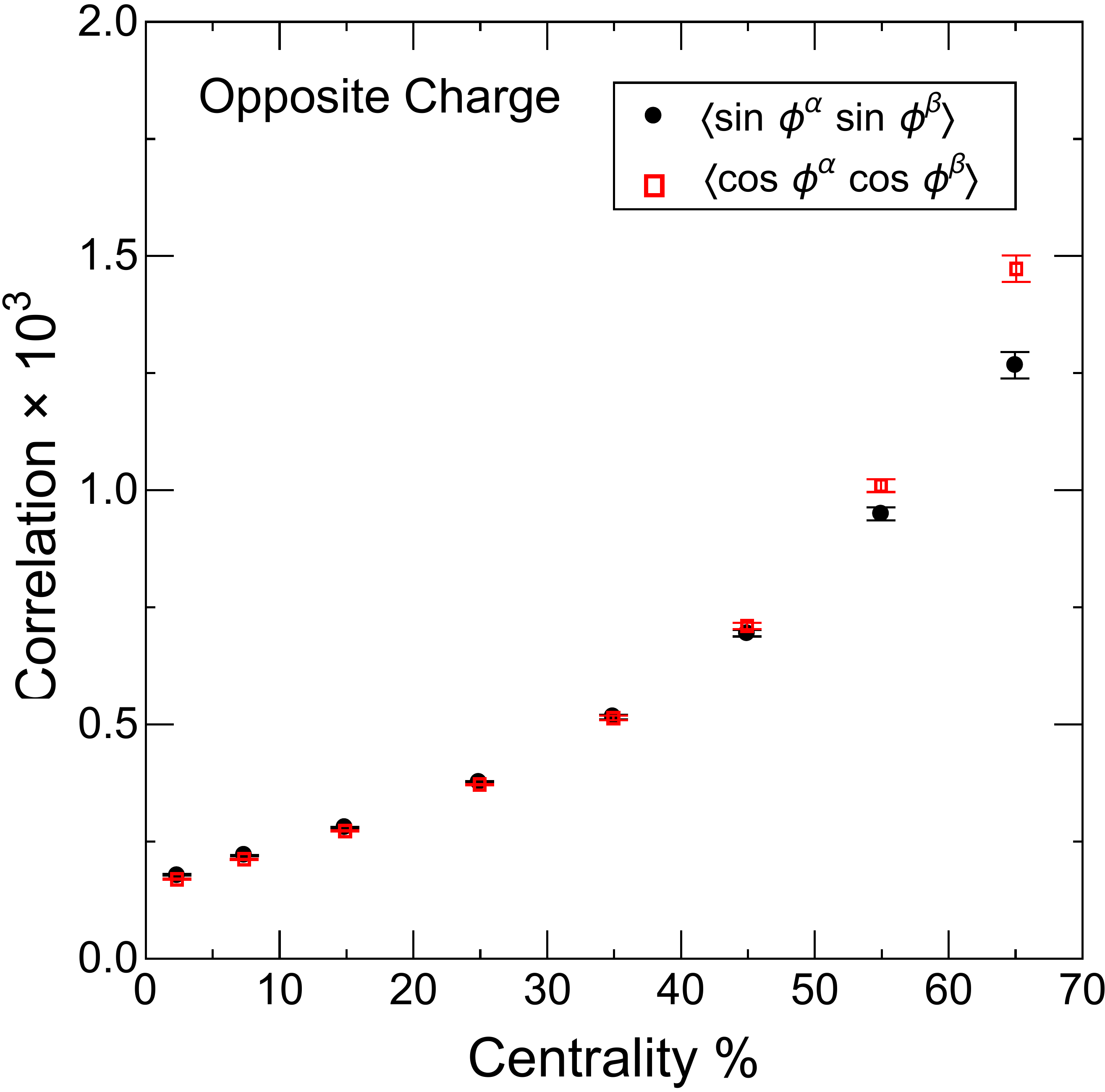}
\caption{ The in-plane and out-of-plane projected azimuthal correlations from STAR for AuAu collisions at $200 \rm GeV$. Figures from \cite{Bzdak:2009fc}.  } \label{fig:jl:ccss} 
\end{center}
\end{figure*}

If the CME were to be dominant contribution, its extraction would have been easy. As it turns out, however, the measured $\gamma$-correlator from STAR at RHIC energy (with data shown in Sec. \ref{subsec_6_chirality}) is dominated by background contributions, as already pointed out in \cite{Bzdak:2009fc} shortly after the data came out.  This situation can be recognized by a joint analysis  of the  $\gamma$ and $\delta$ 
correlators as defined in Eqs.~\eqref{eq:jl:gamma} and \eqref{eq:jl:delta}, both of which were measured, as discussed in Sec.~\ref{subsec_5_3}.
Given these two correlators, one can define  in-plane and out-of-plane projected azimuthal correlations:   $\langle \cos\phi^\alpha \cos\phi^\beta \rangle = (\delta^{\alpha\beta} + \gamma^{\alpha\beta})/2 $ and $\langle \sin\phi^\alpha \sin\phi^\beta \rangle = (\delta^{\alpha\beta} - \gamma^{\alpha\beta})/2 $. These projections are shown in Fig.~\ref{fig:jl:ccss} for the STAR data at 200~GeV. In a scenario of CME dominance, one would expect the out of plane projection to be larger than the in-plan projection. Specifically, one expects same side, out-of plane correlation for same charge pairs, i.e. $\langle \sin\phi^\pm \sin\phi^\pm \rangle >0 $, and opposite side out-of plane correlations for opposite charge pairs  $\langle \sin\phi^+ \sin\phi^- \rangle <0 $.  At the same time, the in-plane correlation should vanish,  $\langle \cos\phi^\alpha \cos\phi^\beta \rangle  \simeq 0 $ for both same and opposite charge pairs.
The data obviously deviates from such a pattern, in fact demonstrating a strong presence of background contributions. Indeed, the sizable $\langle \cos\phi^\alpha \cos\phi^\beta \rangle$ can only be background correlations which most likely would also make a sizable contribution toward the out-of-plane $\langle \sin\phi^\alpha \sin\phi^\beta \rangle$ correlation. This however does {\em not} exclude the possible presence of CME, and actually may still hint at certain  CME contribution. For example, let us take a closer look at the same charge correlations: there appears a sizable negative $\langle \cos\phi^\alpha \cos\phi^\beta \rangle$ correlation in contrast to   the $\langle \sin\phi^\alpha \sin\phi^\beta \rangle$ correlation that appears nearly zero and becomes slightly positive in the relatively peripheral region. Such a significant ``mismatch'' between in-plane and out-of-plane correlations could have two possible explanations. It could either be due to a negative (i.e. back-to-back type) correlation that somehow occurs {\em only} in-plane. However, at present it is not clear what  mechanism may cause such a background. Or, it could  be due to a positive out-of-plane-only correlation, such as the  CME signal, on top of a back-to-back bulk background, which induces comparable negative contributions to both in-plane and out-of-plane correlations. In any case, the above discussion shows that the situation is rather involved, and the challenge is to extract a small signal immersed in a dominant background. 
 
A number of specific sources of background correlations have been identified (see e.g. reviews \cite{Bzdak:2012ia,Kharzeev:2015znc} for detailed discussions). We briefly discuss them here: \\
{\em Transverse momentum conservation (TMC) ---} This leads to a back-to-back correlations. To see this, consider the extreme case of producing only two particles. Obviously, their momenta must be balanced in the transverse direction, $\vec{\bf p}$ and $-\vec{\bf p}$, resulting in a back-to-back correlation. For a system of N particles, this correlation is ``diluted'' by $1/N$. A quantitative analysis of TMC can be found in \cite{Bzdak:2010fd}. Of course, if the transverse momentum conservation is confined to certain local scale, then the effect would be  amplified. Nevertheless, transverse momentum conservation does not distinguish between charges of particles and thus the correlations are the same for  same-sign and opposite-sign pairs. Therefore,  one can remove their influence by considering the difference between same and opposite charged correlations, $\gamma^{OS} -\gamma^{SS}$. \\
{\em Local charge conservation (LCC) ---} This effect arises from possible charge neutrality over small size spatial domains upon freeze-out. Due to strong collective expansion, particles produced in close spatial proximity tend to have their momentum directions collimated on average. The neutrality enforces a companion negative charge for every positive charge produced from the same local domain. This leads to a near-side azimuthal correlations for opposite-sign pairs, resulting in positive contributions to  both in-plane and out-of-plane correlations with the former slightly stronger. Quantitative analysis~\cite{Pratt:2010gy,Schlichting:2010na,Pratt:2010zn,Schenke:2019ruo} suggests that the LCC could account for a substantial amount of the observed opposite-sign correlation patterns. However the precise mechanism of such local neutrality is unclear and the spatial size of such neutral domains, which is a key parameter controlling the strength of this correlation, remains poorly constrained.  
\\ 
{\em Resonance decay ---} Hadrons produced from resonance, or more generally cluster, decay will have their momenta correlated in azimuthal angle, inheriting the parent resonance's momentum~\cite{Wang:2009kd}. A most relevant example for the $\gamma$-correlator background is the neutral $\rho$ decay into charged pions, i.e. $\rho^0 \to \pi^+ \pi^-$, which makes a considerable contribution to a near-side correlation for opposite-sign hadron pairs and has similar characteristics to the LCC contribution. The hadronic decay contributions could be quantified by hybrid bulk evolution models which include a hadronic cascade stage, e.g., implemented via UrQMD simulations.

All the background sources discussed roughly scale inversely with the multiplicity. Therefore, they should increase as one decreases the centrality, i.e. considers more peripheral collisions. They all induce similar angular correlations for both in-plane and out-of-plane projected components, however with the in-plane component larger in magnitude due to stronger in-plane collective expansion. Therefore their contributions to the $\gamma$-correlator, being a difference between the in-plane and out-of-plane components, are roughly proportional to $v_2$ as in Eq.~\eqref{eq:jl:FH} and are absorbed together into the $F$-term there.  Finally let us note, that centrality dependence of the same-sign out-of-plane correlation, left panel of Fig.~\ref{fig:jl:ccss}, is much weaker as compared to the other  correlations.

To wrap up this discussion, the current understanding is that the measured $\gamma$-correlator, while being sensitive to CME contributions, is dominated by strong non-CME background correlations but also bears hints of nonzero CME-like signal at RHIC energy region. To convincingly extract a possible CME signal, progress needs to be made in three directions. First, one needs to get the backgrounds under control  by quantitatively understanding and computing them. Above we discussed the present status but obviously more needs to be done. Second, one needs to be able to quantitatively predict the CME signal. Here  major progress has been  and continues to be made, as we shall discuss next. Lastly, one needs to develop new experimental approaches that may be more sensitive to the  CME and may allow suppression/separation/subtraction of the backgrounds: a lot of interesting developments are ongoing along this line, which we discuss in the last two parts of this subsection.

\subsubsection{Quantitative modeling of anomalous chiral transport}
\label{sec:7.7.2}

To address the difficulty in the experimental detection of CME signals, it is imperative to develop a quantitative modeling framework that simulates the anomalous chiral transport while accurately accounts for the realistic environment in heavy ion collisions. In the past few years a matured framework, called the Anomalous-Viscous Fluid Dynamics (AVFD), has been developed~\cite{Jiang:2016wve,Shi:2017cpu,Shi:2018sah}. We note in passing that there have been phenomenological study of CME based on kinetic transport models as well~\cite{Sun:2018idn,Sun:2016mvh,Sun:2016nig,Deng:2016knn,Deng:2018dut,Zhao:2019ybo,Zhao:2019crj,Shou:2014zsa}. 

\begin{figure*}[hbt!]
\begin{center} 
\includegraphics[scale=0.22]{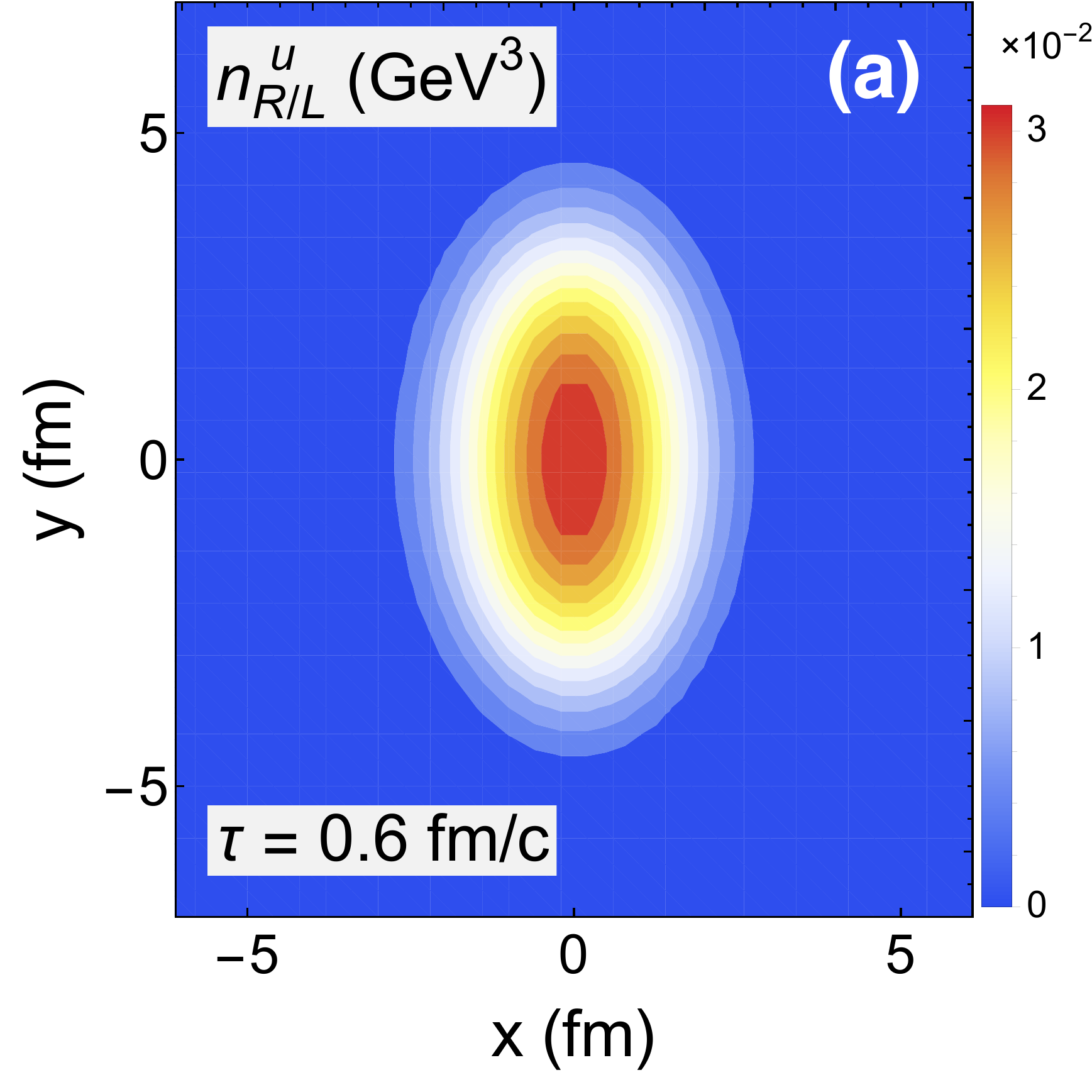} 
\includegraphics[scale=0.22]{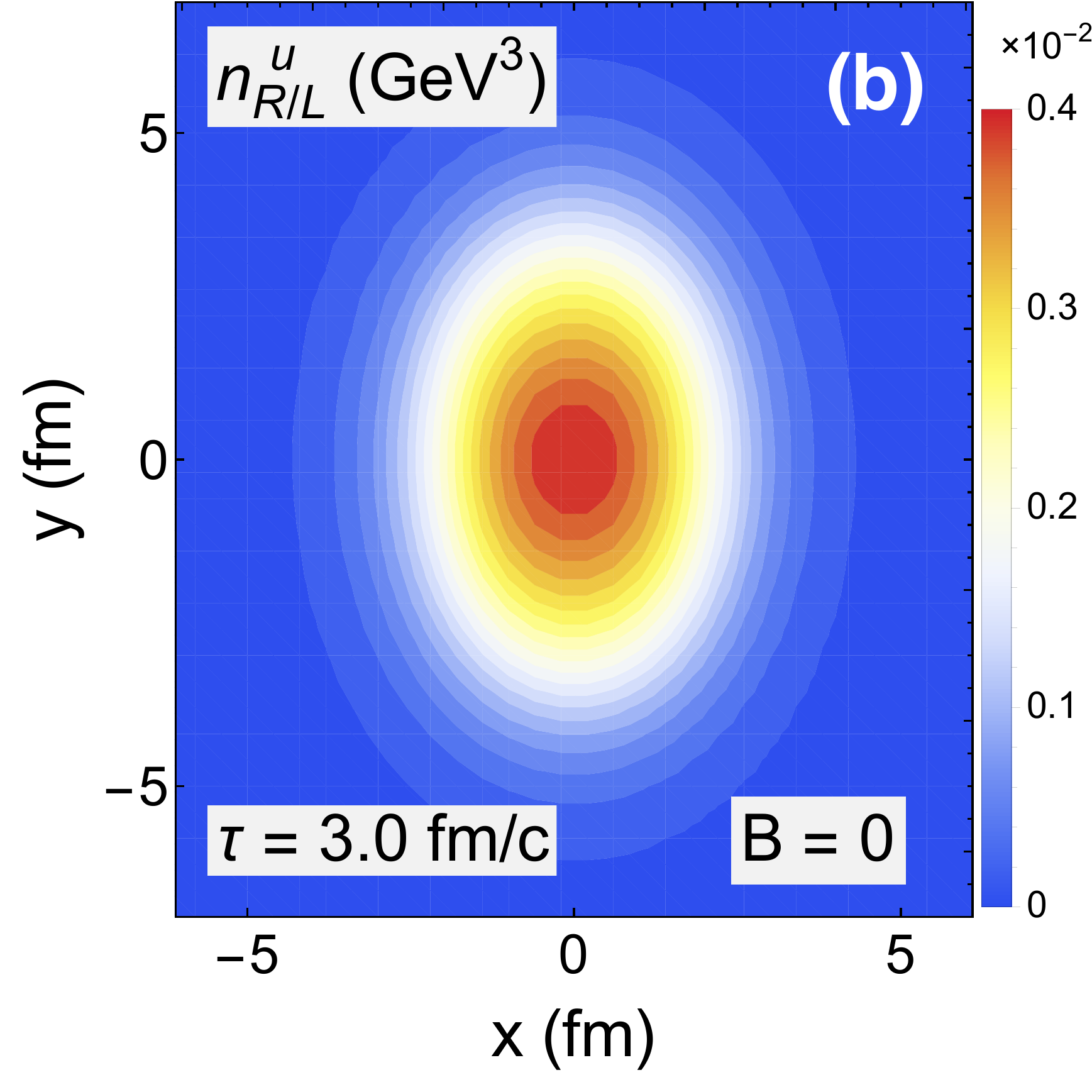}
\includegraphics[scale=0.22]{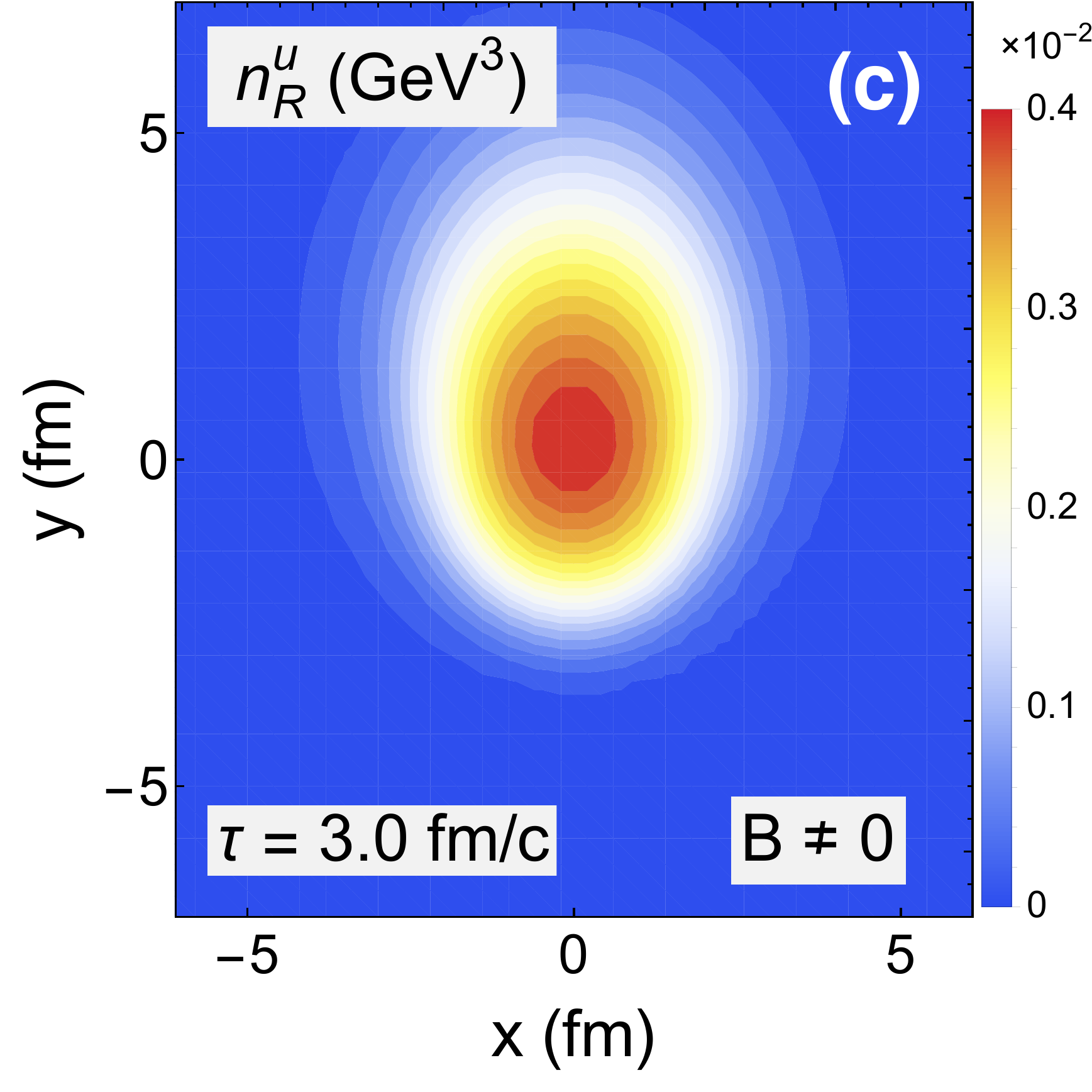} 
\includegraphics[scale=0.22]{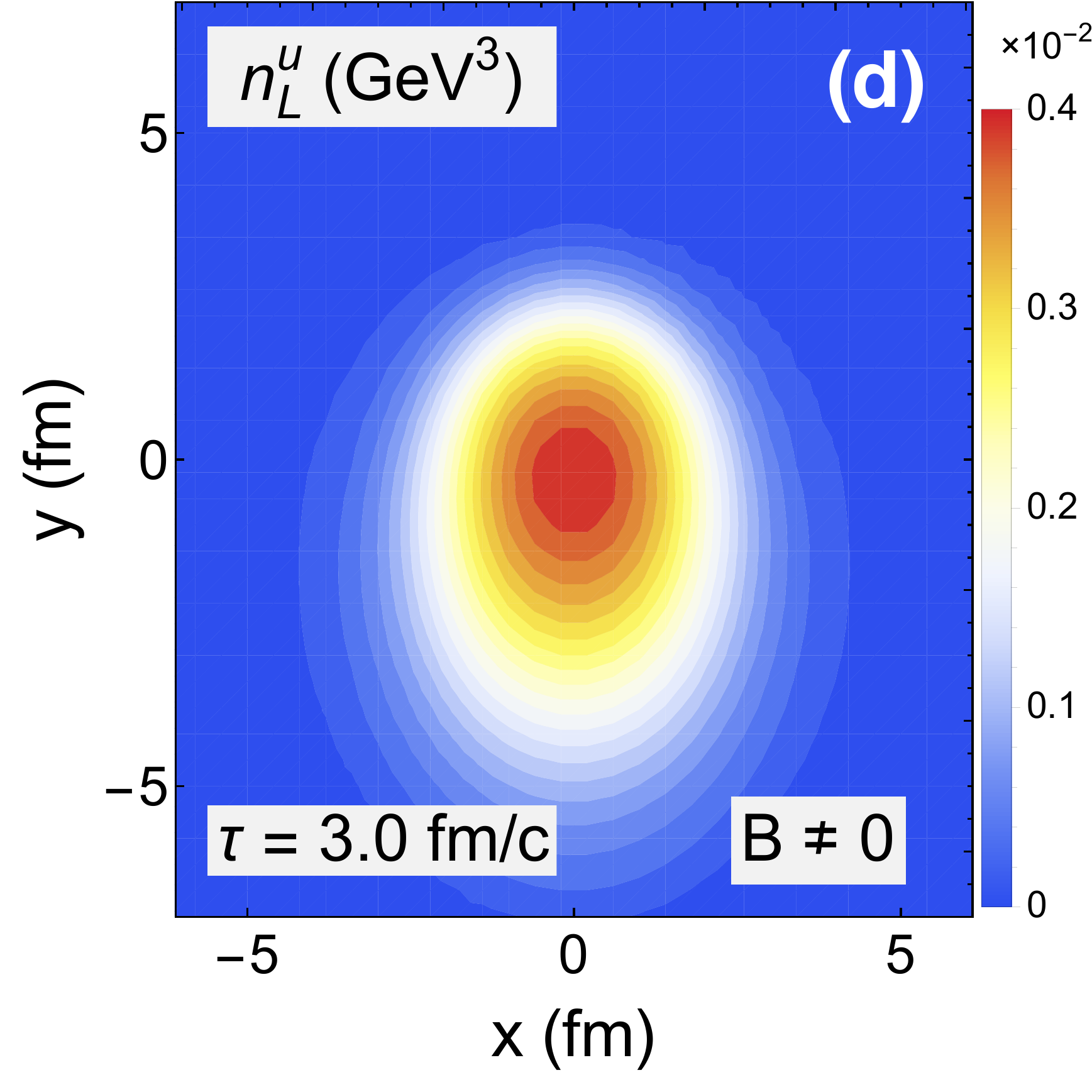}
\caption{The evolution of up-quark densities calculated with AVFD for a given initial charge density distribution. The  initial charge density at the initial time $\tau_0=0.60$~fm/c is depicted in panel-(a). Starting from  the same initial condition, three cases of evolution are studied: Evolution without magnetic field (panel-(a)), evolution with nonzero ${\bf B}$ field along positive y-axis for right-handed quarks (panel-(c)), and left-handed quarks (panel-(c)). All densities in panels (b)-(c) are taken at a time $\tau = 3$~fm/c. The figure is adapted from \cite{Jiang:2016wve,Shi:2017cpu}.} \label{fig:jl:f6} 
\end{center}
\end{figure*}

The main idea of AVFD is to extend the hydrodynamic bulk evolution~\cite{Shen:2014vra} to properly include the evolution of the fermion currents following the anomalous fluid dynamics framework described in Sec.~\ref{subsec_5_1_fluid}. In AVFD the evolution of fermion currents is calculated on top of a viscous hydrodynamics background. This approach accounts for both anomalous and normal viscous transport effects, however, it neglects the feedback of the anomalous current onto the bulk-evolution. The most important feature of AVFD is the evolution of the ${\bf B}$-field driven anomalous current $J^{\mu, (1)}_{anomalous} $,  Eqs. (\ref{eq:jl:Cfluid}) and (\ref{eq:jl:Cfluid2}), where the left-handed and the right-handed components contribute with opposite sign. In Fig. \ref{fig:jl:f6} we demonstrate the effect of such chiral transport within AVFD. Specifically we consider the evolution of up-quark densities starting from  a given initial  density distribution. The  initial up-quark density at time $\tau_0=0.60$~fm/c is depicted in panel-(a).  In panels (b)-(d) we show the  up-quark densities after the system  has evolved to the time $\tau=3.00$~fm/c. In panel-(b) we see the density in absence of a magnetic field. In this case right-handed and left handed quarks show the same distribution. In panels (c) and (d) we show the distributions  with a non-zero magnetic field ${\bf B}$ along positive y-axis for right-handed (panel (c)) and left-handed (panel(d)) up-quarks. The effect of the anomalous transport is clearly visible: the right-handed quarks are predominantly  transported along the direction of the magnetic field, i.e. towards the positive $y$, while the left-handed  quarks are transported in the opposite direction. For the negatively charged down-quarks one would see just the opposite pattern: Right-handed d-quarks will predominantly propagate towards negative $y$ while left-handed ones propagate to positive $y$.

\begin{figure}[hbt!]
\begin{center} 
\includegraphics[scale=0.4]{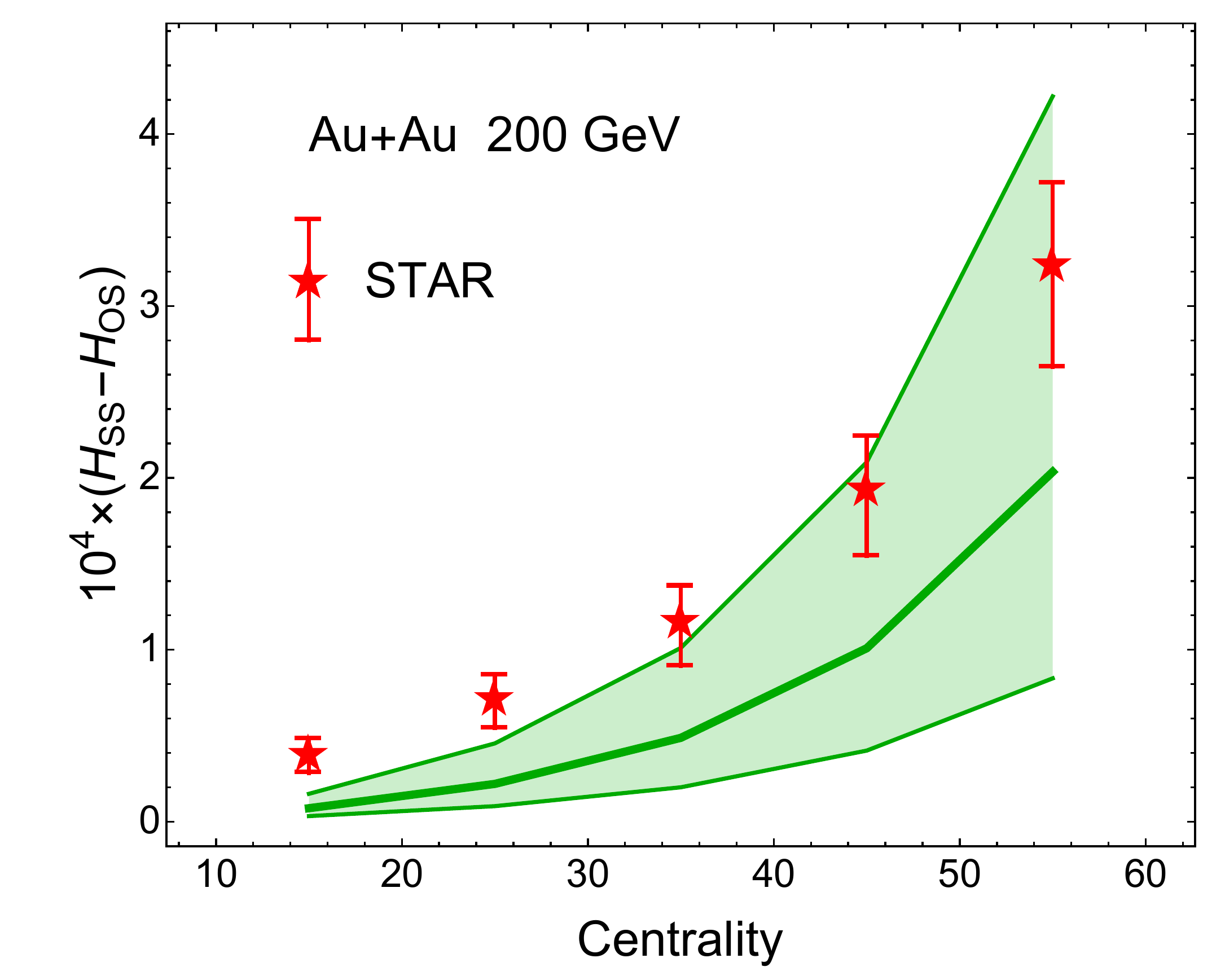}
\caption{The CME-induced H-correlator obtained by the AVFD simulations and compared with STAR data. Figure from \cite{Jiang:2016wve,Shi:2017cpu}. } \label{fig:jl:avfd} 
\end{center}
\end{figure}

Within the AVFD framework in a heavy ion collision  the charge separation arises then as follows.
The $\vec B$ field points along the out-of-plane direction, which we will denote by the $y$-direction. If initially we have a nonzero axial charge density, i.e. an imbalance between right-handed and left-handed quarks, we will have a situation shown similar to those shown in panels (c) or (d) of Fig. \ref{fig:jl:f6}.  For example, if we have more right-handed then left-handed quarks, the positively-charged up-quark density would move more towards the positive $y$-direction while, at the same time, the negatively charged d-quark density would move towards the negative y-direction. As a result we have a net charge current parallel to the magnetic field. If, on the other hand we have more left-handed than right-handed quarks, the charge current would be anti-parallel to the magnetic field. These currents cause a separation of electric charge across the reaction plane which can be quantified by the coefficient in 
Eq. (\ref{eq:jl:separation}). AVFD, thus, allows for a quantitative understanding of the generation and evolution of a CME-induced charge separation signal during  the hydrodynamic stage  as well as its dependence on various theoretical ingredients (for detail see \cite{Shi:2017cpu}). With reasonable estimates of key parameters, such as the strength and duration of the magnetic field, the results of the AVFD model for the CME-induced values of the H-correlator, defined in Eq.~\eqref{eq:jl:FH}, are shown in Fig.~\ref{fig:jl:avfd} and found to be  consistent  with those extracted from the STAR measurements.

More recently this framework has been further developed into event-by-event simulations (EBE-AVFD)~\cite{Shi:2018sah}, which is the state-of-the-art tool for quantifying anomalous chiral transport in heavy ion collisions. The EBE-AVFD includes fluctuating initial conditions as well as a hadronic cascade stage via UrQMD after the hydrodynamic evolution which accounts for the influence of hadronic re-scatterings and resonance decays. Further improvements of the framework shall address a number of issues, including more accurate determination of axial charge initial conditions~\cite{Mace:2016svc,Mace:2016shq}, the fluctuation and dissipation of axial charge during the hydrodynamic evolution~\cite{Iatrakis:2014dka,Guo:2016nnq,Hou:2017szz,Lin:2018nxj}, as well as the possible generation of the CME current in the pre-equilibrium stage~\cite{Huang:2017tsq}. 

%First, the initial conditions for the vector charge densities (of each flavor of quarks) as well as the axial charge densities arising from gluonic topological fluctuations need to be determined. So far there has been a lack of sophisticated models for estimating such initial conditions, posing a most severe theoretical uncertainty. The evolution of the axial charges suffers extra complications due to gluonic topological fluctuations as well as finite mass corrections which  need to be incorporated as stochastic noise for the axial current evolution equations: for a state-of-the-art discussion, see e.g.~\cite{Iatrakis:2015fma}.

%Second, there is short pre-equilibrium stage, on the order of $(0.5\sim 1)\, \rm fm/c$ time, before the fireball reaches the near-equilibrium hydrodynamic stage. The anomalous transport during this stage could be studied via either the classical statistical field simulations or the chiral kinetic theory. Several attempts have been made along both directions and such efforts are under active development. It may be anticipated that such pre-equilibrium modeling shall then be easily coupled to the subsequent hydrodynamic evolution. 

\subsubsection{Developing new  measurement methods}

Recognizing the limitations of the $\gamma$-correlator, it is natural to look for alternative measurement methods that would help extracting  possible CME signals. As can be seen from the two-component decomposition in Eq.~\eqref{eq:jl:FH}, a useful strategy is to  ``dial'' certain external control conditions so as to vary one component while keep the other component stay constant, or to vary the two components in opposite ways, i.e. allowing one to increase while the other to decrease.   A number of new approaches were put forward in recent years, all of which follow more or less that philosophy.
As we discussed in Section~\ref{subsec_5_3}, the signal is driven by the magnetic field whereas the background is driven by the elliptic flow, $v_{2}$. Therefore, in order to disentangle signal and background, it would be helpful to vary the magnetic field  while keeping the elliptic flow constant or vice versa. The  event-shape-engineering approach, already discussed in Sec.~\ref{subsec_6_chirality}, is an attempt to vary the elliptic flow, while keeping the magnetic field more or less constant~\cite{Bzdak:2011np,Acharya:2017fau}. Alternatively, one may want to keep the elliptic flow constant and change the magnetic field. This is the idea behind the isobar run which we discuss at the end of this subsection. Let us turn to the event-shape-engineering approach first. 
%The event-shape-engineering approach, already mentioned in Sec.~\ref{subsec_6_chirality}, is an example of such approach~\cite{Bzdak:2011np}, by varying the flow-driven backgrounds while maintaining similar CME signal~\cite{Acharya:2017fau}. (The isobaric collisions, to be discussed at length in the next part, follows the opposite approach, i.e. by varying the magnetic-field-driven signal while maintaining identical bulk backgrounds.)  
Within a given centrality class of events, the background contributions can be considerably varied by selecting the events' observed elliptic flow coefficient $v_2$. This could allow a subtraction of the $v_2$-dependent portion from the inclusive $\gamma$-correlator and, therefore, provide a potential CME signal that is independent of $v_2$. The working assumption of this method is that the magnetic-field-driven CME signal is  uncorrelated with the geometric shape fluctuations and stays constant across the whole centrality class. This however is not entirely true and the magnetic field component projected to the event-plane (which is the relevant component   for measured CME signal)  would actually change with the event selection. There is thus the danger of over-subtraction in the $\gamma$-correlator. There could also be issues with potential selection bias and event plane resolution. 

A different approach is to   compare the  $\gamma$-correlator measurements with respect to the event plane (EP) and that with respect to the reaction plane (RP)~\cite{Xu:2017qfs,Zhao:2018blc}. Event-by-event fluctuations in the initial conditions cause a de-correlation in azimuthal angles among  event-wise elliptic anisotropy, the magnetic field orientation  and the reaction plane. The event-plane, by definition is the direction maximizing the elliptic anisotropy $v_2$ and therefore the EP-measured $v_2$ value is greater than that from RP method. On the other hand, the magnetic field direction is more correlated with RP than with EP.  Thus the RP-projected magnetic field value is found to be greater than the EP-projected value. Therefore,  the $v_2$-driven background and the magnetic-field-driven signal would vary differently between EP and RP based measurements. This would allow the possibility of inferring the signal/background ratio by contrasting EP-measured and RP-measured $\gamma$-correlator. A main caveat with this approach is that reaction plane is a theoretical concept and can only be approximately inferred, e.g., from the so-called ZDC plane at RHIC. This uncertainty requires careful study.  

Another category of new measurements focuses on characterizing the background components by examining the correlations in a situation where the backgrounds will be present while the CME contribution is expected to be absent. For example, one can measure the $\gamma$-correlator in a small colliding system (e.g. pPb at LHC or dAu at RHIC) in which the magnetic field is decorrelated with event plane and thus the CME does not contribute to the $\gamma$-correlator. Or, one can measure a modified $\gamma$-correlator that is not defined with respect to the elliptic event plane but to the triangular event plane: in this case again the magnetic field is decorrelated with event plane and, again, the CME does not contribute to the $\gamma$-correlator. In both cases, the various background correlations should still be present and thus can be determined from measurements.  These approaches were developed by CMS and successfully used to put a rather stringent limit on the existence of CME signal in PbPb collisions at $5.02$~TeV \cite{Khachatryan:2016got,Sirunyan:2017quh}. Recently, the  STAR collaboration also applied these analysis methods for constraining background contributions at RHIC. 

Yet another strategy aims to find observables in which the background correlations could be significantly suppressed. For example, one may apply an invariant mass selection on the $\pi^+ \pi^-$ pairs which dominate   the opposite-sign pairs used in the $\gamma$-correlator measurements~\cite{Zhao:2017nfq,Zhao:2018blc}. The aforementioned resonance decay contributions, e.g. $\rho^0 \to \pi^+ \pi^-$ could be suppressed   by an invariant mass selection away from the $\rho^0$ mass region. This method helps control resonance decay contributions but can not suppress other background correlations. The CME signal would be mostly produced in the soft regime with low invariant mass, where recent measurements suggest a nonzero signal.  

There is also a new proposal of the so-called $R$-correlator method~\cite{Magdy:2017yje,Magdy:2018lwk}. To construct this observable, one first examines the distribution $N(\Delta S_\parallel)$ of the event-wise  out-of-plane dipole charge separation $\Delta S_\parallel$ for   a given set of events. The next step is to construct a ratio $C(\Delta S_\parallel)$ via normalizing $N(\Delta S_\parallel)$  by a baseline  distribution $N_{shuffled}$ obtained from the same set of events after random reassignment of the charge of each particle in an event. This ratio is to be further contrasted with a similarly constructed ratio for the in-plane separation $C(\Delta S_\perp)$, leading to the $R$-correlator: $R(\Delta S) = C(\Delta S_\parallel)/C(\Delta S_\perp)$. Since CME-induced charge separation will only influence the out-of-plane distribution, the $R$-correlator is expected to be sensitive to the CME. Indeed AVFD simulations unambiguously demonstrate such sensitivity, with the so-constructed $R(\Delta S) $ distribution becoming more and more concave-shaped when the CME signal is increased. The step of normalizing the charge separation distributions by shuffled baseline distributions also appears to effectively suppress background contributions to certain extent. A  preliminary STAR analysis~\cite{Ye:2018oqi} based on this method shows that at comparable multiplicity, a rather flat distribution of the $R$-correlator for p+Au and d+Au collisions (where pure backgrounds prevail) while a strongly concave-shape distribution is seen for peripheral Au+Au collisions, providing an indication of CME signal. At present, potential issues with this method mainly concern the lack of a full understanding on the behavior (e.g. convex versus concave shape) of the $R$-correlator in response to various backgrounds~\cite{Bozek:2017plp}  as well as the influence of factors like kinematic cuts, multiplicity fluctuations and event-plane resolution. Finally we note in passing a newly proposed observable to look for charge separation via signed balance function based on the idea of examining the momentum ordering of charged pairs along the out-of-plane direction~\cite{Tang:2019pbl}.

Clearly each of the new methods discussed above, just like the $\gamma$-correlator, has its own advantages and drawbacks. The best strategy to move forward is perhaps to look for a consistent  pattern regarding the presence of a CME signal by a global analysis of the various methods and observables. A first attempt at such a global analysis was recently carried out by the STAR collaboration and reported at the Quark Matter 2018 conference~\cite{Ye:2018oqi}. 
 The STAR results suggest that a CME signal could likely be at the 10\% level of the measured inclusive $\gamma$-correlator. However, the currently large uncertainty as well as the discrepancy among different measurement methods would not rule out a signal in the range of $0\sim20$\%. A rigorous conclusion, therefore, requires future experimental and theoretical effort to fully understand and improve each approach's methodology, to carefully examine and address potential issues with them, and to meaningfully combine all analyses together.

 \subsubsection{The isobaric collisions}
 \label{sec7:isobar}

As can be seen from the above discussions, while new proposals for measurement methods are under active development,   each of them may face difficulties of one sort or another. 
The theoretical uncertainties and experimental limitations have made it very challenging to  quantitatively separate the possible signal from the backgrounds. 
 In the past few years, a novel idea emerged and matured, based on contrasting collisions of the {\em isobaric nuclei pairs}~\cite{Skokov:2016yrj,Kharzeev:2019zgg}. In one sentence, this idea could be summarized as follows: the bulk collective dynamics  controls the background correlations, while the magnetic field controls the CME signal, and the strategy is to create and compare two colliding systems that have identical  bulk properties but different magnetic field. 

\begin{figure}[hbt!]
\begin{center} 
\includegraphics[scale=0.27]{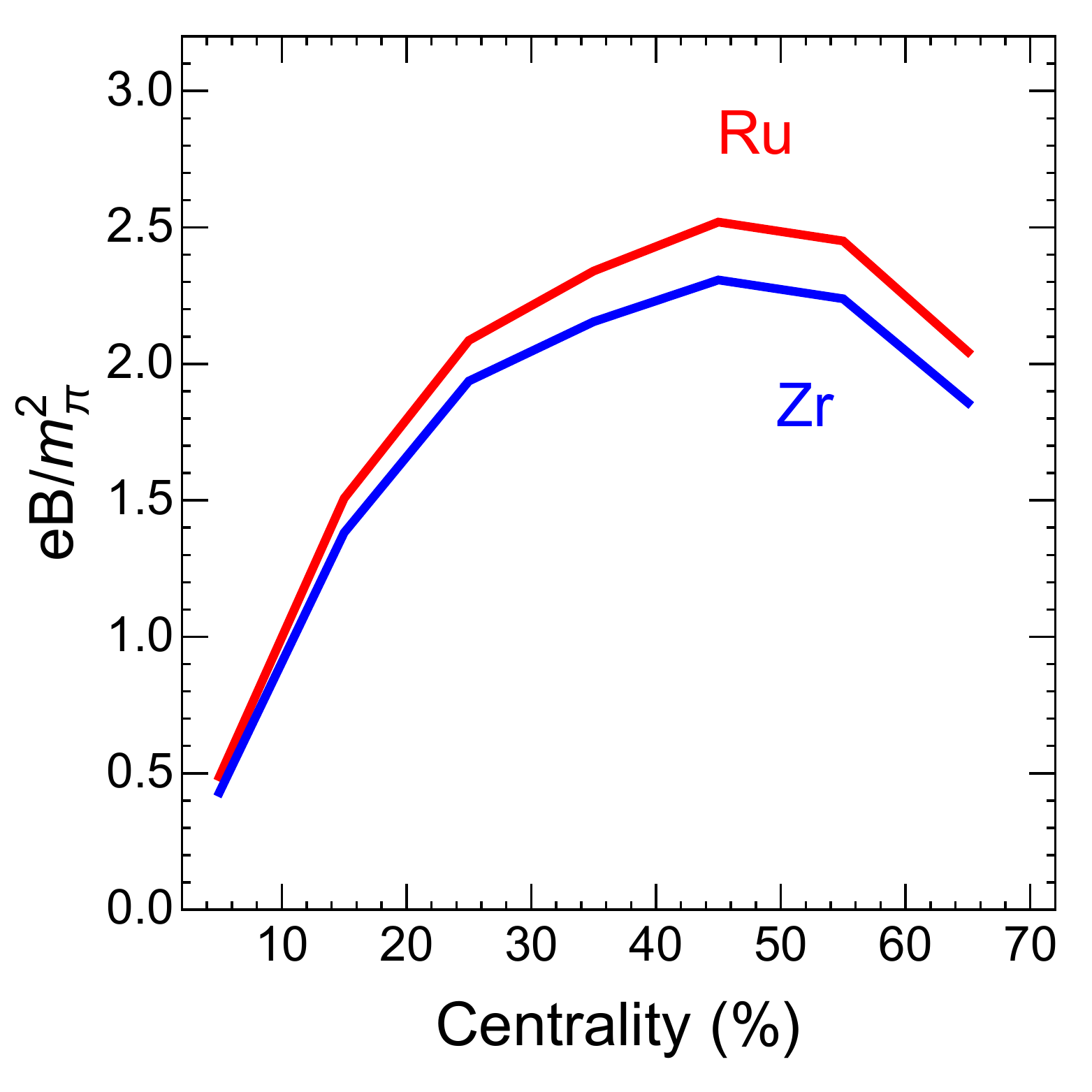}
\includegraphics[scale=0.27]{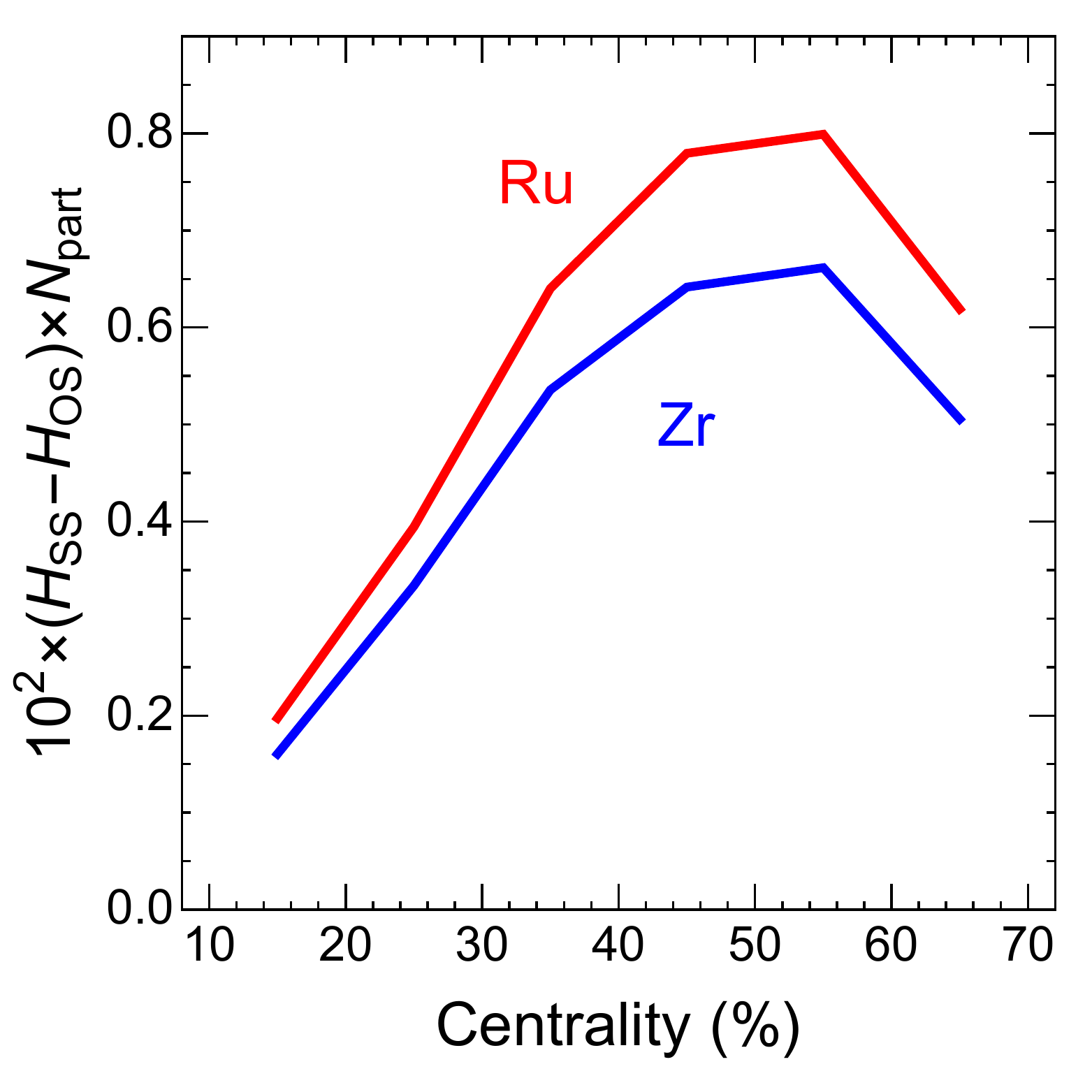}
\includegraphics[scale=0.34]{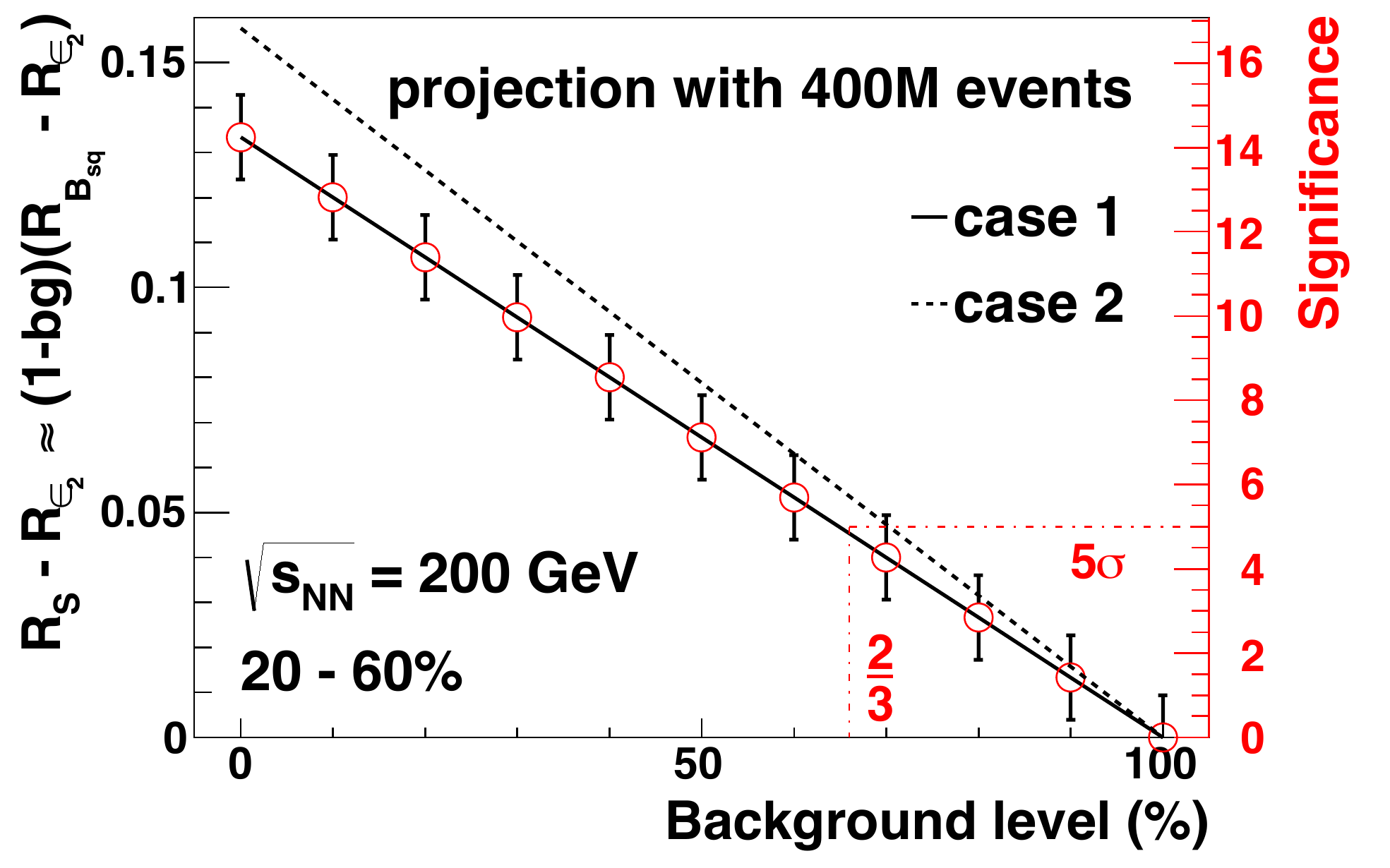}
\caption{The magnetic fields (left) and the AVFD-computed H-correlator signals (middle) for the RuRu and ZrZr collision systems at $200$ GeV. The right panel demonstrates the projected differentiation capability with 400-million events (vertical axis) of the contrast measurement versus the background level in $\gamma$-correlator (horizontal axis) for the isobaric collisions at RHIC. Left and middle panels are from \cite{Shi:2017cpu} while the right panel is from \cite{Skokov:2016yrj,Deng:2016knn}.} \label{fig:jl:isobar} 
\end{center}
\end{figure}

 Let us explain how this idea works in some details.  A pair of isobars are two different nuclei that have the same mass number A but different electric charge (i.e. proton number) Z. Specifically for our discussions, we consider Ruthenium (Ru) and Zirconium (Zr), both having A=96 nucleons while Z=44 for Ru and Z=40 for Zr. One can then perform experiments with Ru+Ru collisions and with Zr+Zr collisions at the same beam energy, and look for the difference of desired observables between the two colliding systems. The 10\% difference in the electric charge implies a difference in the magnetic field of also about 10\%. This can also be seen from the computed magnetic field values with event-by-event simulations in Fig.~\ref{fig:jl:isobar} (left).
Since the CME current  is proportional to the magnetic field, see Eq.~\eqref{eq:jl:cme}, and since the  
 $\gamma$-correlator measures the variance of the CME current, the $\gamma$-correlator scales with the square of the magnetic field $\gamma \sim B^{2}$. Therefore, a 10\% difference in the magnetic field translates into a 20\% difference in the CME-induced contribution to the $\gamma$-correlator. This can also be seen in the results from the AVFD simulation shown in the middle panel of Fig.~\ref{fig:jl:isobar}.  
%This would imply a difference at about 20\% level in the CME-induced correlation signals (i.e. the H-correlator) between the two, as demonstrated by the AVFD simulation results in Fig.~\ref{fig:jl:isobar} (middle). 

The key for success of the isobar run is ensure  that the background correlations will be exactly the same. This requires that one compares the $\gamma$-correlator for events from Ru+Ru and Zr+Zr, which have identical bulk properties, in particular multiplicity and elliptic flow $v_2$. Since the two isobars have the same number of nucleons, naively one would expect that a simple centrality (or multiplicity) selection would ensure
this requirement.
% ensure one would just select events based on centrality, which is essentially a multiplicity selection,  and then compares Ru+Ru events with Zr+Zr events in the same centrality bin.  Since the two isobars have the same number of nucleons, one would naively expect the produced bulk matter to behave in about the same way. i.e. one expects more or less the same $v_{2}$ for the two systems with the same centrality.
However, uncertainties in the initial nucleon distributions in the two isobar nuclei may actually bring percent-level difference in the elliptic flow between the two systems~\cite{Xu:2017zcn} for a given centrality.  Sine the  CME signal in the $\gamma$-correlator is expected to be rather small, this presents a serious problem which needs to be controlled. As shown in  \cite{Shi:2018sah}, this problem can be overcome by selecting events on {\em both multiplicity and elliptic-flow}. Event-by-event simulations~\cite{Shi:2018sah} for both initial conditions and for the final state observables have demonstrated that the joint-multiplicity-ellipticity cut would be effective and sufficient for this purpose.

Experimentally one compares the measured $\gamma$-correlator between Ru+Ru and Zr+Zr, for which the signal differs by about 20\% while the backgrounds are (ideally speaking) identical. The actual difference in the $\gamma$-correlator depends of course   on the background level in $\gamma$-correlator. This is shown in the right panel of Fig.~\ref{fig:jl:isobar}, where we show the results of calculations Refs.~\cite{Skokov:2016yrj,Deng:2016knn} for the expected differentiation capability with 400-million events (vertical axis) of the contrast measurement versus the background level in $\gamma$-correlator (horizontal axis) for the isobaric collisions at RHIC.  For example, if the background level were to be less than 2/3 of the $\gamma$-correlator, then the isobar experiment would allow an observation of CME signal at $5\sigma$ level or beyond with just 400-million events.  

The isobaric collision experiment was successfully completed in the RHIC 2018 with more than twice the planned 400-million events collected by the STAR detector. With this improved statistics and with the latest estimate of from the STAR analysis of a CM signal of about 10\%, one could hope for a signal of at least $2 \sigma$ significance.  Furthermore, there are now an array of new observables besides just the $\gamma$-correlator, as discussed in the preceding part. One may therefore still anticipate a fairly reasonable chance to have a statistically significant statement on the presence of CME signal based on a synthetic analysis of multiple observables. 

Clearly, the isobaric collision experiment provides a unique opportunity to detect the possible presence of CME  in heavy ion collisions. A conclusive observation of the CME signal  would directly establish, for the first time, the existence of chirality-flipping topological  transitions that stem from the vacuum structure of QCD as well as provide  evidence for the restoration of chiral symmetry in the high temperature quark-gluon plasma. Therefore,  given the characteristic beam energy dependence of the CME signal as discussed in Sec.~\ref{subsec_5_4}, a positive CME signal from the isobar run  at $200$ GeV would strongly suggest future measurements at lower beam energies. A measurement in the intermediate BES energy region would possibly detect a maximal CME signal while one in lowest BES energy region may help establish the expected disappearance of the CME signal.

%\subsubsection{The beam energy dependence}

%\subsubsection{Vorticity driven effects}

%Given the nonzero averaged vorticity in non-central collisions, one may also expect anomalous transport current induced by the CVE in Eq.(\ref{eq:jl:cve}). This may become especially relevant toward lower beam energy collisions in which both the averaged vorticity and the baryonic chemical potential would be more significant. A possible signature would be the separation of baryons/anti-baryons along the out-of-plane direction, and preliminary efforts were made to look for it. We refer interested readers to \cite{Kharzeev:2015znc} for a more detailed discussion.  

%=====================================================================================
%=====================================================================================
%

%%% Local Variables:
%%% mode: latex
%%% TeX-master: "BES_Main_current"
%%% End:

%-------------Section 8-----------------------------------
%
%=====================================================================================
%=====================================================================================
\section{Summary and Outlook}
\label{sec8}
We close this review by summarizing the present status as well as discussing open questions. We then finish with discussions of experimental upgrades at current experimental facilities as well as future opportunities at upcoming new facilities.  

\subsection{Present status}
The measurements taken during the first phase of the  RHIC Beam Energy Scan (BES-I) have provided
rather intriguing results, with hints at potentially very interesting physics.

\begin{enumerate}
\item Global observables such as elliptic flow  and particle yields suggest that the systems created
  in the energy range of the RHIC BES are (locally) equilibrated and that their bulk dynamics is
  controlled by the collision geometry and collective expansion.  

\item Several observables, however, exhibit  quite an intriguing energy dependence
  below $\sqrt{s_{NN}} \lesssim 20 \gev$. The directed flow as well as the higher order cumulants
  ($\kappa_{3} $ and $\kappa_{4}$) of
   net-protons show a non-monotonic behavior which may indicate dramatic changes in the equation of
  state which, for example, may be due to a critical point or first-order phase
  transition. In addition femtoscopy measurements seem to indicate a maximum 
  emission time and minimum expansion velocity at the same beam energy (about $\sqrt{s_{NN}} \simeq 20 \gev$),  
  suggesting that the system
  might cross the softest point of the EoS at this energy.
  A closer analysis of the measured net-proton cumulants finds that below
  $\sqrt{s_{NN}}\lesssim 15\gev$ the system exhibits significant four-particle correlations which
  increase in strength rapidly with decreasing collision energy.
  Model calculations, such as UrQMD,  as well as general arguments based on baryon
  number conservation predict the cumulants (within fixed rapidity window $\Delta y=1$) to
  decrease toward lower beam energy.
    
\item The charge asymmetry of the azimuthal correlator, $\gamma$, which is sensitive to the
  anomalous chiral transport current, seems to show features in line with expectations from the
  Chiral Magnetic Effect at high collision energy while it vanishes for energies below
  $\sqrt{s_{NN}}\lesssim 10\gev$. This may be an indication  that at this energy the 
  system created in these collisions changes from one dominantly in the phase with restored
  chiral symmetry to one dominantly in the phase with spontaneously broken chiral symmetry. 
  Similarly, analyses of the charge asymmetry by means of the H-correlator as well as the slope
  parameter of the charge-dependent elliptic flow, which is sensitive to the Chiral Magnetic Wave,
  show a positive signal at high energy which disappears for energies below $\sqrt{s_{NN}}\lesssim 10\gev$.

\end{enumerate}

We should of course point out that the above findings are partially based on preliminary data and may depend upon model inputs in certain aspects. Also the presently available statistics, especially for the cumulant measurements as well as those for the charge dependent correlator $\gamma$, is rather limited. Much better statistics would be required in order to draw any firm conclusions. This situation will improve dramatically in the second phase of the RHIC beam energy scan (BES-II). 
Aside from the limited statistics, there are a number of open questions which need to be addressed. 

\subsection{Open questions}
\begin{enumerate}
\item Criticality and net-proton cumulants:
  \begin{itemize}
  \item
  The critical point signatures described in Section~\ref{sec4} are
characterized by specific patterns of non-monotonous behavior of
various cumulants some of which (such as $\kappa_4$)
appear to be intriguingly matched by experimental data described in Section~\ref{sec6}. Before drawing
conclusions, it is important to keep in mind that the predictions in
Section~\ref{sec4} are based on an assumption of approximate {\em
  static} equilibrium. This assumption has to be relaxed in order to
address the effects of the expansion dynamics on critical point
signatures, some of which, such as memory effects, are
anticipated. The hydrodynamic framework incorporating fluctuations and
critical slowing down needs to be developed and compared to
experimental measurements.

  \item The proton cumulant measurements at the lowest BES energies show that the system exhibits
    large and positive four-particle correlations and sizable three-particle correlations which are
    negative. Such a pattern cannot be explained by simply invoking attractive interactions among
    the protons.  Similarly, the directed flow, $v_{1}$, does not support a simple picture based on
    attractive interactions either.  This raises the question of what dynamics can explain both the net-proton cumulants and the directed flow. Can the observed patterns be related to a change in
    the equation of state and if not, what other dynamical effect plays a role?
      %\item 
      Clearly the steep increase of the fourth-order net-proton cumulant ratio $\cum{4}/\cum{2}$
    calls for measurements at even lower  energies or equivalently higher baryon densities. In
    addition, of course, higher statistics for all energies below $\sqrt{s_{NN}}\lesssim 20\gev $ are needed.  
    
  \item While not discussed in detail in this review, it would be equally interesting to
    experimentally explore and establish the existence of a  crossover transition at $\mu_{B}\simeq
    0$. As discussed in Section \ref{sec4:baryon:cumulants},  to achieve this goal future high energy
    collider
    experiments should pursue the measurement of
    cumulants of the net-proton distribution to sixth, $\kappa_6/\kappa_2$, eigth,
    $\kappa_8/\kappa_2$, or even higher order. Naturally, for these measurements to be successful
    high statistics data sets as well as excellent control of the systematics are
    mandatory.

\end{itemize}
\item Chirality and chiral magnetic effect (CME):
\begin{itemize}
\item A crucial but missing piece of information for a quantitative understanding of the CME
  is the time evolution of the in-medium magnetic field. Can it be quantitatively
  computed including its collision energy dependence? Can it be connected with the fluid rotation in some 
  way? Can the lifetime of the magnetic field be quantitatively connected with certain other
  experimental observables, such as the polarization difference between particles and antiparticles
  or the di-electron transverse momentum broadening?
 
\item A major obstacle in the search for the Chiral Magnetic Effect is the characterization of non-CME physics backgrounds 
  and not all of them  are well understood. Could combining various experimental methods together
  help constrain, extract, or remove the backgrounds? Can background effects be quantified and
  implemented into dynamical models so that they are able to predict also the energy dependence of
  both the signal and expected backgrounds? What additional theoretical developments and
  measurements are required to achieve this?

\item At present, the most promising experimental strategy to isolate the CME signal is the isobar
  run (see Sec. \ref{sec7:isobar}) which has just been completed at RHIC where data have been taken
  at a collision energy of $\sqrt{s_{NN}}=200\gev$. If the analysis of the data show a 
  positive CME signal, a definitive discovery will require additional measurements at lower
  beam energies. These measurements are also necessary to unambiguously establish that chiral
  symmetry is actually restored in the system created in these collisions.
 \end{itemize}

\end{enumerate}

\subsection{Theory efforts}
    {To address these questions one needs not only additional measurements but also progress in theory. In particular, a definitive answer to the questions of the existence of the  QCD critical point and anomalous transport requires a coherent theory effort aiming at a quantitative understanding of the dynamical evolution of the system created in a heavy ion collision. Essential to this is the development of a quantitative dynamical framework which describes the entire evolution of the system from the initial collision to the final state. Such a framework should build on the progress achieved over the last decade which resulted in a quantitative description of heavy ion collisions at the highest RHIC and LHC energies. However to address the physics discussed in this review, several important further developments are needed:
\begin{itemize}
    \item The initial conditions for the hydrodynamic evolution are substantially different at lower energies, where the penetration time of the nuclei could be  potentially as long as the equilibration time. Also the stopping of the incoming nucleons needs to be understood very well as they are the source of net-baryon density at midrapidity. Some progress in this direction has already been made \cite{Shen:2017bsr}.
    \item The (viscous) hydrodynamic evolution needs to be extended to propagate all conserved currents as well as the anomalous current. In addition, it needs to account for and propagate the fluctuations which occur in the vicinity of the critical point. Also, a definitive calculation of the CME will require magneto-hydrodynamics to properly determine the time evolution of the magnetic field and its impact on the anomalous currents. And, as already discussed in Section~\ref{sec:4:map_eos}, an equation of state with  critical point which is consistent with lattice QCD results at $\mu_B=0$ is needed~\cite{Parotto:2018pwx}. 
    
    Considerable progress has been made on many fronts: First results for the  the propagation of conserved and anomalous currents have been obtained \cite{Shen:2017bsr,Jiang:2016wve,Shi:2017cpu} (see also Sect.\ref{sec:7.7.2}). The treatment of fluctuations  within stochastic hydrodynamics \cite{Kapusta:2011gt,Bluhm:2018plm,Singh:2018dpk,Hirano:2018diu,Nahrgang:2018afz} and their coupling to critical modes \cite{Nahrgang:2011mg,Herold:2016uvv,Sakaida:2017rtj} have been studied. Alternatively, the time evolution of hydrodynamic fluctuations  by means of deterministic propagation of correlations functions have  been explored~\cite{Akamatsu:2016llw,An:2019osr}, including the self-consistent description of critical fluctuations and critical slowing down \cite{Stephanov:2017ghc,Rajagopal:2019xwg}. First calculations have applied magneto-hydrodynamics to study the time evolution of the magnetic field \cite{Inghirami:2016iru,Gursoy:2018yai}. 
    \item A transformation of the hydrodynamic fields to actual particles needs to preserve the fluctuations and all conserved and anomalous currents \cite{Oliinychenko:2019zfk}. In addition, as the critical point and phase transition affect also the hadronic phase, the kinetic evolution of the hadronic phase needs to account for this~\cite{Stephanov:2009ra}. 
\end{itemize}
All these essential pieces still need to be put together in one common framework which can be used to perform a quantitative comparison with experimental data.
Clearly there are many open conceptual as well as technical problems and an ever growing community of theorists worldwide is actively working to address these challenges. %and solve them.
}

%\vspace{-0.15in}
\begin{figure}[!hbt] 
\begin{center}
\includegraphics[scale=0.36]{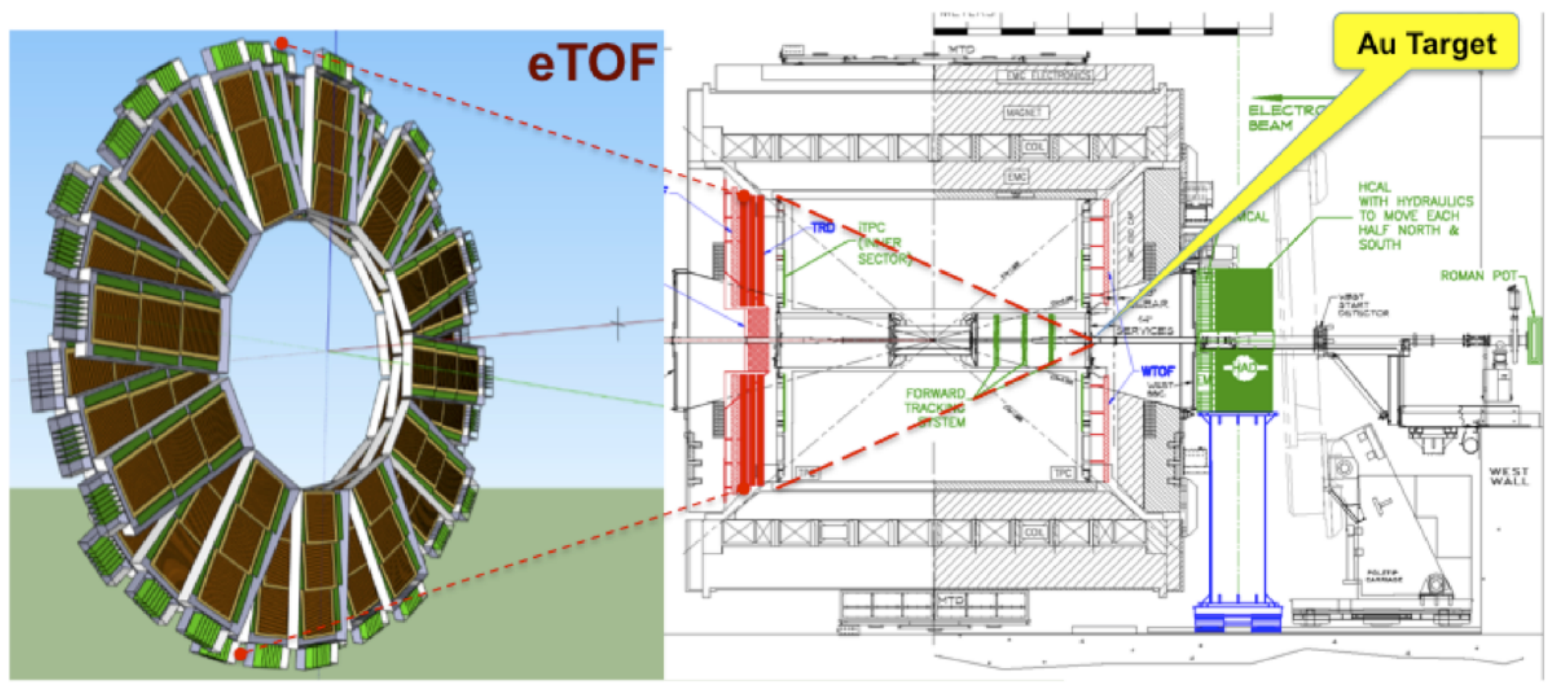} 
%\vspace{-0.1in}
\caption{STAR fixed-target setup during the BES-II at RHIC.}
 \label{fig:exp:fxt}
%\vspace{-0.15in}
\end{center}
\end{figure}

\subsection{Future Upgrades and Physics Program at RHIC and the global picture}
\label{sec8:future}

During the years of 2019 - 2021, the STAR experiment at RHIC will
undertake the second beam energy scan campaign (BES-II), with the goal to dramatically reduce both
statistical and systematical uncertainties for all of the observables and to map the QCD phase
diagram with highest possible precision. As shown in Table~\ref{tab:exp:bes}, BES-II will cover the
energy range $7.7 \gev \le \sqrt{s_{NN}} \le 19.6 \gev$ corresponding to
$420 \mev \ge \mu_B \ge 200 \mev$~\cite{STAR_NOTE_25}. The expected statistical uncertainties for the
observables relevant for the search for anomalous transport and the QCD critical point are
illustrated in Fig.~\ref{fig:exp:f15} (green band) for the cumulants and in Fig. \ref{fig:exp:f12} (pink band) for the $\gamma$ correlator.
In addition to the increased statistics STAR detector upgrades
such as that of the inner TPC, the end-cap time of flight detector, and the event plane detector allow for increased rapidity coverage,
better event plane resolution, and centrality selection and particle identification
at forward rapidities. These new
capabilities will  enable STAR to address questions concerning the range of the proton correlations,
more precise measurements of the $\gamma$ correlator, and a suppression of auto-correlations for the
fluctuation measurements.

Both the fourth-order proton cumulants, which have the largest value at the lowest available RHIC
collider energy, as well as the CME observable, $\gamma^{OS}-\gamma^{SS}$, which vanishes right at
the lowest energy, call for measurements at even lower energies. To address this issue, the STAR
collaboration has developed a fixed-target (FXT) program, which uses a gold-target at the one end of
the TPC accompanied with a time of flight wall at the other end of the TPC (see
Fig.~\ref{fig:exp:fxt}). With the fixed target setup, STAR will be able to measure particle
productions and correlations in Au+Au collisions in the energy range $3 \gev \leq \sqrt{s_{NN}} \leq
19.6 \gev$, extending maximum accessible value of the chemical potential 
from  $\mu_B \simeq 400\mev$ to $\mu_B \simeq 750 \mev$.
Importantly for the control of systematics, the highest center-of-mass energy of the FXT mode
overlaps with the lowest energy from colliding mode at $\sqrt{s_{NN}}= 7.7 \gev$, while its lower
range overlaps with the  collision energies explored by future experiments, such as CBM at
FAIR~\cite{STAR_NOTE_26} and MPD at NICA.

\begin{figure}[!hbt] 
\begin{center}
\includegraphics[scale=0.66]{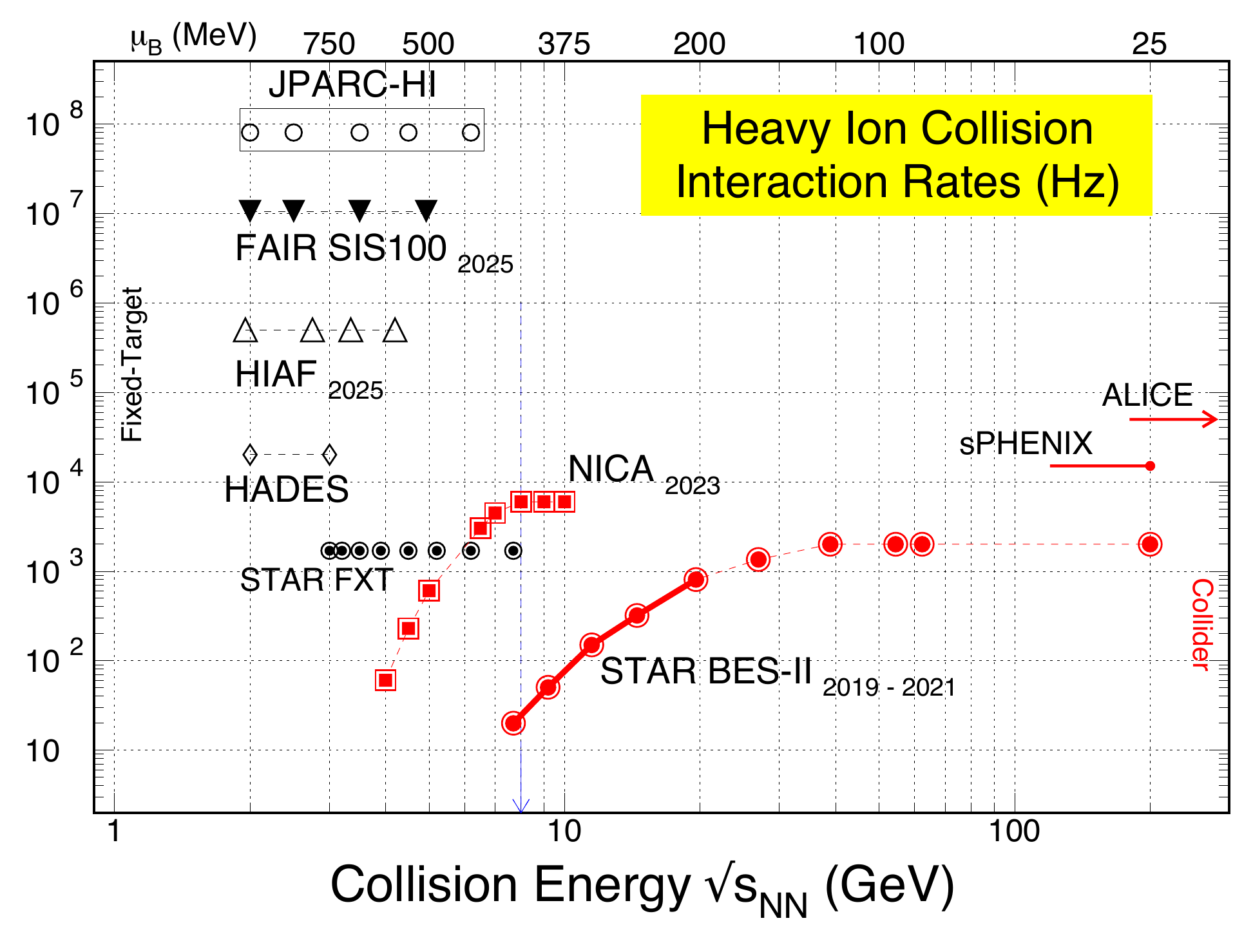} 
%\vspace{-0.1in}
\caption{Interaction rates for high-energy nuclear collision facilities: the second phase RHIC beam energy scan (BES-II, $7.7 < \sqrt{s_{NN}} < 19.6$ GeV, open-circles), NICA (open-squares) as well as  the fixed-target projects including HADES (diamonds),   HIAF (open-triangles), FAIR (filled-triangles) and JPARC-HI (open-circles). STAR fixed-target range is indicated with filled circles.  
  }
 \label{fig:exp:f20}
%\vspace{-0.15in}
\end{center}
\end{figure}

The exciting physics presented in this review has sparked many new experimental initiatives
worldwide.  In addition to BESII and the STAR FXT program, there are a number of approved
experiments coming online in the near future, which will be able to address the physics discussed
here.  They are the compressed baryonic matter (CBM) experiment at the FAIR facility, the
multi-purpose-detector (MPD) at NICA and the
CSR-external target experiment (CEE) at HIAF. In addition to those there are plans to have an experiment at J-PARC in
Japan. And, last but not least, there is the HADES experiment at GSI which has already taken data
for Au+Au collisions at $\sqrt{s_{NN}}\simeq 2.2 \gev$ and is presently analyzing proton
cumulants which, if successful, will provide and important low energy reference point for the fluctuation
measurements. The energy range together with the (expected) interaction rate of the various
facilities  are shown Fig.~\ref{fig:exp:f20}, which 
demonstrate  the worldwide
excitement about the physics presented in this review.

%%% Local Variables:
%%% mode: latex
%%% TeX-master: "BES_Main_current"
%%% End:

\section*{Acknowledgements}

This material is based upon work partly supported by the U.S. Department of Energy, 
Office of Science, Office of Nuclear Physics, under contract number 
DE-AC02-05CH11231, contract number DE-AC03-76SF00098, contract number DE-FG0201ER41195, as well as within the framework of the Beam Energy Scan Theory (BEST) Topical Collaboration. 
The work is also supported in part by MOST of China 973-project No. 2015CB856900 and NSFC Grant No. 11735007, by Ito Science Foundation (2017) and JSPS KAK-ENHI Grant No. 25105504, as well as by NSF Grant No. PHY-1913729.  
AB also acknowledges partial support by the Polish Ministry of Science and Higher Education and the National Science Centre  Grant No. 2018/30/Q/ST2/00101. JL also acknowledges hospitality during sabbatical visits at CCNU and SCNU where part of this work was completed. The authors thank Dmitri Kharzeev, Swagato Mukherjee, Shuzhe Shi and Paul Sorensen for useful discussions and communications.

%-------------Appendix-----------------------------------
\appendix

\section{Introduction to Cumulants}
\label{sec:ab:cumulant-intro}

%\textbf{(Equations will be checked at the end)}

In this Appendix we present a general discussion of the multiplicity distribution $P(N)$
and its various characteristics. $P(N)$ is the probability to produce $N$
particles and by definition $0\leq P(N)\leq 1$, and $\sum_{N}P(N)=1$. For
example $P(N)$ could describe the probability to produce $N$ protons in
Au+Au collisions. In this review we discuss several distributions: the
Poisson distribution, the binomial distribution and the negative binomial
distribution. Here we mostly consider a simplified case where $P(N)$ depends
on one variable only but we also generalize to the higher number
of variables.

\subsection*{Factorial moments}
It is convenient to define the generating function%
\begin{equation}
H(z)=\sum_{N}P(N)z^{N},\qquad H(1)=1,  \label{eq:ab:H}
\end{equation}%
where the value of $H(z)$ at $z=1$ is determined by the normalization condition
$\sum_{N}P(N)=1$. Having $H(z)$ we can readily calculate the average number of particles, the
average number of pairs, triplets etc. In general, it is easy to see that%
\begin{equation}
F_{k}\equiv \left\langle \frac{N!}{(N-k)!}\right\rangle =\left. \frac{d^{k}}{%
dz^{k}}H(z)\right| _{z=1},  \label{eq:ab:Fk}
\end{equation}%
where $F_{k}$ are called the factorial moments and $k\geq 1$. $%
F_{1}=\left\langle N\right\rangle $ is the average number of particles, $%
F_{2}=\left\langle N(N-1)\right\rangle $ is the average number of pairs, $%
F_{3}=\left\langle N(N-1)(N-2)\right\rangle $ is the average number of
triplets etc.
Let us add here that $F_{k}$ are directly related to typically measured
multi-particle rapidity\footnote{%
In this Section we will usually discuss rapidity distributions but our
discussion is general and applies to any other variable or variables.} (or
any other variable) distributions, see, e.g., Ref. \cite{Bialas:1999tv} 
\begin{equation}
F_{k}=\int dy_{1}\cdots dy_{k}\,\rho _{k}(y_{1},...,y_{k}),
\label{eq:ab:Fk-rhok}
\end{equation}%
where $\rho _{1}(y)$ is a single-particle rapidity distribution\footnote{%
In this review we will usually denote it by $\rho (y)$.}, $\rho
_{2}(y_{1},y_{2})$ is the two-particle density\footnote{%
To be more precise, $\rho _{2}(y_{1},y_{2})dy_{1}dy_{2}$ is the number of
pairs of particles such that one particle is located in a rapidity interval $%
y_{1}-dy_{1}/2<y<y_{1}+dy_{1}/2$ and another one is located in $%
y_{2}-dy_{2}/2<y<y_{2}+dy_{2}/2$.} (distribution of pairs) etc.

In the case of two species of particles we  have%
\begin{equation}
H(z,\bar{z})=\sum_{N,\bar{N}}P(N,\bar{N})z^{N}z^{\bar{N}},\qquad H(1,1)=1,
\end{equation}%
where $P(N,\bar{N})$ is, e.g., the probability to produce $N$ protons and $%
\bar{N}$ antiprotons. Here the factorial moments are given by%
\begin{equation}
F_{i,k}\equiv \left\langle \frac{N!}{(N-i)!}\frac{\bar{N}!}{(\bar{N}-k)!}%
\right\rangle =\left. \frac{d^{i}}{dz^{i}}\frac{d^{k}}{d\bar{z}^{k}}H(z,\bar{%
z})\right| _{z=\bar{z}=1},
\end{equation}%
where $F_{1,1}=$ $\left\langle N\bar{N}\right\rangle $, $F_{2,1}=$ $%
\left\langle N(N-1)\bar{N}\right\rangle $ etc.
If $P(N,\bar{N})=P(N)P(\bar{N})$, that is, the number of, say, protons is
independent from the number of, say, antiprotons we obtain 
\begin{equation}
H(z,\bar{z})=H(z)H(\bar{z}),
\end{equation}%
and $F_{i,k}=F_{i}F_{k}$.
As we will discuss later, the factorial moments are very useful in
experimental analysis since they are easy to correct for detector efficiency %
\cite{Bzdak:2012ab,Luo:2017faz}.

\subsection*{Factorial cumulants}
Another useful characteristics of $P(N)$ is given by the factorial
cumulants, which are generated from the following function 
\begin{equation}
G(z)=\ln \left( H(z)\right) =\ln \left( \sum_{N}P(N)z^{N}\right) ,
\label{eq:ab:G}
\end{equation}%
and the factorial cumulants are given by%
\begin{equation}
\hat{C}_{k}=\left. \frac{d^{k}}{dz^{k}}G(z)\right| _{z=1}.
\label{eq:ab:Ckhat}
\end{equation}
It is instructive to list the first three $\hat{C}_{k}$. $\hat{C}%
_{1}=\left\langle N\right\rangle $ is the average number of particles and%
\begin{eqnarray}
\hat{C}_{2} &=&\left\langle N(N-1)\right\rangle -\left\langle N\right\rangle
^{2}  \notag \\
&=&F_{2}-F_{1}^{2},  \label{eq:ab:C2hat}
\end{eqnarray}%
\begin{eqnarray}
\hat{C}_{3} &=&\left\langle N(N-1)(N-2)\right\rangle -\left\langle
N\right\rangle ^{3}-3\left\langle N\right\rangle \hat{C}_{2}  \notag \\
&=&F_{3}-F_{1}^{3}-3F_{1}\hat{C}_{2},  \label{eq:ab:C3hat}
\end{eqnarray}%
where we expressed $\hat{C}_{2}$ and $\hat{C}_{3}$ through the factorial
moments and $\hat{C}_{2}$. In general it is straightforward to express $%
\hat{C}_{k}$ through $F_{i}$ by taking appropriate derivatives of $\ln
\left( H(z)\right) $, see Eq. (\ref{eq:ab:G}), and using Eq. (\ref{eq:ab:Fk}%
).

We note that $\hat{C}_{k}$ are directly related to the integrals of the
multi-particle correlation functions 
\begin{equation}
\hat{C}_{k}=\int dy_{1}\cdots dy_{k}C_{k}(y_{1},\ldots ,y_{k}),
\label{eq:ab:Ckhat-corr-funct}
\end{equation}%
where $C_{k}(y_{1},\ldots ,y_{k})$ is the $k$-particle genuine correlation
function. The integration is performed over the same region of phase-space
where $P(N)$ is measured. For example, the two-particle rapidity density 
\begin{equation}
\rho _{2}(y_{1},y_{2})=\rho (y_{1})\rho (y_{2})+C_{2}(y_{1},y_{2}),
\label{eq:ab:rho2}
\end{equation}%
is given by a product of two single-particle distributions, $\rho
(y_{1})\rho (y_{2})$, plus the two-particle correlation function $%
C_{2}(y_{1},y_{2})$. Taking the integral of Eq. (\ref{eq:ab:rho2}) over a
given interval in rapidity we obtain%
\begin{eqnarray}
F_{2} &=&F_{1}^{2}+\int C_{2}(y_{1},y_{2})dy_{1}dy_{2}  \notag \\
&=&F_{1}^{2}+\hat{C}_{2},
\end{eqnarray}%
which is the same as Eq. (\ref{eq:ab:C2hat}). Analogously, for three
particles we have \cite{Botet:2002gj} (see also, e.g., Ref. \cite%
{Bzdak:2015dja}) 
\begin{eqnarray}
\rho _{3}(y_{1},y_{2},y_{3}) &=&\rho (y_{1})\rho (y_{2})\rho (y_{3})+\rho
(y_{1})C_{2}(y_{2},y_{3})+\rho (y_{2})C_{2}(y_{3},y_{1})  \notag \\
&&+\rho (y_{3})C_{2}(y_{1},y_{2})+C_{3}(y_{1},y_{2},y_{3}),
\end{eqnarray}%
where $C_{3}(y_{1},y_{2},y_{3})$ is the three-particle genuine correlation
function in rapidity. Calculating the integral over a given bin in rapidity
we obtain%
\begin{equation}
F_{3}=F_{1}^{3}+3F_{1}\hat{C}_{2}+\hat{C}_{3},  \label{eq:ab:F3}
\end{equation}%
which is the same as Eq. (\ref{eq:ab:C3hat}). In Ref. \cite{Bzdak:2015dja}
explicit formulas for $\rho _{k}(y_{1},...,y_{k})$ up to $k=6$ are presented.

The factorial cumulants are very useful in analysis of physical processes
since they are directly related to correlations present in the system and
the order of the factorial cumulant is related to the genuine correlation of
the same order, see, e.g., \cite{Ling:2015yau,Bzdak:2016sxg,Kitazawa:2017ljq}%
. This is not the case for the factorial moments, where, e.g., as seen from
Eq. (\ref{eq:ab:F3}), $F_{3}$ is a combination of factorial cumulants of
different orders.

In the case of two species of particles we have%
\begin{equation}
G(z,\bar{z})=\ln \left( H(z,\bar{z})\right) =\ln \left( \sum_{N,\bar{N}%
}P(N,\bar{N})z^{N}\bar{z}^{\bar{N}}\right) ,  \label{eq:ab:Gzzbar}
\end{equation}%
and 
\begin{eqnarray}
\hat{C}_{n+m}^{(n,m)} &=&\left. \frac{d^{n}}{dz^{n}}\frac{d^{m}}{d\bar{z}^{m}%
}G(z,\bar{z})\right| _{z=\bar{z}=1}  \notag \\
&=&\int dy_{1}\cdots dy_{n}d\bar{y}_{1}\cdots d\bar{y}%
_{m}C_{n+m}(y_{1},...,y_{n};\bar{y}_{1},...,\bar{y}_{m}),
\label{eq:ab:Cnmhat}
\end{eqnarray}%
where $C_{n+m}(...)$ is the $n+m$ genuine correlation function involving $n$
and $m$ particles of different species. For example%
\begin{equation}
\hat{C}_{2}^{(1,1)}=\left\langle N\bar{N}\right\rangle -\left\langle
N\right\rangle \left\langle \bar{N}\right\rangle ,
\end{equation}%
which is zero if, e.g., the number of protons $N$ is independent from the
number of antiprotons $\bar{N}$.

Again, the mixed factorial cumulants $\hat{C}_{n+m}^{(n,m)}$ are directly
related to the correlations present in the system. Suppose that $P(N,\bar{N}%
)=P(N)P(\bar{N})$, that is the number of, say, protons is independent from
the number of, say, antiprotons. In this case 
\begin{equation}
G(z,\bar{z})=\ln \left( \sum_{N}P(N)z^{N}\sum_{\bar{N}}P(\bar{N})z^{\bar{N}%
}\right) =\ln (H(z))+\ln (H(\bar{z})),
\end{equation}%
and according to Eq. (\ref{eq:ab:Cnmhat}), the mixed factorial cumulants $%
\hat{C}_{n+m}^{(n,m)}=0$ for $n\geq 1$ and $m\geq 1$.

\subsection*{Cumulants}
The cumulants of a given multiplicity distribution $P(N)$ are generated from
the cumulant generating function $K(t)$ given as 
\begin{equation}
K(t)=G\left( e^{t}\right) =\ln \left( H(e^{t})\right) =\ln \left(
\sum_{N}P(N)e^{tN}\right) ,  \label{eq:ab:K}
\end{equation}%
and the cumulants read%
\begin{equation}
\kappa _{i}=\left. \frac{d^{i}}{dt^{i}}K(t)\right| _{t=0}.
\end{equation}
As already mentioned in Section~\ref{sec6:exp:cumulants}, the STAR Collaboration
measured cumulants of net-proton \cite{Adamczyk:2013dal}, net-charge \cite%
{Adamczyk:2014fia} and net-kaon \cite{Adamczyk:2017wsl} distributions in
Au+Au collisions, see also Refs. \cite{Luo:2015ewa,Luo:2017faz}. The
definition of $K(t,\bar{t})$ for two species of particles is analogous to
Eq. (\ref{eq:ab:Gzzbar}), where we sum over $P(N,\bar{N})e^{tN}e^{\bar{t}%
\bar{N}}$. The net-proton cumulants, for example, are given by 
\begin{equation}
\kappa _{i}=\left. \frac{d^{i}}{dt^{i}}K(t,-t)\right| _{t=0},\qquad
K(t,-t)=\ln \left( \sum_{N,\bar{N}}P(N,\bar{N})e^{t(N-\bar{N})}\right) ,
\end{equation}%
where the corresponding $K(t,\bar{t})$ is taken at $\bar{t}=-t$. In this
case the cumulants are expressed through the moments of $N-\bar{N}$
distribution. Clearly, one could define also the cumulants of the sum $N+%
\bar{N}$ by taking appropriate derivatives of $K(t,+t)$ at $t=0$.

We note here that if $P(N,\bar{N})=P(N)P(\bar{N})$, that is, the number of,
say, protons, is independent from the number of, say, antiprotons we obtain%
\begin{equation}
K(t,-t)=K(t)+K(-t),
\end{equation}%
an the cumulants of, say, net-proton number can be expressed as a sum of
proton cumulants plus (even cumulants) or minus (odd cumulants) antiproton
cumulants.
Performing straightforward derivatives of $K(t,-t)$ we obtain the net-proton
(or baryon, charge, strangeness, etc.) cumulants%
\begin{eqnarray}
\kappa _{1} &=&\left\langle N-\bar{N}\right\rangle ,\quad  \notag \\
\kappa _{2} &=&\left\langle \left( \delta \lbrack N-\bar{N}]\right)
^{2}\right\rangle ,  \notag \\
\kappa _{3} &=&\left\langle \left( \delta \lbrack N-\bar{N}]\right)
^{3}\right\rangle ,  \notag \\
\kappa _{4} &=&\left\langle \left( \delta \lbrack N-\bar{N}]\right)
^{4}\right\rangle -3\left\langle \left( \delta \lbrack N-\bar{N}]\right)
^{2}\right\rangle ^{2},\quad \label{eq:ab:kappa-net}
\end{eqnarray}%
where $\delta \lbrack N-\bar{N}]=N-\bar{N}-\left\langle N-\bar{N}%
\right\rangle $.
As discussed in Section~\ref{sec4:baryon:cumulants}, 
the cumulants appear
naturally in statistical physics as derivatives of the partition function.

It is useful to express the cumulants through the factorial cumulants $%
\hat{C}_{k}$. Let us start with one species of particles. In this case $%
K(t)=G(x(t))$, where $x(t)=e^{t}$, see Eq. (\ref{eq:ab:K}), and the second
order cumulant reads 
\begin{eqnarray}
\kappa _{2} &=&\left. \frac{d^{2}}{dt^{2}}G(x(t))\right| _{t=0}  \notag \\
&=&\left[ \frac{d^{2}G(x)}{dx^{2}}\left( \frac{dx(t)}{dt}\right) ^{2}+\frac{%
dG(x)}{dx}\frac{d^{2}x(t)}{dt^{2}}\right] _{t=0,x=1}  \notag \\
&=&\left\langle N\right\rangle +\hat{C}_{2},
\end{eqnarray}%
where, according to Eq. (\ref{eq:ab:Ckhat}), $dG(x)/dx$ and $d^{2}G(x)/dx^{2}
$ at $x=1$ equal $\hat{C}_{1}=\left\langle N\right\rangle $ and $\hat{C}_{2}$%
, respectively. Clearly, $dx/dt=e^{t}$ which is one at $t=0$. Performing
straightforward calculations we obtain\footnote{%
Additionally: $\kappa _{5}=\left\langle N\right\rangle +15\hat{C}_{2}+25%
\hat{C}_{3}+10\hat{C}_{4}+\hat{C}_{5}$ and $\kappa _{6}=\left\langle
N\right\rangle +31\hat{C}_{2}+90\hat{C}_{3}+65\hat{C}_{4}+15\hat{C}_{5}+%
\hat{C}_{6}$.} \cite{Ling:2015yau,Bzdak:2016sxg} 
\begin{eqnarray}
\kappa _{2} &=&\left\langle N\right\rangle +\hat{C}_{2},  \notag \\
\kappa _{3} &=&\left\langle N\right\rangle +3\hat{C}_{2}+\hat{C}_{3},  \notag
\\
\kappa _{4} &=&\left\langle N\right\rangle +7\hat{C}_{2}+6\hat{C}_{3}+\hat{C}%
_{4}.  \label{eq:ab:kappa-Chat}
\end{eqnarray}

For two species of particles the formulas are a bit more complicated (see
the Appendix of Ref. \cite{Bzdak:2016sxg})%
\begin{eqnarray}
\kappa _{2} &=&\left\langle N\right\rangle +\left\langle \bar{N}%
\right\rangle +\hat{C}_{2}^{(2,0)}+\hat{C}_{2}^{(0,2)}-2\hat{C}_{2}^{(1,1)} 
\notag \\
\kappa _{3} &=&\left\langle N\right\rangle -\left\langle \bar{N}%
\right\rangle +3\hat{C}_{2}^{(2,0)}-3\hat{C}_{2}^{(0,2)}+\hat{C}_{3}^{(3,0)}-%
\hat{C}_{3}^{(0,3)}-3\hat{C}_{3}^{(2,1)}+3\hat{C}_{3}^{(1,2)}  \notag \\
\kappa _{4} &=&\left\langle N\right\rangle +\left\langle \bar{N}%
\right\rangle +7\hat{C}_{2}^{(2,0)}+7\hat{C}_{2}^{(0,2)}-2\hat{C}%
_{2}^{(1,1)}+6\hat{C}_{3}^{(3,0)}+6\hat{C}_{3}^{(0,3)}-6\hat{C}_{3}^{(2,1)}-6%
\hat{C}_{3}^{(1,2)}+  \notag \\
&&\hat{C}_{4}^{(4,0)}+\hat{C}_{4}^{(0,4)}-4\hat{C}_{4}^{(3,1)}-4\hat{C}%
_{4}^{(1,3)}+6\hat{C}_{4}^{(2,2)},  \label{eq:ab:kappa-Chat-two}
\end{eqnarray}%
which obviously reduces to Eq. (\ref{eq:ab:kappa-Chat}) if $\bar{N}=0$ since
in this case only $C_{k}^{(k,0)}$ may be different than zero. Here $\hat{C}%
_{2}^{(1,1)}$ measures the correlations between, say, protons and
antiprotons, $\hat{C}_{2}^{(0,2)}$ is the factorial cumulant for antiprotons
only etc. It is clear that the cumulants for the net-protons only are not
enough to determine all $\hat{C}_{i+k}^{(i,k)}$.
As seen from Eqs. (\ref{eq:ab:kappa-Chat}) and (\ref{eq:ab:kappa-Chat-two}),
the cumulants mix the factorial cumulants of different orders and thus their
interpretation is not straightforward. We will come back to this problem
shortly.

     {At the end of this part, let us clarify the notation frequently used in, e.g., experimental plots, see for example Fig. \ref{fig:nu:exp:f16}. The standard  deviation squared, $\sigma^2$, is simply the second central moment $\langle (x - \langle x\rangle)^2\rangle$, where $x$ equals to $N$ or $N - \bar{N}$. Clearly
$\sigma^2 = \kappa_2$. The skewness, $S$, is defined as $\langle (x - \langle x\rangle)^3\rangle / \sigma^3$ and thus $S\sigma$ equals to $\langle (x - \langle x\rangle)^3\rangle / \sigma^2 = \kappa_3/\kappa_2$. 
The kurtosis or, more precisely, the excess kurtosis, $\kappa$ is defined as 
$\kappa = \langle (x - \langle x\rangle)^4\rangle/\sigma^4 - 3$ and  therefore  
$\kappa \sigma^2 = \langle (x - \langle x\rangle)^4\rangle/\sigma^2 - 3\sigma^2 = \kappa_{4}/\kappa_{2}$, see Eq. (\ref{eq:ab:kappa-net}).
\footnote{     {For the normal distribution,  
$\langle (x - \langle x\rangle)^4\rangle/\sigma^4=3$ which provides a kind of baseline and therefore commonly used to shift the kurtosis value for defining the excess kurtosis. }}}

\subsection*{Discussion}

Here we discuss the factorial moments, factorial cumulants and cumulants in
different physical situations.
Let us start with one species of particles distributed according to the
Poisson distribution
\begin{eqnarray}
  P(N)=e^{-\ave{N}}\frac{\ave{N}^{N}}{N!}.
  \label{eq:append:poisson}
\end{eqnarray}
In this case, the generating function, given by Eq. (%
\ref{eq:ab:H}), reads%
\begin{eqnarray}
\left. H(z)\right| _{\text{Poisson}} &=&\sum_{N=0}^{\infty }e^{-\left\langle
N\right\rangle }\frac{\left\langle N\right\rangle ^{N}}{N!}%
z^{N}=e^{\left\langle N\right\rangle (z-1)}\sum_{N=0}^{\infty
}e^{-z\left\langle N\right\rangle }\frac{\left( z\left\langle N\right\rangle
\right) ^{N}}{N!}  \notag \\
&=&e^{\left\langle N\right\rangle \left( z-1\right) },  \label{eq:ab:H-Poiss}
\end{eqnarray}%
and consequently, see Eq. (\ref{eq:ab:Fk}), we obtain%
\begin{equation}
F_{k}=\left\langle N\right\rangle ^{k},  \label{eq:ab:Fk-Poiss}
\end{equation}%
that is, the number of pairs, triplets, etc., is simply expressed by a
product of $\left\langle N\right\rangle $. This result corresponds to the
situation where multiparticle rapidity distribution, $\rho _{k}$ for each $k$%
, is given by a product of single particle distributions%
\begin{equation}
\rho _{k}(y_{1},...,y_{k})=\rho (y_{1})\cdots \rho (y_{k}),
\end{equation}%
that is, the multiparticle genuine correlation functions $%
C_{k}(y_{1},...,y_{k})=0$ for each $k$. Indeed, performing integration over
a given bin in acceptance, see Eq. (\ref{eq:ab:Fk-rhok}), we obtain Eq. (\ref%
{eq:ab:Fk-Poiss}).\footnote{%
Strictly speaking, $F_{k}=\left\langle N\right\rangle ^{k}$ means that the
factorial cumulants $\hat{C}_{k}$, being the integrals over $%
C_{k}(y_{1},...,y_{k})$, equal zero, which does not necessary mean $%
C_{k}(y_{1},...,y_{k})=0$.}

Using Eq. (\ref{eq:ab:G}) we obtain $G(z)=\left\langle N\right\rangle \left(
z-1\right) $, and consequently $\hat{C}_{k}=0$ for $k\geq 1$. By the same
token we have $K(t)=$ $\left\langle N\right\rangle \left( e^{t}-1\right) $
and $\kappa _{n}=\left\langle N\right\rangle $. The last relation can be
immediately seen from Eq. (\ref{eq:ab:kappa-Chat}) in the absence of $\hat{C}%
_{k}$. Clearly, for the Poisson distribution all cumulant ratios equals one%
\begin{equation}
\frac{\kappa _{n}}{\kappa _{m}}=1,\text{\quad Poisson distribution.}
\end{equation}

Before we discuss more complicated and interesting examples, we want to
explain why physically the factorial cumulants $\hat{C}_{k}$, being the
integrals of the genuine correlation functions, are zero for the Poisson
distribution. Suppose that we generate some number of particles, $N$, from a
given multiplicity distribution $P(N)$, which we further assume to be the
Poisson distribution. Next we randomly split the particles between the left
and the right bin (for example in rapidity) with probability $1/2$, namely,
for each particle we choose randomly whether it goes to the left or to the
right bin. Now, we can calculate $P(N_{L},N_{R})$, being the joint
probability to observe $N_{L}$ particles in the left interval and $N_{R}$
particles in the right interval. Obviously $\left\langle N_{L}\right\rangle
=\left\langle N_{R}\right\rangle $. We obtain%
\begin{equation}
P(N_{L},N_{R})=P(N_{L}+N_{R})\frac{(N_{L}+N_{R})!}{N_{L}!N_{R}!}\left( \frac{%
1}{2}\right) ^{N_{L}}\left( \frac{1}{2}\right) ^{N_{R}},
\label{eq:ab:P-two-bins}
\end{equation}%
where the multiplicity distribution $P(N)$ is taken at $N=N_{L}+N_{R}$. We
observe that in general $P(N_{L},N_{R})$ cannot be factorized to the product 
$P_{L}(N_{L})P_{R}(N_{R})$ where $P_{L}$ is the multiplicity distribution in
the left bin and $P_{R}$ that for the right bin. The factorization works for
the Poisson distribution, Eq.~\eqref{eq:append:poisson}, since $(N_{L}+N_{R})!$ cancels and we obtain%
\footnote{%
This result can be easily generalized to more than two bins and to arbitrary
values of probabilities of particle distribution between the bins.} \ 
\begin{equation}
P(N_{L},N_{R})=\frac{\left\langle N_{L}\right\rangle ^{N_{L}}}{N_{L}!}%
e^{-\left\langle N_{L}\right\rangle }\frac{\left\langle N_{R}\right\rangle
^{N_{R}}}{N_{R}!}e^{-\left\langle N_{R}\right\rangle },
\end{equation}%
which is the product of the two Poisson distributions. It means that for the
Poisson distribution there is no correlation between the numbers of
particles in the left and the right bins. Obviously any other distribution
would introduce multiplicity correlations. For example, if a source is
producing always (in each event), say, $10$ particles, and we know that,
say, $N_{L}=2$ we immediately know that $N_{R}=8$. In this case $N_{L}$ and $%
N_{R}$ are maximally correlated (or more precisely anti-correlated).

To better understand the cumulants and the factorial cumulants let us
consider a slightly more complicated model. Suppose that clusters (e.g.,
resonances) are distributed according to Poisson distribution $P_{c}(N_{c})$%
, where $N_{c}$ is the number of clusters. Next, each cluster decays into
exactly $m$ final particles, namely, $N_{c}$ clusters result in $N=mN_{c}$
particles. In this case, the final multiplicity distribution is given by 
\begin{equation}
P(N)=P_{c}(N_{c})\,\delta _{N,\,mN_{c}}.
\end{equation}%
The generating function is given by%
\begin{equation}
H(z)=\sum_{N}P(N)z^{N}=\sum_{N_{c}}P_{c}(N_{c})(z^{m})^{N_{c}}=e^{\left%
\langle N_{c}\right\rangle \left( z^{m}-1\right) },
\end{equation}%
where the last equality assumes $P_{c}(N_{c})$ to be the Poisson
distribution with the average number of clusters $\left\langle
N_{c}\right\rangle $. It follows directly from Eq. (\ref{eq:ab:H-Poiss})
where $z$ is replaced by $z^{m}$. Performing suitable calculations we obtain 
$F_{1}=m\left\langle N_{c}\right\rangle =\left\langle N\right\rangle $ and 
\begin{equation}
F_{2}=\left\langle N\right\rangle ^{2}+\left\langle N\right\rangle \left(
m-1\right) ,\qquad F_{3}=\left\langle N\right\rangle ^{3}+3\left\langle
N\right\rangle ^{2}\left( m-1\right) +\left\langle N\right\rangle \left(
m-1\right) \left( m-2\right) ,
\end{equation}%
and more complicated expressions for the higher order factorial moments.
Clearly $F_{i}>\left\langle N\right\rangle ^{i}$ which indicates specific
correlations.

On the other hand, the factorial cumulants, generated from $\ln
(H)=\left\langle N_{c}\right\rangle \left( z^{m}-1\right) $, read $\hat{C}%
_{1}=m\left\langle N_{c}\right\rangle =\left\langle N\right\rangle $ and for 
$k\geq 2$ we have \cite{Bzdak:2016jxo} 
\begin{equation}
\hat{C}_{k}=\left\{ 
\begin{array}{c}
\left\langle N_{c}\right\rangle \frac{m!}{(m-k)!},\quad k\leq m \\ 
\text{ \ \ \ \ \ }0,\quad \text{\ \ \ \ \ \ \ \ }k>m%
\end{array}%
\right. .  \label{eq:ab:Ckhat-clust}
\end{equation}%
In other words, the factorial cumulants vanish if the order of the factorial
cumulant is larger than $m$ - the number of particle from a single cluster.
This is not surprising. The factorial cumulants are directly related to
correlations present in the system. Having clusters decaying exactly into $m$
particles we introduce two- and up to $m$-particle correlations but not $%
(m+1)$-particle correlations. Consequently $\hat{C}_{m+1}$, $\hat{C}_{m+2}$
etc. equal zero. It is worth mentioning that in our simple example clusters
themselves do not bring any correlations since their multiplicity
distribution is given by Poisson. However, any other distribution different
than Poisson would in general result in $\hat{C}_{k}\neq 0$ also for $k>m$.\footnote{The interpretation of factorial cumulants can at times be rather tricky. 
For example, one may
have n-particle correlations even if one has less that $n$ particles in the system. This is best
illustrated by the extreme example, where we have a source that in each
event produces exactly one particle. In this case $P(N=1)=1$ and $H(z)=z$.
Obviously $F_{1}=1$ and $F_{i}=0$ for $i\geq 2$. The same relations hold for
the cumulants. However, the factorial cumulants are $\hat{C}%
_{i}=(-1)^{i-1}(i-1)!$. For example, the two particle density is zero (we
have only one particle) and thus following Eq. (\ref{eq:ab:rho2}), we obtain 
$C_{2}(y_{1},y_{2})=-\rho (y_{1})\rho (y_{2}),$ which integrates to $\hat{C}%
_{2}=-1$. Therefore, even though we have only one particle we have 
integrated n-particle correlations
for all $n$. This seemingly paradoxical situation is best resolved by realizing that the factorial cumulants, $\hat{C}_{k}$, ``measure'' the deviation from Poisson distribution. In our
example the distribution with exactly one particle deviates considerably from Poisson with $\ave{N}=1$. It is these deviations which are encoded in the $\hat{C}_{k}$. 
This example also nicely demonstrates that interactions between produced particles should not be confused with correlations since both are not necessarily related, as is evident by a source producing one particle.      
}

As seen from Eq. (\ref{eq:ab:kappa-Chat}), the cumulants $\kappa _{i}$ are
influenced by the factorial cumulants up to the $i$-th order. In particular,
if $m=2$, that is we have only two-particle correlations in the system, all
cumulants are influenced, which makes the interpretation of the cumulants
rather tricky. Performing straightforward calculations we obtain%
\begin{equation}
\kappa _{i}=m^{i}\left\langle N_{c}\right\rangle ,\quad i\geq 1,
\end{equation}%
which follows directly from the cumulant generating function, which in our
simple model reads $K(t)=\left\langle N_{c}\right\rangle \left(
e^{mt}-1\right) $. For example, taking $m=2$ we obtain $\kappa _{4}/\kappa
_{2}=4$ even though the system has only two particle correlations. This demonstrates that
the higher order cumulants do not necessarily carry information about
nontrivial multiparticle correlations,  and that factorial cumulants seem to be
better suited for this purpose.

Another example worth considering in more detail is the bi-modal or two-component distribution
discussed in Sec.~\ref{sec:6:bimodal}. In this case the multiplicity distribution is given by
Eq.~\eqref{eq:two_component}
\begin{equation}
P(N)=(1-\alpha )P_{(a)}(N)+\alpha P_{(b)}(N),
\end{equation}
so that the factorial moment generating function is given by
\begin{equation}
H(z)=(1-\alpha )H_{(a)}(z)+\alpha H_{(b)}(z).
\end{equation}
Here $H_{(a,b)}(z)$ are the generating functions for the distributions $P_{(a,b)}(N)$. The factorial
cumulant generating function is then
\begin{equation}
  G(z)=\ln\left[  (1-\alpha )H_{(a)}(z)+\alpha H_{(b)}(z) \right]= G_{(a)}(z)+\ln\left[ (1-\alpha)
  + \alpha e^{G_{(b)}(z)-G_{(a)}(z)}\right],
\end{equation}
where $G_{(a,b)}(z)=\ln[H_{(a,b)}(z)]$ are the factorial cumulant generating functions for the distributions
$P_{(a,b)}(N)$.
Noting, that the second term in the above equation resembles the cumulant generating function
$K_{Bn}(t)$ of the Bernoulli distribution with the probability $p=\alpha$, which is given by
\begin{equation}
  K_{Bn}(t) = \ln\left[ 1 - \alpha +\alpha e^{t} \right],
\end{equation}
we have
\begin{equation}
  G(z)= G_{(a)}(z)+K_{Bn} \left( G_{(b)}(z)-G_{(a)}(z) \right) .
\end{equation}
Let us now assume that both, $P_{(a)}(N)$ and $P_{(b)}(N)$ are Poisson distributions so that all but the
first factorial cumulants vanish,
\begin{eqnarray}
  \hat{C}^{(a,b)}_{1} &=& \left. \frac{d}{d z} G_{(a,b)}(z) \right|_{z=1}= \ave{N_{(a,b)}} \non
  \hat{C}^{(a,b)}_{k} &=& \left. \frac{d^{k}}{d z^{k}} G_{(a,b)}(z)\right|_{z=1}= 0; \quad k>1.
\end{eqnarray}
In this case the factorial cumulants for the bi-modal distribution $P(N)$ are simply given by
\begin{eqnarray}
  \hat{C}_{1} &=& \ave{N_{(a)}} - \kappa_{Bn,1}\left( \ave{N_{(a)}} - \ave{N_{{(b)}}} \right) \non
  \hat{C}_{k} &=&  (-1)^{k} \kappa_{Bn,k}\left( \ave{N_{(a)}} - \ave{N_{{(b)}}} \right)^{k}; \quad k>1.
\end{eqnarray}
Here, $\kappa_{Bn,k}$ denote the cumulants of the Bernoulli distribution with $\kappa_{Bn,1}=\alpha$,
$\kappa_{Bn,2}=\alpha ( 1-\alpha)$,  $\kappa_{Bn,3}=\alpha(1-\alpha)(1-2\alpha)$ and so on.
For the bi-modal distribution discussed in Sec.~\ref{sec:6:bimodal}, which fits the measured STAR cumulants,
we have $\alpha\ll1 $ and $\ave{N_{(a)}} - \ave{N_{(b)}}\simeq 15$ so that, to a good
approximation, the higher order factorial cumulants are given by
\begin{equation}
\hat{C}_{k} \simeq \alpha \, (-1)^{k}\, \left( \ave{N_{(a)}} - \ave{N_{{(b)}}} \right)^{k}; \quad k>1. 
\end{equation}
Therefore, with increasing order the factorial cumulants increase in magnitude (by a factor of
$\sim 15$ in case of the fit to the STAR data ) and alternate in sign. 
Most interestingly, we obtain large factorial cumulants even though in
each event particles are produced from a single Poisson distribution, which by itself has vanishing
factorial cumulants.  What happens in the case of the two-component
model can be easily understood following the discussion 
around Eq. (\ref{eq:ab:P-two-bins}). Suppose we split particles originating
from $P(N)$ between the left and the right bin. Having a large number of
particles in the left bin it is more likely that they are originating from $%
P_{(a)}(N)$, since in our example $ \ave{N_{(a)}} > \ave{N_{(b)}}$.  Consequently, we  expect a large number of
particles in the right bin as well. Similarly, heaving a small number of particles
in the left bin indicates that $P_{(b)}(N)$ is active
and we expect a smaller number of particles in the right bin. In other
words, having a superposition of two Poisson distributions with different
means results in $P(N_{L},N_{R})$ being different than $%
P(N_{L})P(N_{R})$, which indicates multiplicity correlations.

The final example we want to discuss is related to the multiplicity
distribution of charged particles, which is typically described with the
negative binomial distribution (see, e.g., Ref. \cite{Ansorge:1988kn})%
\begin{equation}
P(N;\left\langle N\right\rangle ,k)=\frac{\Gamma \left( N+k\right) }{\Gamma
\left( N+1\right) \Gamma \left( k\right) }\left( \frac{\left\langle
N\right\rangle }{k}\right) ^{N}\left( 1+\frac{\left\langle N\right\rangle }{k%
}\right) ^{-N-k},
\end{equation}%
where $k$ measures deviation from the Poisson distribution.\footnote{%
The negative binomial distribution goes to the Poisson distribution if $%
k\rightarrow \infty $ at fixed $\left\langle N\right\rangle $. For example $%
\left\langle N^{2}\right\rangle -\left\langle N\right\rangle
^{2}=\left\langle N\right\rangle \left( 1+\frac{\left\langle N\right\rangle 
}{k}\right) $ which goes to $\left\langle N^{2}\right\rangle -\left\langle
N\right\rangle ^{2}=\left\langle N\right\rangle $.} In this case, the
generating function reads%
\begin{equation}
H(z)=\left( 1+\frac{\left\langle N\right\rangle }{k}\left( 1-z\right)
\right) ^{-k}.  \label{eq:ab:H-NBD}
\end{equation}%
The first factorial moment is obviously $F_{1}=\left\langle N\right\rangle $
and for $i\geq 2$ we have 
\begin{equation}
F_{i}=\left\langle N\right\rangle ^{i}\frac{(k+1)(k+2)\cdots (k+i-1)}{k^{i-1}%
},
\end{equation}%
which goes to the Poisson limit, $F_{i}=\left\langle N\right\rangle ^{i}$,
if $k\rightarrow \infty $ at fixed $\left\langle N\right\rangle $. Taking
the logarithm of Eq. (\ref{eq:ab:H-NBD}) and using Eq. (\ref{eq:ab:Ckhat})
we obtain the factorial cumulant%
\begin{equation}
\hat{C}_{i}=\frac{(i-1)!\left\langle N\right\rangle ^{i}}{k^{i-1}},
\end{equation}%
which goes to zero if $k\rightarrow \infty $ at fixed $\left\langle
N\right\rangle $, being the proper Poisson limit. Finally, the cumulants are
given by a sum of the factorial cumulants as shown in Eq. (\ref%
{eq:ab:kappa-Chat}). The negative binomial distribution is usually defined
with $k>0$ and thus the factorial cumulants are all positive and
consequently the cumulants are larger than $\left\langle N\right\rangle $,
and for example 
\begin{equation}
\frac{\kappa _{4}}{\kappa _{2}}=1+6\left\langle N\right\rangle \frac{%
\left\langle N\right\rangle +k}{k^{2}},
\end{equation}%
which is larger then one (unless $k\rightarrow \infty $).

%%% Local Variables:
%%% mode: latex
%%% TeX-master: "BES_Main_current"
%%% End:

%\include{appen2}



\vspace{0.5in}
%\section*{References}
\bibliographystyle{elsarticle-num} 
\bibliography{BES_BIB.bib}%,Extra_BIB.bib}


%\begin{thebibliography}{00}


%\bibitem{}

%\end{thebibliography}
\end{document}